\documentclass[rmp,twocolumn,nofootinbib,superscriptaddress,eqsecnum]{revtex4}
\pdfoutput=1
\usepackage{graphics}
\usepackage{epsfig}
\usepackage{longtable}

\usepackage{color}

\usepackage{amssymb}
\usepackage{amsmath}
\usepackage{graphicx}
\usepackage{float}

\def \be{\begin{equation}}
\def \ee{\end{equation}}
\def \bew{\begin{widetext}\begin{equation}}
\def \eew{\end{equation}\end{widetext}}
\def \bmlett{\begin{mathletters}}
\def \emlett{\end{mathletters}}

\def \ind{\hspace{0.5 cm}}

\def \r{{\bf r}}

\def \p{{\bf p}}

\def \pd{\phantom{\dagger}}

\def \FF{{\mathcal F}}

\def \clS{ {\mathcal S} }	
\def \smat{ s}			

\def \tV{\tilde{V}}
\def \tI{\tilde{I}}

\def \Re{\textrm{Re }}

\def \ts{\tilde{s}}

\def \dtild{\tilde{\delta}}

\def \bS{\bar{S}}

\def \bS{\bar{S}}

\def \ua{\uparrow}
\def \da{\downarrow}
\def \ra{\rightarrow}
\def \la{\leftarrow}

\def \hx{\hat{x}}

\def \hp{\hat{p}}
\def \hF{\hat{F}}
\def \hI{\hat{I}}
\def \hQ{\hat{Q}}
\def \hV{\hat{V}}
\def \ha{\hat{a}}
\def \hb{\hat{b}}
\def \hq{\hat{q}}
\def \hu{\hat{u}}
\def \hv{\hat{v}}
\def \hFF{\hat{\FF}}
\def \htV{\hat{\tilde{V}}}
\def \htI{\hat{\tilde{I}}}
\def \hz{\hat{z}}

\def \hd{\hat{d}}
\def\hddag{{\hat d}^\dagger}

\def \hain{\hat{a}_{{\rm in}}}
\def \hbin{\hat{b}_{{\rm in}}}
\def \haout{\hat{a}_{{\rm out}}}
\def \hbout{\hat{b}_{{\rm out}}}
\def \huin{\hat{u}_{{\rm in}} }
\def \hvin{\hat{v}_{{\rm in}} }

\def\hbr{\hat{b}^\rightarrow}
\def\hBr{\hat{B}^\rightarrow}
\def\hbrd{\hat{b}^{\dagger\rightarrow}}
\def\hBrd{\hat{B}^{\dagger\rightarrow}}

\def \hdbout{{\hat b}_{\rm out}^\dagger}

\def\ap{\hat{a}_{\rm P}}
\def\as{\hat{a}_{\rm S}}
\def\ai{\hat{a}_{\rm I}}

\def \ain{a_{\rm in}}
\def\aout{a_{\rm out}}
\def \bin{b_{\rm in}}
\def \bout{b_{\rm out}}

\def\Aout{A_{\rm out}}
\def\Bout{B_{\rm out}}

\def\Ain{A_{\rm in}}
\def\Bin{B_{\rm in}}

\def\vin{v_{\rm in}}

\def \VN{\tV }
\def \IN{\tI }

\def\oscchi{\chi_{xx} }
\def \chiIF{\chi_{IF}}
\def \chiFI{\chi_{FI}}
\def \tchiIF{ \tilde{\chi}_{IF} }
\def \ychi{\chi_{yy} }

\def\kbt{k_{\rm B}T}
\def\be{\begin{equation}}
\def\ee{\end{equation}}

\def\w01{\omega_{01}}
\def\r0{R_0}
\def\nb{n_{\rm B}}

\def\omegar{\omega_{\rm c}} 
\def\tomegar{{\tilde\omega}_{\rm c}}
\def \Qc{Q_{\rm c}}

\def\chiR{\chi_{\rm c}} 
\def\chiM{\chi_{\rm M}} 
\def\omegaq{\omega_{\rm q}}
\def\omegal{\omega_{\rm L}}

\def\omegap{\omega_{\rm P}}
\def\omegas{\omega_{\rm S}}
\def\omegai{\omega_{\rm I}}

\def\ks{\kappa_{\rm S}}
\def\ki{\kappa_{\rm I}}

\def\omegam{\Omega}	

\def\barbin{{\bar b}_{\rm in}}

\def\hc{\hat{c}}
\def\xrms{x_{\rm ZPF}}

\def\hxi{\hat{\xi}}
\def\hdxi{\hat{\xi}^\dagger}
\def\hxidag{\hat{\xi}^\dagger}

\def\SVV{S_{\rm V V}}
\def\SNN{S_{{\dot N} {\dot N}}  }
\def\Snn{S_{nn}}
\def\Ndotbar{{\overline{{\dot N}}}}
\def\Sthetatheta{S_{\theta\theta}}

\def\Sxx{S_{xx}}
\def\Sxxsym{ \bS_{x x}}
\def\SIxx{S^{\rm I}_{xx}}

\def \SIzz{S^{\rm I}_{zz}}
\def \SFFz{S_{F_z F_z}}

\def \hFz{\hat{F}_z}

\def\SFF{S_{FF}}
\def\twd{t_{\rm WD}}
\def\ttwd{{\tilde{t}}_{\rm WD}}
\def\nb{n_{\rm B}}

\def\aIF{\ha_{\rm IF}}
\def\aLO{\ha_{\rm LO}}
\def\au{\ha_{\rm u}}
\def\al{\ha_{\rm l}}
\def\omegaLO{\omega_{\rm LO}}

\def\bSxtot{{\bS}_{xx, {\rm tot}} }
\def\bSxadd{{\bS}_{xx, {\rm add}} }
\def\bSxeq{{\bS}_{xx, {\rm eq}} }
\def\Zin{Z_{\rm{in}}}
\def\Zout{Z_{\rm{out}}}
\def\Zs{Z_{\rm{s}}}
\def\Vin{V_{\rm in}}
\def\Vout{V_{\rm{out}}}
\def\TN{T_{\rm{N}}}
\def\ZN{Z_{\rm{N}}}

\def\kb{k_{\rm B}}
\def\gammam{\gamma_0}

\def\Zc{Z_{\rm c}}
\def\Vc{V_{\rm c}}
\def\VR{V_{\rm R}}
\def\vp{v_{\rm p}}

\def\XAin{X_A^{\rm in}}
\def\XBin{X_B^{\rm in}}
\def\YAin{Y_A^{\rm in}}
\def\YBin{Y_B^{\rm in}}

\def\XAout{X_A^{\rm out}}
\def\XBout{X_B^{\rm out}}
\def\YAout{Y_A^{\rm out}}
\def\YBout{Y_B^{\rm out}}

\def\dW{d\widehat W}
\def\dWd{d\widehat W^\dagger}

\begin{document}


\title{Introduction to Quantum Noise, Measurement and Amplification}
\author{A.A. Clerk}
\affiliation{Department of Physics, McGill University, 3600 rue
University\\
Montr\'eal, QC Canada H3A 2T8}
\email{clerk@physics.mcgill.ca}

\author{M.H. Devoret}
\affiliation{Department of Applied Physics, Yale University\\
PO Box 208284, New Haven, CT 06520-8284}
\author{S.M. Girvin}
\affiliation{Department of Physics, Yale University\\
PO Box 208120, New Haven, CT 06520-8120}
\author{Florian Marquardt}
\affiliation{Department of Physics, Center for NanoScience, and Arnold Sommerfeld Center for Theoretical Physics, Ludwig-Maximilians-Universit\"at
M\"unchen\\ Theresienstr. 37, D-80333 M\"unchen, Germany}
\author{R.J. Schoelkopf}
\affiliation{Department of Applied Physics, Yale University\\
PO Box 208284, New Haven, CT 06520-8284}

\date{April 15, 2010}

\begin{abstract}

The topic of quantum noise has become extremely timely due to the rise of quantum information
physics and the resulting interchange of ideas between the condensed matter and atomic, molecular,
opticalÐquantum optics communities. This review gives a pedagogical introduction to the physics of
quantum noise and its connections to quantum measurement and quantum amplification. After
introducing quantum noise spectra and methods for their detection, the basics of weak continuous
measurements are described. Particular attention is given to the treatment of the standard quantum
limit on linear amplifiers and position detectors within a general linear-response framework. This
approach is shown how it relates to the standard Haus-Caves quantum limit for a bosonic amplifier
known in quantum optics and its application to the case of electrical circuits is illustrated, including
mesoscopic detectors and resonant cavity detectors.
\end{abstract}

\maketitle

\tableofcontents



%


\section{Introduction}

Recently, several advances have led to a renewed interest in the quantum
mechanical aspects of noise in mesoscopic electrical circuits,
detectors and amplifiers.
One motivation is that such systems can operate simultaneously at high frequencies and at low temperatures, entering the regime where $\hbar \omega > \kbt$.  As such, quantum zero-point fluctuations will play a more dominant role in determining their behaviour than the more familiar thermal fluctuations. A second motivation
comes from the relation between quantum noise and quantum measurement.  There exists an ever-increasing number of experiments in mesoscopic electronics where one is forced
to think about the quantum mechanics of the detection process, and about
fundamental quantum limits which constrain the performance of the detector or amplifier used.

Given the above, we will focus in this review on discussing what is known as the ``standard quantum limit" (SQL) on both displacement detection and amplification. To preclude any possible confusion, it is worthwhile to state explicitly from the start that there is no limit to how well one may resolve the position of a particle in an {\em instantaneous} measurement. Indeed, in the typical Heisenberg microscope setup, one would scatter photons off an electron, thereby detecting its position to an accuracy set by the wavelength of photons used. The fact that its momentum will suffer a large uncontrolled perturbation, affecting its future motion, is of no concern here. Only as one tries to further increase the resolution will one finally encounter relativistic effects (pair production) that set a limit given by the Compton wavelength of the electron.
The situation is obviously very different if one attempts to observe the whole trajectory of the particle. As this effectively amounts to measuring both position and momentum, there has to be a tradeoff between the accuracies of both, set by the Heisenberg uncertainty relation. The way this is enforced in practice is by the uncontrolled perturbation of the momentum
during one position measurement adding to the noise in later measurements, a phenomenon known as "measurement back-action".

Just such a situation is encountered in ``weak measurements" \cite{Braginsky92}, where one integrates the signal over time, gradually learning more about the system being measured; this review will focus on such measurements.  There are many good reasons why one may be interested in doing a weak measurement, rather than an instantaneous, strong projective measurement. On a practical level, there may be limitations to the strength of the coupling between the system and the detector, which have to be compensated by integrating the signal over time. One may also deliberately opt not to disturb the system too strongly, e.g.~to be able to apply quantum feedback techniques for state control. Moreover, as one reads out an oscillatory signal over time, one effectively filters away noise (e.g.~of a technical nature) at other frequencies. Finally, consider an example like detecting the collective coordinate of motion of a micromechanical beam. Its zero-point uncertainty (ground state position fluctuation) is typically on the order of the diameter of a proton. It is out of the question to reach this accuracy in an instantaneous measurement by scattering photons of such a small wavelength off the structure, since they would instead resolve the much larger position fluctuations of the individual atoms comprising the beam (and induce all kinds of unwanted damage), instead of reading out the center-of-mass coordinate.
The same holds true for other collective degrees of freedom.

The prototypical example we will discuss several times is that of a weak measurement detecting the motion of a harmonic oscillator (such as the mechanical beam). The measurement then actually follows the slow evolution of amplitude and phase of the oscillations (or, equivalently, the two quadrature components), and the SQL derives from the fact that these two observables do not commute. It essentially says that the measurement accuracy will be limited to resolving both quadratures down to the scale of the ground state position fluctuations, within one mechanical damping time. Note that, in special applications, one might be interested only in one particular quadrature of motion. Then the Heisenberg uncertainty relation does not enforce any SQL and one may again obtain unlimited accuracy, at the expense of renouncing all knowledge of the other quadrature.

Position detection by weak measurement essentially amounts to amplifying the quantum signal up to a classically accessible level. Therefore, the theory of quantum limits on displacement detection is intimately connected to limits on how well an amplifier can work. If an amplifier does not have any preference for any particular phase of the oscillatory signal, it is called ``phase-preserving", which is the case relevant for amplifying and thereby detecting both quadratures equally well\footnote{In the literature this is often referred to as a 'phase insensitive' amplifier. We prefer the term 'phase-preserving' to avoid any ambiguity.}.
We will derive and discuss in great detail the SQL for phase-preserving linear amplifiers \cite{Haus62,Caves82}. Quantum mechanics demands that such an amplifier adds noise that corresponds to half a photon added to each mode of the input signal, {\em in the limit of high 
photon-number gain} $G$. In contrast, for small gain, the minimum number of added noise quanta, $(1-1/G))/2$, can become arbitrarily small as the gain is reduced down to $1$ (no amplification). One might ask, therefore, whether it shouldn't be possible to evade the SQL by being content with small gains? The answer is no, since high gains $G\gg1$ are needed to amplify the signal to a level where it can be read out (or further amplified) using classical devices without their noise having any further appreciable effect, converting $1$ input photon into $G\gg1$ output photons. In the words of Caves,  it is necessary to generate an output that ``we can lay our grubby, classical hands on" \cite{Caves82}. It is a simple exercise to show that feeding the input of a first, potentially low-gain amplifier into a second amplifier results in an overall bound on the added noise that is just the one expected for the product of their respective gains. Therefore, as one approaches the classical level, i.e. large overall gains, the SQL in its simplified form of half a photon added always applies.

Unlike traditional discussions of the amplifier SQL, we will devote considerable attention
to a general linear response approach based on the quantum 
relation between susceptibilities and noise.   This approach treats the amplifier or detector as a black box with an input port coupling to the signal source and an output port to access the amplified signal. It is more suited for mesoscopic systems than the quantum optics scattering-type approach, and it leads us to the quantum noise inequality: 
a relation between the noise added to the output and the back-action noise feeding back to the signal source. In the ideal case (what we term a ``quantum-limited detector"), the product of these two contributions reaches the minimum value allowed by quantum mechanics.
We will show that optimizing this inequality on noise is a necessary pre-requisite for having
a detector achieve the quantum-limit on a specific measurement task, such as linear ampification.

There are several motivations for understanding in principle, and realizing in
practice, amplifiers whose noise reaches this minimum quantum
limit.  Reaching the quantum limit on continuous position detection
has been one of the goals of many recent experiments
on quantum electro-mechanical
\cite{Cleland02, Knobel03, LaHaye04, Naik06, FlowersJacobs07, Regal08, Poggio08, Etaki08}
and opto-mechanical systems
\cite{Arcizet2006, Gigan2006,Kippenberg07,Harris08,2009_05_MarquardtGirvin_OptomechanicsReview}.
As we will show, having a near-quantum limited detector would allow one
to continuously monitor the quantum zero-point fluctuations of a mechanical resonator.
Having a quantum limited detector is also necessary for such tasks as single-spin NMR detection \cite{Rugar04},
as well as gravitational wave detection \cite{Abramovici92}.  The topic of quantum-limited detection is also directly
relevant to recent activity exploring feedback control
of quantum systems \cite{Wiseman93, Wiseman94, Doherty00, Korotkov01b, Ruskov02}; such schemes {\it necessarily} need a
close-to-quantum-limited detector.

This review is organized as follows. We start in
Sec.~\ref{sec:QuantumNoiseSpectra} 
by providing a short review
of the basic statistical properties of quantum noise, including its
detection. In Sec.~\ref{sec:QuantumMeasurements} we turn to quantum measurements,
and give a basic introduction to weak, continuous measurements.  To make things concrete,
we discuss heuristically measurements of both a qubit and an oscillator using a simple resonant cavity detector, giving an idea of the origin of the quantum limit in each case.  Sec.~\ref{sec:GenLinResponseTheory} is devoted to a more rigorous treatment of quantum constraints on noise arising from general quantum linear response theory.  The heart of the review is contained in Sec.~\ref{sec:QLAmplifiers}, where we give a thorough discussion of quantum limits on amplification and continuous position detection. We also briefly discuss various methods for beating the usual quantum limits on added noise using back-action evasion techniques.  We are careful to distinguish two very distinct modes of amplifier operation (the ``scattering" versus ``op amp" modes);
we expand on this in Sec.~\ref{sec:ScatteringAmp}, where we discuss both modes of operation in a simple two-port bosonic amplifier.  Importantly, we show that an amplifier can be quantum limited in one mode of operation, but fail to be quantum limited in the other mode of operation.  Finally, in Sec.~\ref{sec:QLinpractice}, we highlight a number of practical considerations that one must keep in mind when trying to perform a quantum limited measurement.    Table I provides a synopsis of the main results discussed in the text, as well as definitions of symbols used.

In addition to the above, we have supplemented the main text with several pedagogical appendices
which cover basic background topics. Particular attention is given to the quantum mechanics of transmission
lines and driven electromagnetic cavities, topics which are especially relevant given recent experiments making use of microwave stripline resonators.
These appendices appear as a separate on-line only supplement to the published article \cite{ClerkEPAPS}, but are included in this arXiv version of the article.
In Table~\ref{table:Appendices}, we list the contents of these appendices.
Note that while some aspects of the
topics discussed in this review have been studied in the quantum optics and
quantum dissipative systems communities and are the subject of
several comprehensive books \cite{Braginsky92, Weiss99, Haus00,
Gardiner00}, they are somewhat newer to the condensed matter physics
community; moreover, some of the technical machinery developed
in these fields is not directly applicable to the study of
quantum noise in quantum electronic systems.
Finally, note  that while this article is a review, there is considerable new material presented, especially in our discussion of quantum amplification (cf.~Secs.~\ref{subsec:TwoKindsAmps},\ref{sec:ScatteringAmp}).



\begin{longtable*}{ll}
\caption{Table of symbols and main results.} \\


\hline \hline \\[-2ex]
	\multicolumn{1}{l}{Symbol} &
	\multicolumn{1}{l}{Definition / Result} \\ \hline
\\[-1.8ex]
\endhead

\multicolumn{2}{c}{\em General Definitions} \\

$f[\omega]$			& Fourier transform of the function or operator $f(t)$, defined via
					$f[\omega] = \int_{-\infty}^{\infty} dt f(t) e^{i \omega t}$ \\
					& (Note that for operators, we use the convention
					$\hat{f}^{\dag}[\omega] = 
						\int_{-\infty}^{\infty} dt \hat{f}^{\dag}(t) e^{i \omega t}$, implying
					$\hat{f}^{\dag}[\omega] = \left( \hat{f}[-\omega] \right)^{\dag}$ ) \\

$\clS_{FF}[\omega]$     	& Classical noise spectral density or power spectrum:
                        			$\clS_{FF}[\omega] = \int_{-\infty}^{+\infty} dt\, e^{i \omega t} \langle F(t) F(0) \rangle$ \\

$S_{FF}[\omega]$         	& Quantum noise spectral density:
                        			$S_{FF}[\omega] = \int_{-\infty}^{+\infty} dt\, e^{i \omega t} \langle \hF(t) \hF(0) \rangle$ \\

$\bS_{FF}[\omega]$       	& Symmetrized quantum noise spectral density
					$\bS_{FF}[\omega] = \frac{1}{ 2} ( S_{FF}[\omega] + S_{FF}[-\omega] )
					= \frac{1}{2} \int_{-\infty}^{+\infty} dt\, e^{i \omega t} \langle \{ \hF(t), \hF(0) \} \rangle$  \\

$\chi_{AB}(t)$			& General linear response susceptibility describing the response of $A$ to a perturbation which couples to $B$; \\
					& 	\ind in the quantum case, given by the
					Kubo formula $\chi_{AB}(t) = -\frac{i}{\hbar} \theta(t) \langle [ \hat{A}(t), \hat{B}(0)] \rangle$ [Eq. (\ref{OscLF})] \\

$A$                   			& Coupling constant (dimensionless) between measured system and detector/amplifier, \\\
					& \ind e.g. ${\hat V}=A F(t) {\hat \sigma}_x, {\hat V}=A{\hat x}{\hat F},$ or ${\hat V}=A \hbar \omegar {\hat \sigma}_z {\hat a}^{\dagger} {\hat a}$ \\

$M, \Omega$           	& Mass and angular frequency of a mechanical harmonic oscillator.\\
$\xrms$           			& Zero point uncertainty of a mechanical oscillator, $\xrms = \sqrt{ \frac{\hbar}{ 2 M \Omega} }$. \\
$\gamma_0$              	& Intrinsic damping rate of a mechanical oscillator due to coupling to a bath via ${\hat V}=A{\hat x}{\hat F}$ :\\
                      			& \ind $\gamma_0 = \frac{A^2}{ 2M\hbar \Omega} (S_{FF}[\Omega] - S_{FF}[-\Omega])$ [Eq. (\ref{OscGamma})] \\
			
$\omegar$			& Resonant frequency of a cavity \\
$\kappa, \Qc$			& Damping, quality factor of a cavity: $\Qc = \omegar/\kappa$ \\

\\
	
\multicolumn{2}{c}{\em Sec.~\ref{sec:QuantumNoiseSpectra} Quantum noise spectra} \\

$T_{\rm eff}[\omega]$ 	& Effective temperature at a frequency $\omega$ for a given quantum noise spectrum, defined via \\
                      			& \ind  $\frac{S_{FF}[\omega]}{ S_{FF}[-\omega]} = \exp\left(\frac{ \hbar \omega}{ \kb T_{\rm eff}[\omega]} \right)$ [Eq. (\ref{OscTeff})]\\

		&	Fluctuation-dissipation theorem relating the symmetrized noise spectrum to the dissipative part \\
                      &  \ind for an equilibrium bath: $\bS_{FF}[\omega] = \frac{1}{ 2} \coth(\frac{\hbar \omega}{ 2 \kb T}) (S_{FF}[\omega] - S_{FF}[-\omega])$
                      	[Eq. (\ref{eq:FDFinal})]\\
 & \\

\multicolumn{2}{c}{\em Sec.~\ref{sec:QuantumMeasurements} Quantum Measurements} \\

	&	Number-phase uncertainty relation for a coherent state: \\
	&		\ind $ \Delta N \Delta \theta \geq \frac{1}{2} $  [Eq.~(\ref{eq:numberphaseuncertainty}), (\ref{eq:AppNumberPhase})] \\

$\dot{N}$			&	Photon number flux of a coherent beam \\
$\delta \theta$		&	Imprecision noise in the measurement of the phase of a coherent beam\\

	&	Fundamental noise constraint for an ideal coherent beam: \\
	&	 \ind $ \SNN\Sthetatheta = \frac{1}{4} $  
	[Eq.~(\ref{eq:numberphasenoiseproduct}), (\ref{eq:AppNumberPhaseNoise})] \\

$\Sxxsym^0(\omega)$			& symmetrized spectral density of zero-point position fluctuations of a damped harmonic oscillator \\

$\bSxtot(\omega)$				& total output noise spectral density (symmetrized) of a linear position detector, referred back to the oscillator \\
$\bSxadd(\omega)$				& added noise spectral density (symmetrized) of a linear position detector, referred back to the oscillator \\

\\

\multicolumn{2}{c}{\em Sec.~\ref{sec:GenLinResponseTheory}:
	General linear response theory} \\

$\hat x$              & Input signal \\
$\hat F$              & Fluctuating force from the detector, coupling to ${\hat x}$ via ${\hat V}=A{\hat x}{\hat F}$ \\
$\hat I$              & Detector output signal\\

		&	General quantum constraint on the detector output noise, backaction noise and gain: \\
               	& 	\ind $\bS_{II}[\omega] \bS_{FF}[\omega] - \left| \bS_{IF}[\omega] \right|^2  \geq
                         \left| \frac{\hbar \tchiIF [\omega]}{2} \right|^2
                         \left( 1 +
                         \Delta\left[ \frac{\bS_{IF}[\omega]}{\hbar
			     \tilde{\lambda}[\omega]/2} \right] \right)$ [Eq. (\ref{NoiseConstraint})] \\
                      & \ind where $\tchiIF[\omega] \equiv  \chiIF[\omega] - \left[ \chiFI[\omega] \right]^*$ and  $\Delta[z]  = ( \left| 1 + z^2 \right| - \left(1 + |z|^2 \right))/2$. \\
                      & \ind [Note: $1+\Delta[z]\geq0$ and $\Delta=0$ in most cases of relevance, see discussion around Eq.~(\ref{CorrQLCondition})]\\

$\alpha$              & Complex proportionality constant characterizing a quantum-ideal detector: \\
                      	& \ind $| \alpha |^2 =  \bS_{II} / \bS_{FF}$ and $\sin \left( \arg \alpha [\omega] \right) = \frac{ \hbar  | \lambda[\omega] |  / 2} { \sqrt{\bS_{II}[\omega] \bS_{FF}[\omega]} }$ [Eqs. (\ref{AlphaDefn},\ref{eq:AlphaPhase})] \\

& \\
$\Gamma_{\rm meas}$   & Measurement rate (for a QND qubit measurement) [Eq.~\ref{eq:GammaMeasDefn}]\\

$\Gamma_{\varphi}$    & Dephasing rate (due to measurement back-action) [Eqs.~(\ref{eq:cavitydephasingrate}),(\ref{eq:DephRate2})] \\

		&	Constraint on weak, continuous QND qubit state detection : \\
		&	\ind $\eta = \frac{ \Gamma_{\rm meas} }{\Gamma_{\varphi}} \leq 1$  [Eq. (\ref{eq:EfficiencyRatio})]\\

 & \\

\multicolumn{2}{c}{\em Sec.~\ref{sec:QLAmplifiers}: Quantum Limit on  Linear Amplifiers and Position Detectors} \\

$G$                   	& 	Photon number (power) gain, e.g. in Eq. (\ref{eq:fullb}) \\
			&	Input-output relation for a bosonic scattering amplifier: $\hat{b}^{\dagger}  =  \sqrt{G} \hat{a}^{\dagger} + \mathcal{\hat{F}}^{\dagger}$ [Eq.(\ref{eq:fullb})]  \\
$(\Delta a)^2$ 	& 	Symmetrized field operator uncertainty for the scattering description of a bosonic amplifier:\\
			& 		\ind $\left(\Delta a \right)^2 \equiv \frac{1}{2} \left \langle \{  \hat{a}, \hat{a}^{\dag} \} \right \rangle  - \left| \langle a \rangle \right|^2 $  \\

			&	Standard quantum limit for the noise added by a phase-preserving bosonic scattering amplifier \\
			&	\ind in the high-gain limit, $G\gg 1$, where $\langle (\Delta a)^2 \rangle_{\rm ZPF}=\frac{1}{ 2}$: \\
  			& 	\ind $\frac{\left( \Delta b \right)^2}{G}  \geq   \left( \Delta a \right)^2 + \frac{1}{2}$ [Eq. (\ref{CavesQL})] \\

 & \\





$G_P[\omega]$         & Dimensionless power gain of a linear position detector or voltage amplifier  \\
				 & \ind (maximum ratio of the power delivered 
				 by the detector output to a load, vs. the power fed into signal source): \\
                      &   \ind $  G_P[\omega] = \frac{ | \chiIF[\omega] |^2}{4 \textrm{Im } \chi_{FF}[\omega]
                 			\cdot \textrm{Im } \chi_{II} [\omega]]}  $ [Eq. (\ref{GPDefn})]\\
                      & \ind	For a quantum-ideal detector, in the high-gain limit: $G_P \simeq \left[ \frac{\textrm{Im } \alpha }{ |\alpha| } \frac{4 \kb T_{\rm eff}}{\hbar \omega }\right]^2$ [Eq. (\ref{eq:GPTeff})] \\


$\bS_{xx,{\rm eq}}[\omega,T]$    & Intrinsic equilibrium noise $\bS_{xx,{\rm eq}}[\omega,T]   =
         \hbar
         \coth\left( \frac{\hbar \omega}{2 \kb T} \right)
         \left[ -\textrm{Im } \oscchi[\omega] \right]$ [Eq. (\ref{SxEquilib})] \\


$A_{\rm opt}$           & Optimal coupling strength of a linear position detector which minimizes the added noise at frequency $\omega$: \\
                        & \ind $A^4_{\rm opt}[\omega] = \frac{ \bS_{II}[\omega]}{ |\lambda[\omega] \oscchi[\omega]|^2 \bS_{FF}[\omega] }$ [Eq. (\ref{AOptOffRes})] \\

$\gamma[A_{\rm opt}]$   & Detector-induced damping of a quantum-limited linear position detector at optimal coupling, fulfills \\
                        & \ind $\frac{ \gamma[A_{\rm opt}]}{\gamma_0 + \gamma[A_{\rm opt}]} =
     \left| \frac{\textrm{Im } \alpha}{\alpha}  \right|
     \frac{1}{\sqrt{G_P[\Omega]}}
     = \frac{\hbar \Omega}{4 \kb T_{\rm eff}} \ll 1$ [Eq. (\ref{OptA2})] \\

		&	Standard quantum limit for the added noise spectral density of a linear position detector (valid at each frequency $\omega$): \\
  		& 	\ind	$S_{xx,{\rm add}}[\omega] \geq \lim_{T\ra0}S_{xx,{\rm eq}}[w,T]$ [Eq.~(\ref{eq:SxxSQL})] \\

		&	Effective increase in oscillator temperature due to coupling to the detector backaction,  \\
		&	\ind for an ideal detector, with $\hbar \Omega / \kb \ll T_{{\rm bath}} \ll T_{\rm eff}$: \\
            	&  	\ind	$T_{{\rm osc}} \equiv \frac{ \gamma \cdot T_{\rm eff} +
         \gamma_0 \cdot T_{{\rm bath}}}
     { \gamma  + \gamma_0}
     \ra
     \frac{\hbar \Omega}{4 \kb} + T_{{\rm bath}}$     [Eq.~(\ref{eq:QLheating})] \\
\\

$\Zin, \Zout$		&	Input and output impedances of a linear voltage amplifier \\
$Z_{\rm s}$		&	Impedance of signal source attached to input of a voltage amplifier \\
$\lambda_V$		&	Voltage gain of a linear voltage amplifier \\
$\VN(t)$			&	Voltage noise of a linear voltage amplifier \\
				&	\ind (Proportional to the intrinsic output noise of the generic linear-response detector [Eq.~(\ref{eq:SVsub})] ) \\
$\IN(t)$			&	Current noise of a linear voltage amplifier \\
				&	\ind (Related to the back-action force noise of the generic linear-response detector [Eqs.~(\ref{eq:SIsub})] ) \\
$\TN$			&	Noise temperature of an amplifier [defined in Eq.~(\ref{eq:TNDefinition})] \\
$\ZN$			&	Noise impedance of a linear voltage amplifier [Eq. \ref{ZOpt1})] \\
				&	Standard quantum limit on the noise temperature of a linear voltage amplifier:  \\
				&	\ind $\kb \TN[\omega] \geq \frac{\hbar \omega}{2}$
					[Eq.(\ref{eq:TNQL})] \\
\\

\multicolumn{2}{c}{\em Sec.~\ref{sec:ScatteringAmp}: Bosonic Scattering Description of a Two-Port Amplifier} \\
			
$\hV_a (\hV_b)$			&	Voltage at the input (output) of the amplifier \\	
						&		\ind Relation to bosonic mode operators: Eq.~(\ref{eq:Vdefn}) \\
$\hI_a (\hI_b)$  			&	Current drawn at the input (leaving the output) of the amplifier \\
						&		\ind Relation to bosonic mode operators: Eq.~(\ref{eq:Idefn}) \\
$\lambda'_I$				&	Reverse current gain of the amplifier \\
$s[\omega]$				& 	Input-output $2 \times 2$ scattering matrix of the amplifier [Eq.~(\ref{eq:sdefn})] \\
						&		\ind Relation to op-amp parameters $\lambda_V, \lambda'_I, \Zin, \Zout$:
								Eqs.~(\ref{eq:OpAmpParams})\\
$\htV (\htI)$   				&	Voltage (current) noise operators of the amplifier \\
$\hFF_a[\omega]$, 
	$\hFF_b[\omega]$		&	Input (output) port noise operators in the scattering description [Eq.~(\ref{eq:sdefn})]  \\
						&	\ind Relation to op-amp noise operators $\htV, \htI$: Eq.~(\ref{eq:OpAmpNoises})  \\

\end{longtable*}

\begin{table*}[t]
	\caption{Contents of online appendix material.  Page numbers refer to the supplementary material.}
	\label{table:Appendices}
	 
	\begin{tabular*}{0.75\textwidth}{@{\extracolsep{\fill}} ll}
\hline
	Section &
	Page 
\\
\hline
	A.  Basics of Classical and Quantum Noise &
		1 \\
	B.  Quantum Spectrum Analyzers: Further Details &
		4 \\
	C.  Modes, Transmission Lines and Classical Input-Output Theory &
		8 \\
	D.  Quantum Modes and Noise of a Transmission Line &
		15 \\
	E.  backaction and Input-Output Theory for Driven Damped Cavities &
		18 \\
	F.  Information Theory and Measurement Rate &
		29 \\
	G.  Number Phase Uncertainty &
		30 \\
	H.  Using Feedback to Reach the Quantum Limit &
		31 \\
	I.  Additional Technical Details &
		34 \\
\hline
	
\end{tabular*}	
\end{table*}

\section{Quantum Noise Spectra}
\label{sec:QuantumNoiseSpectra}

\subsection{Introduction to quantum noise}
\label{subsec:IntroQuantumNoise}


In classical physics, the study of a noisy time-dependent quantity invariably involves its spectral density $\clS[\omega]$.  The spectral density tells us the
intensity of the noise at a given frequency, and is directly related to the auto-correlation function of the noise.\footnote{%
For readers unfamiliar with the basics of classical noise, 
a compact review is given in Appendix \ref{sec:ReviewClassicalNoise}.}  
In a similar fashion, the study of quantum noise involves
quantum noise spectral densities.  These are defined in a manner which mimics the classical case:  
\be
S_{xx}[\omega]=\int_{-\infty}^{+\infty} dt\, e^{i\omega t}
\langle \hat x(t) \hat x(0)\rangle.
\ee
Here $\hat x$ is a quantum operator (in the Heisenberg
representation) whose noise we are interested in, and the angular brackets indicate the quantum
statistical average evaluated using the quantum density matrix.
Note that we will use $\clS[\omega]$ throughout this review to denote the
spectral density of a classical noise, while $S[\omega]$ will denote a quantum
noise spectral density.  

As a simple introductory example illustrating important differences from the classical
limit, consider the position noise of a simple
harmonic oscillator having mass $M$ and frequency $\Omega$. The
oscillator is maintained in equilibrium with a large heat bath at
temperature $T$ via some infinitesimal coupling which we will
ignore in considering the dynamics.  The
solutions of the Heisenberg equations of motion are the same as
for the classical case but with the initial position $x$ and momentum $p$
replaced by the corresponding quantum operators.
It follows that the position autocorrelation
function is
\begin{eqnarray}
G_{xx}(t) &=& \langle \hat x(t) \hat x(0)\rangle 
\label{eq:autocorrxx}
\\
&=& \langle \hat x(0)\hat x(0)\rangle\cos(\Omega t)  + \langle
\hat p(0) \hat x(0)\rangle\frac{1}{M\Omega}\sin(\Omega
t).\nonumber
\end{eqnarray}
Classically the second term on the RHS vanishes because in thermal
equilibrium $x$ and $p$ are uncorrelated random
variables.   As we will see shortly below for the quantum case,
the symmetrized (sometimes called the `classical') correlator
vanishes in thermal equilibrium, just as it does classically:
$\langle \hat x \hat p + \hat p \hat x\rangle =0$.
Note however that in the quantum case, the canonical commutation
relation between position and momentum implies there must be some
correlations between the two, namely $\langle\hat x(0)\hat p(0)\rangle - \langle \hat p(0) \hat
x(0)\rangle=i\hbar$.
These correlations are easily evaluated by writing $\hx$ and $\hp$ in terms of harmonic oscillator
ladder operators.  We find that in thermal equilibrium:
$\langle \hat p(0) \hat x(0)\rangle =
-i\frac{\hbar}{2}$
and $\langle \hat x(0)\hat p(0) \rangle = +i\frac{\hbar}{2}$.
Not only are the position and momentum correlated, but their
correlator is imaginary!\footnote{Notice that this occurs because
the product of two non-commuting hermitian operators is not itself
an hermitian operator.}  This means that, despite the fact that
the position is an hermitian observable with real eigenvalues, its
autocorrelation function is complex and given from
Eq.~(\ref{eq:autocorrxx}) by:
\be
G_{xx}(t) =x_{\rm ZPF}^2\left\{\nb(\hbar\Omega) e^{+i\Omega t}
+ [\nb(\hbar\Omega)+1]e^{-i\Omega t}\right\},
\label{eq:complexGxx}
\ee
where $x_{\rm ZPF}^2 \equiv \hbar / 2M\Omega$ is the RMS zero-point uncertainty of $x$ in
the quantum ground state, and  $\nb$ is the Bose-Einstein occupation factor.  The complex
nature of the autocorrelation function follows from the fact that
the operator $\hat x$ does not commute with itself at different
times.

Because the correlator is complex it follows that the spectral
density is not symmetric in frequency:
\begin{eqnarray}
&&S_{xx}[\omega] = 2\pi x_{\rm ZPF}^2
\label{eq:Sxxquantum}
\\
&\times&\left\{\nb(\hbar\Omega) \delta(\omega + \Omega) +
[\nb(\hbar\Omega)+1]\delta(\omega - \Omega)\right\}\nonumber
\end{eqnarray}
In contrast, a classical autocorrelation function is always real, and hence a classical
noise spectral density is always symmetric in frequency.    Note that in the high temperature limit
$\kbt\gg\hbar\Omega$ we have
$\nb(\hbar\Omega)\sim \nb(\hbar\Omega)+1\sim
\frac{\kbt}{\hbar\Omega}$.  Thus, in this limit the $S_{xx}[\omega]$ becomes symmetric 
in frequency as expected classically, and coincides with the classical expression for 
the position spectral density (cf.~Eq.~(\ref{eq:Sxxclassical})).  
  
\begin{figure}[t]
\begin{center}
\includegraphics[width=0.9 \columnwidth]{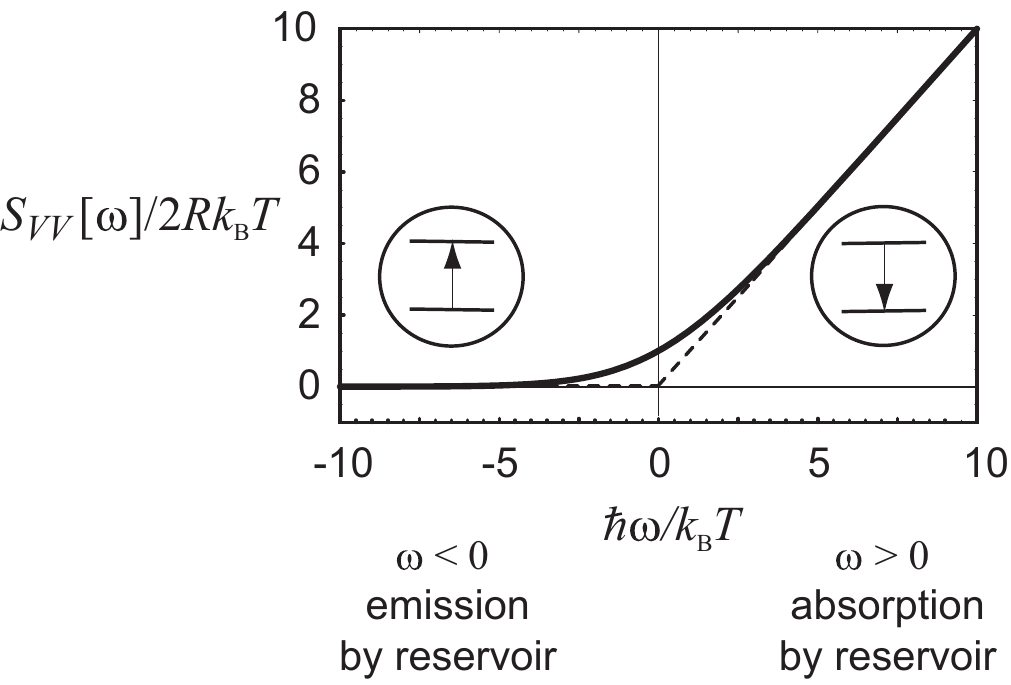}
\caption{Quantum noise spectral density of voltage fluctuations across a resistor (resistance $R$) as a function of frequency at zero temperature (dashed line) and finite temperature (solid line).
\label{fig:resistornoiseplot}}
\end{center}
\end{figure}

The Bose-Einstein factors suggest a way to understand the frequency-asymmetry of 
Eq.~(\ref{eq:Sxxquantum}):  the positive frequency part
of the spectral density has to do with stimulated emission of
energy {\em into} the oscillator and the negative frequency part
of the spectral density has to do with emission of energy {\em by}
the oscillator. That is, the positive frequency part of the
spectral density is a measure of the ability of the oscillator to
{\em absorb} energy, while the negative frequency part is a
measure of the ability of the oscillator to {\em emit} energy.
As we will see, this is generally true, even for non-thermal states.  
Fig.~\ref{fig:resistornoiseplot} illustrates this idea for the case of the
voltage noise spectral density of a resistor (see Appendix
\ref{subapp:ResistorNoise} for more details).  Note that the result Eq.~(\ref{eq:Sxxquantum})
can be extended to the case of a bath of many harmonic oscillators.  As described in Appendix
\ref{app:QuantumResistor} a resistor can be modeled as an
infinite set of harmonic oscillators and from this model the
Johnson/Nyquist noise of a resistor can be derived.


\subsection{Quantum spectrum analyzers}
\label{subsec:QuantumSpectrumAnalyzers}

The qualitative picture described in the previous subsection can be confirmed by considering simple systems which act as effective
spectrum analyzers of quantum noise.  The simplest such example
is a quantum-two level system (TLS) coupled to a 
quantum noise source \cite{Aguado00, Gavish00, Schoelkopf03}.  Describing the TLS as a fictitious spin-1/2
particle with spin down (spin up) representing the ground state (excited state), its Hamiltonian is 
${\hat H}_{0}=\frac{\hbar \w01}{2}{\hat \sigma}_{z}$, where $\hbar \omega_{01}$ is the energy splitting between the two states.  The TLS is then coupled to an external noise source via an additional term in the Hamiltonian
\begin{equation}
{\hat V}=A \hF {\hat \sigma} _{x},
\label{eq:TLSCoupling}
\end{equation}
where $A$ is a coupling constant, and the operator $\hF$ represents the external noise source.  
The coupling Hamiltonian $\hV$ can lead to the exchange of energy between the two-level system and noise source, and hence transitions between its two eigenstates.  The corresponding Fermi Golden Rule transition rates can be compactly expressed in terms of the quantum noise spectral density of $\hF$, $S_{FF}[\omega]$:
\begin{subequations}
\begin{eqnarray}
\Gamma _{\uparrow } & = & \frac{A^{2}}{\hbar ^{2}}S_{FF}[-\omega _{01}]
\label{gamup} \\
\Gamma _{\downarrow } & = & \frac{A^{2}}{\hbar ^{2}}S_{FF}[+\omega
_{01}]. \label{gamdown}
\end{eqnarray}
\label{eqs:FGRRates}
\end{subequations}
Here, $\Gamma_{\ua}$ is the rate at which the qubit is excited from its ground to excited state; $\Gamma_{\da}$ is the corresponding rate for the opposite, relaxation process.  As expected, positive (negative) frequency noise corresponds to absorption (emission) of energy by the noise source.  Note that if the noise source is in thermal equilibrium at temperature $T$, the transition rates of the TLS must satisfy the detailed balance relation $\Gamma_{\ua} / \Gamma_{\da} =  e^{-\beta \hbar \omega_{01}}$, where $\beta = 1/k_B T$.  This in turn implies that in thermal equilibrium, the quantum noise spectral density must satisfy:
\begin{equation} S_{FF}[+\w01 ]=e^{\beta \hbar \w01 }S_{FF}[-\w01].
\label{DetailedBalance}
\end{equation}
 The more general situation is where the noise source is {\it not} in thermal
equilibrium; in this case, no general detailed balance relation
holds. However, if we are concerned only with a single particular
frequency, then it is always possible to {\it define} an
`effective temperature' $T_{\rm eff}$ for the noise using
Eq.~(\ref{DetailedBalance}), i.e.
\begin{equation}
     \kb T_{\rm eff}[\omega] \equiv
     \frac{\hbar \omega}
     {\log \left[\frac{S_{FF}[\omega]}{S_{FF}[-\omega]} \right]}
     \label{OscTeff}
\end{equation}
Note that for a non-equilibrium system, $T_{\rm eff}$ will in general be frequency-dependent.     
In NMR language, $T_{\rm eff}$
will simply be the `spin temperature' of
our TLS spectrometer once it reaches steady state after being
coupled to the noise source. 

Another simple quantum noise spectrometer is a harmonic oscillator
(frequency $\Omega$, mass $M$, position $x$) coupled 
to a noise source (see e.g.~\textcite{Schwinger61, Dykman78}).  The coupling Hamiltonian is now:
\begin{equation}
     {\hat V} = A \hat{x} \hat{F} =  A \left[
         \xrms
         (\hat{a} + \hat{a}^{\dagger}) \right] \hat{F}
         \label{eq:SHOCoupling}
\end{equation}
where $\ha$ is the oscillator annihilation operator, $\hat{F}$ is the operator describing the fluctuating noise,
and $A$ is again a coupling constant.  We see that $(-A \hF)$ plays the role of a fluctuating force acting on the oscillator.
In complete analogy to the previous subsection, noise in $\hat{F}$ at the oscillator
frequency $\Omega$ can cause transitions between the oscillator energy eigenstates.
The corresponding Fermi Golden Rule transition rates are again simply related to the noise spectrum $S_{FF}[\omega]$.
Incorporating these rates into a simple master equation describing the probability to find the oscillator in a particular energy state, 
one finds that the stationary state of the oscillator is a Bose-Einstein distribution evaluated at the effective temperature $T_{\rm eff}[\Omega]$
defined in Eq.~(\ref{OscTeff}).  Further, one can use the master equation to 
derive a very classical-looking equation for the average energy 
$\langle E \rangle$ of the oscillator (see Appendix \ref{sec:SHOQNOISE}):
\begin{equation}
     \frac{d}{dt} \langle E \rangle =
         P - \gamma \langle E \rangle
     \label{OscEnergy}
\end{equation}
where
\begin{eqnarray}
     P & = &  
                 \frac{A^2}{4 M}
         \left( S_{FF}[\Omega] + S_{FF}[-\Omega] \right)
         \equiv 
                \frac{A^2   \bS_{FF}[\Omega]}{2 M}
                           \label{OscD} \\
     \gamma & = & 
         \frac{A^2\xrms^2}{\hbar^2}
         \left( S_{FF}[\Omega] - S_{FF}[-\Omega] \right)
             \label{OscGamma}
\end{eqnarray}
The two terms in Eq.~(\ref{OscEnergy}) describe, respectively,
heating and damping of the oscillator by the noise source.  The
heating effect of the noise is completely analogous to what
happens classically: a random force causes the oscillator's
momentum to diffuse, which in turn causes $\langle E \rangle$ to grow linearly in time at rate
proportional to the force noise spectral density.
In the quantum case, Eq.~(\ref{OscD}) indicates that it is the
{\it symmetric-in-frequency} part of the noise spectrum
, $\bS_{FF}[\Omega]$, which is
responsible for this effect, and which thus plays the role of a
classical noise source.
This is another reason why $\bS_{FF}[\omega]$ 
is often referred to as the ``classical" part of the
noise.\footnote{%
Note that with our definition, $\langle \hF^2 \rangle =  \int_{-\infty} ^{\infty} 
(d \omega / 2 \pi) \bS_{FF}[\omega]$.  It is common in engineering
contexts to define so-called ``one-sided" classical spectral densities, which 
are equal to two times our definition.}
In contrast, we see that the {\it asymmetric-in-frequency}
part of the noise spectrum is responsible for the damping.  This also has
a simple heuristic interpretation:  damping is caused by the net tendency 
of the noise source to absorb, rather than emit, energy from the oscillator.    


The damping induced by the noise source may equivalently be attributed
to the oscillator's motion inducing an average value to $\langle F \rangle$ which is 
out-of-phase with $x$, i.e.  
     $  \delta \langle A \cdot F(t) \rangle = - M \gamma \dot{x}(t)$.
Standard quantum linear response theory yields:
\begin{equation}
     \delta \langle A \cdot \hat{F}(t) \rangle =
         A^2 \int dt' \chi_{FF}(t-t') \langle {\hat x}(t') \rangle 
         \label{OscLR}
\end{equation}
where we have introduced the susceptibility
\begin{equation}
     \chi_{FF}(t) = \frac{-i}{\hbar} \theta(t) \left \langle
         [\hat{F}(t),\hat{F}(0)] \right \rangle
         \label{OscLF}
\end{equation}
Using the fact that the oscillator's motion only involves the frequency $\Omega$, we thus have:
\begin{eqnarray}
     \gamma &=& \frac{2A^2\xrms^2}{\hbar} \left[
          - \textrm{Im} \chi_{FF}[\Omega] \right] 
         \label{eq:GammaLambdaF}
\end{eqnarray}
A straightforward manipulation of Eq.~(\ref{OscLF})
for $\chi_{FF}$ shows that this expression for $\gamma$ is {\it
exactly} equivalent to our previous expression, Eq.~(\ref{OscGamma}).


In addition to giving insight on the meaning of the symmetric and asymmetric
parts of a quantum noise spectral density, the above example also directly yields the quantum version of the
fluctuation-dissipation theorem \cite{Callen51}. As we saw earlier, if our
noise source is in thermal equilibrium, the positive and negative
frequency parts of the noise spectrum are strictly related to one
another by the condition of detailed balance (cf.\
Eq.~(\ref{DetailedBalance})). This in turn lets us link the
classical, symmetric-in-frequency part of the noise to the
damping (i.e.~the asymmetric-in-frequency part of the noise).
Letting $\beta = 1/(\kb T)$ and making use of Eq.~(\ref{DetailedBalance}), we have:
\begin{eqnarray}
     \bS_{FF}[\Omega] & \equiv & \frac{S_{FF}[\Omega] + S_{FF}[-\Omega]}{2}
             \nonumber 
              \\
     & = & \frac{1}{2} \coth (\beta \hbar \Omega / 2)
         \left( S_{FF}[\Omega] - S_{FF}[-\Omega] \right)
                 \nonumber \\
     & = & \coth (\beta \hbar \Omega / 2) \frac{\hbar \Omega
     M}{A^2} \gamma[\Omega]
         \label{eq:FDFinal}
     \label{FDT}
\end{eqnarray}
Thus, in equilibrium, the condition that noise-induced transitions
obey detailed balance immediately implies that noise and damping
are related to one another via the temperature.  For $T \gg \hbar
\Omega$, we recover the more familiar classical version of the
fluctuation dissipation theorem:
\begin{equation}
     A^2 \bS_{FF}[\Omega] = 2 \kb T M \gamma
\end{equation}
Further insight into the fluctuation dissipation theorem
is provided in Appendix \ref{subsec:TLClassicalStatMech}, where we discuss it
in the simple but instructive context of a transmission line terminated by an impedance
$Z[\omega]$. 

We have thus considered two simple examples of how one can measure quantum
noise spectral densities.  Further details, as well as examples of other quantum noise spectrum analyzers, are
given in Appendix \ref{app:QSAFurtherDetails}.

\section{Quantum Measurements}
\label{sec:QuantumMeasurements}

Having introduced both quantum noise and
quantum spectrum analyzers, we are now in a position to
introduce the general topic of quantum measurements.
All practical measurements are affected by noise.
Certain quantum measurements remain limited
by quantum noise {\it even though} they use completely ideal apparatus.  As we will see,
the limiting noise here is associated with the fact that canonically
conjugate variables are incompatible observables in quantum mechanics.

The simplest, idealized description of a quantum measurement,
introduced by von Neumann
\cite{vonNeumann32,Bohm89,WheelerZurek84,HarocheRaimond2006},
postulates that the measurement process instantaneously collapses
the system's quantum state onto one of the eigenstates of the
observable to be measured. As a consequence, any initial
superposition of these eigenstates is destroyed and the values of
observables conjugate to the measured observable are perturbed.
This perturbation is an intrinsic feature of quantum mechanics and
cannot be avoided in any measurement scheme, be it of the
{}``projection-type'' described by von Neumann or rather a weak,
continuous measurement to be analyzed further below.

\begin{figure}[t]
\begin{center}
\includegraphics[width=3.45in]{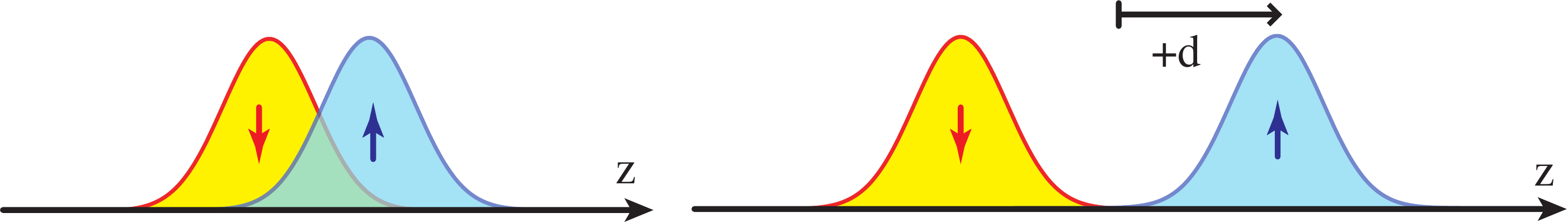}
\caption{(Color online) Schematic illustration of position distributions of an
atom in the detector plane of a Stern-Gerlach apparatus whose
field gradient is in the $z$ direction.  For small values of the
displacement $d$ (described in the text), there is significant
overlap of the distributions and the spin cannot be unambiguously
inferred from the position.  For large values of $d$ the spin is
perfectly entangled with position and can be inferred from the
position.  This is the limit of strong projective measurement.}
\label{fig:SGpositiondist}
\end{center}
\end{figure}

To form a more concrete picture of quantum measurement, we begin
by noting that every quantum measurement apparatus consists of a macroscopic
`pointer' coupled to the microscopic system to be measured. (A
specific model is discussed in \textcite{Nieuwenhuizen2001}.) This
pointer is sufficiently macroscopic that its position can be read
out `classically'.  The interaction between the microscopic system
and the pointer is arranged so that the two become strongly
correlated. One of the simplest possible examples of a quantum
measurement is that of the Stern-Gerlach apparatus which measures
the projection of the spin of an $S=1/2$ atom along some chosen
direction.  What is really measured in the experiment is the final
position of the atom on the detector plate.  However, the magnetic
field gradient in the magnet causes this position to be
perfectly correlated (`entangled') with the spin projection so
that the latter can be inferred from the former.  Suppose for
example that the initial state of the atom is a product of a
spatial wave function $\xi_0(\vec r)$ centered on the entrance to
the magnet, and a spin state which is the superposition of up and
down spins corresponding to the eigenstate of ${\hat\sigma}_x$:
\be
|\Psi_0\rangle =
\frac{1}{\sqrt{2}}\left\{|\uparrow\rangle+|\downarrow\rangle
\right\} |\xi_0\rangle.
\ee
After passing through a magnet with field gradient in the $z$
direction, an atom with spin up is deflected upwards and an atom
with spin down is deflected downwards.  By the linearity of
quantum mechanics, an atom in a spin superposition state thus ends
up in a superposition of the form
\be
|\Psi_1\rangle = \frac{1}{\sqrt{2}}\left\{|\uparrow\rangle
|\xi_+\rangle + |\downarrow\rangle |\xi_-\rangle \right\},
\ee
where $\langle\vec r|\xi_\pm\rangle = \psi_1(\vec r\pm d\hat z)$
are spatial orbitals peaked in the plane of the detector.  The
deflection $d$ is determined by the device geometry and the
magnetic field gradient.  The $z$-direction position distribution
of the particle for each spin component is shown in
Fig.~\ref{fig:SGpositiondist}. If $d$ is sufficiently large
compared to the wave packet spread then, given the position of the
particle, one can unambiguously determine the distribution from
which it came and hence the value of the spin projection of the
atom.  This is the limit of a strong `projective' measurement.

In the initial state one has
$\langle\Psi_0|{\hat\sigma}_x|\Psi_0\rangle=+1$,
but in the final state one has
\be
\langle\Psi_1|{\hat\sigma}_x|\Psi_1\rangle=\frac{1}{2}\left\{
\langle\xi_-|\xi_+\rangle + \langle\xi_+|\xi_-\rangle  \right\}
\label{eq:SternGerlachoverlap}
\ee
For sufficiently large $d$ the states $\xi_\pm$ are orthogonal and
thus the act of ${\hat\sigma}_z$ measurement destroys the spin
coherence
\be
\langle\Psi_1|{\hat\sigma}_x|\Psi_1\rangle\rightarrow 0.
\label{eq:SternGerlachoverlap2}
\ee
This is what we mean by projection or wave function `collapse'.
The result of measurement of the atom position will yield a random
and unpredictable value of $\pm\frac{1}{2}$ for the $z$ projection
of the spin.  This destruction of the coherence in the transverse
spin components by a strong measurement of the longitudinal spin
component is the first of many examples we will see of the
Heisenberg uncertainty principle in action. Measurement of one
variable destroys information about its conjugate variable. We
will study several examples in which we understand microscopically
how it is that the coupling to the measurement apparatus causes
the `backaction' quantum noise which destroys our knowledge of
the conjugate variable.

In the special case where the eigenstates of the
observable we are measuring are also stationary states (i.e.~energy eigenstates),
measuring the observable a second time would reproduce
\emph{exactly} the same measurement result, thus providing a way
to confirm the accuracy of the measurement scheme.
These optimal
kinds of \emph{repeatable} measurements are called
{}``\emph{Quantum Non-Demolition}'' (QND) measurements
\cite{Braginsky80,Braginsky92,Braginsky96,AsherPeres}.   A simple example
would be a sequential pair of Stern-Gerlach devices oriented in
the same direction.  In the absence of stray magnetic
perturbations, the second apparatus
 would always yield the same answer as the
first.  
In practice, one terms a measurement QND if the observable being measured is an eigenstate of the {\it ideal} Hamiltonian of the measured system (i.e.~one ignores any couplings between this system and sources of dissipation).  This is reasonable if such couplings give rise to dynamics on timescales longer than what is needed to complete the measurement.  This point is elaborated in Sec.~\ref{sec:QLinpractice}, where we discuss practical considerations related to the quantum limit.  We also discuss in that section the fact that the repeatability of QND measurements is of fundamental
practical importance in overcoming detector inefficiencies
\cite{Gambetta07}.

A common confusion is to think that a QND measurement has
no effect on the state of the system being measured.  While this
is true if the initial state is an eigenstate of the observable,
it is \emph{not} true in general. 
 Consider again our example of a
spin oriented in the $x$ direction.
  The result of the first $\hat\sigma_z$ measurement will be
that the state randomly and completely unpredictably collapses
onto one of the two $\hat\sigma_z$ eigenstates:
the state is indeed altered by the measurement.
However all
subsequent measurements using the same orientation for the
detectors will always agree with the result of the first
measurement.  Thus QND measurements may affect the state of the
system, but never the value of the observable (once it is
determined). Other examples of QND measurements include: (i)
measuring the electromagnetic field energy stored inside a cavity
by determining the radiation pressure exerted on a moving piston
\cite{Braginsky92}, (ii) detecting the presence of a photon in a
cavity by its effect on the phase of an atom's superposition state
\cite{Haroche99,HarocheRaimond2006}, and (iii) the
{}``dispersive'' measurement of a qubit state by its effect on the
frequency of a driven microwave resonator \cite{Blais04,Wallraff04,Lupascu07},
which is the first canonical example we will describe below.

In contrast to the above, in non-QND measurements, the back-action
of the measurement will affect the observable being studied. The
canonical example we will consider below is the position
measurement of a harmonic oscillator.  Since the position operator
does not commute with the Hamiltonian, the QND criterion is not
fulfilled. Other examples of non-QND measurements include: (i)
photon counting via photo-detectors that absorb the photons, (ii)
continuous measurements where the observable does not commute with
the Hamiltonian, thus inducing a time-dependence of the
measurement result, (iii) measurements that can be repeated only
after a time longer than the energy relaxation time of the
system (e.g.~for a qubit, $T_1$) .

\subsection{Weak continuous measurements}
\label{subsec:WeakMeasExamples}

In discussing ``real" quantum measurements, another key notion to introduce
is that of \emph{weak, continuous measurements}
\cite{Braginsky92}. Many measurements in practice take an extended
time-interval to complete, which is much longer than the
{}``microscopic'' time scales (oscillation periods etc.) of the
system. The reason may be quite simply that the coupling strength
between the detector and the system cannot be made arbitrarily
large, and one has to wait for the effect of the system on the
detector to accumulate.  For example, in our Stern-Gerlach
measurement suppose
 that we are only able to achieve small magnetic field gradients
 and that consequently, the displacement $d$ cannot be made large compared to the
 wave packet spread (see Fig.~\ref{fig:SGpositiondist}).
 In this case the states $\xi_\pm$
 would have non-zero overlap and it would not be possible to
 reliably distinguish them:  we thus would only have a ``weak" measurement.
However, by cascading together a series of such measurements and taking
 advantage of the fact that they are QND, we can eventually achieve
 an unambiguous strong projective measurement:
  at the end of the cascade, we 
 are certain of which $\hat \sigma_z$ eigenstate the spin is in.
   During this
 process, the overlap of $\xi_\pm$ would gradually fall to zero
 corresponding to a smooth continuous loss of phase coherence in
 the transverse spin components.  
  At the end of the process, the QND nature
 of the measurement ensures that the
 probability of measuring $\sigma_z = \ua$ or $\da$ will accurately
 reflect the initial wavefunction.
 Note that it is only in this case of weak
 continuous measurements that makes sense to define a
 measurement rate in terms of a rate of gain of information about
 the variable being measured, and a corresponding dephasing
 rate, the rate at which information about the conjugate variable
 is being lost. We will see that these rates are intimately
 related via the Heisenberg uncertainty principle.

While strong projective measurements may seem to be the ideal, 
there are many cases where one may intentionally desire a weak continuous measurement; this was already discussed in the introduction.
There are many practical examples of weak,
continuous measurement schemes.  These include:
(i) charge measurements, where the current through a
device (e.g.\  quantum point contact or single-electron
transistor) is modulated by the presence/absence of a nearby
charge, and where it is necessary to wait for a sufficiently long
time to overcome the shot noise and distinguish between the two
current values, (ii) the weak dispersive qubit measurement
discussed below, (iii) displacement detection of a nano-mechanical
beam (e.g.\  optically or by capacitive coupling to a charge
sensor), where one looks at the two quadrature amplitudes of the
signal produced at the beam's resonance frequency.

Not surprisingly, quantum noise plays a crucial role in determining the
properties of a weak, continuous quantum measurement.  For such measurements, noise both determines the back-action effect of the measurement on the measured
system, as well as how quickly information is acquired in the measurement process.
Previously we saw that a crucial feature of quantum noise is the asymmetry between
positive and negative frequencies; we further saw that this
corresponds to the difference between absorption and emission
events.  For measurements, another key aspect of quantum noise
will be important:  as we will discuss extensively,
\emph{quantum mechanics places constraints on the
noise of any system capable of acting as a detector or amplifier}.
These constraints in turn place restrictions on any \emph{weak,
continuous} measurement, and lead directly to quantum limits on how well
one can make such a measurement.


In the rest of this section, we give an introduction to how one
describes a weak, continuous quantum measurement, considering the
specific examples of using parametric coupling to a resonant
cavity for QND detection of the state of a qubit and the
(necessarily non-QND) detection of the position of a harmonic
oscillator.  In the following section
(Sec.~\ref{sec:GenLinResponseTheory}), we give a derivation of  a very general
quantum mechanical constraint on the noise of any system capable
of acting as a detector, and show how this constraint {\it
directly} leads to the quantum limit on qubit detection. Finally,
in Sec.~\ref{sec:QLAmplifiers}, we will turn to the important but
slightly more involved case of a quantum linear amplifier or
position detector.  We will show that the basic quantum noise
constraint derived Sec.~\ref{sec:GenLinResponseTheory}
 again leads to a quantum limit;
here, this limit is on how small one can make the added noise of a
linear amplifier.

Before leaving this introductory section, it is worth pointing out
that the theory of weak continuous measurements is sometimes described
in terms of some set of auxiliary systems which are sequentially and
momentarily weakly coupled to the system being measured. (See
Appendix \ref{app:drivencavity}.)
%
%
One then envisions a
sequence of projective von Neumann measurements on the auxiliary
variables. The weak entanglement between the system of interest
and one of the auxiliary variables leads to a kind of partial
collapse of the system wave function (more precisely the density
matrix) which is described in mathematical terms not by projection
operators, but rather by POVMs (positive operator valued
measures).  We will not use this and the related `quantum
trajectory' language here, but direct the reader to the literature
for more information on this important approach.
\cite{Brun02,AsherPeres,Korotkov06,HarocheRaimond2006}

\subsection{Measurement with a parametrically coupled resonant cavity}
\label{subsec:CavityDetector}

A simple yet experimentally practical
example of a quantum detector consists of a
resonant optical or RF cavity parametrically coupled to the
 system being measured.  Changes in the variable being measured
 (e.g.~the state of a qubit or the position of an oscillator) shift the cavity frequency
 and produce a varying phase shift in the carrier signal reflected from the cavity.  This
 changing phase shift can be converted (via homodyne interferometry)
 into a changing intensity; this can then be detected using diodes or photomultipliers.

In this subsection, we will analyze weak, continuous measurements made using such a
parametric cavity detector; this will serve as a good introduction to the more general
approaches presented in later sections.  We will show that this detector
is capable of reaching the `quantum-limit', meaning that
it can be used
to make a weak, continuous measurement {\it as optimally} as is allowed by quantum mechanics.
This is true for both the (QND) measurement of the state of a qubit, and the (non-QND)
measurement of the position of a harmonic oscillator.
Complementary analyses of
weak, continuous qubit measurement are given in
\textcite{Makhlin00, Makhlin01} (using a single-electron transistor) and in
\textcite{Korotkov01b, Korotkov01c, Gurvitz97, Pilgram02, Clerk03} (using a quantum point
contact).  
We will focus here on a high-$Q$ cavity detector; weak qubit measurement with a low-$Q$
cavity was studied in \cite{Johansson06}.

It is worth noting the widespread usage of cavity detectors in experiment.
One important current realization is a microwave cavity used to
read out the state of a superconducting qubit 
\cite{Illichev03,Illichev04, Mooij04, Mooij05, Blais04,Wallraff04,Schuster05,Delsing05,Hakonen05}.
Another class of examples are optical cavities used to measure mechanical degree of freedom.
Examples of such systems include those where one
of the cavity mirrors is mounted on a
cantilever \cite{Gigan2006,Arcizet2006,KlecknerandBouwmeester2006}.
Related systems involve a freely suspended mass \cite{Abramovici92,Corbitt2007},
an optical cavity with a thin transparent membrane in the middle \cite{Harris08} and,
more generally, an elastically deformable whispering gallery mode
resonator \cite{2006_11_Kippenberg_RadPressureCooling}.
Systems where a microwave cavity is coupled to a mechanical element are also
under active study \cite{Blencowe07,Regal08,KonradCooling2008}.
 
We start our discussion with a general observation.  The cavity uses interference and the \emph{wave} nature of light
to convert the input signal to a phase shifted wave.  For small
phase shifts we have a weak continuous measurement.
Interestingly, it is the complementary \emph{particle} nature of
light which turns out to 
limit the measurement.
As we will see, it both limits the rate at which we can make a measurement
(via photon shot noise in the output beam) and also controls the
backaction disturbance of the system being measured
(due to photon shot noise inside the cavity acting on the system being
measured).  These two dual aspects are an important part
of any weak, continuous quantum measurement;  hence, understanding both the
output noise (i.e.~the measurement imprecision) and back-action noise of detectors will be crucial.

All of our discussion of 
noise in the cavity system will be framed in terms of the number-phase uncertainty relation for coherent states.  
A coherent photon state
contains a Poisson distribution of the number of photons, implying that the 
fluctuations in photon number obey $(\Delta N)^2={\bar N}$, where 
$\bar N$ is the mean number of photons.
Further, coherent states are over-complete and states of different phase
are not orthogonal to each other; 
this directly implies (see Appendix~\ref{app:numberphase})
that there is an uncertainty in any measurement of the phase.
For large $\bar N$, this is
given by:
\be
( \Delta\theta)^2 = \frac{1}{4\bar N}.
 \label{eq:phasevar}
\ee
Thus, large-$\bar N$ coherent states obey the number-phase uncertainty relation
\be
\Delta N  \Delta\theta = \frac{1}{2}
\label{eq:numberphaseuncertainty}
\ee
analogous to the position-momentum uncertainty relation.

Eq.~(\ref{eq:numberphaseuncertainty}) can also be usefully formulated in terms of noise spectral densities
associated with the measurement.
Consider a continuous photon beam carrying an average photon flux $\Ndotbar$.
The variance in the number of photons detected grows linearly in time and can
be represented as $(\Delta N)^2=\SNN t$, where $\SNN$ is the  
white-noise spectral density of photon-flux fluctuations. On a physical level, it describes
photon shot noise, and is given by $\SNN = \Ndotbar$.

Consider now the phase of the beam.
Any homodyne
measurement of this phase will be subject to
the same photon shot noise fluctuations discussed above (see Appendix~\ref{app:numberphase} for more details).
Thus, if the phase of the beam has some
nominal small value $\theta_0$, the output signal from the
homodyne detector integrated up to time $t$ will be of the form
$I=\theta_0t + \int_0^td\tau\, \delta\theta(\tau)$,
where $\delta\theta$ is a noise representing the imprecision in
our measurement of $\theta_0$ due to the photon shot noise in the
output of the homodyne detector.  
An unbiased estimate of the phase obtained from $I$ is
$\theta = I/t$, which obeys
$\langle \theta\rangle = \theta_0$.  Further, one has
$(\Delta\theta)^2=\Sthetatheta/t$,
where $\Sthetatheta$ is the spectral density of the $\delta\theta$
white noise.
Comparison with Eq.~(\ref{eq:phasevar}) yields
\be
\Sthetatheta = \frac{1}{4\Ndotbar}.
\ee
The results above lead us to the fundamental wave/particle
relation for ideal coherent beams
\be
\sqrt{\SNN\Sthetatheta} = \frac{1}{2}
\label{eq:numberphasenoiseproduct}
\ee


Before we study the role that these uncertainty relations play in
measurements with high $Q$ cavities, let us consider the simplest
case of reflecting light from a mirror without a cavity.  The
phase shift of the beam (having wave vector $k$) when the mirror
moves a distance $x$ is $2kx$. Thus, the uncertainty in the phase
measurement corresponds to a position imprecision which can again
be represented in terms of a noise spectral density
 $\SIxx = \Sthetatheta/4k^2$.
 Here the superscript I refers to the fact that this is noise representing imprecision in the measurement, not actual fluctuations in the position.  We also need to
 worry about backaction:  each photon hitting the mirror transfers a momentum $2\hbar k$ to the mirror, so photon shot noise corresponds to a random backaction force noise spectral density
$ \SFF = 4\hbar^2k^2 \SNN$
 Multiplying these together we have the central result for
 the product of the backaction force noise and the imprecision
\be
\SFF\SIxx = \hbar^2\SNN\Sthetatheta=\frac{\hbar^2}{4}
\label{eq:SFFSIxxproduct}
\ee
or in analogy with Eq.~(\ref{eq:numberphaseuncertainty})
\be
\sqrt{\SFF\SIxx}=\frac{\hbar}{2}.
\label{eq:sqrtforceimprecisionproduct}
\ee
Not surprisingly, the situation considered here is as ideal as possible.  Thus, the RHS above is actually a {\it lower bound} on the product of imprecision and back-action noise for {\it any} detector capable of significant amplification; we will prove this rigorously in Sec.~\ref{subsec:NoiseConstraint}.
Eq.~(\ref{eq:sqrtforceimprecisionproduct}) thus represents the quantum-limit on the noise of our detector.  As we will see shortly, having a detector with quantum-limited
noise is a prerequisite for reaching the quantum limit on various different measurement tasks (e.g.~continuous position detection  of an oscillator and QND qubit state detection).  Note that in general, a given detector will have {\it more} noise than the quantum-limited value;
we will devote considerable effort in later sections to determining the conditions needed to achieve the lower bound of Eq.~(\ref{eq:sqrtforceimprecisionproduct}).

We now turn to the story of measurement using a high $Q$ cavity;
it will be similar to the above discussion, except that we have to 
account for the filtering of the noise by the cavity response.
We relegate relevant calculational details related to  
Appendix \ref{app:drivencavity}.  The cavity
is simply described as a single bosonic mode coupled weakly to
electromagnetic modes outside the cavity.  The Hamiltonian of
the system is given by:
\begin{equation}
     {\hat H} = H_0 +  \hbar \omegar \left(1  + A
      {\hz} \right) {\hat a}^{\dagger} {\hat a} + {\hat H}_{\rm
      envt}.
     \label{CavityHam1}
\end{equation}
Here, $H_0$ is the unperturbed Hamiltonian of the system whose
variable $\hz$ (which is not necessarily a position)
is being measured, ${\hat a}$ is the annihilation operator for the
cavity mode, and $\omegar$ is the cavity resonance
frequency in the absence of the coupling $A$.  We will take both $A$ and
$\hz$ to be dimensionless. The term ${\hat
H}_{\rm envt}$ describes the electromagnetic modes outside the
cavity, and their coupling to the cavity; it is responsible for
both driving and damping the cavity mode.  The damping is
parameterized by rate $\kappa$, which tells us how quickly energy
leaks out of the cavity; we consider the case of a high
quality-factor cavity, where $\Qc \equiv \omegar/\kappa \gg 1$.

Turning to the interaction term in Eq.~(\ref{CavityHam1}), we see
that the parametric coupling strength $A$ determines the change in
frequency of the cavity as the system variable $\hz$ changes.
We will assume for simplicity that the dynamics of $\hz$
is slow compared to $\kappa$.
In this limit the reflected phase shift
simply varies slowly in time adiabatically following the
instantaneous value of $\hz$.  We will also assume that the
coupling $A$ is small enough that the phase shifts are always very
small and hence the measurement is weak. Many photons will have to
pass through the cavity before much information is gained about
the value of the phase shift and hence the value of $\hz$.

We first consider the case of a `one-sided' cavity where only one
of the mirrors is semi-transparent, the other being perfectly
reflecting. 
In this
case, a wave incident on the cavity (say, in a one-dimensional
waveguide) will be perfectly reflected, but with a phase shift
$\theta$ determined by the cavity and the value of $\hz$.  
The reflection coefficient at the bare cavity frequency $\omegar$ is simply given
by \cite{Walls94}
\be
r = -\frac{1+2i A \Qc \hz  }{1-2i A \Qc \hz  }.
\label{eq:onesidedcavityreflection}
\ee
 Note that $r$ has unit
magnitude because all photons which are incident are reflected if
the cavity is lossless.  For weak coupling we can write the
reflection phase shift as
$r= - e^{i\theta}$,
where
\be
\theta \approx 4 \Qc A \hz = (A \omegar \hz) \twd 
\label{eq:cavityphaseshift}
\ee
 We see that the scattering phase
shift is simply the frequency shift caused by the parametric
coupling multiplied by the Wigner delay time \cite{Wigner55}
\be
t_{\rm WD} = {\rm Im} \frac{\partial \ln
r}{\partial\omega}=4/\kappa.
\ee
 Thus the measurement
imprecision noise power for a given photon flux $\Ndotbar$
incident on the cavity is given by
\be
\SIzz = \frac{1}{(A \omegar \twd)^2}\Sthetatheta.
\label{eq:SIxxcavity}
\ee
The random part of the generalized backaction force conjugate to $\hz$ is
from Eq.~(\ref{CavityHam1})
\be
\hFz \equiv -\frac{\partial \hat H}{\partial \hz}= -A \hbar \omegar \,
\delta\hat n
\label{eq:FzDefn}
\ee
where, since $\hz$ is dimensionless, $\hFz$ has units of
energy.  Here $\delta\hat n = \hat n - \bar n = \ha^\dagger \ha -
\langle \ha^\dagger \ha \rangle$ represents the photon number
fluctuations around the mean $\bar n$ inside the cavity. The
backaction force noise spectral density is thus
\be
\SFFz = (A \hbar \omegar)^2 \Snn
\label{eq:1sidedcavityforcenoise}
\ee
As shown in Appendix \ref{app:drivencavity}, the cavity filters
the photon shot noise so that at low frequencies $\omega\ll
\kappa$ the number fluctuation spectral density is simply
\be
\Snn = {\bar n}\twd.
\label{eq:1sidedcavitySnn}
\ee
The mean photon number in the cavity is found to be
$\bar n = \Ndotbar \twd$,
where again $\Ndotbar$ the mean photon flux incident on the
cavity. From this it follows that
\be
\SFFz = (A \hbar \omegar  \twd)^2\SNN.
\label{eq:SFFcavity}
\ee
Combining this with Eq.~(\ref{eq:SIxxcavity}) again yields the
same result as Eq.~(\ref{eq:sqrtforceimprecisionproduct}) obtained
without the cavity.  The parametric cavity detector (used in this way) is thus
a quantum-limited detector, meaning that
the product of its noise spectral densities achieves the ideal
minimum value.
%


We will now examine how the quantum limit on the noise of our detector
directly leads to quantum limits on different measurement tasks.  In particular,
we will consider the cases of continuous position detection and QND qubit state
measurement.

\subsubsection{QND measurement of the state of a qubit using a resonant cavity}

\begin{figure}[t]
\begin{center}
\includegraphics[width=3.45in]{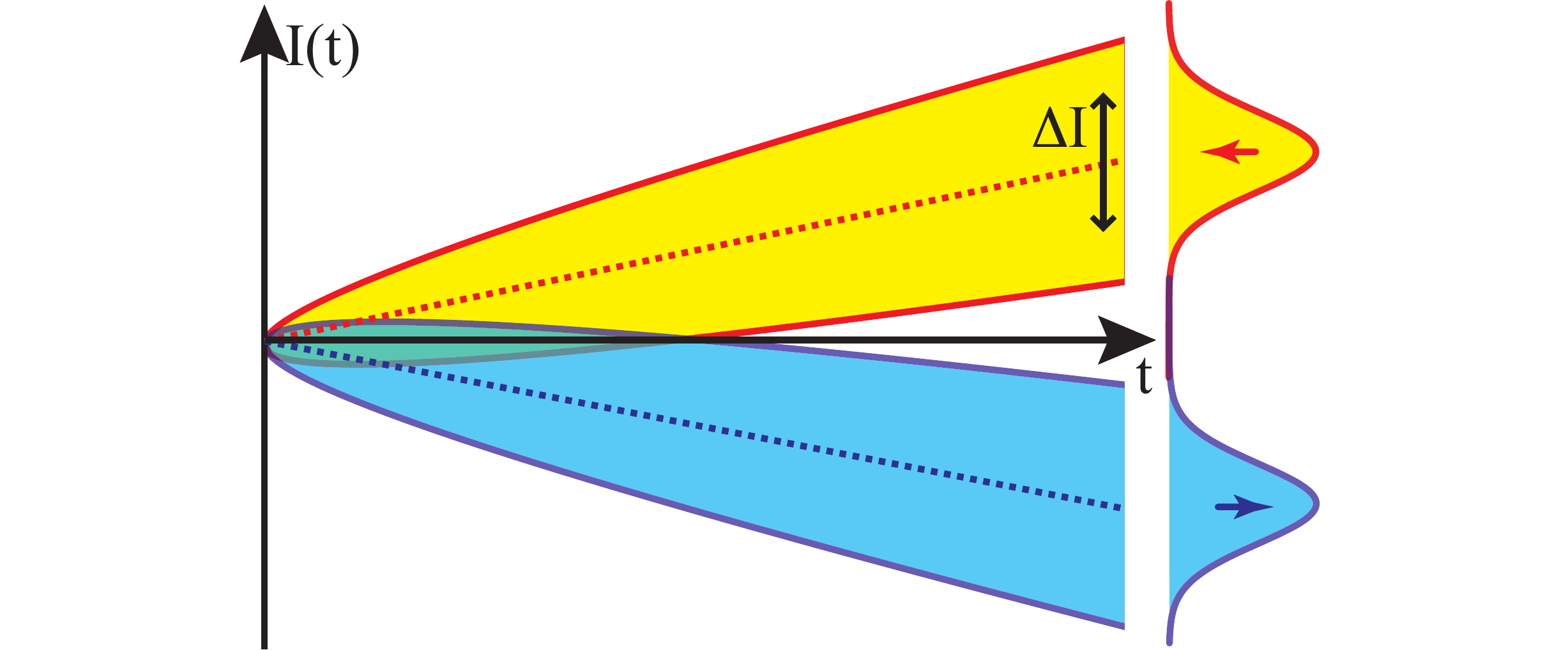}
\caption{(Color online) Distribution of the integrated output for the cavity detector, $I(t)$,
for the two different qubit states.  The separation of the means of the distributions
grows linearly in time, while the width of the distributions only grow as the $\sqrt{t}$.}
\label{fig:IntegratedOutput}
\end{center}
\end{figure}

Here we specialize to the case where the system operator $\hz ={\hat\sigma}_z$ represents the state of a spin-1/2 quantum bit.
Eq.~(\ref{CavityHam1}) becomes
\begin{equation}
     {\hat H} = \frac{1}{2}\hbar\omega_{01}{\hat \sigma}_z 
     + \hbar \omegar \left(1 +A {\hat \sigma}_z \right){\hat a}^{\dagger} {\hat
     a} + {\hat H}_{\rm envt}
     \label{eq:qubitcavityHam}
\end{equation}
We see that ${\hat\sigma}_z$ commutes with all terms in the
Hamiltonian and is thus a constant of the motion (assuming that
${\hat H}_{\rm envt}$ contains no qubit decay terms so that
$T_1=\infty$) and hence the measurement will be QND. From
Eq.~(\ref{eq:cavityphaseshift}) we see that the two states of the
qubit produce phase shifts $\pm\theta_0$ where
\be
\theta_0=A \omegar \twd.
\ee
As $\theta_0 \ll 1$, it will take many reflected photons before we
are able to determine the state of the qubit.  This is a direct
consequence of the unavoidable photon shot noise in the output of
the detector, and is a basic feature of weak measurements--
information on the input is only acquired gradually in time.

Let $I(t)$ be the homodyne signal for the wave reflected from the
cavity integrated up to time $t$.  Depending on the state of the
qubit the mean value of $I$ will be
$\langle I \rangle = \pm \theta_0 t$,
and the RMS gaussian fluctuations about the mean will be
$\Delta I = \sqrt{\Sthetatheta t}$.
As illustrated in Fig.~\ref{fig:IntegratedOutput} and discussed
extensively in \textcite{Makhlin01}, the integrated signal is
drawn from one of two gaussian distributions which are better and
better resolved with increasing time (as long as the measurement
is QND). The state of the qubit thus becomes ever more reliably
determined.
%
%
The signal energy to noise energy ratio becomes
\be
{\rm SNR} = \frac{\langle I \rangle^2}{(\Delta I)^2}
	= \frac{\theta_0^2}{\Sthetatheta} t
\label{eq:cavitySNR}
\ee
which can be used to define the measurement rate via
\be
\Gamma_{\rm meas} \equiv \frac{ {\rm SNR} }{2} = \frac{\theta_0^2}{2\Sthetatheta}
= \frac{1}{2 \SIzz}.
\label{eq:cavitymeasurementrate}
\ee
There is a certain arbitrariness in the scale factor of $2$ appearing in the
definition of the measurement rate; this particular choice
is motivated by precise information theoretic grounds 
(as defined, $\Gamma_{\rm meas}$ is the rate at which
the `accessible information' grows, c.f 
Appendix~\ref{app:measurementrateinformationtheory}).

While Eq.~(\ref{eq:qubitcavityHam}) makes it clear that the state
of the qubit modulates the cavity frequency, we can easily
re-write this equation to show that this same interaction term is
also responsible for the {\it back-action} of the measurement
(i.e.~the disturbance of the qubit state by the measurement
process):
\begin{equation}
     {\hat H} = \frac{\hbar }{2} \left(\omega_{01} + 2 A \omegar {\hat
     a}^{\dagger} {\hat a} \right)
     {\hat \sigma}_z + \hbar \omegar  {\hat a}^{\dagger} {\hat
     a} + {\hat H}_{\rm envt}
\end{equation}
We now see that the interaction can also be viewed as providing a
`light shift' (i.e.~ac Stark shift) of the qubit splitting
frequency \cite{Blais04,Schuster05} which contains a constant part
$2 A {\bar n}A \omegar$ plus a randomly fluctuating part
$\Delta\omega_{01}=2\hFz/\hbar$
which depends on ${\hat n} = {\hat a}^{\dagger} {\hat a}$, the
number of photons in the cavity.  During a measurement, ${\hat n}$
will fluctuate around its mean and act as a fluctuating
back-action `force' on the qubit. In the present QND case, noise
in ${\hat n} = {\hat a}^{\dagger} {\hat a}$ cannot cause
transitions between the two qubit eigenstates.  This is the
opposite of the situation considered in Sec.~\ref{subsec:QuantumSpectrumAnalyzers},
where we wanted to use the qubit as a spectrometer. Despite the
lack of any noise-induced transitions, there still is a
back-action here, as noise in ${\hat n}$ causes the effective
splitting frequency of the qubit to fluctuate in time. For weak
coupling, the resulting phase diffusion leads to
measurement-induced dephasing of superpositions in the qubit
\cite{Blais04,Schuster05} according to
\begin{eqnarray}
\left\langle e^{-i\varphi}\right\rangle &=& \left\langle
e^{-i\int_0^td\tau\,\Delta\omega_{01}(\tau)} \right\rangle.
\end{eqnarray}
For weak coupling the dephasing rate is slow and thus we are
interested in long times $t$.  In this limit the integral is a sum
of a large number of statistically independent terms and thus we
can take the accumulated phase to be gaussian distributed.  Using
the cumulant expansion we then obtain
\begin{eqnarray}
	\left\langle e^{-i\varphi}\right\rangle &=&
	\exp\left( -\frac{1}{2} \left\langle
		\left[ \int_0^td\tau\,\Delta\omega_{01}(\tau) \right]^2
		\right\rangle \right) \nonumber\\
&=& \exp\left(-\frac{2}{\hbar^2}\SFFz t \right).
\label{eq:BAdephasing}
\end{eqnarray}
Note also that the noise correlator
above is naturally symmetrized-- the quantum asymmetry of the noise
plays no role for this type of coupling. Eq.~(\ref{eq:BAdephasing}) yields the 
dephasing rate
\be
\Gamma_\varphi = \frac{2}{\hbar^2}\SFFz = 2\theta_0^2\SNN.
\label{eq:cavitydephasingrate}
\ee

Using Eqs.~(\ref{eq:cavitymeasurementrate}) and (\ref{eq:cavitydephasingrate}), we find the 
interesting conclusion that the dephasing rate and measurement rates coincide:
\be
	\frac{\Gamma_\varphi}{\Gamma_{\rm meas}}
	=	\frac{4}{\hbar^2} \SIzz \SFFz =
 4  \SNN\Sthetatheta = 1.
\label{eq:cavityreachesquantumlimit}
\ee
As we will see and prove rigorously, this represents the ideal, quantum-limited case
for QND qubit detection:  the best one can do is measure as quickly as one dephases.
In keeping with our
earlier discussions, it represents
the enforcement of the
Heisenberg uncertainty principle. The faster you gain information
about one variable, the faster you lose information about the
conjugate variable.  Note that in general, the ratio $\Gamma_\varphi / \Gamma_{\rm meas}$ will be larger than one, as an arbitrary detector
will not reach the quantum limit on its noise spectral densities.  Such a non-ideal
detector produces excess back-action beyond what is required quantum mechanically.

In addition to the quantum noise point of view presented above,
there is a second
complementary way in which to understand the origin of
measurement induced dephasing \cite{Stern90} which is analogous to
our description of loss of transverse spin coherence in the
Stern-Gerlach experiment in Eq.~(\ref{eq:SternGerlachoverlap}).
The measurement takes the incident wave, described by a coherent
state $| \alpha \rangle$, to a reflected wave described by a
(phase shifted) coherent state $| r_{\ua} \cdot \alpha \rangle$ or
$| r_{\da} \cdot \alpha \rangle$, where $r_{\ua/\da}$ is the
qubit-dependent reflection amplitude given in
Eq.~(\ref{eq:onesidedcavityreflection}). Considering now the full
state of the qubit plus detector, measurement results in a state change:
\begin{eqnarray}
     \frac{1}{\sqrt{2}} \Big(
         | \ua \rangle +  | \da \rangle
         \Big) \otimes | \alpha \rangle &\ra&
     \frac{1}{\sqrt{2}} \Big(
     e^{+i \omega_{01} t/2}| \ua \rangle \otimes |r_{\ua} \cdot \alpha
\rangle
\nonumber\\
     &+&
     e^{-i \omega_{01} t/2}
     | \da \rangle \otimes |r_{\da} \cdot \alpha \rangle
     \Big)
     \label{EntangledState}
\end{eqnarray}
As $| r_{\ua} \cdot \alpha \rangle \neq | r_{\da} \cdot \alpha
\rangle$, the qubit has become entangled with the detector:  the
state above cannot be written as a product of a qubit state times
a detector state. 
 To assess the coherence of the final qubit state (i.e.~the relative
phase between $\ua$ and $\da$), one looks at the off-diagonal
matrix element of the qubit's reduced density matrix:
\begin{eqnarray}
     \rho_{\da \ua} & \equiv & \textrm{Tr }_{\textrm {detector}}
         \langle \da | \psi \rangle
         \langle \psi | \ua \rangle \\
     & = &
         \frac{e^{+i \omega_{01} t}}{2} \langle r_{\ua} \cdot \alpha |
             r_{\da} \cdot \alpha \rangle
             \label{RhoSup2} \\
     & = &
         \frac{e^{+i \omega_{01} t}}{2}
         \exp\left[
             - |\alpha|^2 \left(1 - r_{\ua}^{*} r_{\da} \right)
             \right]
              \label{RhoSup3}
\end{eqnarray}
In Eq.~(\ref{RhoSup2}) we have used the usual expression for the
overlap of two coherent states.  We see that the measurement
reduces the magnitude of $\rho_{\ua \da}$: this is dephasing.  The
amount of dephasing is directly related to the overlap between the
different detector states that result when the qubit is up or
down; this overlap can be straightforwardly found using
Eq.~(\ref{RhoSup3}) and $|\alpha|^2 = \bar N = \Ndotbar t$, where
$\bar N$ is the mean number of photons that have reflected from
the cavity after time $t$. We have
\begin{equation}
         \left|
         \exp\left[
             - |\alpha|^2 \left(1 - r_{\ua}^{*} r_{\da} \right)
             \right] \right|
         =
             \exp \left[
                 - 2 \bar N \theta_0^2 \right]
         \equiv \exp\left[ -\Gamma_{\varphi} t \right]
\end{equation}
with the dephasing rate $\Gamma_{\varphi}$ being given by:
\begin{equation}
     \Gamma_{\varphi} = 2 \theta_0^2
         \Ndotbar
         \label{eq:OverlapDephRate}
\end{equation}
in complete agreement with the previous result in
Eq.(\ref{eq:cavitydephasingrate}).


\subsubsection{Quantum limit relation for QND qubit state detection}
\label{subsubsec:CavityQubitQL}

We now return to the ideal quantum limit relation of Eq.~(\ref{eq:cavityreachesquantumlimit}).
As previously stated, this is a lower bound:
quantum mechanics enforces
the constraint that in a QND qubit measurement {\it the best you can possibly do is measure
as quickly as you dephase} \cite{Averin00b,Makhlin01, Devoret00, Clerk03,
Averin03, Korotkov01c}:
\begin{equation}
     \Gamma_{\rm meas} \leq \Gamma_{\varphi}
     \label{QLCond}
\end{equation}
While a detector with quantum limited noise has an equality above,
most detectors will be very far from this ideal limit, and will dephase
the qubit faster than they acquire information about its state.
We provide a proof of Eq.~(\ref{QLCond}) in
Sec.~\ref{subsec:QLimitProof}; for now, we note that its heuristic
origin rests on the fact that both measurement and dephasing rely
on the qubit becoming entangled with the detector. Consider again
Eq.~(\ref{EntangledState}), describing the evolution of the
qubit-detector system when the qubit is initially in a
superposition of $\ua$ and $\da$.  To say that we have truly
measured the qubit, the two detector states $|r_{\ua} \alpha
\rangle$ and $|r_{\da} \alpha \rangle$ need to correspond to
different values of the detector output (i.e.~phase shift $\theta$
in our example); this necessarily implies they are orthogonal.
This in turn implies that the qubit is completely dephased:
$\rho_{\ua \da} = 0$, just as we saw in
Eq.~(\ref{eq:SternGerlachoverlap2}) in the Stern-Gerlach example.
Thus, {\it measurement implies dephasing}. The opposite is not
true. The two states $|r_{\ua} \alpha \rangle$ and $|r_{\da}
\alpha \rangle$ could in principle be orthogonal without them
corresponding to different values of the detector output
(i.e.~$\theta$).  For example, the qubit may have become entangled
with extraneous microscopic degrees of freedom in the detector.
Thus, on a heuristic level, the origin of Eq.~(\ref{QLCond}) is
clear.

Returning to our one-sided cavity system, we see from
Eq.~(\ref{eq:cavityreachesquantumlimit}) that the one-sided cavity
detector reaches the quantum limit.  It is natural to now ask {\it
why} this is the case: is there a general principle in action here
which allows the one-sided cavity to reach the quantum limit? The
answer is yes: {\it reaching the quantum limit requires that there
is no `wasted' information in the detector} \cite{Clerk03}. There
should not exist any unmeasured quantity in the detector which
could have been probed to learn more about the state of the qubit.
In the single-sided cavity detector, information on the state of
the qubit is {\it only} available in (that is, is entirely
encoded in) the phase shift of the reflected beam; thus, there is
no `wasted' information, and the detector does indeed reach the
quantum limit. 

To make this idea of `no wasted information' more concrete, 
we now consider a simple detector
which fails to reach the quantum limit precisely due to
`wasted' information.  Consider again a 1D cavity system
where now {\it both} mirrors are slightly transparent. Now, a wave
incident at frequency $\omega_{\rm R}$ on one end of the cavity
will be partially reflected and partially transmitted. If the
initial incident wave is described by a coherent state $| \alpha
\rangle$, the scattered state can be described by a tensor product
of the reflected wave's state and the transmitted wave's state:
\begin{equation}
     | \alpha \rangle \ra | r_{\sigma} \cdot \alpha \rangle | t_{\sigma}
     \cdot \alpha \rangle
\end{equation}
where the qubit-dependent reflection and transmission amplitudes
$r_{\sigma}$ and $t_{\sigma}$ are given by \cite{Walls94}:
\begin{eqnarray}
     t_{\da} & = & \frac{1}{1 + 2 i A \Qc } \\
     r_{\da} & = & \frac{2 i \Qc A }{1 + 2 i A \Qc }
     \label{TwoSidedR}
\end{eqnarray}
with $t_{\ua} = (t_{\da})^*$ and $r_{\ua} = (r_{\da})^*$.  Note
that the incident beam is almost perfectly transmitted:
$|t_{\sigma}|^2 = 1 - O(A \Qc )^2$.

Similar to the one-sided case, the two-sided cavity could be used
to make a measurement by monitoring the phase of the transmitted
wave. Using the expression for $t_{\sigma}$ above, we find that
the qubit-dependent transmission phase shift is given by:
\begin{equation}
     \tilde{\theta}_{\ua/\da} = \pm \tilde{\theta}_0
     = \pm 2 A \Qc
\end{equation}
where again the two signs correspond to the two different qubit
eigenstates.  The phase shift for transmission is only half as
large as in reflection so the Wigner delay time associated with
transmission is
\be
\ttwd = \frac{2}{\kappa}.
\ee
Upon making the substitution of $\ttwd$ for $\twd$, the one-sided
cavity 
Eqs.~(\ref{eq:SIxxcavity}) and (\ref{eq:1sidedcavityforcenoise})
remain valid.  However the internal
cavity photon number shot noise remains fixed so that
Eq.~(\ref{eq:1sidedcavitySnn}) becomes
\be
\Snn = 2{\bar n}\ttwd.
\label{eq:2sidedcavitySnn}
\ee
which means that
\be
\Snn = 2\Ndotbar \ttwd^2 = 2\SNN \ttwd^2
\ee
and
\be
\SFFz = 2\hbar^2 A^2 \omegar^2 \ttwd^2\SNN.
\label{eq:2sidedSFFcavity}
\ee
As a result the backaction dephasing doubles relative to the
measurement rate and we have
\be
\frac{\Gamma_{\rm meas}}{\Gamma_\varphi}= 2\SNN\Sthetatheta =
\frac{1}{2}.
\label{eq:2sidedcavityreachesquantumlimit}
\ee
Thus the two-sided cavity fails to reach the quantum limit by a
factor of 2.

  Using
the entanglement picture, we may again alternatively calculate the
amount of dephasing from the overlap between the detector states
corresponding to the qubit states $\ua$ and $\da$ (cf.\
Eq.~(\ref{RhoSup2})). We find:
\begin{eqnarray}
     e^{-\Gamma_{\varphi} t} & = &
         \Big| \langle t_{\ua} \alpha | t_{\da} \alpha \rangle
          \langle r_{\ua} \alpha | r_{\da} \alpha \rangle \Big|
         \\
     & = &
         \exp\left[- |\alpha|^2
             \left(1 - (t_{\ua})^* t_{\da} - (r_{\ua})^* r_{\da}
             \right) \right]
         \label{TwoSidedOverlap}
\end{eqnarray}
Note that {\it both} the change in the transmission and reflection
amplitudes contribute to the dephasing of the qubit.  Using the
expressions above, we find:
\begin{equation}
     \Gamma_{\varphi}t  =
              4 (\tilde{\theta}_0)^2 |\alpha|^2
         =
             4 (\tilde{\theta}_0)^2 \bar N
         =
             4 (\tilde{\theta}_0)^2 \Ndotbar t
         = 2 \Gamma_{\rm meas}t.
\end{equation}
Thus, in agreement with the quantum noise result, the two-sided
cavity misses the quantum limit by a factor of two.

Why does the two-sided cavity fail to reach the quantum limit? The
answer is clear from Eq.~(\ref{TwoSidedOverlap}):  even though we
are not monitoring it, there is information on the state of the
qubit available in the phase of the reflected wave.  
Note from
Eq.~(\ref{TwoSidedR}) that the magnitude of the reflected wave is
weak ($\propto A^2$), but (unlike the transmitted wave) the
difference in the reflection phase associated with the two qubit
states is large ($\pm \pi/2$).   The `missing information' in the
reflected beam makes a direct contribution to the dephasing rate
(i.e.~the second term in Eq.~(\ref{TwoSidedOverlap})), making it
larger than the measurement rate associated with measurement of
the transmission phase shift. In fact, there is an equal amount of
information in the reflected beam as in the transmitted beam, so
the dephasing rate is doubled.  We thus have a concrete example of 
the general principle connecting a failure to reach the quantum limit to the presence of
`wasted information'.  Note that the application of this principle
to generalized quantum point contact detectors is found in \textcite{Clerk03}.

Returning to our cavity detector, we note in closing 
that it is often technically easier to work with the transmission 
of light through a two-sided cavity, rather than reflection from a one-sided cavity.  
One can still reach the quantum limit in the two-sided cavity case if on uses an 
asymmetric cavity in which the input mirror has much less
transmission than the output mirror.  Most photons are reflected
at the input, but those that enter the cavity will almost
certainly be transmitted.  The price to be paid is that the input
carrier power must be increased.


\subsubsection{Measurement of oscillator position using a resonant
cavity}
\label{subsec:measurementofoscillatorposition}


The qubit measurement discussed in the previous subsection was an example of a QND measurement: the back-action did not affect the observable being measured.  We now consider the simplest example of a non-QND measurement, namely the weak continuous measurement of the position of a harmonic oscillator.  The detector will again be a parametrically-coupled
resonant cavity, where the position of the oscillator $x$ changes the frequency of the
cavity as per Eq.~(\ref{CavityHam1}) (see, e.g., \textcite{Tittonen1999}).  Similar to the qubit case, for a sufficiently weak coupling the phase shift of the
reflected beam from the cavity will depend linearly on the position $x$ of the oscillator (cf.~Eq.~(\ref{eq:cavityphaseshift})); by reading out this phase, we may thus measure
$x$.  The origin of 
backaction noise is the same as before, namely photon shot noise
in the cavity.  Now however this represents a random force which
changes the momentum of the oscillator.  During the subsequent time
evolution these random force perturbations will reappear as random
fluctuations in the position. Thus the measurement is \emph{not}
QND.  This  will mean that the minimum uncertainty of even an
ideal measurement is larger (by exactly a factor of 2) than the
`true' quantum uncertainty of the position (i.e.~the ground state uncertainty).  This is known
as the standard quantum limit on weak continuous position detection.  It is also
an example of a general principle 
that a linear
`phase-preserving' amplifier necessarily adds noise, and that the
minimum added noise exactly doubles the output noise for the case
where the input is vacuum (i.e.~zero-point) noise.  
A more general discussion of the quantum limit on amplifiers and position detectors 
will be presented in Sec.~\ref{sec:QLAmplifiers}.


We start by emphasizing that we are speaking here of a \emph{weak} continuous measurement of
the oscillator position.  
The measurement is sufficiently weak
that the position undergoes many cycles of oscillation before
significant information is acquired.  Thus we are not talking
about the instantaneous position but rather the overall amplitude
and phase, or more precisely the two quadrature amplitudes
describing the smooth envelope of the motion,
\be
{\hat x}(t)= \hat{X}(t)\cos(\Omega t) + \hat{Y}(t)\sin(\Omega t).
\label{eq:twoquadraturesdefined}
\ee
One can easily show that for an oscillator, the two
quadrature amplitudes $\hat X$ and $\hat Y$ are canonically
conjugate and hence do not commute with each other
\be
	[\hat X,\hat Y] = \frac{i\hbar}{M\Omega}=2i\xrms^2.
	\label{eq:QuadsConjugate}
\ee
As the measurement is both weak and continuous, it will yield information on both 
$\hat X$ and $\hat Y$.  As such, one is effectively trying to simultaneously measure two incompatible observables.  This basic fact is intimately related to the property mentioned above, that even a completely ideal weak continuous position measurement will have a total uncertainty which is twice the zero-point uncertainty.


We are now ready to start our heuristic analysis of position
detection using a cavity detector; relevant 
calculational details presented in Appendix
\ref{subapp:CavityPositionDetector}. Consider first the
mechanical oscillator we wish to measure. We take it to be a
simple harmonic oscillator of natural frequency $\Omega$ and
mechanical damping rate $\gammam$.
 For weak damping, and at zero coupling to the detector, the spectral density of the oscillator's
 position fluctuations is given by Eq.~(\ref{eq:Sxxquantum}) with the delta function replaced by a
 Lorentzian\footnote{This form is valid only for weak damping because we are assuming that the
 oscillator frequency is still sharply defined.  We have evaluated the Bose-Einstein factor exactly
 at frequency $\Omega$ and we have assumed that the Lorentzian centered at positive (negative) frequency
 has negligible weight at negative (positive) frequencies.}
\begin{eqnarray}
&&\Sxx[\omega] =  x_{\rm ZPF}^2 \bigg\{\nb(\hbar\Omega)
\frac{\gammam}{(\omega+\Omega)^2+(\gammam/2)^2}
\nonumber\\
&+&
[\nb(\hbar\Omega)+1]\frac{\gammam}{(\omega-\Omega)^2+(\gammam/2)^2}\bigg\}.
\label{eq:Sxxquantumdamped}
\end{eqnarray}

\begin{figure}[t]
\begin{center}
\includegraphics[width=3.45in]{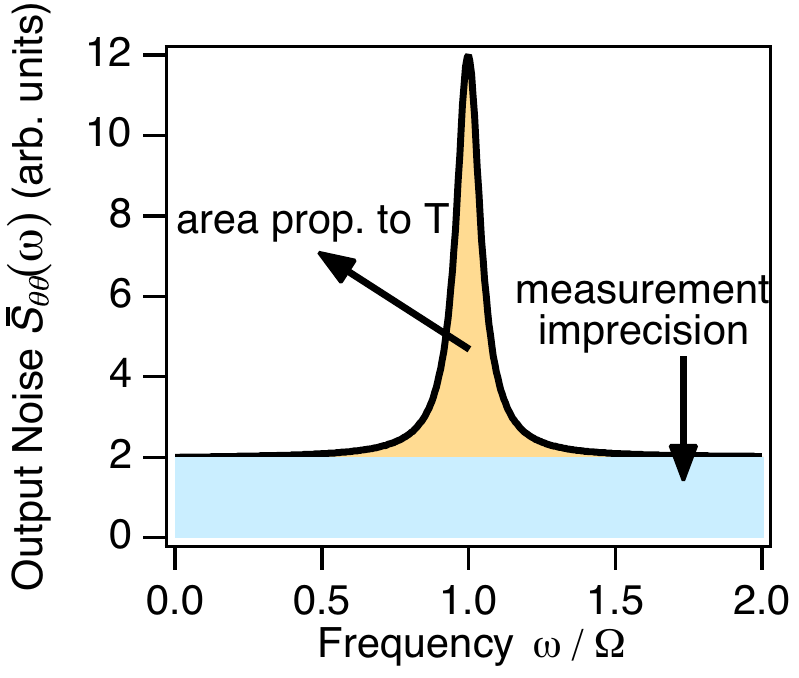}
\caption{(Color online)
Spectral density of the symmetrized output noise $\bS_{\theta \theta}[\omega]$
of a linear position detector.  The oscillator's noise appears as a Lorentzian sitting
above a noise floor (i.e. the measurement imprecision).  As discussed in the text,
the width of the peak is proportional to the oscillator damping rate $\gammam$, while
the area under the peak is proportional to temperature.  This latter fact can be used
to calibrate the response of the detector.}
\label{fig:oscclassicalspecdensity}
\end{center}
\end{figure}

When we now weakly couple the oscillator to the cavity (as per
Eq.~(\ref{CavityHam1}), with $\hz = \hx / \xrms$) and drive the
cavity on resonance, the phase shift $\theta$ of the reflected
beam will be proportional to $x$ (i.e. $\delta \theta(t) = [d
\theta / dx] \cdot x(t)$).  As such, the oscillator's position
fluctuations will cause additional fluctuations of the phase
$\theta$, over and above the intrinsic shot-noise induced phase
fluctuations $\Sthetatheta$. We consider the usual case where
the noise spectrometer being used to measure the noise in
$\theta$ (i.e.~the noise in the homodyne current) measures the
symmetric-in-frequency noise spectral density; as such, it is
the symmetric-in-frequency position noise that we will detect.
In the classical limit $\kbt \gg \hbar\Omega$, this is just
given by: \begin{eqnarray} \Sxxsym[\omega] &\equiv& \frac{1}{2}
\left( 	\Sxx[\omega] + \Sxx[-\omega] \right)\nonumber\\
&\approx&\frac{\kbt}{2 M\Omega^2}
\frac{\gammam}{(|\omega|-\Omega)^2+(\gammam/2)^2}
\end{eqnarray} 
If we ignore back-action
effects, we expect to see this Lorentzian profile riding on top
of the background imprecision noise floor; this is illustrated
in Fig.~\ref{fig:oscclassicalspecdensity}.

Note that additional stages of amplification
would also add noise, and would thus further
augment this background noise floor.  If we
subtract off this noise floor, the FWHM of the
curve will give the damping parameter
$\gammam$, and the area under the experimental
curve \be \int_{-\infty}^\infty
\frac{d\omega}{2\pi}\Sxxsym[\omega] =
\frac{\kbt}{M\Omega^2} \ee measures the
temperature.  What the experimentalist actually
plots in making such a curve is the output of
the entire detector-plus-following-amplifier
chain.   Importantly, if the temperature is
known, then the area of the measured curve can
be used to calibrate the coupling of the
detector and the gain of the total overall
amplifier chain (see, e.g.,
\onlinecite{LaHaye04, FlowersJacobs07}). One
can thus make a calibrated plot where the measured output noise
is referred back
to the oscillator position. 

Consider now the case where the oscillator is at zero temperature.
Eq.~(\ref{eq:Sxxquantumdamped}) then yields
 for
the symmetrized noise spectral density
\be
\Sxxsym^0[\omega] = \xrms^2
\frac{\gammam/2}{(|\omega|-\Omega)^2+(\gammam/2)^2}.
\label{eq:zeroTsymmspecden}
\ee
One might expect that one could see this Lorentzian directly in the output
noise of the detector (i.e.~the $\theta$ noise), sitting above the measurement-imprecision
noise floor.  However, this neglects the effects of measurement backaction.
From the classical equation of motion
we expect the response of the oscillator to the backaction
force $F=F_z / \xrms$ (cf.~Eq.~(\ref{eq:FzDefn})) at frequency $\omega$ to produce an additional displacement
$\delta x[\omega] = \oscchi[\omega] F[\omega]$,
where $\oscchi[\omega]$ is the mechanical susceptibility
\be
\oscchi[\omega] \equiv \frac{1}{M}\frac{1}{\Omega^2-\omega^2
-i\gammam \omega}.
\ee
These extra oscillator fluctuations will show up as additional fluctuations in the output of
the detector. For simplicity, let
us focus on this noise at the oscillator's resonance frequency $\Omega$.
As a result of the detector's back-action, the total measured position noise
(i.e.~inferred spectral density) at the  frequency $\Omega$ is given by:
\begin{eqnarray}
	\bSxtot[\Omega] &=&
	 	\Sxxsym^0[\Omega] +
	 	\frac{|\oscchi[\Omega]|^2}{2}
	 		\left[ \SFF[+\Omega]+\SFF[-\Omega] \right ]\nonumber\\
 	&+&
		\frac{1}{2} \left[\SIxx[+\Omega] + \SIxx[-\Omega] \right]
 		\label{eq:SxxHeuristic} \\
 & = & \Sxxsym^0[\Omega] + \bSxadd[\Omega]
 	\label{eq:SxxHeuristic2}
\end{eqnarray}
The first term here is just the intrinsic zero-point noise of the oscillator:
\be
\Sxxsym^0[\Omega]=\frac{2\xrms^2}{\gammam} = \hbar |\oscchi[\Omega]|.
	\label{eq:SxxT}
\ee
The second term $\bSxadd$ is the total noise added by the measurement,
and includes {\it both} the measurement imprecision
$\SIxx \equiv \SIzz \xrms^2$ and the extra fluctuations
caused by the backaction.  We stress that $\bSxtot$ corresponds to a position
noise spectral density {\it inferred} from the output of the detector:  one simply
scales the spectral density of total output fluctuations
$\bS_{\theta \theta,{\rm tot}}[\Omega]$ by $( d\theta / dx )^2$.

Implicit in Eq.~(\ref{eq:SxxT}) is the assumption that the back
action noise and the imprecision noise are uncorrelated and
thus add in quadrature.  It is not obvious that this is
correct, since in the cavity detector
the backaction noise and
output shot noise are both caused by the vacuum noise in the
beam incident on the cavity.
It turns out there are
indeed correlations, however the \emph{symmetrized} (i.e.\ `classical') correlator $\bS_{\theta F}$ 
does vanish for our choice of a resonant cavity drive.
Further, Eq.~(\ref{eq:SxxHeuristic})
 assumes that the measurement does not change the damping rate of the oscillator.
 Again, while this will not be true for an arbitrary detector, it is the case
 here for the cavity detector when (as we have assumed) it is driven on resonance.
Details justifying both these statements are given in Appendix \ref{app:drivencavity};
the more general case with non-zero noise correlations and back-action damping is discussed in 
Sec.~\ref{subsec:PositionDetector}.

\begin{figure} \begin{center}
\includegraphics[width=3.45in]{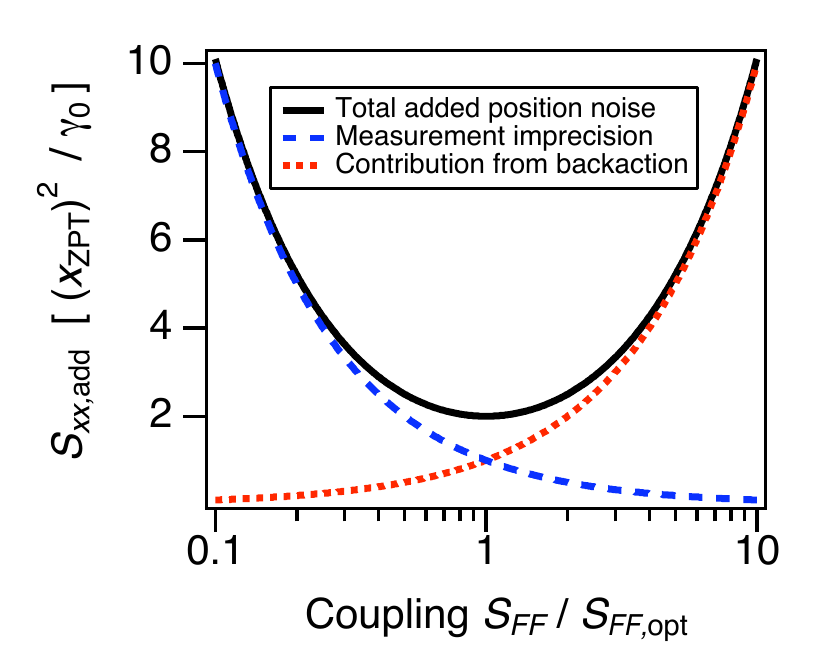} 
\caption{(Color online) Noise
power of the added position noise of a linear position
detector, evaluated at the oscillator's resonance frequency
($\bSxadd[\Omega]$), as a function of the magnitude of the
back-action noise spectral density $S_{FF}$. $S_{FF}$ is
proportional to the oscillator-detector coupling, and in the
case of the cavity detector, is also proportional to the power
incident on the cavity. The optimal value of $S_{FF}$ is given
by $S_{FF,{\rm opt}} = \hbar M \Omega \gammam /2$ (cf.
Eq.~(\ref{eq:optimalbackaction})). We have assumed that there
are no correlations between measurement imprecision noise and
back-action noise, as is appropriate for the cavity detector.}
\label{fig:optimalcoupling} \end{center}
\end{figure}

Assuming we have a quantum-limited detector that obeys
Eq.~(\ref{eq:SFFSIxxproduct}) (i.e.~$\SIxx \SFF = \hbar^2/4$)
and that the shot noise is symmetric in frequency, the added
position noise spectral density at resonance (i.e.~second term
in Eq.~(\ref{eq:SxxHeuristic2})) becomes:
\be 	
\bSxadd[\Omega] =
\left[|\chi[\Omega]|^2\SFF+\frac{\hbar^2}{4}\frac{1}{\SFF}\right].
\ee
Recall from Eq.~(\ref{eq:SFFcavity}) that the backaction
noise is proportional to the coupling of the oscillator to the
detector and to the intensity of the drive on the cavity. The
added position uncertainty noise is plotted in
Fig.~\ref{fig:optimalcoupling} as a function of $\SFF$.  We see
that for high drive intensity, the backaction noise dominates
the position uncertainty, while for low drive intensity, the
output shot noise (the last term in the equation above)
dominates.

 The added noise (and hence the total noise $\bSxtot[\Omega]$) is minimized when the drive intensity is tuned so that $\SFF$ is equal to $S_{FF,{\rm opt}}$, with:
\be
S_{FF,{\rm opt}} = \frac{\hbar}{2|\oscchi[\Omega]|} =
\frac{\hbar}{2}M\Omega\gammam.
\label{eq:optimalbackaction}
\ee
The more heavily damped is the oscillator, the less
susceptible it is to backaction noise and hence the higher is the
optimal coupling. At the optimal coupling strength, the measurement imprecision
noise and back-action noise each make equal contributions to the added noise, yielding:
\be
	\bSxadd[\Omega] =
	\frac{\hbar}{M\Omega\gammam}= \Sxxsym^0[\Omega]
	\label{eq:SxxaddQL}
\ee
Thus, the spectral density of the added position noise is {\it
exactly} equal to the noise power associated with the
oscillator's zero-point fluctuations.  This represents a {\it
minimum} value for the added noise of any linear position
detector, and is referred to as the standard quantum limit on
position detection. 
Note that this limit only involves
the added noise of the detector, and thus has nothing to do
with the initial temperature of the oscillator.  

We emphasize that to reach the above quantum
limit on weak continuous position detection, one needs the detector itself to be quantum limited,
i.e.~the product $\SFF \SIxx$ must be as small as is allowed by quantum mechanics, namely
$\hbar^2/4$.  Having a quantum-limited detector however is not enough:  in addition,
one must be able to achieve sufficiently strong coupling to reach
the optimum given in Eq.~(\ref{eq:optimalbackaction}). Further, 
the measured output noise
must be dominated by the output noise of the cavity, not by the
added noise  of following amplifier stages. 

A related, stronger quantum limit refers to the {\it total} inferred
position noise from the measurement, $\bSxtot[\omega]$.  It
follows from Eqs.~(\ref{eq:SxxaddQL}),(\ref{eq:SxxHeuristic2}) that at resonance, the
smallest this can be is {\it twice} the oscillator's zero point
noise:
\be
\bSxtot[\Omega]= 2\Sxxsym^0[\Omega].
\label{eq:twicevacnoise}
\ee
Half the
noise here is from the oscillator itself, half is from the
added noise of the detector.  Reaching this quantum limit is
even more challenging:  one needs both to reach the quantum
limit on the added noise {\it and} cool the oscillator to its
ground state.  

\begin{figure}[t]
\begin{center}
\includegraphics[width=3.45in]{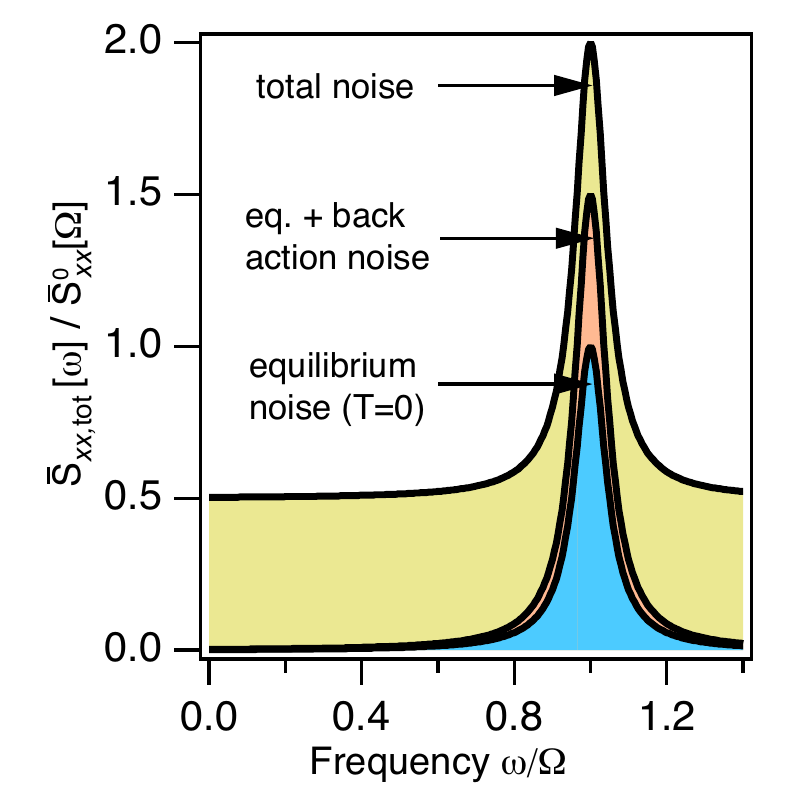}
\caption{(Color online) Spectral density of measured position fluctuations of a harmonic oscillator,
$\bSxtot[\omega]$, as a function of frequency $\omega$, for a detector which reaches the quantum limit at the oscillator frequency $\Omega$.  We have assumed that without the coupling to the detector, the oscillator would be in its ground state.  The y-axis has been normalized by the zero-point position-noise spectral density $\bS^0_{xx}[\omega]$, evaluated at  $\omega = \Omega$.  One clearly sees that the total noise at $\Omega$ is twice the zero-point value, and that the peak of the Lorentzian rises a factor
of three above the background.  This background represents the measurement-imprecision, and is equal to $1/2$ of $\bS^0_{xx}(\Omega)$.}
\label{fig:zeroTSHOlineshape}
\end{center}
\end{figure}

Finally, we emphasize that the optimal value of the coupling
derived above was specific to the choice of minimizing the
total position noise power at the resonance frequency.  If a
different frequency had been chosen, the optimal coupling would
have been different; one again finds that the minimum possible
added noise corresponds to the ground state noise at that
frequency.  It is interesting to ask what the total position noise
would be as a function of frequency, assuming that the coupling
has been optimized to minimize the noise at the resonance
frequency, and that the oscillator is initially in the ground
state.  From our results above we have \begin{eqnarray}
&&\bSxtot[\omega]\nonumber\\ &=&
\xrms^2\frac{\gammam/2}{(|\omega|-\Omega)^2+(\gammam/2)^2}+
\frac{\hbar}{2} \left[
\frac{|\oscchi[\omega]|^2}{|\oscchi[\Omega]|} +
|\oscchi[\Omega]| \right]\nonumber\\ &\approx&
\frac{\xrms^2}{\gammam}\left\{
1+3\frac{(\gammam/2)^2}{(|\omega|-\Omega)^2+(\gammam/2)^2}
\right\} \end{eqnarray} which is plotted in
Fig.~\ref{fig:zeroTSHOlineshape}.  Assuming that the detector
is quantum limited, one sees that the Lorentzian peak rises
above the constant background by a factor of three when the
coupling is optimized to minimize the total noise power at
resonance.  This represents the best one can do when continuously monitoring
zero-point position fluctuations.  
Note that the value of this peak-to-floor ratio
is a direct consequence of two simple facts which hold for an optimal coupling, at the quantum limit: i) the total added noise at resonance (back-action plus measurement imprecision) is equal to the zero-point noise, and ii) back-action and measurement imprecision make {\it equal} contributions to the total added noise.  Somewhat surprisingly, the {\it same} maximum peak-to-floor ratio is obtained when one tries to continuously monitor coherent qubit oscillations with 
a linear detector which is transversely coupled to the qubit \cite{Korotkov01c}; this is also a
non-QND situation.  Finally,
if one only wants to detect the noise peak (as
opposed to making a continuous quantum-limited measurement), one could use two 
independent detectors coupled to the oscillator and look at the cross-corelation between the
two output noises:  in this case, there need not be any noise floor \cite{Jordan05b, Doiron07}.


In Table \ref{table:QLExperiments}, 
we give a summary of recent experiments which approach the quantum limit on 
weak, continuous position detection of a mechanical resonator.  Note that in many of these experiments, 
the effects of detector back-action were not seen.  This could either be the result of too low of a detector-oscillator coupling, or due to the presence of excessive thermal noise.  As we have shown, the back-action force noise serves to slightly heat the oscillator.  If it is already at an elevated temperature due to thermal noise, this additional heating can be very hard to resolve.

In closing, we stress that this subsection has given only a
very rudimentary introduction to the quantum limit on position
detection.  A complete discussion which treats the important
topics of back-action damping, effective temperature, noise
cross-correlation and power gain is given in
Sec.~\ref{subsec:PositionDetector}.


\begin{table*}[t]
	\caption{\label{table:QLExperiments}Synopsis of recent experiments 
	approaching the quantum limit on continuous position detection of a mechanical resonator.
	The second column corresponds to the best measurement imprecision
	noise spectral density $\bS^I_{xx}$ achieved in the experiment.  This value is 
	compared against the zero-point position noise spectral density
	$\bar{S}_{xx}^0$, calculated using the total measured resonator damping
	(which may include a back-action contribution).  
	All spectral densities are at
	the oscillator's resonance frequency $\Omega$.  
	As discussed in the text, there is no quantum
	limit on how small one can make $\bS^I_{xx}$; for an ideal detector,
	one needs to tune the detector-resonator coupling so that
	$\bS^I_{xx} = \bar{S}_{xx}^0/2$ in order to reach the quantum limit on position detection.  
	The third column presents the product of the measured imprecision noise 
	(unless otherwise noted) and measured back-action noises, divided
	by $\hbar/2$; this quantity must be one to achieve the quantum limit on the added noise.} 
	\begin{tabular}{ |c|c|c|c|}
\hline
	Experiment &
	Mechanical &
	Imprecision noise & 
	Detector Noise Product
	\footnote{A blank value in this column indicates that back-action was not measured in the experiment.}
\\
	&
	frequency [Hz] &
	vs. zero-point noise &
	$\sqrt{ \bS^I_{xx} \bS_{FF} } / (\hbar/2)$
\\
	&	
	$ \Omega / (2 \pi) $ &
	$ \sqrt{ \bS^I_{xx} / \Sxxsym^0[\Omega] } $ &
\\
\hline

\textcite{Cleland02} (quantum point contact) &
	$1.5 \times 10^6$ &
	$4.2 \times 10^4$  &
\\
\hline

\textcite{Knobel03} (single-eletron transistor) &
	$1.2 \times 10^8$ &
	$1.8 \times 10^2$ &
\\
\hline

\textcite{LaHaye04} (single-electron transistor) &
	$ 2.0 \times 10^7$ &
	$5.4$ &
\\
\hline

\textcite{Naik06} (single-electron transistor) &
	$ 2.2 \times 10^7$ &
	$5.3$
	\footnote{Note that back-action effects dominated the mechanical $Q$ in this
	measurement, lowering it from  $1.2 \times 10^5$ to $\sim 4.2 \times 10^2$.
	If one compares the imprecision against the zero-point noise of the uncoupled 
	mechanical resonator, one finds  
	$\sqrt{ \bS^I_{xx} / \Sxxsym^0[\Omega] } \sim 0.33 $.}  &
	$8.1 \times 10^2$ 
\\
 (if $\bS^I_{xx}$ had been limited by SET shot noise) &
	&
	&
	$1.5 \times 10^1$
\\
\hline

\textcite{Arcizet2006} (optical cavity) &
	$ 8.1 \times 10^5$ &
	$0.87$ &
\\
\hline

\textcite{FlowersJacobs07} (atomic point-contact) &
	$ 4.3 \times 10^7$ &
	$29$ &
	$1.7 \times 10^3$ 
\\
\hline

\textcite{Regal08} (microwave cavity) &
	$ 2.4 \times 10^5$ &
	$ 21 $ &
\\
\hline

\textcite{Kippenberg07} (optical cavity) &
	$ 4.1 \times 10^7$ &
	$0.50$  &
\\
\hline

\textcite{Poggio08} (quantum point contact) &
	$ 5.2 \times 10^3$ &
	$ 63 $ &
\\
\hline

\textcite{Etaki08} (d.c. SQUID) &
	$ 2.0 \times 10^6$ &
	$ 47 $ &
\\
\hline

\textcite{Aspelmeyer09} (optical cavity) &
	$ 9.5 \times 10^2$ &
	$0.57$  &
\\
\hline


\textcite{Kippenberg09} (optical cavity) &
	$ 6.5 \times 10^7$ &
	$5.5$  &
	$1.0 \times 10^2$
\\
\hline

\textcite{Lehnert09} (microwave cavity) &
	$ 1.0 \times 10^6$ &
	$ 0.63 $  &  
\\

\hline
\end{tabular}	
\end{table*}



\section{General Linear Response Theory}
\label{sec:GenLinResponseTheory}

\subsection{Quantum constraints on noise}
\label{subsec:NoiseConstraint}

In this section, we will further develop the connection
between quantum limits and noise discussed in previous sections, focusing now on a more general approach.  As before,
we will emphasize the idea that reaching the quantum limit requires a detector having ``quantum-ideal" noise properties.  The approach here is different from typical treatments in the
quantum optics literature \cite{Haus00, Gardiner00}, and uses
nothing more than features of quantum linear response.  Our
discussion here will expand upon \textcite{Clerk03, Clerk04c};
somewhat similar approaches to quantum measurement are also
discussed in \textcite{Braginsky92} and \textcite{Averin03}.

In this subsection, we will start by heuristically sketching how
constraints on noise
(similar to Eq.~(\ref{eq:SFFSIxxproduct}) for the cavity detector) can
emerge directly from the Heisenberg uncertainty
principle.  We then present a rigorous and general quantum constraint
on noise.  We introduce both the notion of a generic
linear response detector, and the basic quantum constraint on
detector noise.  In the next subsection (C), we will discuss how this noise constraint
leads to the quantum limit on QND state detection of a qubit.  The quantum limit on a linear amplifier (or a position detector) is discussed in detail in the
next section.

\subsubsection{Heuristic weak-measurement noise constraints}

As we already stressed in the introduction, there is no fundamental
quantum limit on the accuracy with which a given observable can be
measured, at least not within the framework of non-relativistic
quantum mechanics. For example, one can, in principle, measure the
position of a particle to arbitrary accuracy in the course of a
projection measurement.
However, the situation is different when we specialize
to continuous, non-QND measurements. Such a measurement can be
envisaged as a series of instantaneous measurements, in the limit
where the spacing between the measurements $\delta t$ is taken to
zero. Each measurement in the series has a limited resolution and
perturbs the conjugate variables, thereby affecting the subsequent
dynamics and measurement results. Let us discuss this briefly for
the example of a series of position measurements of a free
particle.

After initially measuring the position with an accuracy $\Delta
x$, the momentum suffers a random perturbation of size $\Delta
p\geq\hbar/(2\Delta x)$. Consequently, a second position
measurement taking place a time $\delta t$ later will have an
additional uncertainty of size $\delta t (\Delta p /m) \sim \hbar
\delta t /(m \Delta x)$. Thus, when trying to obtain a good
estimate of the position by averaging several such measurements,
it is not optimal to make $\Delta x$ too small, because otherwise
this additional perturbation, called the {}``back-action'' of the
measurement device, will become large. The back-action can be
described as a random force $\Delta F=\Delta p/ \delta t$. A
meaningful limit $\delta t\rightarrow0$ is obtained by keeping
both $\Delta x^{2} \delta t \equiv \bar{S}_{xx}$ and $\Delta p^{2}/
\delta t \equiv \bar{S}_{FF}$ fixed. In this limit, the deviations
$\delta x(t)$ describing the finite measurement accuracy and the
fluctuations of the back-action force $F$ can be described as
white noise processes, $\left\langle \delta x(t)\delta
x(0)\right\rangle =\bar{S}_{xx}\delta(t)$ and $\left\langle
F(t)F(0)\right\rangle =\bar{S}_{FF}\delta(t)$. The Heisenberg
uncertainty relation $\Delta p\Delta x\geq\hbar/2$ then implies
$\bar{S}_{xx}\bar{S}_{FF}\geq\hbar^{2}/4$ \cite{Braginsky92}.
Note this is completely analogous to the relation Eq.~(\ref{eq:SFFSIxxproduct})
we derived for the resonant cavity detector using the fundamental number-phase
uncertainty relation.  In this section, we will derive rigorously more general
quantum limit relations on noise spectral densities of this form.

\subsubsection{Generic linear-response detector}
\label{subsubsec:GenericLinear}

\begin{figure}[t]
\begin{center}
\includegraphics[width=3.45in]{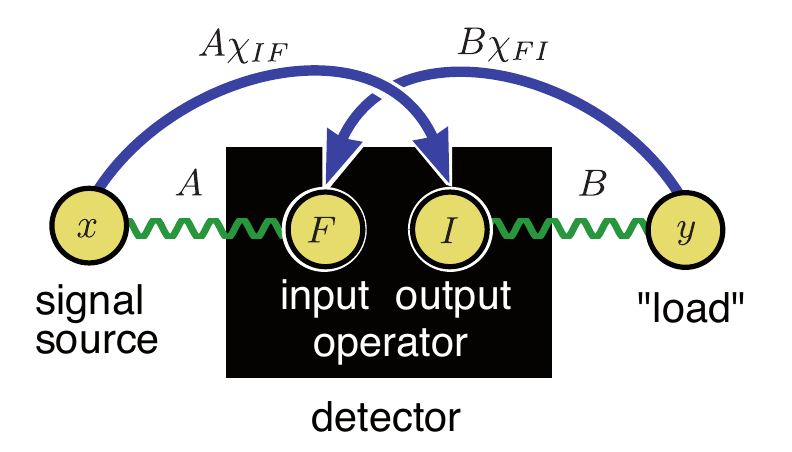}
\caption{(Color online) Schematic of a generic linear response detector.}
\label{fig:generaldetector}
\end{center}
\end{figure}

To rigorously discuss the quantum limit, we would like to start
with a description of a detector which is as general as possible.
To that end, we will think of a detector as some physical system
(described by some unspecified Hamiltonian ${\hat H}_{det}$ and
some unspecified  density matrix ${\hat \rho}_0$) which is
time-independent in the absence of coupling to the signal source.
The detector has both an input port, characterized by an operator
$\hat{F}$, and an output port, characterized by an operator
$\hat{I}$ (see Fig.~\ref{fig:generaldetector}). The output
operator $\hat{I}$ is simply the quantity which is read-out at the
output of the detector (e.g., the current in a single-electron
transistor, or the phase shift in the cavity detector of the
previous section). The input operator $\hat{F}$ is the detector
quantity which directly couples to the input signal (e.g., the
qubit), and which causes a back-action disturbance of the signal
source; in the cavity example of the previous section, we had $\hF
= \hat{n}$, the cavity photon number. As we are interested in weak
couplings, we will assume a simple bilinear form for the
detector-signal interaction Hamiltonian:
\begin{equation}
     {\hat H}_{int} = A \hat{x} \hat{F}
     \label{HIntGen}
\end{equation}
Here, the operator $\hat{x}$ (which is not necessarily a position operator)
carries the input signal.  Note that because
$\hat x$ belongs to the signal source, it necessarily commutes
with the detector variables $\hat I, \hat F$.

We will always assume the coupling strength $A$ to be small enough that we
can accurately describe the output of the detector using linear
response.\footnote{The precise conditions for when linear response breaks down will
depend on specific details of the detector.  For example, in the cavity detector discussed in 
Sec.~\ref{subsec:CavityDetector}, 
one would need the dimensionless coupling $A$ to satisfy $A \ll 1/(\Qc \langle \hz \rangle)$ to ensure that the non-linear dependence of the phase shift $\theta$ on the  signal 
$\langle \hz \rangle$ is negligible.  This translates to the signal modulating the cavity frequency by an amount
much smaller than its linewidth $\kappa$.}
We thus have:
\begin{equation}
     \langle \hat{I}(t) \rangle =
         \langle \hat{I} \rangle_0 + A \int dt' \chiIF(t-t')
             \langle \hat{x}(t') \rangle,
             \label{LinResp}
\end{equation}
where $\langle \hat{I} \rangle_0$ is the input-independent value
of the detector output at zero-coupling, and $\chiIF(t)$ is the
linear-response susceptibility or gain of our detector.
Note that in \textcite{Clerk03} and \textcite{Clerk04c}, this gain coefficient
is denoted $\lambda$.
Using standard time-dependent perturbation theory in the coupling
$\hat{H}_{int}$, one can easily derive Eq.~(\ref{LinResp}), with
$\chiIF(t)$ given by a Kubo-like formula:
\begin{equation}
     \chiIF(t) = -\frac{i}{\hbar} \theta(t)
         \left\langle \left[ \hI(t), \hF(0) \right]
             \right\rangle_0
             \label{eq:gain}
\end{equation}
Here (and in what follows), the operators $\hI$ and $\hF$ are
Heisenberg operators with respect to the detector Hamiltonian, and
the subscript $0$ indicates an expectation value with respect to
the density matrix of the uncoupled detector.

As we have already discussed, there will be unavoidable noise in
both the input and output ports of our detector.  This noise is
subject to quantum mechanical constraints, and its presence is
what limits our ability to make a measurement or amplify a signal.
We thus need to quantitatively characterize the noise in both
these ports.  Recall from the discussion in
Sec.~\ref{subsec:QuantumSpectrumAnalyzers} that it is the symmetric-in-frequency
part of a quantum noise spectral density which plays a role akin
to classical noise.  We will thus want to characterize the {\it
symmetrized} noise correlators of our detector (denoted as always with a
bar).  Redefining these operators so that their average value is
zero at zero coupling (i.e.~$\hF \ra \hF - \langle \hF \rangle_0$,
$\hI \ra \hI - \langle \hI \rangle_0$), we have:
\begin{subequations}
\begin{eqnarray}
     \bS_{FF}[\omega] & \equiv & \frac{1}{2}
         \int_{-\infty}^{\infty} dt
                 e^{i \omega t}
         \langle \{ \hF(t),\hF(0) \} \rangle_0
     \label{SFSymm} \\
     \bS_{II}[\omega] & \equiv & \frac{1}{2}
         \int_{-\infty}^{\infty} dt
                 e^{i \omega t}
         \langle \{ \hI(t),\hI(0) \} \rangle_0
     \label{SISymm} \\
     \bS_{IF}[\omega] & \equiv & \frac{1}{2}
         \int_{-\infty}^{\infty} dt
                 e^{i \omega t}
         \langle \{ \hI(t),\hF(0) \} \rangle_0
     \label{SIFSymm}
\end{eqnarray}
\end{subequations}
where $\{,\}$ indicates the anti-commutator, $\bS_{II}$ represents
the intrinsic noise in the output of the detector, and $\bS_{FF}$
describes the back-action noise seen by the source of the input
signal.  In general, there will be some correlation between these
two kinds of noise; this is described by the cross-correlator
$\bS_{IF}$.


Finally, we must also allow for the possibility that our detector could operate
in reverse (i.e.~with input and output ports playing opposite roles).  
We thus introduce the reverse gain
$\chiFI$ of our detector.  This is the response coefficient
describing an experiment where we couple our input signal to the
{\it output} port of the detector (i.e.~$\hat{H}_{int} = A \hx
\hI$), and attempt to observe something at the {\it input} port
(i.e.~in $\langle \hat{F}(t) \rangle$).  In complete analogy to
Eq.~(\ref{LinResp}), one would then have:
\begin{equation}
     \langle \hat{F}(t) \rangle =
         \langle \hat{F} \rangle_0 + A \int dt' \chiFI(t-t')
             \langle \hat{x}(t') \rangle
             \label{RevLinResp}
\end{equation}
with:
\begin{equation}
     \chiFI(t) = -\frac{i}{\hbar} \theta(t)
         \langle \left[ \hF(t), \hI(0) \right] \rangle_0
         \label{ReverseGain}
\end{equation}
Note that if our detector is in a time-reversal symmetric, thermal equilibrium state, then Onsager reciprocity relations would imply
either $\chiIF = \chiFI^*$ (if $I$ and $F$ have the same parity under time-reversal) or $\chiIF = -\chiFI^*$ 
(if $I$ and $F$ have the opposite parity under time-reversal)  (see, e.g.~\textcite{Pathria}).  
Thus, if the detector is in equilibrium, the presence of gain necessarily implies the presence of reverse gain.  
Non-zero reverse gain is also found in many standard classical electrical amplifiers such as op amps \cite{Boylestad06}.

The reverse gain is something that we must worry about even if we are not interested in 
operating our detector in reverse.  To see why, note that 
to make a measurement of the output operator $\hI$, we must
necessarily couple to it in some manner.  If $\chiFI \neq 0$,
the noise associated with this coupling could in turn lead to
additional back-action noise in the operator $\hF$.  Even if the
reverse gain did nothing but amplify vacuum noise entering the
output port, this would heat up the system being measured at the
input port and hence produce excess backaction.
Thus, the ideal situation is to have $\chiFI=0$, implying a high
asymmetry between the input and output of the detector, and requiring
the detector to be in a state far from thermodynamic equilibrium.  
We note that almost all mesoscopic detectors that have been studied in
detail (e.g.~single electron transistors and generalized quantum
point contacts) have been found to have a vanishing reverse gain:
$\chiFI = 0$ \cite{Clerk03}.
For this reason, we will
often focus on the ideal (but experimentally relevant) situation where $\chiFI = 0$ in what
follows.

Before proceeding, it is worth emphasizing that there is a
relation between the detector gains $\chiIF$ and $\chiFI$ and
the {\it unsymmetrized} $I$-$F$ quantum noise correlator,
$S_{IF}[\omega]$. This spectral density, which need not be
symmetric in frequency, is defined as:
\begin{eqnarray}
     S_{I F}[\omega] = \int_{-\infty}^{\infty} dt
     e^{i \omega t} \langle \hat{I}(t) \hat{F}(0) \rangle_0
    \label{eq:SIFunsymm}
\end{eqnarray}
Using the definitions, one can easily show that:
\begin{subequations}
\begin{eqnarray}
     \bS_{I F}[\omega] & = & \frac{1}{2} \left[
             S_{I F}[\omega] + S_{I F}[-\omega]^{*}
     \right]
     \label{eq:SIFBar}  \\
                \chiIF[\omega] - \chiFI[\omega]^*
         &  = &
         -\frac{i}{\hbar} \left[
             S_{I F}[\omega] - S_{I F}[-\omega]^*
        \right]
     \label{eq:LambdaSIF}
\end{eqnarray}
\end{subequations}
Thus, while $\bS_{IF}$ represents the classical part of the
$I$-$F$ quantum noise spectral density, the gains $\chiIF,
\chiFI$ are determined by the quantum part of this spectral
density.  This also demonstrates that though the gains have an
explicit factor of $1/\hbar$ in their definitions, they have a
well defined $\hbar \ra 0$ limit, as the asymmetric-in-frequency
part of $S_{IF}[\omega]$ vanishes in this limit.

\subsubsection{Quantum constraint on noise}

Despite having said nothing about the detector's Hamiltonian or
state (except that it is time-independent), we can nonetheless
derive a very general {\it quantum} constraint on its noise
properties.  Note first that for purely classical noise spectral
densities, one always has the inequality:
\begin{eqnarray}
    \bS_{II}[\omega] \bS_{FF}[\omega] - \left| \bS_{IF}[\omega] \right|^2  & \geq & 0
\end{eqnarray}
This simply expresses the fact that the correlation between two
different noisy quantities cannot be arbitrarily large; it follows immediately from
the Schwartz inequality.  In the
quantum case, this simple constraint becomes modified whenever there is an asymmetry
between the detector's gain and reverse gain.  This asymmetry is parameterized by the quantity
$\tchiIF[\omega]$:
\begin{eqnarray}
    \tchiIF[\omega]
        & \equiv & \chiIF[\omega] - \left( \chiFI[\omega] \right)^* 
\end{eqnarray}
We will show below that the following quantum noise inequality
(involving {\it symmetrized} noise correlators) is always valid (see also Eq. (6.36) in \textcite{Braginsky96}): 
\begin{eqnarray}
    \bS_{II}[\omega] \bS_{FF}[\omega] - \left| \bS_{IF}[\omega] \right|^2  & \geq & \nonumber \\
        \left| \frac{\hbar \tchiIF[\omega]}{2} \right|^2
        \left( 1 +
            \Delta\left[ \frac{\bS_{IF}[\omega]}{\hbar
                \tchiIF[\omega]/2} \right] \right) & &  \label{NoiseConstraint}  
\end{eqnarray}
where 
\begin{eqnarray}
    \Delta[z] & = &
        \frac{
            \left| 1 + z^2 \right| - \left(1 + |z|^2 \right)
        }{2},
    \label{eq:DeltaFactor}
\end{eqnarray}

To interpret the quantum noise inequality Eq.~(\ref{NoiseConstraint}), note that
$1 + \Delta[z] \geq 0$.  
Eq.~(\ref{NoiseConstraint}) thus implies that if our detector has gain and does not have a perfect symmetry between input and output (i.e. $\chiIF \neq \chiFI^*$), then it must in general have a minimum amount of back-action and output noise; moreover, these two noises cannot be perfectly anti-correlated.   As we will show in the following sections, this constraint on the noise of a detector directly leads to quantum limits on various different measurement tasks.  Note that in the zero-frequency limit, $\tchiIF$ and $\bS_{IF}$ are both real, implying that the term involving $\Delta$ in Eq.~(\ref{NoiseConstraint}) vanishes.  The result is a simpler looking inequality found elsewhere in the literature \cite{Clerk03, Averin03, Clerk04c}.  


While Eq.~(\ref{NoiseConstraint}) may appear vaguely reminiscent to the
standard fluctuation dissipation theorem, its origin is quite
different: in particular, the quantum noise constraint applies
irrespective of whether the detector is in equilibrium.
Eq.~(\ref{NoiseConstraint}) instead follows directly from
Heisenberg's uncertainty relation applied to the frequency
representation of the operators ${\hat I}$ and ${\hat F}$. In its
most general form, the Heisenberg uncertainty relation gives a
lower bound for the uncertainties of two observables in terms of
their commutator and their noise correlator \cite{1966_Gottfried_QuantumMechanics}: 
\begin{equation}
(\Delta A)^{2}(\Delta B)^{2}\geq\frac{1}{4}\left\langle \left\{
\hat{A},\hat{B}\right\} \right\rangle
^{2}+\frac{1}{4}\left|\left\langle
\left[\hat{A},\hat{B}\right]\right\rangle
\right|^{2}\,.\label{eq:Heisenberg}\end{equation} Here we have
assumed $\langle \hat{A} \rangle = \langle \hat{B} \rangle =0$.
Let us now choose the Hermitian operators $\hat{A}$ and $\hat{B}$
to be given by the cosine-transforms of $\hat{I}$ and $\hat{F}$,
respectively, over a finite time-interval $T$:
\begin{subequations}
\begin{eqnarray}
    \hat{A} & \equiv & \sqrt{\frac{2}{T}}\int_{-T/2}^{T/2}dt\,\cos(\omega t + \delta)\,\hat{I}(t)\\
    \hat{B} & \equiv & \sqrt{\frac{2}{T}}\int_{-T/2}^{T/2}dt\,\cos(\omega t)\,\hat{F}(t)
\end{eqnarray}
\end{subequations}
Note that we have phase shifted the transform of $\hI$ relative to
that of $\hF$ by a phase $\delta$. In the limit
$T\rightarrow\infty$ we find, at any finite frequency
$\omega\neq0$:
\begin{subequations}
\begin{eqnarray}
    (\Delta A)^{2} & = & \bar{S}_{II}[\omega],\,(\Delta B)^{2}=\bar{S}_{FF}[\omega]
        \label{eq:HeisNoise1} \\
    \left\langle \left\{ \hat{A},\hat{B}\right\} \right\rangle  & = &
        2{\rm Re}\,e^{i \delta} \bar{S}_{IF}[\omega]\\
    \left\langle \left[\hat{A},\hat{B}\right]\right\rangle  & = & \int_{-\infty}^{+\infty}dt\,
        \cos(\omega t + \delta) \,\left\langle \left[\hat{I}(t),\hat{F}(0)\right]\right\rangle
    \nonumber \\
    & = &
            i \hbar \Re \left[e^{i \delta}
            \left( \chiIF[\omega] -
            \left( \chiFI[\omega]  \right)^* \right) \right]
        \label{eq:HeisNoise3}
\end{eqnarray}
\end{subequations}
In the last line, we have simply made use of the Kubo formula
definitions of the gain and reverse gain
(cf.~Eqs.~(\ref{eq:gain}) and (\ref{ReverseGain})).  As a
consequence of
Eqs.~(\ref{eq:HeisNoise1})-(\ref{eq:HeisNoise3}), the
Heisenberg uncertainty relation (\ref{eq:Heisenberg}) directly
yields:
\begin{eqnarray}
     \bS_{II}[\omega] \bS_{FF}[\omega] & \geq &
         \left[\textrm{Re }
         \left( e^{i \delta} \bS_{IF}[\omega] \right) \right]^2 +
         \label{DeltaNoiseConstraint}
          \\
         &&
         \frac{\hbar^2}{4}
         \left[ \textrm{Re }
            e^{i \delta}
         \left(\chiIF[\omega] -
         \left( \chiFI[\omega]\right)^* \right)
         \right]^2
         \nonumber
\end{eqnarray}
Maximizing the RHS of this inequality over $\delta$ then yields
the general quantum noise constraint of
Eq.~(\ref{NoiseConstraint}).

With this derivation, we can now interpret the quantum noise
constraint Eq.~(\ref{NoiseConstraint}) as stating that the {\it
noise at a given frequency} in two observables, $\hI$ and $\hF$,
is bounded by the value of their commutator at that frequency. The
fact that $\hI$ and $\hF$ do not commute is necessary for the
existence of linear response (gain) from the detector, but also
means that the noise in both $\hI$ and $\hF$ cannot be arbitrarily
small.  A more detailed derivation, yielding additional important
insights, is described in Appendix \ref{app:noiseineq}.

Given the quantum noise constraint of Eq.~(\ref{NoiseConstraint}),
we can now very naturally define a ``quantum-ideal" detector (at a
given frequency $\omega$) as one which minimizes the LHS of
Eq.~(\ref{NoiseConstraint})-- a quantum-ideal detector has a
minimal amount of noise at frequency $\omega$.  We will often be
interested in the ideal case where there is no reverse gain (i.e.
measuring $\hI$ does not result in additional back-action noise in
$\hF$); the condition to have a quantum limited detector thus
becomes:
\begin{eqnarray}
     && \bS_{II}[\omega] \bS_{FF}[\omega] - \left| \bS_{IF}[\omega] \right|^2
       =
         \label{CorrQLCondition}
      \\
      && \left| \frac{\hbar \chiIF[\omega]}{2} \right|^2
      \left(1 +
                \Delta\left[ \frac{\bS_{IF}[\omega]}{\hbar
                \chiIF[\omega]/2} \right] \right)
                 \nonumber
\end{eqnarray}
where $\Delta[z]$ is given in Eq.~(\ref{eq:DeltaFactor}).  Again,
as we will discuss below, in most cases of interest (e.g.~zero
frequency and/or large amplifier power gain), the last term on the RHS
will vanish.  In the following sections, we will demonstrate that
the ``ideal noise" requirement of Eq.~(\ref{CorrQLCondition}) is
necessary in order to achieve the quantum limit on QND detection
of a qubit, or on the added noise of a linear amplifier.

Before leaving our general discussion of the quantum noise
constraint, it is worth emphasizing that achieving
Eq.~(\ref{CorrQLCondition}) places a very strong constraint on the
properties of the detector. In particular, there must exist a
tight connection between the input and output ports of the
detector-- in a certain restricted sense, the operators $\hI$ and
$\hF$ must be proportional to one another (see
Eq.~(\ref{PropCond}) in Appendix \ref{app:noiseineq}). As is
discussed in Appendix \ref{app:noiseineq}, this proportionality
immediately tells us that {\it a quantum-ideal detector cannot be
in equilibrium}. The proportionality exhibited by a quantum-ideal
detector is parameterized by a single complex valued number
$\alpha[\omega]$, whose magnitude is given by:
\begin{eqnarray}
     | \alpha[\omega] |^2 & = &  \bS_{II}[\omega] / \bS_{FF}[\omega]
     \label{AlphaDefn}
\end{eqnarray}
While this proportionality requirement may seem purely formal, it
does have a simple heuristic interpretation; as is discussed in
\textcite {Clerk03}, it may be viewed as a formal expression of
the principle that a quantum-limited detector must not contain any `wasted information'  (cf.~Sec.~\ref{subsubsec:CavityQubitQL}).

\subsubsection{Evading the detector quantum noise inequality}
\label{subsubsec:NoiseCrossCorr}

We now turn to situations where the RHS of Eq.~(\ref{NoiseConstraint}) vanishes, implying that there is no additional quantum constraint on the noise of our detector beyond what exists classically. In such situations, one could have a detector with perfectly correlated back-action and output noises (i.e. 
$\bS_{FF} \bS_{II} = | \bS_{IF} |^2$), or even with a vanishing back-action $\bS_{FF}=0$.  Perhaps not
surprisingly, these situations are not of much utility.  As we will now show, in cases where the RHS of Eq.~(\ref{NoiseConstraint}) vanishes,
the detector may be low noise, but will necessarily be deficient in another important regard:  {\it it will not be good enough that we can ignore the extra noise associated with the measurement of the detector output $\hI$}.  As we have already discussed, reading-out $\hI$ will invariably involve coupling the detector output to some other physical system.  In the ideal case, this coupling will not generate any additional back-action on the system coupled to the detector's input port.  In addition,  the signal at the detector output should be large enough that any noise introduced in measuring $\hI$ is negligible; we already came across this idea in our discussion of the resonant cavity detector (see comments following Eq.~(\ref{eq:SxxaddQL})).
This means that we need our detector to truly {\it amplify} the input signal, not simply reproduce it at the output with no gain in energy.  As we now show, a detector which evades the quantum constraint of Eq.~(\ref{NoiseConstraint}) by making the RHS of the inequality zero will {\it necessarily} fail in one or both of the above requirements.

 The most obvious case where the quantum noise constraint vanishes is for a detector 
 which has equal forward and reverse gains, $\chiFI = \chiIF^*$.  As we have already mentioned, this relation will necessarily hold if the the detector is time-reversal symmetric and in equilibrium, and $\hI$ and $\hF$ have the same parity under time-reversal.  In this case, the relatively large reverse gain implies that  in analyzing a given measurement task, it is not sufficient to just consider the noise of the detector: one {\it must necessarily} also consider the noise associated with whatever system is coupled to $\hI$ to readout the detector output, as this noise will be fed back to the detector input port, causing additional back-action; we give an explicit example of this in the next sub-section, when we discuss QND qubit detection.  
Even more problematically, when $\chiFI = \chiIF^*$, there is never any amplification by the detector.
 As we will discuss in Sec.~\ref{subsubsec:PowerGain}, the proper metric of the detector's ability
to amplify is its dimensionless power gain:   what is the power supplied at the output of the detector versus the amount of power drawn at the input from the signal source?  When $\chiFI = \chiIF^*$, one has negative feedback, with the result that the power gain cannot be larger than $1$ (cf.~Eq.~\ref{eq:GPReverse}).  There is thus no amplification when $\chiFI = \chiIF^*$.  Further, if one also insists that the noise constraint of Eq.~(\ref{NoiseConstraint}) is optimized, then one finds the power gain must be exactly $1$; this is explicitly demonstrated in Appendix \ref{subapp:NoAmplification}.  The detector thus will simply act as a transducer, reproducing the input signal at the output without any increase in energy.  We have here a specific example of a more general idea that will be discussed extensively in Sec.~\ref{sec:QLAmplifiers}:  if a detector only acts as a transducer, it need not add any noise.  

%

At finite frequencies, there is a second way to make the RHS of
the quantum noise constraint of Eq.~(\ref{NoiseConstraint}) vanish:  one needs the 
quantity $\bS_{IF}[\omega] / \tchiIF$ to be purely imaginary, and larger in magnitude than $\hbar/2$.  In this case, it would again be possible to have the LHS of the noise constraint of Eq.~(\ref{NoiseConstraint}) equal to zero.  However, one again finds
that in such a case, the dimensionless power gain of the detector is at most equal to one; it thus
does not amplify.  This is shown explicitly in Appendix \ref{subapp:NoAmplification}.  
%
An important related statement is that a quantum-limited detector with a large power gain must have the quantity $\bS_{IF} / \chiIF$ be real.  Thus, {\it at the quantum limit, correlations between
  the back-action force and the intrinsic output noise fluctuations must have the same phase as the
  gain $\chiIF$}.  As we discuss further in Sec.~\ref{subsec:VoltageAmp}, this requirement can 
  be interpreted in terms of the principle of `no wasted information' introduced in Sec.~\ref{subsubsec:CavityQubitQL}.

\subsection{Quantum limit on QND detection of a qubit}
\label{subsec:QLimitProof}

In Sec.~\ref{subsec:CavityDetector},
we discussed the quantum limit on QND qubit detection in the
specific context of a resonant cavity detector.  We will now show how the full
quantum noise constraint of Eq.~(\ref{NoiseConstraint})
directly leads to this quantum limit for an arbitrary weakly-coupled detector.
Similar to Sec.~\ref{subsec:CavityDetector}, we couple the input operator of our generic linear-response
detector to the $\hat{\sigma}_z$ operator of the qubit we wish to measure
(i.e.~we take $\hat{x} = \hat{\sigma}_z$ in Eq.~(\ref{HIntGen})); we also
consider the QND regime, where $\hat{\sigma}_z$ commutes with the qubit Hamiltonian.
As we saw
in Sec.~\ref{subsec:CavityDetector}, the quantum limit in this
case involves the inequality $\Gamma_{\rm meas} \leq
\Gamma_{\varphi}$, where $\Gamma_{\rm meas}$ is the measurement
rate, and $\Gamma_{\varphi}$ is the back-action dephasing rate.
For the latter quantity, we can directly use the results of our
calculation for the cavity system, where we found the dephasing
rate was set by the zero-frequency noise in the cavity photon
number (cf.~Eq.~\ref{eq:cavitydephasingrate}).
 In complete
analogy, the back-action dephasing rate here will be determined by
the zero-frequency noise in the input operator $\hF$ of our
detector:
\begin{equation}
     \Gamma_{\varphi} = \frac{2 A^2}{\hbar^2} \bS_{FF}
     \label{eq:DephRate2}
\end{equation}
We omit frequency arguments in this subsection, as it is always the zero-frequency  susceptibilities and spectral densities which appear.

The measurement rate (the rate at which information on the state of qubit
is acquired) is also defined in complete analogy to what was done
for the cavity detector.  We imagine we turn the
measurement on at $t=0$ and start to integrate up the output
$I(t)$ of our detector:
\begin{equation}
     \hat{m}(t) = \int_0^t dt' \hI(t')
\end{equation}
The probability distribution of the integrated output $\hat{m}(t)$
will depend on the state of the qubit; for long times, we may
approximate the distribution corresponding to each qubit state as
being gaussian. Noting that we have chosen $\hI$ so that its
expectation vanishes at zero coupling, the average value of
$\langle \hat{m}(t) \rangle$ corresponding to each qubit state is (in
the long time limit of interest):
\begin{eqnarray}
     \langle \hat{m}(t) \rangle_{\ua}  =  A \chiIF t, \quad
     \langle \hat{m}(t) \rangle_{\da}  =  - A \chiIF t
\end{eqnarray}
The variance of both distributions is, to leading order,
independent of the qubit state:
\begin{equation}
     \langle \hat{m}^2 (t) \rangle_{\ua/\da} - \langle \hat{m} (t)
\rangle_{\ua/\da}^2
      \equiv \langle \langle \hat{m}^2(t) \rangle \rangle_{\ua/\da}
      = \bS_{II} t
\end{equation}
For the last equality above, we have taken the long-time limit,
which results in the variance of $\hat{m}$ being determined
completely by the zero-frequency output noise $\bS_{II}[\omega=0]$ of
the detector.  The assumption here is that due to the weakness of
the measurement, the measurement time (i.e.~$1/ \Gamma_{\rm
meas}$) will be much longer than the autocorrelation time of the
detector's noise.

We can now define the measurement rate, in complete analogy to the
cavity detector of the previous section
(cf.~Eq.~(\ref{eq:cavitymeasurementrate})),
by how quickly the resolving power of the measurement grows:\footnote{%
     The strange looking factor of $1/4$ here is purely chosen for
convenience;
     we are defining the measurement rate based on the information theoretic definition given in
     Appendix \ref{app:measurementrateinformationtheory}.  This factor of four is consistent
with the definition used in
     the cavity system.
}
\begin{equation}
     \frac{1}{4}
     \frac{
     \left[
         \langle \hat{m}(t) \rangle_{\ua} - \langle \hat{m}(t) \rangle_{\da}
         \right]^2}
         {\langle \langle \hat{m}^2(t) \rangle \rangle_{\ua} +
             \langle \langle \hat{m}^2(t) \rangle \rangle_{\da}}
             \equiv
             \Gamma_{\rm meas} t.
\end{equation}
This yields:
\begin{equation}
     \Gamma_{\rm meas} = \frac{A^2 \left( \chiIF \right)^2}{2 \bS_{II}}
     \label{eq:GammaMeasDefn}
\end{equation}

Putting this all together, we find that the ``efficiency" ratio
$\eta = \Gamma_{\rm meas} / \Gamma_{\varphi}$ is given by:
\begin{equation}
     \eta \equiv \frac{\Gamma_{\rm meas}}{\Gamma_{\varphi}} =
         \frac{\hbar^2 \left( \chiIF \right)^2}{4 \bS_{II} \bS_{FF}}
         \label{eq:EfficiencyRatio}
\end{equation}
In the case where our detector has a vanishing reverse gain (i.e.~$\chiFI = 0$),
the quantum-limit bound $\eta \leq 1$ follows immediately
from the quantum noise constraint of Eq.~(\ref{NoiseConstraint}).
We thus see that achieving the quantum-limit for QND qubit detection requires both
a detector with quantum-ideal noise properties, as defined by
Eq.~(\ref{CorrQLCondition}), as well as a detector with a
vanishing noise cross-correlator: $\bS_{IF} = 0$.  

If in contrast $\chiFI \neq 0$, it would seem that it is possible to have $\eta \geq 1$.  This is 
of course an invalid inference:  as discussed, $\chiFI \neq 0$ implies that we must necessarily 
consider the effects of extra noise injected into detector's output
port when one measures $\hI$, as the reverse gain will bring this noise back to the qubit, causing extra dephasing.  
The result is that one can do no better than $\eta = 1$.  To see this explicitly, consider the
extreme case $\chiIF = \chiFI$ and $\bS_{II} = \bS_{FF} = 0$, and suppose we use a second detector to read-out the output $\hI$ of the first detector.  This second detector
has input and output operators $\hF_2$, $\hI_2$; we also take it to have a vanishing reverse gain, so that we do not have to also worry about how its output is read-out.  Coupling the detectors linearly in the standard way (i.e.~$H_{\rm int,2} = \hI \hF_2$), the overall gain  of the two detectors in series is $\chi_{I_2 F_2} \cdot \chi_{IF}$, while the back-action driving the qubit dephasing is described by the spectral density $(\chi_{FI})^2 S_{F_2 F_2}$.  Using the fact that our second detector must itself satisfy the quantum noise inequality, we have:
 \begin{eqnarray}
 	\left[ \left(\chiFI\right)^2 \bS_{F_2 F_2} \right] \bS_{I_2 I_2} \geq
		\frac{\hbar^2}{4} \left(\chi_{I_2F_2} \cdot \chiIF \right)^2
\end{eqnarray}
Thus, the overall chain of detectors satisfies the usual, zero-reverse gain quantum noise inequality, implying that we will still have $\eta \leq 1$.


\section{Quantum Limit on Linear Amplifiers and Position Detectors}
\label{sec:QLAmplifiers}

In the previous section, we established the fundamental quantum
constraint on the noise of any system capable of acting as a
linear detector; we further showed that this quantum noise
constraint directly leads to the ``quantum limit" on
non-demolition qubit detection using a weakly-coupled detector. In
this section, we turn to the more general situation where our
detector is a phase-preserving quantum linear amplifier:  the
input to the detector is described by some time-dependent operator
$\hx(t) $ which we wish to have amplified at the
output of our detector.  As we will see, the quantum limit in this
case is a limit on how small one can make the noise added by the
amplifier to the signal.  The discussion in this section both furthers and generalizes
the heuristic discussion of position detection using a cavity detector presented in
Sec.~\ref{subsec:CavityDetector}.

In this section, we will start by presenting a heuristic
discussion of quantum constraints on amplification.  We will then
demonstrate explicitly how the previously-discussed quantum noise
constraint leads directly to the quantum limit on the added noise
of a phase-preserving linear amplifier; we will examine both the
cases of a generic linear position detector and a generic voltage
amplifier, following the approach outlined in \textcite{Clerk04c}.
We will also spend time explicitly connecting the linear response
approach we use here to the bosonic scattering formulation of the
quantum limit favoured by the quantum optics community
\cite{Haus62, Caves82, Grassia98, Courty99}, paying particular
attention to the case of two-port scattering amplifier.  We will
see that  there are some important subtleties involved in
converting between the two approaches.  In particular,
there exists a crucial difference between the case where
the input signal is tightly coupled to the input of the amplifier (the case usually considered in the quantum optics community), versus the case where, similar to an ideal op-amp, the input signal is only weakly coupled to the input of the amplifier (the case usually
considered in the solid state community).

\subsection{Preliminaries on amplification}
\label{subsec:AmpPrelims}

What exactly does one mean by `amplification'?  As we will see
(cf.~Sec.~\ref{subsubsec:PowerGain}), a precise definition
requires that the energy provided at the output of the amplifier
be much larger than the energy drawn at the input of the
amplifier-- the ``power gain"  of the amplifier must be larger than
one.  For the moment, however, let us work with the cruder
definition that amplification involves making some time-dependent
signal `larger'. To set the stage, we will first consider an
extremely simple classical analogue of a linear amplifier.  Imagine
the ``signal"  we wish to amplify is the coordinate $x(t)$ of a
harmonic oscillator; we can write this signal as:
\begin{equation}
x(t)=x(0) \cos(\omega_{S}t)+\frac{p(0)}{M\omega_{S}}
\sin(\omega_{S}t)
\end{equation}
Our signal has two quadrature amplitudes, i.e.  the amplitude of
the cosine and sine components of $x(t)$.  To ``amplify" this signal,
we start at $t=0$ to parametrically drive the
oscillator by changing its frequency $\omega_{S}$ periodically in
time: $\omega_{S}(t)=\omega_{0}+\delta\omega\sin(\omega_{P}t)$,
where we assume $\delta\omega\ll\omega_{0}$. The well-known
physical example is a swing whose motion is being excited by
effectively changing the length of the pendulum at the right
frequency and phase. For a {}``pump frequency'' $\omega_{P}$
equalling twice the {}``signal frequency'',
$\omega_{P}=2\omega_{S}$, the resulting dynamics will lead to an
amplification of the initial oscillator position, with the energy
provided by the external driving:
\begin{equation}
x(t)=x(0)e^{\lambda
t}\cos(\omega_{S}t)+\frac{p(0)}{M\omega_{S}}e^{-\lambda
t}\sin(\omega_{S}t)
\label{eq:ClassDPA}
\end{equation}
Thus, one of the quadratures is amplified exponentially, at a rate
$\lambda=\delta\omega/2$, while the other one decays. In a
quantum-mechanical description, this produces a squeezed state out
of an initial coherent state. Such a system is called a
{}``degenerate parametric amplifier'', and we discuss its quantum
dynamics in more detail in Sec.~\ref{subsec:BackactionEvasion}.
We will see that such an
amplifier, which only amplifies a single quadrature, is not
required quantum mechanically to add any noise
\cite{Caves80b,Caves82, Braginsky92}.

Can we now change this parametric amplification scheme slightly in
order to make \emph{both} signal quadratures grow with time? It
turns out this is impossible, as long as we restrict ourselves to
a driven system with a single degree of freedom. The reason in
classical mechanics is that Liouville's theorem requires phase
space volume to be conserved during motion. More formally,
this is related to the conservation of Poisson brackets, or, in
quantum mechanics, to the conservation of commutation relations.
Nevertheless, it is certainly desirable to have an amplifier that
acts equally on both quadratures (a so-called
{}``phase-preserving'' or {}``phase-insensitive'' amplifier),
since the signal's phase is often not known beforehand. The way
around the restriction created by Liouville's theorem is to add
more degrees of freedom, such that the phase space volume can
expand in both quadratures (i.e.~position and momentum) of the
interesting signal degree of freedom, while being compressed in
other directions. This is achieved most easily by coupling the
signal oscillator to another oscillator, the {}``idler mode''. The
external driving now modulates the coupling between these
oscillators, at a frequency that has to equal the sum of the
oscillators' frequencies. The resulting scheme is called a
phase-preserving non-degenerate parametric amplifier (see 
Sec.~\ref{subsec:NDParamps}).

Crucially, there is a price to pay for the introduction of an
extra degree of freedom:  there will be noise associated with the
``idler" oscillator, and this noise will contribute to the noise
in the output of the amplifier.  Classically, one could make the
noise associated with the ``idler" oscillator arbitrarily small by
simply cooling it to zero temperature.  This is not possible
quantum-mechanically; there are always zero-point fluctuations of
the idler oscillator to contend with.  It is this noise which sets
a fundamental quantum limit for the operation of the amplifier.
We thus have a heuristic accounting for why there is a
quantum-limit on the added noise of a phase-preserving linear
amplifier:  one needs extra degrees of freedom to amplify both
signal quadratures, and such extra degrees of freedom invariably
have noise associated with them.

\subsection{Standard Haus-Caves derivation of the quantum limit on a bosonic
amplifier}
\label{subsec:CavesArg}

We now make the ideas of the previous subsection more precise by
quickly sketching the standard derivation of the quantum limit on
the noise added by a phase-preserving amplifier.  This derivation
is originally due to \textcite{Haus62}, and was both clarified and extended by
\textcite{Caves82}; the amplifier quantum limit was also motivated
in a slightly different manner by  
\textcite{Heffner62}.\footnote{Note that \textcite{Caves82} provides a thorough discussion of why the derivation of the amplifier
quantum limit given in \textcite{Heffner62} is not rigorously correct} 
While extremely compact,
the Haus-Caves derivation can
lead to confusion when improperly applied; we will discuss this
in Sec.~\ref{subsec:TwoKindsAmps}, as well as in Sec.~\ref{sec:ScatteringAmp}, where
we apply this argument carefully to the important case of a
two-port quantum voltage amplifier, and discuss the connection to the general linear-response formulation of Sec.~\ref{sec:GenLinResponseTheory}.

The starting assumption of this derivation is that both the input and output ports of
the amplifier can be described by sets of bosonic modes.  If we
focus on a narrow bandwidth centered on frequency $\omega$, we can
describe a classical signal $E(t)$ in terms of a complex number $a$ defining
the amplitude and phase of the signal (or equivalently the two
quadrature amplitudes) \cite{Haus62,Haus00}
\be
E(t) \propto i[a e^{-i\omega t} - a^* e^{+i\omega t}].
\ee
In the quantum case, the two signal quadratures of $E(t)$
(i.e.~the real and imaginary parts of $a(t)$)
cannot be measured simultaneously because they are canonically conjugate; this is
in complete analogy to a harmonic oscillator (cf.~Eq.~(\ref{eq:twoquadraturesdefined})).
As a result $a,a^*$ must be elevated to the status of photon ladder operators:
 $a \ra \hat{a}, a^* \ra a^{\dagger}$ .

Consider the simplest case, where there is only a single mode at
both the input and output, with corresponding 
operators $\hat{a}$ and $\hat{b}$.\footnote{ To relate this to the
linear response detector of Sec.~\ref{subsec:NoiseConstraint}, one
could naively write $\hat{x}$, the operator carrying the input signal, as,
e.g., $\hat{x} = \hat{a} + \hat{a}^{\dag}$, and the output
operator $\hI$ as, e.g., $\hI = \hat{b} + \hat{b}^{\dagger}$ (we
will discuss how to make this correspondence in more detail in
Sec.~\ref{sec:ScatteringAmp})}
It follows that the input signal into the amplifier is described
by the expectation value $\langle \hat{a} \rangle$, while the
output signal is described by $\langle \hat{b} \rangle$.
Correspondingly, the symmetrized noise in both these quantities is
described by:
\begin{eqnarray}
     \left(\Delta a \right)^2  \equiv  \frac{1}{2} \left \langle \{  \hat{a}, \hat{a}^{\dag} \} \right \rangle
        - \left| \langle \ha \rangle \right|^2
\end{eqnarray}
with an analogous definition for $(\Delta b)^2$.

To derive a quantum limit on the added noise of the amplifier, one
uses two simple facts. First, both the input and the output
operators must satisfy the usual commutation relations:
\begin{eqnarray}
     \left[  \hat{a}, \hat{a}^{\dag} \right] = 1, \quad
     \left[ \hat{b}, \hat{b}^{\dag} \right]    =  1
\end{eqnarray}
Second, the linearity of the amplifier and the fact that it is
phase preserving (i.e.~both signal quadratures are amplified the
same way) implies a simple relation between the output operator
$\hat{b}$ and the input operator $\hat{a}$:
\begin{eqnarray}
     \hat{b}  =  \sqrt{G} \hat{a}, \quad
     \hat{b}^{\dagger} & = & \sqrt{G} \hat{a}^{\dagger}
     \label{eq:naiveinputoutput}
\end{eqnarray}
where $G$ is the dimensionless {\em ``photon number gain"} of the
amplifier.  It is immediately clear however this expression
cannot possibly be correct as written because it violates the
fundamental bosonic commutation relation $[ \hat{b},
\hat{b}^{\dagger} ] = 1$. We are therefore forced to write
\begin{eqnarray}
     \hat{b}  =  \sqrt{G} \hat{a} + \mathcal{\hat{F}}, \quad
     \hat{b}^{\dagger}  =  \sqrt{G} \hat{a}^{\dagger} + \mathcal{\hat{F}}^{\dagger}
     \label{eq:fullb}
\end{eqnarray}
where $\mathcal{\hat{F}}$ is an operator representing additional
noise added by the amplifier.  Based on the discussion of the
previous subsection, we can anticipate what $\mathcal{\hat{F}}$
represents:  it is noise associated with the additional degrees of
freedom which must invariably be present in a phase-preserving
amplifier.

As $\mathcal{\hat{F}}$ represents noise, it has a vanishing
expectation value; in addition, one also assumes that this noise
is uncorrelated with the input signal, implying $[
\mathcal{\hat{F}}, \hat{a} ] = [ \mathcal{\hat{F}}, \hat{a}^{\dag}
] = 0$ and $\langle \mathcal{\hat{F}} \hat{a} \rangle = \langle
\mathcal{\hat{F}} \hat{a}^{\dag} \rangle = 0$. Insisting that $[
\hat{b}, \hat{b}^{\dagger} ] = 1$  thus yields:
\begin{equation}
     \left[ \mathcal{\hat{F}}, \mathcal{\hat{F}}^{\dagger} \right] = 1 - G
     \label{eq:addednoisecommutator}
\end{equation}

The question now becomes how small can we make the noise described
by $\mathcal{\hat{F}}$? Using Eqs. (\ref{eq:fullb}),
  the noise at the amplifier output $\Delta b$ is given by:
\begin{eqnarray}
     \left( \Delta b \right)^2 & = &
         G \left( \Delta a \right)^2 + \frac{1}{2} \left \langle
         \{\mathcal{\hat{F}}, \mathcal{\hat{F}}^{\dag} \} \right \rangle\nonumber \\
     & \geq &
         G \left( \Delta a \right)^2 + \frac{1}{2}
             \left| \left \langle [ \mathcal{\hat{F}}, \mathcal{\hat{F}}^{\dag} ]
\right \rangle
             \right|\nonumber\\
             &\geq&
                 G \left( \Delta a \right)^2 + \frac{ |G-1|}{2}
                 \label{eq:FinalCavesIneq}
\end{eqnarray}
We have used here a standard inequality to bound the expectation
of $\{ \mathcal{\hat{F}}, \mathcal{\hat{F}}^{\dag} \}$. The first
term here is simply the amplified noise of the input, while the
second term represents the noise added by the amplifier. Note that
if there is no amplification (i.e.~$G=1$), there need not be any
added noise. However, in the more relevant case of large
amplification ($G \gg 1$), the added noise cannot vanish.  It is
useful to express the noise at the output as an equivalent noise
at (``referred to") the input by simply dividing out the photon
gain $G$. Taking the large-$G$ limit, we have:
\begin{equation}
     \frac{\left( \Delta b \right)^2}{G}  \geq   \left( \Delta a
\right)^2 + \frac{1}{2}
     \label{CavesQL}
  \end{equation}
  Thus, we have a very simple demonstration that {\em an amplifier
with a large photon gain must add at least
  half a quantum of noise to the input signal}.  Equivalently, the
minimum value of the added noise is
  simply equal to the zero-point noise associated with the
input mode; the total output noise (referred to the input) is at
least twice the zero point input noise.
Note that both these conclusions are identical to what we found in our analysis of the resonant cavity position detector in Sec.~\ref{subsec:measurementofoscillatorposition}.
We will discuss in later sections how this conclusion can also be reached using the general linear-response language of Sec.~\ref{sec:GenLinResponseTheory} (cf.~Sec.~\ref{subsec:PositionDetector}, \ref{subsec:VoltageAmp}).

As already discussed, the added noise operator
$\mathcal{F}$ is associated with additional degrees of freedom
(beyond input and output modes) necessary for phase-preserving
amplification.  To see this more concretely, note that every
linear amplifier is inevitably a {\em non-linear} system
consisting of an energy source and a `spigot' controlled by the
input signal which redirects the energy source partly to the
output channel and partly to some other channel(s).   Hence there
are inevitably other degrees of freedom involved in the
amplification process beyond the input and output channels. An
explicit example is the quantum parametric amplifier, to be discussed 
in the next subsection.
Further insights into amplifier added noise and its connection
to the fluctuation-dissipation theorem can be obtained by considering a simple
model where a transmission line is terminated by an effective negative impedance; we discuss this model in Appendix \ref{subsec:NegativeResistance}.

To see explicitly the role of the additional degrees of freedom,
note first that for $G>1$ the RHS of
Eq.~(\ref{eq:addednoisecommutator}) is negative. Hence the
simplest possible form for the added noise is
\begin{eqnarray}
\mathcal{\hat{F}} = \sqrt{G-1}\hat{d}^\dagger, \quad
\mathcal{\hat{F}^\dagger} = \sqrt{G-1}\hat{d}
\label{eq:addednoise}
\end{eqnarray}
where $\hat{d}$ and $\hat{d}^\dagger$ represent a single
additional mode of the system.  This is the minimum number of
additional degrees of freedom that must inevitably be involved in
the amplification process. Note that for this case,  the
inequality in Eq.~(\ref{eq:FinalCavesIneq}) is satisfied as an
equality, and the added noise takes on its minimum possible value.
If instead we had, say, two additional modes (coupled
inequivalently):
  \be
\mathcal{\hat{F}} = \sqrt{G-1}(\cosh\theta \hat{d}_1^\dagger +
\sinh\theta \hat{d}_2)
  \ee
  it is straightforward to show that the added noise is inevitably
larger than the minimum.  This again can be interpreted in terms
of wasted information, as the extra degrees of freedom are not
being monitored as part of the measurement process and so
information is being lost.

\subsection{Non-degenerate parametric amplifier}
\label{subsec:NDParamps}

\subsubsection{Gain and added noise}

\begin{figure}[t]
\begin{center}
\includegraphics[width=0.5\columnwidth]{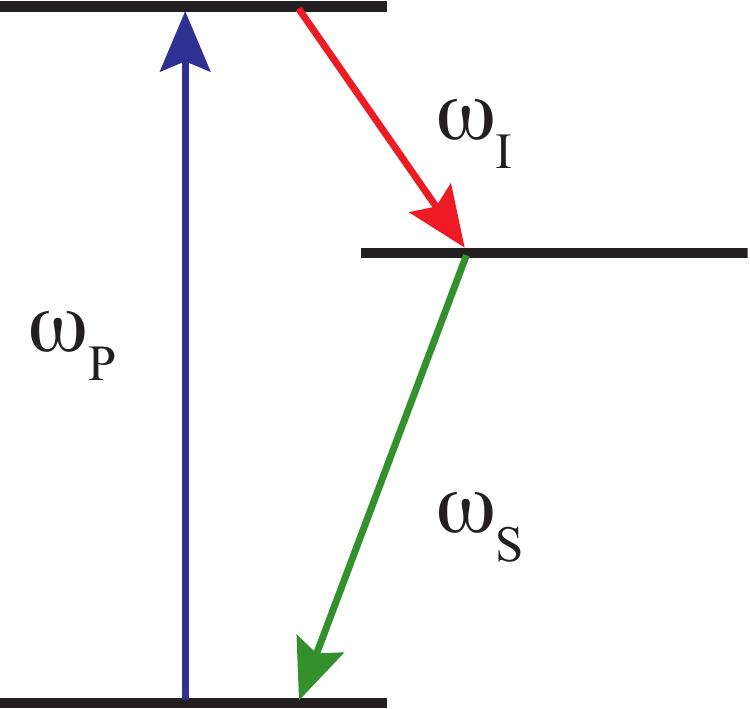}
\caption{(Color online) Energy level scheme of the non-degenerate (phase
preserving) parametric oscillator.}
\label{fig:paramp_level_scheme}
\end{center}
\end{figure}

Before we start our discussion of how the Haus-Caves formulation of the quantum limit connects to the general linear-response approach of Sec.~\ref{sec:GenLinResponseTheory}, it is useful to consider a specific example.  To that end, we analyze here a non-degenerate parametric amplifier, a linear, phase-preserving amplifier which reaches the quantum limit on its added noise \cite{Gordon63,Louisell61,Mollow67a,Mollow67b} 
and directly realizes the ideas of the previous sub-section.
One possible realization \cite{Yurke89} is a 
cavity with three internal resonances that are coupled together by
a non-linear element (such as a Josephson junction) whose symmetry
permits three-wave mixing.  The three modes are called the pump,
idler and signal and their energy level structure (illustrated in
Fig.~\ref{fig:paramp_level_scheme}) obeys
$\omegap=\omegai+\omegas$.
The system Hamiltonian
is then
\begin{eqnarray}
\hat{H}_{\rm sys} &=& \hbar\left(\omegap \ap^\dagger \ap + \omegai
\ai^\dagger\ai + \omegas \as^\dagger\as \right)\nonumber\\
&+& i\hbar \eta \left(\as^\dagger\ai^\dagger\ap -
\as\ai\ap^\dagger\right).
\label{eq:parmapHam}
\end{eqnarray}
We have made the rotating wave approximation in the three-wave
mixing term, and without loss of generality, we take the
non-linear susceptibility $\eta$ to be real and positive. The
system is driven at the pump frequency and the three wave mixing
term permits a single pump photon to split into an idler photon
and a signal photon.  This process is stimulated by signal photons
already present and leads to gain.
 A typical mode of operation
would be the negative resistance reflection mode
in which the input signal is reflected from a
non-linear cavity and the reflected beam extracted using a
circulator \cite{Yurke89, 2008_Devoret_JosephsonRingModulator}. 

The non-linear EOMs become tractable if we assume the pump has
large amplitude and can be treated classically by making the
substitution
\be
\ap = \psi_{\rm P}e^{-i\omegap t}=\psi_{\rm
P}e^{-i(\omegai+\omegas) t},
\ee
where without loss of generality we take $\psi_{\rm P}$ to be real
and positive.  We note here the important point that if this
approximation is not valid, then our amplifier would in any case
not be the linear amplifier which we seek.  With this
approximation we can hereafter forget about the dynamics of the
pump degree of freedom and deal with the reduced system
Hamiltonian
\begin{eqnarray}
\hat{H}_{\rm sys}= \hbar\left(\omegai \ai^\dagger\ai + \omegas
\as^\dagger\as \right) + i\hbar\lambda
\nonumber\\
\left(\as^\dagger\ai^\dagger e^{-i(\omegai+\omegas) t}- \as\ai
e^{+i(\omegai+\omegas) t}   \right)
\end{eqnarray}
where $\lambda\equiv \eta \psi_{\rm P}$. Transforming to the
interaction representation we are left with the following
time-independent quadratic Hamiltonian for the system
\be
\hat{V}_{\rm sys} = i\hbar\lambda \left(\as^\dagger\ai^\dagger -
\as\ai \right)
\label{eq:nondegintrep}
\ee

To get some intuitive understanding of the physics, let us
temporarily ignore the damping of the cavity modes that would result
from their coupling to modes outside the cavity.
We now have a pair of coupled
EOMs for the two modes
\begin{eqnarray}
{\dot{\hat{a}}}_{\rm S} &=& +\lambda \hat{a}_{\rm I}^\dagger\nonumber\\
{\dot{\hat{a}}}_{\rm I}^\dagger &=& +\lambda \as
\end{eqnarray}
for which the solutions are
\begin{eqnarray}
\as(t) &=& \cosh(\lambda t) \as(0) + \sinh(\lambda t) \hat{a}_{\rm
I}^\dagger(0)\nonumber\\
\hat{a}_{\rm I}^\dagger(t)&=& \sinh(\lambda t)\, \as(0) +
\cosh(\lambda t)\, \hat{a}_{\rm I}^\dagger(0)
\end{eqnarray}
We see that the amplitude in the signal channel grows
exponentially in time and that the effect of the time evolution is
to perform a simple unitary transformation which mixes $\as$ with
$\ai^\dagger$ in such a way as to preserve the commutation
relations.  Note the close connection with the form found
from very general arguments in
Eqs.~(\ref{eq:fullb})-(\ref{eq:addednoise}).

We may now tackle the full system
which includes the coupling between the cavity modes and modes external to the cavity.
Such a coupling is of course necessary in order to feed the input signal into the cavity, as well as extract the amplified output signal.  It will also result in the damping
of the cavity modes, which will cut off the exponential growth found above and yield a fixed amplitude gain.  We will present and motivate the main results in this subsection, relegating details to how one treats the bath modes (so-called input-output theory \cite{Walls94}) to 
Appendix \ref{app:drivencavity}.   Working in the standard Markovian limit, we obtain the following EOMs in the interaction representation:
\begin{eqnarray}
{\dot{\hat{a}}}_{\rm S} &=& -\frac{\ks}{2}\,\as +\lambda
\ai^\dagger -\sqrt
{\ks}\, {\hat b}_{\rm S,in}, \nonumber\\
{\dot{\hat{a}}}_{\rm I}^\dagger &=& -\frac{\ki}{2}\, \ai^\dagger
+\lambda\, \as -\sqrt{\ki}\, \hat{b}^\dagger_{\rm I, in}.
\label{eq:nondegEOMs}
\end{eqnarray}
Here $\ks$ and $\ki$ are the respective damping rates of the
cavity signal mode and the idler mode.  The coupling to extra-cavity modes also lets signals and noise enter the cavity from the baths:
this is described by the bosonic operators $\hb_{\rm S, in}$ and
$\hb_{\rm I, in}$ which drive (respectively) the signal and idler modes.   
$\hb_{\rm S, in}$ describes both the input signal to be amplified plus vacuum noise entering from the bath coupled to the signal mode, whereas $\hb_{\rm I, in}$ simply describes vacuum noise.\footnote{%
Note that the $\hb$ operators are not dimensionless, as 
$\hb^\dag \hb$ represents a photon flux.  This is discussed fully in Appendix \ref{app:drivencavity}.}

Let us fix our attention on signals inside a frequency window
$\delta \omega$ centered on $\omegas$ (hence zero frequency in the
interaction representation).  For simplicity, we will first consider the case where
the signal bandwidth $\delta \omega$ is almost infinitely narrow (i.e.~much much smaller than the damping rate of the cavity modes).  It then suffices to find the steady state solution of these EOMs
\begin{eqnarray}
 \as&=&\frac{2\lambda}{\ks}\,\ai^\dagger-\frac{2}{\sqrt{\ks}}\,{\hat b}_{\rm S,in}\\
 \ai^\dagger
 &=&\frac{2\lambda}{\ki}\,\as-\frac{2}{\sqrt{\ki}}\,{\hat b}_{\rm I,in}^\dagger.
\end{eqnarray}

The output signal of the non-degnerate paramp is the signal leaving the cavity signal mode
and entering the external bath modes; it is described by an operator $\hb_{\rm S, out}$.  
The standard input-output theory treatment of the extra-cavity modes  \cite{Walls94}, presented
in Appendix \ref{app:drivencavity}, 
 yields the simple relation (cf.~Eq.~(\ref{eq:boutput})):
\begin{eqnarray}
	\hb_{\rm S,out} & = & \hb_{\rm S,in} + \sqrt{\kappa_{\rm S} } \ha_{\rm S}.
	\label{eq:NDPAInputOutput}
\end{eqnarray}
The first term corresponds to the reflection of the signal and noise incident on the cavity from the bath, while
the second term corresponds to radiation from the cavity mode into the bath.  Using this, we find that
the output signal from the cavity is given by
\be
\hat{b}_{\rm S,out} = \frac{Q^2+1}{Q^2-1} \,\hat{b}_{\rm S,in} +
\frac{2Q}{Q^2-1}\, \hat{b}^\dagger_{\rm I,in},
\ee
where $Q\equiv \frac{2\lambda}{\sqrt{\ki\ks}}$ 
is proportional to
the pump amplitude and inversely proportional to the cavity decay
rates. 
We have to require $Q^2<1$ to make sure that the parametric amplifier does not 
settle into self-sustained oscillations, i.e.~it works below threshold. Under that condition, we can define the 
photon-number gain $G_0$ via
\begin{eqnarray}
	-\sqrt{G_0}=(Q^2+1)/(Q^2-1),
	\label{eq:ParampG}
\end{eqnarray}
such that
\be
\hat{b}_{\rm S,out} = -\sqrt{G_0}\, \hat{b}_{\rm S,in} - \sqrt{G_0
- 1}\, \hat{b}^\dagger_{\rm I,in}.
\label{eq:nondegparampout}
\ee
In the ideal case, the noise associated with $\hat{b}_{\rm I,in},\hat{b}_{\rm S,in}$
is simply vacuum noise.  As a result, the input-output relation Eq.~(\ref{eq:nondegparampout}) is 
precisely of the Haus-Caves form Eq.~(\ref{eq:addednoise}) for an ideal, quantum-limited amplifier.  It demonstrates that the non-degenerate
parametric amplifier reaches the quantum limit for minimum
added noise. In the limit of large gain the output noise
(referred to the input) for a vacuum input signal is precisely
doubled. 

\subsubsection{Bandwidth-gain tradeoff}

The above results neglected the finite bandwidth $\delta \omega$ of the input signal to the amplifier.  The gain $G_0$ given in Eq.~(\ref{eq:ParampG}) is only the gain at precisely the mean signal frequency $\omega_{\rm S}$; for a finite bandwidth, we also need to understand how the power gain varies as a function of frequency over the entire signal bandwidth.  As we will see, a parametric amplifier suffers from the fact that as one increases the overall magnitude of the gain at the center frequency $\omega_{\rm S}$ (e.g. by increasing the pump amplitude), one simultaneously narrows the frequency range over which the gain is appreciable.  Heuristically, this is because parametric amplification involves using the pump energy to decrease
the damping and hence
increase the quality factor of the signal mode resonance.  This increase in quality factor leads to amplification, 
but it also reduces the bandwidth over which $\as$ can respond to the input signal $\hat{b}_{\rm S, in}$.

To deal with a finite signal bandwidth, one simply Fourier transforms Eqs.~(\ref{eq:nondegEOMs}).  The resulting equations are easily solved and substituted into
Eq.~(\ref{eq:NDPAInputOutput}), resulting in a frequency-dependent generalization of the input-output relation given in Eq.~(\ref{eq:nondegparampout}):
\begin{eqnarray}
	\hb_{\rm S,out}[\omega] & = &
		-g[\omega] \hb_{\rm S,in}[\omega]
		- g'[\omega] \hb^{\dag}_{\rm I,in}[\omega]
		\nonumber \\
\end{eqnarray}
Here, $g[\omega]$ is the frequency-dependent gain of the amplifier, and $g'[\omega]$ satisfies $|g'[\omega]|^2 = | g[\omega] |^2 -1$.  In the relevant limit where $G_0 = |g[0]|^2 \gg 1$ (i.e.~large gain at the signal frequency), one has to an excellent approximation:
\begin{eqnarray}
	g[\omega] & = &
		\frac{
			\sqrt{G_0} - i \left( \frac{ \ks - \ki}{\ki + \ks} \right) (\omega / D)}
			{1 - i (\omega / D)},
\end{eqnarray}
with 
\begin{eqnarray}
	D = \frac{1}{\sqrt{G_0}}  \frac{\ks \ki}{\ks + \ki}.
\end{eqnarray}
As always, we work in an interaction picture where the signal frequency has been shifted to zero.  $D$ represents the effective operating bandwidth of the amplifier.  Components of the signal with frequencies (in the rotating frame) $|\omega| \ll D$ are strongly amplified, while components with frequencies $| \omega | \gg D$ are not amplified at all, but can in fact be slightly attenuated.  As we already anticipated, the amplification bandwidth $D$ becomes progressively smaller as the pump power and $G_0$ are increased, with the product $\sqrt{G_0} D$ remaining constant.  In a parametric amplifier increasing the gain via increasing the pump strength comes with a price:  the effective operating bandwidth is reduced.  

\subsubsection{Effective temperature}


Recall that in Sec.~\ref{subsec:QuantumSpectrumAnalyzers}, we introduced the concept of an effective temperature of a non-equilibrium 
system, Eq.~(\ref{OscTeff}).  As we will discuss, this concept plays an important role in quantum-limited amplifiers; the degenerate paramp gives us a first example of this.
Returning to the behaviour of the paramp at the signal frequency, we note that
Eq.~(\ref{eq:nondegparampout}) implies that even for vacuum
input to both the signal and idler ports, the
output will contain a real photon flux.  To quantify this in a simple way, 
it is useful to introduce temporal modes which describe the input and output 
fields during a particular time interval $[j \Delta t, (j+1) \Delta t]$ (where $j$ is an integer):
\begin{eqnarray}
	\hat{B}_{{\rm S, in},j} = 
		\frac{1}{\sqrt{\Delta t} } \int_{j \Delta t}^{(j+1)\Delta t} d \tau \hb_{\rm S, in}(\tau),
\end{eqnarray}
with the temporal modes $\hat{B}_{{\rm S,out},j}$ and $\hat{B}_{{\rm I,in},j}$ being defined analogously.
These temporal modes are discussed further in Appendix \ref{subapp:WindowedFT}, where we discuss the windowed Fourier transform (cf.~Eq.~(\ref{eq:C13})).  

With the above definition, we find that the output mode will have a real occupancy even if the input mode is empty:
\begin{eqnarray}
	{\bar n}_{\rm S,out} &=& 
		\langle 0|
			\hat{B}^\dagger_{{\rm S,out},j} \hat{B}_{{\rm S,out},j}
		|0\rangle \nonumber\\
	&=&
		G_0 \langle 0|
			\hat{B}^\dagger_{{\rm S,in},j}\hat{B}_{{\rm S,in},j}
		|0\rangle  +
		\nonumber \\
			&& 
		(G_0-1)\langle 0|
			\hat{B}_{{\rm I,in},j}\hat{B}^\dagger_{{\rm I,in},j}|0\rangle\nonumber\\
&=&G_0-1.
\end{eqnarray}
The dimensionless mode occupancy $\bar{n}_{\rm S,out}$ is best thought of as a photon flux per unit bandwidth
(cf.~Eq.(\ref{eq:WindowedModeOccupancy})).
This photon flux is equivalent to the photon flux that would
appear in equilibrium at the very high effective temperature
(assuming large gain $G_0$)
\be
T_{\rm eff} \approx \hbar\omega_{\rm S} G_0.
\label{eq:Tefflargegain}
\ee
This is an example of a more general principle, to be discussed in 
Sec.~\ref{subsubsec:SimplificationsQuantumIdealDetector}:
a high gain amplifier must have associated with it a large effective
temperature scale.  Referring this total output noise back to the
input, we have (in the limit $G_0 \gg 1$):
\be
	\frac{T_{\rm eff}}{G_0} = 
		\frac{\hbar\omega_{\rm S}}{2} + \frac{\hbar\omega_{\rm S}}{2}
		= \frac{\hbar\omega_{\rm S}}{2} + T_{\rm N}.
\ee
This corresponds to the half photon of vacuum noise associated with the signal source, plus the added noise of a half photon of our phase preserving amplifier (i.e.~the noise temperature $T_N$ is equal to its quantum limited value).  Here, the added noise is simply the vacuum noise associated with the idler port. 

The above argument is merely suggestive that the output noise
looks like an effective temperature.  In fact, it is possible to
show that the photon number distribution of the output is
precisely that of a Bose-Einstein distribution at temperature
$T_{\rm eff}$.  From Eq.~(\ref{eq:parmapHam}) we see that the
action of the paramp is to destroy a pump photon and create a
pair of new photons, one in the signal channel and one in the
idler channel.  Using the SU(1,1) symmetry of the quadratic
hamiltonian in Eq.~(\ref{eq:nondegintrep}) it is possible to
show that, for vacuum input, the output of the paramp is a
so-called `two-mode squeezed state' of the form
\cite{Gerry85, Caves85,KnightBuzek04}
\be
\left|\Psi_{\rm out}\right\rangle = Z^{-1/2}e^{\alpha
b^\dagger_{\rm S}b^\dagger_{\rm I}}\left|0\right\rangle
 \ee
where $\alpha$ is a constant related to the gain and, to
simplify the notation, we have dropped the `out' labels on the
operators.  The normalization constant $Z$ can be worked out by
expanding the exponential and using
\be
(b^\dagger_{\rm S})^n|0,0\rangle = \sqrt{n!}|n,0\rangle
\ee
 to obtain
\be
\left|\Psi_{\rm out}\right\rangle = Z^{-1/2}\sum_{n=0}^\infty
\alpha^n |n,n\rangle
\ee
and hence
\be
Z=\frac{1}{1-|\alpha|^2}
\ee
so the state is normalizable only for $|\alpha|^2<1$.

Because this output is obtained by unitary evolution from the
vacuum input state, the output state is a pure state with zero
entropy.  In light of this, it is interesting to consider the
reduced density matrix obtain by tracing over the idler mode.
The pure state density matrix is:
\begin{eqnarray}
\rho&=&\left|\Psi_{\rm out}\right\rangle \left\langle\Psi_{\rm
out}\right|\nonumber\\
&=&\sum_{m,n=0}^\infty \left|n,n\right\rangle \frac{\alpha^n \alpha*^m}{Z}
\left\langle m,m\right|
\end{eqnarray}
If we now trace over the idler mode we are left with the
reduced density matrix for the signal channel
\begin{eqnarray}
\tilde\rho_{\rm S} = 
	{\rm Tr}_{\rm Idler}\,\left\{ \rho\right\}
& = &
\sum_{n_{\rm S}=0}^\infty \left|n_{\rm
S}\right\rangle \frac{|\alpha|^{2n_{\rm S}}}{Z} \left\langle n_{\rm
S}\right|
\nonumber\\
& \equiv & 
\frac{1}{Z}e^{-\beta\hbar\omega_{\rm S}a^\dagger_{\rm S}a_{\rm
S}}
\end{eqnarray}
which is a pure thermal equilibrium distribution with effective
Boltzmann factor
\be
e^{-\beta\hbar\omega_{\rm S}}=|\alpha|^2<1.
\ee
The effective temperature can be obtained from the requirement
that the signal mode occupancy is $G_0-1$
\be
\frac{1}{e^{\beta\hbar\omega_{\rm S}}-1}=G_0-1
\ee
which in the limit of large gain reduces to
Eq.~(\ref{eq:Tefflargegain}).

This appearance of finite entropy in a subsystem even when the
full system is in a pure state is a purely quantum effect.
Classically the entropy of a composite system is at least as
large as the entropy of any of its components. Entanglement
among the components allows this lower bound on the entropy to be
violated in a quantum system.\footnote{This paradox has
prompted Charles Bennett to remark that a classical house is at
least as dirty as its dirtiest room, but a quantum house can be
dirty in every room and still perfectly clean over all.}  In
this case the two-mode squeezed state has strong entanglement
between the signal and idler channels (since their photon
numbers are fluctuating identically).

\subsection{Scattering versus op-amp modes of operation}
\label{subsec:TwoKindsAmps}

\begin{figure}[t]
\begin{center}
\includegraphics[width=3.45in]{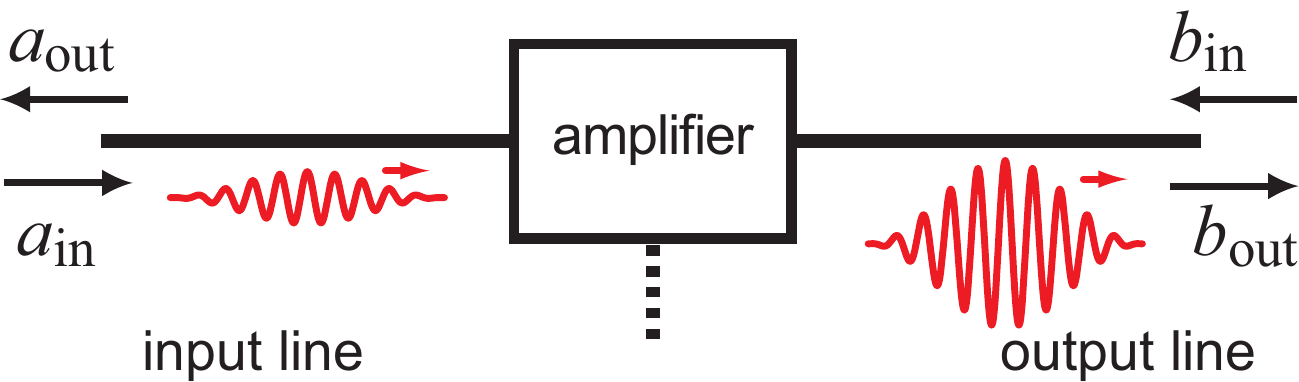}
\caption{(Color online) Schematic of a two-port bosonic amplifier.  Both the input
and outputs of the amplifier are attached to transmission lines.  The
incoming and outgoing wave amplitudes in the input (output) transmission
line are labelled $\hain, \haout$ ($\hbin, \hbout$) respectively.  The voltages
at the end of the two lines ($\hV_a, \hV_b$) are linear combinations of incoming and
outgoing wave amplitudes.}
\label{fig:TwoPortAmp}
\end{center}
\end{figure}
\begin{table}
\caption{\label{tab:TwoAmps} Two different amplifier modes of operation.}
\begin{ruledtabular}
\begin{tabular}{c|c|c}
		Mode &
			Input Signal &
			Output Signal  \\
		&
			$s(t)$ &
			$o(t)$  \\
	\hline
	\hline
		Scattering &
			$s(t) = \ain(t)$  &
			$o(t) = \bout(t) )$  \\
		 &
			($\ain$ indep. of $\aout$)  &
			($\bout$ indep. of $\bin$)  \\
	\hline
		Op-amp &
			$s(t) = V_a(t)$  &
			$o(t) = V_b(t)$  \\
		 &
			($\ain$ depends on $\aout$)  &
			($\bout$ depends on $\bin$)  \\
\end{tabular}
\end{ruledtabular}
\end{table}

\begin{figure}[t]
\begin{center}
\includegraphics[width=3.45in]{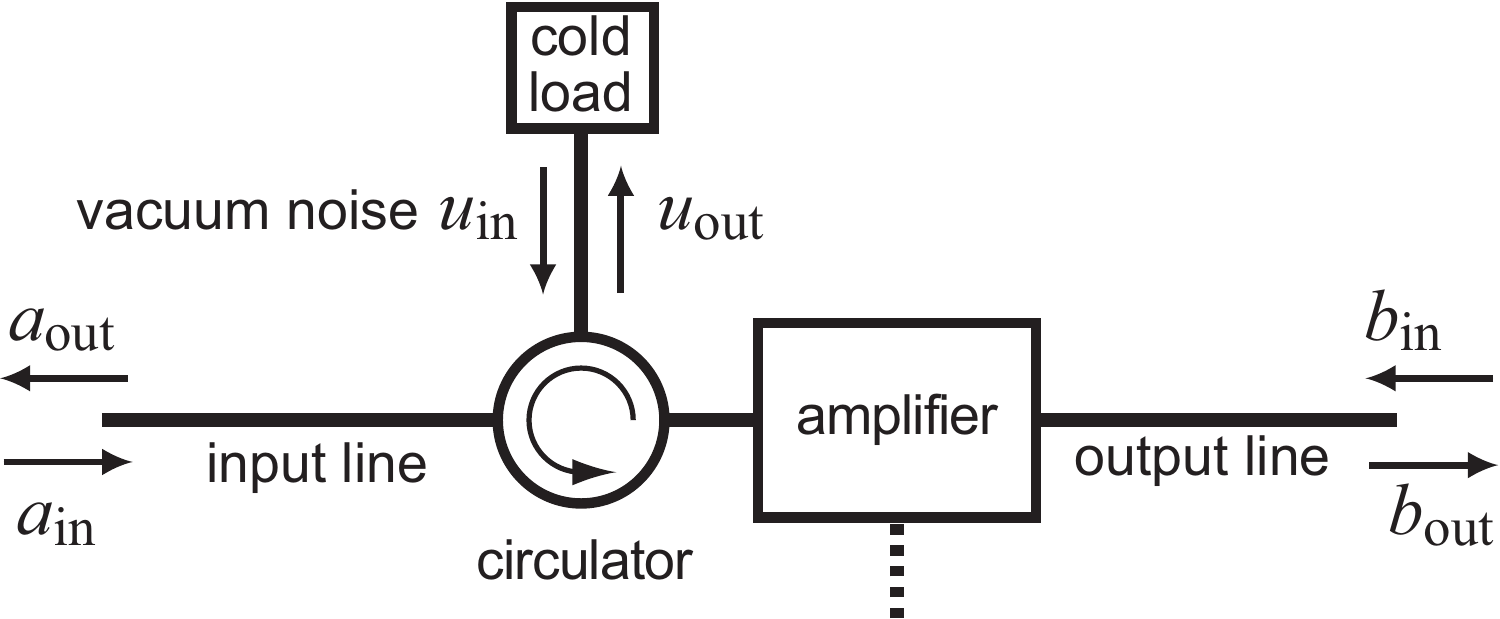}
\caption{Illustration of a bosonic two-port amplifier used in the scattering mode
of operation.  The ``signal" is an incoming wave in the input port of the amplifier, and
does not depend on what is coming out of the amplifier.  This is achieved by connecting the input line to a circulator and a ``cold load" (i.e. a zero temperature resistor):  all that goes back towards the source of the input signal is vacuum noise.}
\label{fig:TwoPortAmpwithcirc}
\end{center}
\end{figure}

We now begin to address the question of how the standard Haus-Caves derivation of the 
amplifier quantum limit presented in Sec.~\ref{subsec:CavesArg} 
relates to the general linear response approach of Sec.~\ref{sec:GenLinResponseTheory}.
Recall that in Sec.~\ref{subsec:measurementofoscillatorposition}, we already used this latter approach to discuss position detection with
a cavity detector, reaching similar conclusions (i.e.~at best, the detector adds noise equal to the zero-point noise).
In that linear-response-based discussion,
we saw that a crucial aspect of
the quantum limit was the trade-off between back-action noise and
measurement imprecision noise.  We saw that reaching the quantum limit required
both a detector with ``ideal" noise, as well as an optimization of the detector-oscillator coupling strength.
Somewhat disturbingly, none of these ideas appeared explicitly in the Haus-Caves derivation;
this can give the misleading impression that the quantum limit never has anything to do with back-action.
A further confusion comes from the fact that many detectors have input and outputs that cannot be described by a set of bosonic modes.  How does one apply the above arguments to such systems?

The first step in resolving these seeming inconsistencies is to
realize that there are really two different ways in which one
can use a given amplifier or detector. In deciding how to
couple the input signal (i.e.~the signal to be amplified) to
the amplifier, and in choosing what quantity to measure, the
experimentalist essentially enforces boundary conditions; as we
will now show, there are in general two distinct ways in which
to do this.  For concreteness, consider the situation depicted
in Fig.~\ref{fig:TwoPortAmp}: a two-port voltage amplifier
where the input and output ports of the amplifier are attached
to one-dimensional transmission lines (see App.~\ref{app:classicalinputoutput}
for a quick review of quantum transmission lines). Similar to the
previous subsection, we focus on a narrow bandwidth signal
centered about a frequency $\omega$.  At this frequency, there
exists both a right-moving and a left-moving wave in each
transmission line.  We label the corresponding amplitudes in
the input (output) line with $\ain, \aout$ ($\bin, \bout$), as
per Fig.~\ref{fig:TwoPortAmp}.
 Quantum mechanically, these amplitudes become operators, much in the same way that
 we treated the mode amplitude $a$ as an operator in the previous subsection.
We will analyze this two-port bosonic amplifier in detail in Sec.~\ref{sec:ScatteringAmp};
here, we will only sketch its operation to introduce the two different amplifier operation modes.
This will then allow us to understand the subtleties
of the Haus-Caves quantum limit derivation.

In the first kind of setup, the experimentalist arranges things so that $\ain$, the amplitude of the wave incident on the amplifier's input port, is {\it precisely}
equal to the signal to be amplified (i.e.~the input signal),
{\it irrespective} of the amplitude of the wave leaving the input port (i.e.~$\aout$).
Further, the output signal is taken to be the amplitude of the outgoing wave exiting the output of the amplifier (i.e.~$\bout$), again, irrespective of whatever might be entering the output port (see Table \ref{tab:TwoAmps}).  In this situation, the Haus-Caves description of the quantum limit in the previous subsection is almost directly applicable;
we will make this precise in Sec.~\ref{sec:ScatteringAmp}.  Back-action is indeed irrelevant, as the prescribed experimental conditions mean that it plays no role.    We will call this mode of operation the ``scattering mode", as it is most relevant to time-dependent experiments where the experimentalist launches a signal pulse at the input of the amplifier and looks at what exits the output port.
One is usually only interested in the scattering mode of operation in
cases where the source producing the input signal is matched to the input of the amplifier:  only in this case is the input wave $\ain$ perfectly transmitted into
the amplifier.   As we will see in Sec.~\ref{sec:ScatteringAmp}, such a perfect matching requires a relatively strong coupling between the signal source and the input of the amplifier;  as such, the amplifier will strongly enhance the damping of the signal source.

The second mode of linear amplifier operation is what we call
the ``op-amp" mode; this is the mode one usually has in mind
when thinking of an amplifier which is weakly coupled to the
signal source, 
and will be the focus of the next two subsections.  The key difference from the ``scattering" mode
is that here, the input signal {\it is not} simply the
amplitude of a wave incident on the input port of the
amplifier; similarly, the output signal {\it is not} the
amplitude of a wave exiting the output port. As such, the
Haus-Caves derivation of the quantum limit does not directly
apply. For the bosonic amplifier discussed here, the op-amp
mode would correspond to using the amplifier as a voltage
op-amp.  The input signal would thus be the voltage at the end
of the input transmission line. Recall that the
voltage at the end of a transmission line involves the
amplitude of {\it both} left and right moving waves, i.e.
$V_a(t) \propto \textrm{Re }[\ain(t) + \aout(t)]$. At first,
this might seem quite confusing:  if the signal source
determines $V_a(t)$, does this mean it sets the value of {\it
both} $\ain(t)$ and $\aout(t)$?  Doesn't this violate
causality?  These fears are of course unfounded.  The signal
source enforces the value of $V_a(t)$ by simply changing
$\ain(t)$ in response to the value of $\aout(t)$.  While there
is no violation of causality, the fact that the signal source
is dynamically responding to what comes out of the amplifier's
input port implies that back-action is indeed relevant.

The op-amp mode of operation is relevant to the typical situation of ``weak coupling" between the signal
source and amplifier input. 
 By ``weak coupling", we mean here something stronger than just requiring that the 
amplifier be linear:  we require additionally that the amplifier does not appreciably change the dissipation of the signal source. 
 This is analogous to the situation in an ideal voltage op-amp, where the amplifier input impedance is much larger than the impedance of the signal source.  We stress the op-amp mode and this limit of weak coupling is the relevant situation in most
electrical measurements.

Thus, we see that the Haus-Caves formulation of the quantum limit is not directly relevant to amplifiers or detectors operated in the usual op-amp mode of operation.
We clearly need some other way to describe quantum amplifiers used in this regime.  
As we will demonstrate in the remainder of this section, the general linear-response
approach of Sec.~\ref{sec:GenLinResponseTheory} is exactly what is needed.  To see this, we will have to expand the discussion of Sec.~\ref{sec:GenLinResponseTheory} to include the concepts of input and output impedance, as well as power gain.  
The linear response approach will allow us to see (similar to  Sec.~\ref{subsec:measurementofoscillatorposition}) that reaching the quantum limit in the op-amp mode does indeed require a trade-off between back-action and measurement imprecision, and requires use of an amplifier with ideal quantum noise properties (cf. Eq.~\ref{NoiseConstraint}).  This approach also has the added benefit of being directly applicable to systems where the input and output of the amplifier are not described by bosonic modes.\footnote{%
Note that the Haus-Caves derivation for the quantum limit of a scattering amplifier has recently been generalized to the case of fermionic operators \cite{Gavish04}.}  
In the next section (Sec.~\ref{sec:ScatteringAmp}), we will return to the scattering description of a two-port voltage amplifier (Fig.~10), and show explicitly how an amplifier can be quantum-limited when used in the scattering mode of operation, but miss the quantum limit when used in the op-amp mode of operation.

\subsection{Linear response description of a position detector}
\label{subsec:PositionDetector}

In this subsection, we will examine the amplifier quantum limit
for a two-port linear amplifier in the usual weak coupling, ``op-amp" regime
of operation.  Our discussion here will make use of the results we have obtained
for the noise properties of a generic linear response detector in
Sec.~\ref{sec:GenLinResponseTheory}, including the fundamental
quantum noise constraint of Eq.~(\ref{NoiseConstraint}).
For simplicity, we will start with the concrete
problem of continuous position detection of a harmonic oscillator.  Our discussion
will thus generalize the discussion of position detection using a cavity detector given in
Sec.~\ref{subsec:measurementofoscillatorposition}.
We start with a generic detector (as introduced in Sec.~\ref{subsec:NoiseConstraint}) coupled
at its input to the position $\hat{x}$ of a harmonic oscillator
(cf.~Eq.~(\ref{HIntGen})).\footnote{For consistency with previous
sections, our coupling Hamiltonian does not have a minus sign.
This is different from the convention of \textcite{Clerk04c},
where the coupling Hamiltonian is written $H_{int} = -A \hat{x}
\cdot \hat{F}$.}
We would like to understand the total output
noise of our amplifier in the presence of the oscillator, and,
more importantly, how small we can make the amplifier's
contribution to this noise.  The resulting lower bound is known
as the standard quantum limit (SQL) on position detection, and is
analogous to the quantum limit on the added noise of a voltage
amplifier (to be discussed in Sec.~\ref{subsec:VoltageAmp}).

\subsubsection{Detector back-action}

We first consider the consequence of noise in the detector input
port. As we have already seen in Sec.~\ref{subsec:QuantumSpectrumAnalyzers}, the
fluctuating back-action force $\hat{F}$ acting on our oscillator
will lead to {\it both} damping and heating of the oscillator.  To
model the intrinsic (i.e.~detector-independent) heating and
damping of the oscillator,  we will also assume that our
oscillator is coupled to an equilibrium heat bath. In the
weak-coupling limit that we are interested in, one can use
lowest-order perturbation theory in the coupling $A$ to describe
the effects of the back-action force $\hat{F}$ on the oscillator.
A full quantum treatment (see~Appendix \ref{app:Langevin}) shows that the oscillator is 
described by an effective classical Langevin equation:\footnote{
    Note that we have omitted a back-action
    term in this equation
    which leads to small renormalizations of the oscillator frequency
    and mass.  These terms are not important for the following
    discussion, so we have omitted them for clarity;
    one can consider $M$ and $\Omega$ in this equation
    to be renormalized quantities.  See
    Appendix \ref{app:Langevin} for more details.}
\begin{eqnarray}
     &&M \ddot{x}(t)  =  -M \Omega^2 x(t) -
     M \gamma_0 \dot{x}(t)+ F_{0}(t)\nonumber\\
     &&- M A^2 \int dt' \gamma(t-t') \dot{x}(t')
      -A \cdot F(t)
     \label{Langevin}
\end{eqnarray}
The position $x(t)$ in the above equation is {\it not} an
operator, but is simply a classical variable whose fluctuations
are driven by the fluctuating forces $F(t)$ and $F_0(t)$.
Nonetheless, the noise in $x$ calculated from Eq.~(\ref{Langevin})
corresponds precisely to $\bS_{xx}[\omega]$, the {\it symmetrized}
quantum mechanical noise in the operator $\hx$. 
The fluctuating force exerted by the detector (which
represents the heating part of the back-action) is described by $A
\cdot F(t)$ in Eq.~(\ref{Langevin}); it has zero mean, and a
spectral density given by $A^2 \bS_{FF}[\omega]$ in Eq.~(\ref{SFSymm}).
The kernel $\gamma(t)$ describes the damping effect of the
detector.  It is given by the {\it asymmetric} part of the
detector's quantum noise, as was derived in
Sec.~\ref{subsec:QuantumSpectrumAnalyzers} (cf.~Eq.~(\ref{OscGamma})). 

Eq.~(\ref{Langevin}) also describes the effects of an equilibrium
heat bath at temperature $T_0$ which models the intrinsic
(i.e.~detector-independent) damping and heating of the oscillator.  The parameter
$\gamma_0$ is the damping arising from this bath, and $F_0$ is the
corresponding fluctuating force.  The spectral density of the
$F_0$ noise is determined by $\gamma_0$ and $T_0$ via the
fluctuation-dissipation theorem (cf.~Eq.~(\ref{FDT})).  $T_0$ and
$\gamma_0$ have a simple physical significance: they are the
temperature and damping of the oscillator when the coupling to the
detector $A$ is set to zero.

To make further progress, we recall from Sec.~\ref{subsec:QuantumSpectrumAnalyzers}
   that even though our detector will in general {\it
not} be in equilibrium, we may nonetheless assign it an effective
temperature $T_{\rm eff}[\omega]$ at each frequency
(cf.~Eq.~(\ref{OscTeff})). 
The effective
temperature of an out-of-equilibrium detector is simply a measure
of the asymmetry of the detector's quantum noise.  We are often
interested in the limit where the internal detector timescales are
much faster than the timescales relevant to the oscillator
(i.e.~$\Omega^{-1}, \gamma^{-1}, \gamma_0^{-1}$). We may then take the
$\omega \ra 0$ limit in the expression for $T_{\rm eff}$,
yielding:
\begin{equation}
     2 \kb T_{\rm eff} \equiv \frac{\bS_{FF}(0)}{M \gamma(0)}
     \label{eq:Teff}
\end{equation}
In this limit, the oscillator position noise calculated from
Eq.~(\ref{Langevin}) is given by:
\begin{eqnarray}
     \bS_{xx}[\omega]
          =    \frac{1}{M}
                 \frac{ 2 (\gamma_0 + \gamma) \kb }
                 {(\omega^2 - \Omega^2)^2 + \omega^2 (\gamma +
\gamma_0)^2}
                 \frac{\gamma_0 T + \gamma T_{\rm eff}}{\gamma_0 +
                 \gamma}
                 \nonumber \\
\end{eqnarray}
This is exactly what would be expected if the oscillator were {\it
only} attached to an equilibrium Ohmic bath with a damping
coefficient $\gamma_{\Sigma} = \gamma_0 + \gamma$ and temperature
$\bar{T} = (\gamma_0 T + \gamma T_{\rm eff}) / \gamma_{\Sigma}$.

\subsubsection{Total output noise}

The next step in our analysis is to link fluctuations in the
position of the oscillator (as determined from
Eq.~(\ref{Langevin})) to noise in the output of the detector. As
discussed in Sec. \ref{subsec:measurementofoscillatorposition},
the output noise consists of the intrinsic output noise of the detector
(i.e.~``measurement imprecision noise") plus the
amplified position fluctuations in the position of the oscillator.
The latter contains both an intrinsic part and a term due to the
response of the oscillator to the backaction.

To start, imagine that we can treat both the oscillator position $x(t)$ and the
detector output $I(t)$ as classically fluctuating quantities.
Using the linearity of the detector's response, we can then write
$\delta I_{\rm total}$, the fluctuating part of the detector's
output, as:
\begin{eqnarray}
     \delta I_{\rm total}[\omega] & = & \delta I_0[\omega] +
     A \chiIF[\omega] \cdot
     \delta x[\omega]
\end{eqnarray}
The first term ($\delta I_0$) describes the intrinsic
(oscillator-independent) fluctuations in the detector output, and
has a spectral density $\bS_{II}[\omega]$.  If we scale this by $| \chiIF |^2$,
we have the measurement imprecision noise discussed in Sec.~\ref{subsec:measurementofoscillatorposition}.  The second term
corresponds to the amplified fluctuations of the oscillator, which
are in turn given by solving Eq.~(\ref{Langevin}):
\begin{eqnarray}
     \delta x[\omega] & = & -\left[ \frac{1/M}{ (\omega^2 - \Omega^2)
+ i
     \omega \Omega / Q[\omega]} \right] (F_0[\omega] - A \cdot
     F[\omega]) \nonumber \\
         & \equiv &
            \oscchi[\omega] (F_0[\omega] - A \cdot
     F[\omega])
     \label{eq:ClassicalDeltax}
\end{eqnarray}
where $Q[\omega] = \Omega / (\gamma_0 + \gamma[\omega])$ is the
oscillator quality factor. It follows that the spectral density of
the total noise in the detector output is given {\it classically}
by:
\begin{eqnarray}
     \clS_{II,{\rm tot}}[\omega] & = &
         \clS_{II}[\omega]\nonumber\\
         & +&
         |\oscchi[\omega] \chiIF[\omega]|^2 \left(A^4 \clS_{FF}[\omega] + A^2
             \clS_{F_0 F_0}[\omega] \right) \nonumber \\
                 &+& 2 A^2 \textrm{Re }
             \left[ \oscchi[\omega] \chiIF[\omega] \clS_{I F}[\omega] \right]
             \label{SItot}
\end{eqnarray}
Here, $\clS_{II}, \clS_{FF}$ and $\clS_{IF}$ are the (classical) detector noise
correlators calculated in the absence of any coupling to the
oscillator.  Note importantly that we have included the fact that
the two kinds of detector noise (in $\hI$ and in $\hF$) may be
correlated with one another.

To apply the classically-derived Eq.~(\ref{SItot}) to our quantum
detector-plus-oscillator system, we recall from
Sec.~\ref{subsec:QuantumSpectrumAnalyzers} that symmetrized quantum noise
spectral densities play the role of classical noise.
The LHS of  Eq.~(\ref{SItot}) thus becomes
$S_{II,{\rm tot}}$,
the total {\it symmetrized} quantum-mechanical output noise of the
detector, while the RHS will now contain the
symmetrized quantum-mechanical detector noise correlators $\bS_{FF}$,
$\bS_{II}$ and $\bS_{IF}$, defined as in Eq.~(\ref{SFSymm}). Though
this may seem rather ad-hoc, one can easily demonstrate that
Eq.~(\ref{SItot}) thus interpreted would be quantum-mechanically
rigorous {\it if} the detector correlation functions obeyed Wick's
theorem. Thus, quantum corrections to Eq.~(\ref{SItot}) will arise
solely from the non-Gaussian nature of the detector noise
correlators. We expect from the central limit theorem that such
corrections will be small in the relevant limit where $\omega$ is
much smaller that the typical detector frequency $\sim \kb T_{\rm
eff} / \hbar$, and neglect these corrections in what follows. Note
that the validity of Eq.~(\ref{SItot}) for a specific model of a
tunnel junction position detector has been explicitly verified in
\textcite{Clerk04b}.

\subsubsection{Detector power gain}
\label{subsubsec:PowerGain}

\begin{figure}[t]
\begin{center}
\includegraphics[width=3.45in]{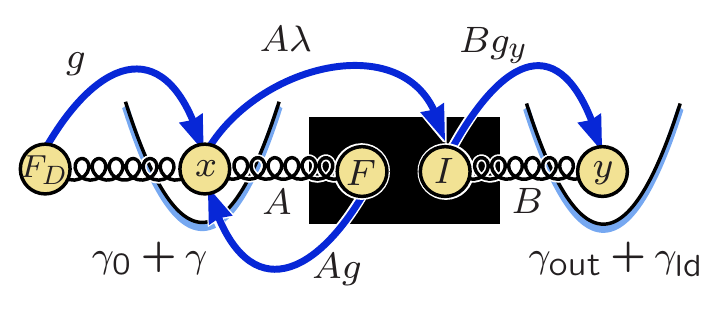}
\caption{(Color online) Schematic of a generic linear-response position detector,
where an auxiliary oscillator $y$ is driven by the detector
output.}
\label{fig:generaldetectorSHO}
\end{center}
\end{figure}

Before proceeding, we need to consider our detector once again in
isolation, and return to the fundamental question of what we mean
by amplification. To be able to say that our detector truly
amplifies the motion of the oscillator, it is not sufficient to
simply say the response function $\chiIF$ must be large (note
that $\chiIF$ is not dimensionless!).  Instead, true
amplification requires that the {\it power} delivered by the
detector to a following amplifier be much larger than the power
drawn by the detector at its input-- i.e., the detector
must have a dimensionless power gain $G_P[\omega]$ much larger
than one.  As we have already discussed, if the power gain was not large, we would need to worry
about the next stage in the amplification of our signal, and how
much noise is added in that process.  Having a large power gain
means that by the time our signal reaches the following amplifier,
it is so large that the added noise of this following amplifier is
unimportant.  The power gain is analogous to the dimensionless
photon number gain  $G$ that appears in the standard Haus-Caves
description of a  bosonic linear amplifier (cf. Eq.~(\ref{eq:fullb})).

To make the above more precise, we start with the idea case of no reverse gain, $\chiFI = 0$.
We will define the power gain $G_P[\omega]$ of our generic
position detector in a way that is analogous to the power
gain of a voltage amplifier.
Imagine we drive the
oscillator we are trying to measure with a force $2 F_{D} \cos \omega t$; this will cause the
output of our detector $\langle \hat{I}(t) \rangle$ to also
oscillate at frequency $\omega$.
To optimally detect this signal
in the detector output, we further couple the detector output $I$ to a second
oscillator with natural frequency $\omega$, mass $M$, and position
$y$: there is a new coupling term in our Hamiltonian, $H'_{int} =
B \hI \cdot \hat{y}$, where $B$ is a coupling strength.  The
oscillations in $\langle I(t) \rangle$ will now act as a driving
force on the auxiliary oscillator $y$ (see Fig~\ref{fig:generaldetectorSHO}).
We can consider the auxiliary oscillator $y$ as a ``load" we are trying to drive with the
output of our detector.

To find the power gain, we need to consider both $P_{\rm out}$, the power supplied to the output oscillator $y$ from
the detector, and $P_{\rm in}$, the power fed into the input of the
amplifier.  Consider first $P_{\rm in}$.  This is simply the time-averaged power
dissipation of the input oscillator $x$ caused by the back-action damping $\gamma[\omega]$.  Using a bar to denote a time average, we have:
\begin{eqnarray}
	P_{\rm in}
		& \equiv &
			M \gamma[\omega] \cdot \overline{\dot{x}^2} 
=
			M \gamma[\omega] \omega^2  |\oscchi[\omega]|^2 F_D^2
\end{eqnarray}
Note that the oscillator susceptibility $\oscchi[\omega]$
depends on both the back-action damping $\gamma[\omega]$ as well as the intrinsic oscillator damping $\gamma_0$ (cf.~Eq.~(\ref{eq:ClassicalDeltax})).

Next, we need to consider the power supplied to the ``load" oscillator $y$ at the detector output.  This oscillator will have some intrinsic, detector-independent damping $\gamma_{\rm ld}$, as well as a back-action damping $\gamma_{\rm out}$.  In the same way that the back-action damping $\gamma$ of the input oscillator $x$ is determined by the
quantum noise in $\hF$ (cf.~Eq.~(\ref{OscLF})-(\ref{OscGamma})), the back-action damping of the load oscillator $y$ is determined by the quantum noise in the output operator $\hI$:
\begin{eqnarray}
     \gamma_{{\rm out}}[\omega]
     & = & \frac{B^2}{M \omega}
     	\left[-	\textrm{Im } \chi_{II}[\omega] \right]
     \nonumber
     \\
         &=&  \frac{B^2}{ M \hbar \omega}
         	\left[ \frac{S_{II}[\omega] - S_{II}[-\omega]}{2} \right]
     \label{gammaout}
\end{eqnarray}
where $\chi_{II}$ is the linear-response susceptibility which determines
how $\langle \hI \rangle$ responds to a perturbation coupling to $\hI$:
\begin{eqnarray}
        \chi_{I I}[\omega] =
        -\frac{i}{\hbar}
         \int_0^{\infty} dt
             \left\langle \left[
                 \hI(t), \hI(0)
                  \right] \right \rangle e^{i \omega t}
\end{eqnarray}
As the oscillator $y$ is being driven on resonance, the relation
between $y$ and $I$ is given by $y[\omega] = \ychi [\omega]
I[\omega]$ with $\ychi[\omega]=-i[\omega M \gamma_{{\rm out}}
[\omega]]^{-1}$.  From conservation of energy, we have that the {\it net} power flow into the output oscillator from the detector is equal to the power dissipated out of the oscillator through the intrinsic damping $\gamma_{\rm ld}$.  We thus have:
\begin{eqnarray}
	P_{\rm out}  & \equiv &
			M \gamma_{\rm ld} \cdot \overline{\dot{y}^2}
				\nonumber \\
		&= &
			M \gamma_{\rm ld} \omega^2  |\ychi[\omega]|^2
			\cdot | B A \chiIF \oscchi[\omega] F_D |^2
				\nonumber \\
		& = &
			\frac{1}{M }
			\frac{ \gamma_{\rm ld}    }
			{\left( \gamma_{\rm ld} + \gamma_{\rm out}[\omega] \right)^2 }
			\cdot | B A \chiIF \oscchi[\omega] F_D |^2
				\nonumber \\
\end{eqnarray}

Using the above definitions, we find that the ratio between $P_{\rm out}$ and
$P_{\rm in}$ is independent of $\gamma_0$, but depends on $\gamma_{\rm ld}$:
\begin{eqnarray}
	\frac{ P_{\rm out}  }
		{ P_{\rm in}  }
	& = &
			\frac{1}{M^2 \omega^2 }
			\frac{A^2 B^2 |\chiIF[\omega] |^2}
			{\gamma_{\rm out}[\omega] \gamma[\omega] }
			\frac{ \gamma_{\rm ld} / \gamma_{\rm out}[\omega]    }
			{\left(1+  \gamma_{\rm ld}/\gamma_{\rm out}[\omega] \right)^2 }
			\nonumber \\
\end{eqnarray}		
We now define the detector power gain $G_P[\omega]$ as the value of this ratio maximized over the choice of $\gamma_{\rm ld}$ .  The maximum occurs for $\gamma_{\rm ld}= \gamma_{\rm out}[\omega]$ (i.e.~the load oscillator is ``matched" to the output of the detector), resulting in:
\begin{eqnarray}
     G_P[\omega]
         & \equiv &
         	\max \left[
             \frac{ P_{{\rm out}} }{ P_{\rm in} }
             \right]
             \nonumber \\
	& = &
			\frac{1}{4 M^2 \omega^2 }
			\frac{A^2 B^2 |\chiIF|^2}{\gamma_{\rm out} \gamma }
			\nonumber \\
         & = &
             \frac{ |\chiIF[\omega]|^2}
                 {4 \textrm{Im } \chi_{FF}[\omega]
                 \cdot \textrm{Im } \chi_{II} [\omega]}
                 \label{GPDefn}
\end{eqnarray}
In the last line, we have used the relation between the damping rates
$\gamma[\omega]$ and $\gamma_{\rm out}[\omega]$ and the linear-response
susceptibilities $\chi_{FF}[\omega]$ and $\chi_{II} [\omega]$
(cf.~Eqs.~(\ref{eq:GammaLambdaF})
and (\ref{gammaout})).  We thus find that
the power gain is a simple dimensionless ratio formed by
the three different response coefficients characterizing the
  detector, and is independent of the coupling constants $A$ and $B$.
As we will see in subsection \ref{subsec:VoltageAmp}, it is
completely analogous to the power gain of a voltage amplifier,
 which is also determined by
three parameters: the voltage gain, the input impedance and the
output impedance.  Note that there are other important measures of power gain commonly in use in the engineering community:  we will comment on these in
Sec.~\ref{subsec:PowerMatching}.


Finally, the above results are easily generalized to the case where the detector's reverse gain
$\chiFI$ is non-vanishing.  For simplicity, we present results for the case where $\beta = \textrm{Re} \left( \chiIF \chiFI \right) / |\chiIF|^2 \geq 0$,
implying that there is no positive feedback.   Maximizing the ratio of $P_{\rm out} / P_{\rm in}$ over choices of 
$\gamma_{ld}$ now yields
\begin{eqnarray}
	G_{P,{\rm rev}} & = &
		\frac{2 G_P}{1 + 2 \beta G_P + \sqrt{1+ 4 \beta G_P} }
		\leq 1 / \beta .
		\label{eq:GPReverse}
\end{eqnarray}
Here, $G_{P,{\rm rev}}$ is the power gain in the presence of reverse gain, while $G_P$ is the zero-reverse gain power gain given by Eq.~(\ref{GPDefn}).  One can confirm that $G_{P,{\rm rev}}$ is a monotonic increasing function of $G_P$, and is bounded by $1/\beta$.
As we have previously noted in 
Sec.\ref{subsubsec:NoiseCrossCorr}, if $\chiFI = \chiIF^*$, there is no additional
quantum noise constraint on our detector beyond what exists classically (i.e.~the RHS of 
Eq.~(\ref{NoiseConstraint}) vanishes).
We now see explicitly that when $\chiFI = \chiIF^*$, the power gain of our detector can be at most $1$, as $\beta=1$.   Thus, while there is no minimum back-action noise required by quantum mechanics in this case, there is also no amplification: at best, our detector would act as a transducer.  Note further that if the detector has $\chiFI = \chiIF^*$ {\it and} optimizes the inequality of Eq.~(\ref{NoiseConstraint}), then one can show $G_{P,{\rm rev}}$ must be one
(cf.~Appendix~\ref{subapp:NoAmplification}): the detector is simply a transducer.  This is in keeping with the results obtained using the Haus-Caves approach, which also yields the conclusion that a noiseless detector is a transducer.

\subsubsection{Simplifications for a quantum-ideal detector}
\label{subsubsec:SimplificationsQuantumIdealDetector}

We now consider the important case where our detector has no reverse gain
(allowing it to have a large power gain), and also has ``ideal"
quantum noise (i.e.~it satisfies the ideal noise condition of
Eq.~(\ref{CorrQLCondition})).  Fulfilling this condition
immediately places some powerful constraints on our detector.

First, note that we have defined in Eq.~(\ref{eq:Teff}) the
effective temperature of our detector based on what happens at the
input port; this is the effective temperature seen by the
oscillator we are trying to measure.  We could also consider the
effective temperature of the detector as seen at the output
(i.e.~by the oscillator $y$ used in defining the power gain).  This
``output" effective temperature is determined by the quantum noise
in the output operator $\hI$:
\begin{eqnarray}
    \kb T_{\rm eff,out}[\omega] \equiv
        \frac{\hbar \omega}
            { \log \left(S_{II}[+\omega] /S_{II}[-\omega] \right) }
\end{eqnarray}
For a general out-of-equilibrium amplifier, $T_{\rm eff,out}$ does
not have to be equal to the input effective temperature $T_{\rm
eff}$ defined by Eq.~(\ref{OscTeff}).  However, for a
quantum-ideal detector, the effective proportionality between
input and output operators (cf.~Eq.~(\ref{PropCond})) immediately
yields:
\begin{eqnarray}
    T_{\rm eff,out}[\omega] = T_{\rm eff}[\omega]
    \label{eq:OneTEff}
\end{eqnarray}
Thus, a detector with quantum-ideal noise necessarily has the same
effective temperature at its input and its output.  This is all
the more remarkable given that a quantum-ideal detector {\it
cannot} be in equilibrium, and thus $T_{\rm eff}$ cannot represent
a real physical temperature.

Another important simplification for a quantum-ideal detector is
the expression for the power gain.  Using the proportionality
between input and output operators (cf.~Eq.~(\ref{PropCond})), one
finds:
\begin{equation}
    G_P[\omega]  =
        \frac{ \left(\textrm{Im } \alpha \right)^2
                    \coth^2 \left( \frac{\hbar \omega}{2 \kb T_{\rm eff}}
                    \right)
        +   \left( \textrm{Re } \alpha \right)^2 }
        {|\alpha|^2 }
        \label{GPTemp}
\end{equation}
where $\alpha[\omega]$ is the parameter characterizing a quantum
limited detector in Eq.~(\ref{AlphaDefn}); recall that
$|\alpha[\omega]|^2$ determines the ratio of $S_{II}$ and $S_{FF}$. It
follows immediately that for a detector with `ideal noise' to also
have a large power gain ($G_P \gg 1$), one absolutely needs $\kb
T_{\rm eff} \gg \hbar \omega$: {\it a large power gain implies a
large effective detector temperature}.  In the large $G_P$ limit,
we have
\begin{equation}
     G_P \simeq \left[ \frac{\textrm{Im } \alpha }{ |\alpha| }
         \frac{\kb T_{\rm eff}}{\hbar \omega/2 }\right]^2
         \label{eq:GPTeff}
\end{equation}
Thus, the effective temperature of a quantum-ideal detector does
more than just characterize the detector back-action-- it also
determines the power gain.

Finally, an additional consequence of the large $G_P[\omega]$,
large $T_{\rm eff}$ limit is that the gain $\chiIF$ and noise
cross-correlator $\bS_{IF}$ are in phase: $\bS_{IF} / \chiIF$ is
purely real, up to corrections which are small as $\omega / T_{\rm
eff}$. This is shown explicitly in Appendix \ref{app:ImPartsGone}.
Thus, we find that  {\it a large power gain detector with ideal
quantum noise cannot have significant out-of-phase correlations
between its output and input noises}.  This last point may be
understood in terms of the idea of wasted information:  if there
were significant out-of-phase correlations between $\hI$ and
$\hF$, it would be possible to improve the performance of the
amplifier by using feedback.  We will discuss this point more
fully in Sec.~\ref{sec:ScatteringAmp}.
Note that as $\bS_{IF} / \chiIF$ is real, the
last term in the quantum noise constraint of
Eq.~(\ref{NoiseConstraint}) vanishes.

\subsubsection{Quantum limit on added noise and noise temperature}

We now turn to calculating the noise added to our signal
(i.e.~$\langle \hx(t) \rangle$) by our generic position detector.  To
characterize this added noise, it is useful to take the total
(symmetrized) noise in the output of the detector, and refer it back to the
input by dividing out the gain of the detector:
\begin{equation}
     \bSxtot[\omega] \equiv \frac{\bS_{II,{\rm tot}}[\omega]}
{A^2 |\chiIF[\omega]|^2}
\end{equation}
$\bSxtot[\omega]$ is simply the frequency-dependent spectral density of position fluctuations
inferred from the output of the detector.
It is this quantity which will directly determine the sensitivity
of the detector-- given a certain detection bandwidth, what is the
smallest variation of $x$ that can be resolved?  The quantity
$\bSxtot[\omega]$ will have contributions both from the
intrinsic fluctuations of the input signal, as well as a
contribution due to the detector.  We first define $\bSxeq[\omega,T]$ to be the
symmetrized equilibrium position noise of our damped oscillator (whose damping is $\gamma_0 + \gamma$) at temperature $T$:
\begin{equation}
     \bSxeq[\omega,T]   =
         \hbar
         \coth\left( \frac{\hbar \omega}{2 \kb T} \right)
         \left[ -\textrm{Im } \oscchi[\omega] \right]
         \label{SxEquilib}
\end{equation}
where the oscillator susceptibility $\oscchi[\omega]$ is defined in Eq.~(\ref{eq:ClassicalDeltax}).  The total inferred position noise may then be written:
\begin{equation}
     \bSxtot[\omega] \equiv
         \left(\frac{\gamma_0}{\gamma_0 + \gamma}\right)
         \cdot    \bSxeq[\omega,T_0] + \bSxadd[\omega]
     \label{SxTot}
\end{equation}
In the usual case where the detector noise can be approximated as being white, this spectral density will consist of a Lorentzian sitting
atop a constant noise floor (cf.~Fig.~\ref{fig:zeroTSHOlineshape})
The first term in Eq.~(\ref{SxTot}) represents position noise arising from the
fluctuating force $\delta F_0(t)$ associated with the intrinsic (detector-independent) dissipation of the oscillator (cf.~Eq.~(\ref{Langevin})).  The prefactor of this term arises because the strength of the intrinsic Langevin force acting on the oscillator is proportional to $\gamma_0$, not to $\gamma_0 + \gamma$.

The second term in Eq.~(\ref{SxTot}) represents the added position noise due
to the detector.  It has contributions from both from the detector's intrinsic output noise $\bS_{II}$ as well as from the detector's back-action noise $\bS_{FF}$, and may be written:
\begin{eqnarray}
     \bSxadd[\omega]   & = &
         \frac{\bS_{II}}{|\chiIF|^2 A^2} + A^2 \left| \oscchi \right|^2 \bS_{FF} +
         \nonumber \\
         &&
         \frac{
         2 \textrm{Re }
             \left[ \chiIF^* \left( \oscchi \right)^*
             \bS_{IF} \right]}
             { |\chiIF|^2}
             \label{Sx}
\end{eqnarray}
For clarity, we have omitted writing the explicit frequency dependence of
the gain $\chiIF$, susceptibility $\oscchi$, and noise correlators; they should all be evaluated at the frequency $\omega$.
Note that the first term on the RHS corresponds to the ``measurement imprecision" noise of our detector,
$\bS_{xx}^I(\omega)$.

We can now finally address the quantum limit on the added noise in this setup.  As discussed in Sec.~\ref{subsec:TwoKindsAmps},
the Haus-Caves derivation of the quantum limit (cf. Sec.~\ref{subsec:CavesArg}) is not
directly applicable to the position detector we are describing here; nonetheless, we
may use its result to guess what form the quantum limit will take here.  The Haus-Caves argument told us that the added noise of a phase-preserving linear amplifier must be
at least as large as the zero point noise.  We thus anticipate
that if our
detector has a large power gain, the spectral density of the noise
added by the detector (i.e.~$\bSxadd[\omega]$) must be at least as big
as the zero point noise of our damped oscillator:
\begin{equation}
     \bSxadd[\omega] \geq
         \lim_{T\ra0}
         \bSxeq[w,T] =
     \left|
         \hbar \textrm{Im } \oscchi[\omega] \right|
         \label{eq:SxxSQL}
\end{equation}
We will now show that the bound above is rigorously
correct at each frequency $\omega$.

The first step is to examine the dependence of the added noise
$\bSxadd[\omega]$ (as given by Eq.~(\ref{Sx})) on the coupling
strength $A$.  If we ignore for a moment the detector-dependent
damping of the oscillator, the situation is the same as the cavity
position detector of Sec.~\ref{subsec:measurementofoscillatorposition}:
there is an optimal
value of the coupling strength $A$ 
which corresponds to a trade-off between
imprecision noise and back-action (i.e.~first and second terms in Eq.~(\ref{Sx})).
We would thus expect $\bSxadd[\omega]$ to attain a minimum
value at an optimal choice of coupling $A = A_{{\rm opt}}$ where
both these terms make equal contributions (see Fig.~\ref{fig:optimalcoupling}).
Defining $\phi[\omega] = \arg \oscchi[\omega]$, we thus have the bound:
\begin{equation}
         \bSxadd[\omega]  \geq  2 | \oscchi[\omega] | \left[
             \sqrt{ \frac{\bS_{II} \bS_{FF}}{ |\chiIF|^2} }
             +   \frac{
                 \textrm{Re }
                     \left[
                         \chiIF^* e^{-i \phi[\omega]} \bS_{IF}
                     \right]}
                 {|\chiIF|^2}
         \right]
         \label{FirstMin}
\end{equation}
where the minimum value at frequency $\omega$ is achieved when:
\begin{equation}
     A^2_{{\rm opt}} = \sqrt{ \frac{\bS_{II}[\omega] }
         {|\chiIF[\omega] \oscchi[\omega]|^2 \bS_{FF}[\omega]}}
     \label{AOptOffRes}
\end{equation}
Using the inequality
$X^2 + Y^2 \geq 2 |X Y|$ we see that this value serves as a lower
bound on $\bSxadd$ even in the presence of detector-dependent damping.
In the case where the detector-dependent damping is negligible,
the RHS of Eq.~(\ref{FirstMin}) is independent of $A$, and thus
Eq.~(\ref{AOptOffRes}) can be satisfied by simply tuning the
coupling strength $A$; in the more general case where there is
detector-dependent damping, the RHS is also a function of $A$
(through the response function $\oscchi[\omega]$), and it may no longer
be possible to achieve Eq.~(\ref{AOptOffRes}) by simply tuning
$A$.\footnote{Note that in the heuristic discussion of position detection
using a resonant cavity detector in Sec.~\ref{subsec:measurementofoscillatorposition},
these concerns did not arise as there was no back-action damping.}

While Eq.~(\ref{FirstMin}) is certainly a bound on the added
displacement noise $\bSxadd[\omega]$, it does not in itself represent
the quantum limit.  Reaching the quantum limit requires more than
simply balancing the detector back-action and intrinsic output
noises (i.e.~the first two terms in Eq.~(\ref{Sx})); {\it one
also needs a detector with ``quantum-ideal" noise properties, that
is a detector which satisfies Eq.~(\ref{CorrQLCondition})}.
Using the
quantum noise constraint of Eq.~(\ref{NoiseConstraint}) to further
bound $\bSxadd[\omega]$, we obtain:
\begin{eqnarray}
         \bSxadd[\omega]  & \geq&
          2 \left| \frac{\oscchi[\omega] }{\chiIF} \right| \times
          \nonumber \\
         &&
             \Bigg[
             \sqrt{
                 \left(\frac{\hbar | \chiIF |}{2} \right)^2
                 \left(1 + \Delta \left[ \frac{ 2 \bS_{IF}}{\hbar \chiIF} \right]
                 \right) +
                 \left| \bS_{IF} \right|^2 }
             \nonumber \\
             && +   \frac{
                 \textrm{Re }
                     \left[
                         \chiIF^* e^{-i \phi[\omega]} \bS_{IF}
                     \right]}
                 {|\chiIF|}
         \Bigg]
         \label{SecondMin}
\end{eqnarray}
where the function $\Delta[z]$ is defined in
Eq.~(\ref{eq:DeltaFactor}). The minimum value of $\bSxadd[\omega]$ in
Eq.~(\ref{SecondMin}) is now achieved when one has {\it both} an
optimal coupling (i.e.~Eq.~(\ref{AOptOffRes})) {\it and} a quantum
limited detector, that is one which satisfies
Eq.~(\ref{NoiseConstraint}) as an equality. 

Next, we consider the relevant case where our detector is a
good amplifier and has a power gain $G_P[\omega] \gg 1$ over
the width of the oscillator resonance. As we have discussed,
this implies that the ratio $\bS_{IF} /  \chiIF$ is purely
real, up to small $\hbar \omega / (\kb T_{\rm eff} )$
corrections  (see Sec.~(\ref{subsubsec:NoiseCrossCorr}) and
Appendix \ref{app:ImPartsGone} for more details).  This in turn
implies that $\Delta[2 \bS_{IF} / \hbar \chiIF] =0 $;  we thus
have:
\begin{eqnarray}
         \bSxadd[\omega]  \geq \nonumber\\
           2 | \oscchi[\omega] |
           \left[
             \sqrt{
                 \left(\frac{\hbar }{2} \right)^2 +
                 \left(  \frac{\bS_{IF}}{ \chiIF }\right)^2}
             +   \frac{
                 \cos \left[\phi[\omega]\right] \bS_{IF}
                     }
                 {\chiIF}
         \right]
         \label{ThirdMin}
\end{eqnarray}
Finally, as there is no further constraint on $ \bS_{IF} /
\chiIF$ (beyond the fact that it is real), we can minimize the
expression over its value.  The minimum $\bSxadd[\omega]$ is
achieved for a detector whose cross-correlator satisfies:
\begin{equation}
     \frac{\bS_{IF}[\omega]}{\chiIF} \Big|_{\rm optimal} =
         -\frac{\hbar}{2} \cot \phi[\omega],
         \label{OptSIF}
\end{equation}
with the minimum value being given by:
\begin{equation}
\bSxadd[\omega] \Big|_{min} = \hbar | \textrm{Im }\oscchi[\omega] | =
     \lim_{T\ra0} \bSxeq[\omega,T]
     \label{CCBound}
\end{equation}
where $\bSxeq[\omega,T]$ is the equilibrium contribution to
$\bSxtot[\omega]$ defined in Eq.~(\ref{SxEquilib}).  Thus,
in the limit of a large power gain, we have that {\em at each
frequency, the minimum displacement noise added by the detector is
precisely equal to the noise arising from a zero temperature
bath}.  This conclusion is irrespective of the strength of the
intrinsic (detector-independent) oscillator damping.

We have thus derived the amplifier quantum limit (in the context
of position detection) for a two-port amplifier used in the ``op-amp" mode
of operation.  Though we reached a conclusion similar to that given
by the Haus-Caves approach, the linear-response, quantum noise approach
used is quite different.  This approach makes explicitly clear what is needed
to reach the quantum limit.  We find that to reach the quantum-limit
on the added displacement noise $\bSxadd[\omega]$ with a large power
gain, one needs:
\begin{enumerate}
\item A quantum limited detector, that is a detector which
satisfies the ``ideal noise" condition of
Eq.~(\ref{CorrQLCondition}), and hence the proportionality
condition of Eq.~(\ref{PropCond}).

\item A coupling $A$ which satisfies Eq.~(\ref{AOptOffRes}).

\item A detector cross-correlator $\bS_{IF}$ which satisfies
Eq.~(\ref{OptSIF}).
\end{enumerate}
Recall that condition (1) is identical to what is required for
quantum-limited detection of a qubit; it is rather demanding, and
requires that there is no ``wasted" information about the input
signal in the detector which is not revealed in the output
\cite{Clerk03}. Also note that $\cot \phi$ changes quickly as a
function of frequency across the oscillator resonance, whereas
$\bS_{IF}$ will be roughly constant; condition (2) thus implies
that it will not be possible to achieve a minimal
$\bSxadd[\omega]$
across the entire oscillator resonance.  A more reasonable goal is
to optimize $\bSxadd[\omega]$ at resonance, $\omega = \Omega$.  As $\oscchi[\Omega]$
is imaginary, Eq.~(\ref{OptSIF}) tells us that $\bS_{IF}$ should
be zero. Assuming we have a quantum-limited detector with a large
power gain ($\kb T_{\rm eff} \gg \hbar \Omega$), the remaining
condition on the coupling $A$ (Eq.~(\ref{AOptOffRes})) may be
written as:
\begin{equation}
     \frac{ \gamma[A_{\rm opt}]}{\gamma_0 + \gamma[A_{\rm opt}]} =
     \left| \frac{\textrm{Im } \alpha}{\alpha}  \right|
     \frac{1}{2 \sqrt{G_P[\Omega]}}
     = \frac{\hbar \Omega}{4 \kb T_{\rm eff}}
     \label{OptA2}
\end{equation}

As $\gamma[A] \propto A^2$ is the detector-dependent damping of
the oscillator, we thus have that {\it to achieve the
quantum-limited value of $\bSxadd[\Omega]$ with a large power gain,
one needs the intrinsic damping of the oscillator to be much
larger than the detector-dependent damping}.  The
detector-dependent damping must be small enough to compensate the
large effective temperature of the detector; if the bath
temperature satisfies $\hbar \Omega / \kb \ll T_{{\rm bath}} \ll
T_{\rm eff}$, Eq.~(\ref{OptA2}) implies that at the quantum limit,
the temperature of the oscillator will be given by:
\begin{equation}
     T_{{\rm osc}} \equiv \frac{ \gamma \cdot T_{\rm eff} +
         \gamma_0 \cdot T_{{\rm bath}}}
     { \gamma  + \gamma_0}
     \ra
     \frac{\hbar \Omega}{4 \kb} + T_{{\rm bath}}
     \label{eq:QLheating}
\end{equation}
Thus, at the quantum limit and for large $T_{\rm eff}$, the
detector raises the oscillator's temperature by $\hbar \Omega / 4
\kb$.\footnote{ If in contrast our oscillator was initially at
zero temperature (i.e. $T_{bath} = 0$), one finds that the effect
of the back-action (at the quantum limit and for $G_P \gg 1$) is
to heat the oscillator to a temperature $\hbar \Omega /  (\kb \ln
5)$.}
As expected, this additional heating is only {\it half} the
zero-point energy; in contrast, the quantum-limited value of
$\bSxadd[\omega]$ corresponds to the full zero-point result, as it
also includes the contribution of the intrinsic output noise of
the detector.

Finally, we return to Eq.~(\ref{SecondMin}); this is the
constraint on the added noise $\bSxadd[\omega]$ {\it before} we
assumed our detector to have a large power gain, and consequently
a large $T_{\rm eff}$.  Note crucially that {\it if} we did not require a
large power gain, then there need not be {\it any} added noise.
Without the assumption of a large power gain, the ratio $\bS_{IF}
/ \chiIF$ can be made imaginary with a large magnitude.  In this
limit, $1+\Delta[ 2 \bS_{IF} / \chiIF] \ra 0$: the quantum
constraint on the amplifier noises (e.g. the RHS of
Eq.~(\ref{NoiseConstraint})) vanishes.  One can then easily use
Eq.~(\ref{SecondMin}) to show that the added noise $\bSxadd[\omega]$
can be zero.  This confirms a general conclusion that we have seen several
times now (cf.~Secs.~\ref{subsubsec:NoiseCrossCorr}, \ref{subsec:CavesArg}):  
if a detector does not amplify (i.e.~the power gain is unity), it need not produce any added noise.

\subsection{Quantum limit on the noise temperature of a voltage amplifier}
\label{subsec:VoltageAmp}

\begin{figure}[t]
\begin{center}
\includegraphics[width=3.45in]{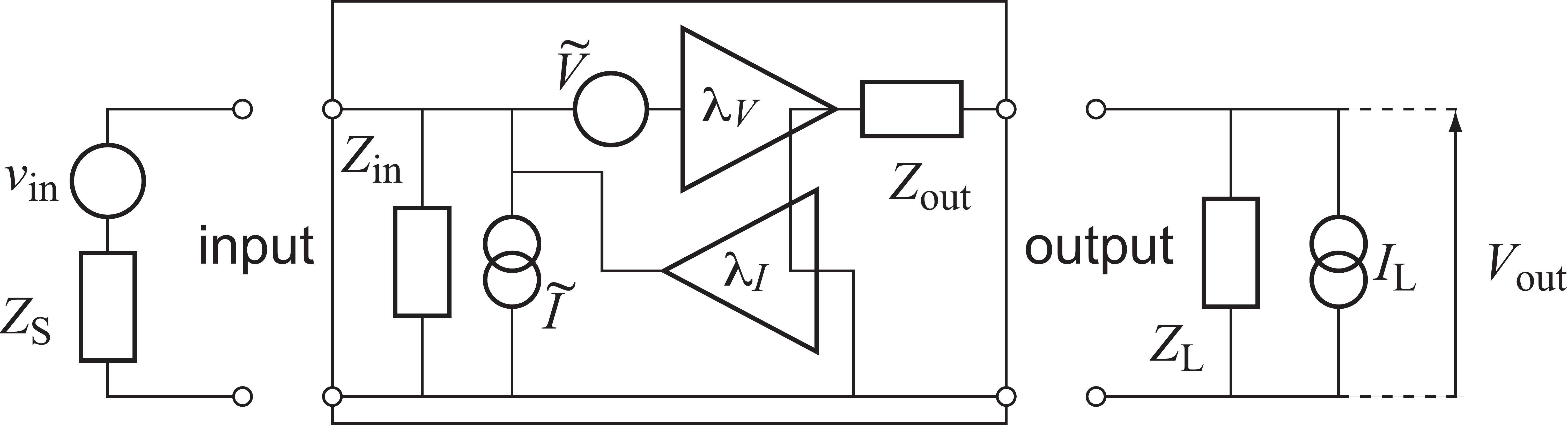}
\caption{Schematic of a linear voltage amplifier, including a 
reverse gain $\lambda_I$.  
$\tilde{V}$ and $\tilde{I}$ represent the standard voltage and current noises
of the amplifier, as discussed further in the text.  The case with reverse gain is discussed in detail
in Sec.~\ref{sec:ScatteringAmp}.}
\label{fig:VoltageAmp}
\end{center}
\end{figure}

We now turn attention to the quantum limit on the
added noise of a generic linear voltage amplifier used in the ``op-amp" mode of
operation (see, e.g.~, \textcite{Devoret00}). For such amplifiers, the added noise is usually
expressed in terms of the ``noise temperature" of the amplifier;
we will define this concept, and demonstrate that, when
appropriately defined, this noise temperature must be bigger than
$\hbar \omega / (2 \kb)$, where $\omega$ is the signal frequency.
Though the voltage amplifier is closely analogous to the position
detector treated in the previous section, its importance makes it
worthy of a separate discussion.
Similar to the last subsection, our discussion here will use the general linear-response approach.
In contrast, in Sec.~\ref{sec:ScatteringAmp}, we will present the bosonic scattering description of a two-port voltage amplifier,
a description similar to that used in formulating the Haus-Caves
proof of the amplifier quantum limit.  We will then be in a good position to contrast the linear-response and scattering approaches, and will clearly see there that the scattering mode of operation and the ``op-amp"
mode of operation discussed here are not equivalent.  
We stress that the general treatment presented here can also be applied directly to the system
discussed in Sec.~\ref{sec:ScatteringAmp}.

\subsubsection{Classical description of a voltage amplifier}

Let us begin by recalling the standard schematic description of a
voltage amplifier, see Fig.~\ref{fig:VoltageAmp}.  
The input voltage to be amplified
$\vin(t)$ is produced by a circuit which has a
Thevenin-equivalent impedance $\Zs$, the source impedance.
We stress that we are considering the ``op-amp" mode of amplifier operation,
and thus the input signal does not correspond to the amplitude of a wave incident upon
the amplifier (see Sec.~\ref{subsec:TwoKindsAmps}).
The
amplifier itself has an input impedance $\Zin$ and an output
impedance $\Zout$, as well as a voltage gain coefficient $\lambda_V$:
assuming no current is drawn at the output (i.e. $Z_{{\rm load}} \ra \infty$ in Fig.~\ref{fig:VoltageAmp}), the output voltage $V_{\rm{out}}(t)$ is simply
 $\lambda_V$
times the voltage across the input terminals of the amplifier.

The added noise of the amplifier is standardly represented by two
noise sources placed at the amplifier input.  There is both a
voltage noise source $\VN(t)$ in series with the input
voltage source,  and a current noise source $\IN(t)$ in
parallel with input voltage source (Fig.~\ref{fig:VoltageAmp}).
The voltage noise produces a
fluctuating voltage $\VN(t)$ (spectral density
$\clS_{\VN \VN}[\omega]$) which simply adds to the signal voltage
at the amplifier input, and is amplified at the output; as such,
it is completely analogous to the intrinsic detector output noise
$\clS_{II}$ of our linear response detector.  In contrast, the current
noise source of the voltage amplifier represents back-action:
this fluctuating current (spectral density $\clS_{\IN \IN}[\omega]$) flows back across the parallel
combination of the source impedance and amplifier input impedance,
producing an additional fluctuating voltage at its input.  The
current noise is thus analogous to the back-action noise $\clS_{FF}$ of
our generic linear response detector.

Putting the above together, the total voltage at the input
terminals of the amplifier is:
\begin{eqnarray}
    v_{\rm{in,tot}}(t) & = &
    \frac{\Zin}{\Zin + \Zs}
        \left[
            \vin(t) + \VN(t)
        \right] - \frac{\Zs \Zin }{ \Zs+ \Zin } \IN(t)
            \nonumber \\
     & \simeq & \vin(t) + \VN(t) -\Zs \IN(t)
     \label{eq:ClassVNoise}
\end{eqnarray}
In the second line, we have taken the usual limit of an ideal
voltage amplifier which has an infinite input impedance (i.e. the
amplifier draws zero current).  The spectral density of the total
input voltage fluctuations is thus:
\begin{eqnarray}
    \clS_{VV,{\rm tot}}[\omega] = \clS_{\vin \vin}[\omega] + \clS_{VV,{\rm add}}[\omega].
\end{eqnarray}
Here $\clS_{v_{{\rm in}} v_{{\rm in}}}$ is the spectral density of the voltage
fluctuations of the input signal $v_{{\rm in}}(t)$, and $S_{VV,{\rm add}}$ is
the amplifier's contribution to the total noise at the input:
\begin{eqnarray}
    \clS_{VV,{\rm add}}[\omega] = \clS_{\VN \VN} +  \left|\Zs\right|^2 \clS_{\IN \IN} -
        2 \textrm{Re} \left[ \Zs^*  \clS_{\VN \IN} \right]
        \nonumber \\
    \label{eq:SVadd}
\end{eqnarray}
For clarity, we have dropped the frequency index for the spectral
densities appearing on the RHS of this equation.

It is useful to now consider a narrow bandwidth input signal at a
frequency $\omega$, and ask the following question:
 {\it if} the signal source was simply an equilibrium resistor at a temperature $T_0$, how much hotter would it
 have to be to produce a voltage noise equal to $\clS_{VV,{\rm tot}}[\omega]$?  The resulting increase in the source
 temperature is defined as the noise temperature $\TN[\omega]$ of the amplifier and is a convenient measure
 of the amplifier's added noise.  It is standard among engineers to define the noise temperature assuming the
 initial temperature of the resistor $T_0 \gg \hbar \omega$.
One may then use the classical expression for the thermal noise of
a resistor, which yields the definition:
\begin{eqnarray}
    2 \textrm{Re } \Zs \cdot \kb \TN [\omega] \equiv \clS_{VV,{\rm tot}}[\omega]
    \label{eq:TNDefinition}
\end{eqnarray}
Writing $\Zs = |\Zs| e^{i \phi}$, we have:
\begin{equation}
     2 \kb \TN =
         \frac{1}{\cos \phi} \left[
             \frac{ \clS_{\VN \VN } } {|\Zs| } + |\Zs| \clS_{ \IN \IN}
             - 2 \textrm{Re } \left(
                e^{-i  \phi} \clS_{\VN \IN} \right) \right]
                 \label{eq:TNeqn}
\end{equation}
It is clear from this expression that $\TN$ will have a minimum
as a function of $|\Zs|$.  For $|\Zs|$ too large, the back-action
current noise of the amplifier will dominate $\TN$, while for
$|\Zs|$ too small, the voltage noise of the amplifier (i.e.~its
intrinsic output noise will dominate).  The situation is
completely analogous to the position detector of the last section;
there, we needed to optimize the coupling strength $A$ to balance
back-action and intrinsic output noise contributions, and thus
minimize the total added noise.  Optimizing the source impedance
thus yields a completely classical minimum bound on $\TN$:
\begin{equation}
     \kb \TN \geq \sqrt{
         \clS_{\VN \VN} \clS_{ \IN \IN} -
            \left[ \textrm{Im  } \clS_{\VN \IN } \right]^2 }  -
             \textrm{Re  } \clS_{\VN \IN}
    \label{eq:TNopt}
\end{equation}
where the minimum is achieved for an optimal source impedance
which satisfies:
\begin{eqnarray}
     |\Zs[\omega]|_{{\rm opt}} & = &
     \sqrt{ \frac
        { \clS_{\VN \VN} [\omega] }
       { \clS_{\IN \IN } [\omega]} }
       \equiv \ZN
     \label{ZOpt1} \\
     \sin \phi[\omega] \big|_{{\rm opt}} & = &
        - \frac{  \textrm{Im } \clS_{\VN  \IN } [\omega]}
         { \sqrt{S_{\VN \VN} [\omega] \clS_{\IN \IN} [\omega]} }
         \label{ZOpt2}
\end{eqnarray}
The above equations define the so-called noise impedance $\ZN$.
We stress again that the discussion so far in this subsection has
been completely classical.

\subsubsection{Linear response description}

It is easy to connect the classical description of a voltage
amplifier to the quantum mechanical description of a generic
linear response detector; in fact, all that is needed is a
``relabeling" of the concepts and quantities we introduced in
Sec.~\ref{subsec:PositionDetector} when discussing a linear
position detector.  Thus, the quantum voltage amplifier will be
characterized by both an input operator $\hQ$ and an output
operator $\hV_{\rm{out}}$; these play the role, respectively, of $\hF$
and $\hI$ in the position detector.  $\hV_{\rm{out}}$ represents the
output voltage of the amplifier, while $\hQ$ is the operator which
couples to the input signal $\vin(t)$ via a coupling
Hamiltonian:
\begin{equation}
     {\hat H}_{int} = \vin(t) \cdot \hQ
     \label{eq:VoltAmpHint}
\end{equation}
In more familiar terms,  $\hat{ \tilde{I}}_{\rm in} -d \hQ / dt$
represents the current flowing into the amplifier.\footnote{
     Note that one could have instead written the coupling Hamiltonian
     in the more traditional form
     $\hat{H}_{int}(t) = \phi(t) \cdot \hat{\tilde{I}}_{\rm{in}}$, where $\phi = \int
dt' \vin(t')$ is the flux
     associated with the input voltage.  The linear response
     results we obtain are exactly the same.  We prefer to work with
the charge $\hat{Q}$ in
     order to be consistent with the rest of the text.}
The voltage gain of our amplifier $\lambda_V$ will again be
given by the Kubo formula of Eq.~(\ref{eq:gain}), with the
substitutions $\hF \ra \hQ, \hI \ra \hV_{{\rm out}}$ (we will
assume these substitutions throughout this section).

We can now easily relate the fluctuations of the input and output
operators to the noise sources used to describe the classical
voltage amplifier.  As usual, symmetrized quantum noise spectral densities $\bS[\omega]$
will play the role of the classical spectral densities $\clS[\omega]$ appearing in the classical
description.  First, as the operator $\hQ$ represents a back-action
force, its fluctuations correspond to the amplifier's current
noise $\IN(t)$:
\begin{eqnarray}
    \clS_{ \IN \IN }[\omega] \leftrightarrow \omega^2 \bS_{QQ}[\omega].
    \label{eq:SIsub}
\end{eqnarray}
Similarly, the fluctuations in the operator $\hV_{\rm{out}}$, when
referred back to the amplifier input, will correspond to the
voltage noise $\VN (t)$ discussed above:
\begin{eqnarray}
    \clS_{\VN \VN}[\omega] \leftrightarrow \frac{ \bS_{V_{\rm{out}} V_{\rm{out}} }[\omega]}{| \lambda_V|^2}
    \label{eq:SVsub}
\end{eqnarray}
A similar correspondence holds for the cross-correlator of these
noise sources:
\begin{eqnarray}
    \clS_{\VN \IN }[\omega] \leftrightarrow + i \omega \frac{ \bS_{V_{\rm{out}} Q} [\omega]}{\lambda_V}
    \label{eq:SVIsub}
\end{eqnarray}

To proceed, we need to identify the input and output impedances of
the amplifier, and then define its power gain.  The first step in
this direction is to assume that the output of the amplifier
($\hV_{\rm out}$) is connected to an external circuit via a term:
\begin{equation}
     \hat{H}'_{int} = q_{{\rm out}}(t) \cdot \hV_{\rm{out}}
     \label{eq:VAmpCoupling}
\end{equation}
where $\tilde{i}_{{\rm out}} = d q_{{\rm out}} / dt$ is the
current in the external circuit. We may now identify the input and
output impedances of the amplifier in terms of the damping at the
input and output.  Using the Kubo formulae for conductance and
resistance yields (cf.~Eq.~(\ref{OscGamma}) and
Eq.~(\ref{gammaout}), with the substitutions $\hF \ra \hQ$ and
$\hI \ra \hV$):
\begin{eqnarray}
     1/ \Zin [\omega] & = &
         i \omega \chi_{QQ}[\omega]
         \label{eq:VAmpZin}\\
     \Zout[\omega] & = &
         \frac{\chi_{VV}[\omega]}{-i \omega}
         \label{eq:VAmpZout}
\end{eqnarray}
i.e.~ $\langle \tilde{I}_{in} \rangle_{\omega} =
\frac{1}{\Zin[\omega]} \vin [\omega]$ and $ \langle V
\rangle_{\omega} = \Zout[\omega] \tilde{i}_{{\rm
out}}[\omega]$, where the subscript $\omega$ indicates the Fourier
transform of a time-dependent expectation value.

We will consider throughout this section the case of no reverse
gain, $\chi_{Q V_{\rm out} } = 0$.  We can define the power gain $G_P$ exactly as we did
in Sec.~\ref{subsubsec:PowerGain} for a linear position detector.  $G_P$ is defined as the ratio of the power delivered to a load attached to the amplifier output divided by the power drawn by the amplifier, maximized over the impedance of the load.  One finds:
\begin{equation}
     G_P
         = \frac{ | \lambda_V |^2 }{4 \textrm{Re }(Z_{{\rm out}})
         \textrm{Re } (1/ \Zin )}
     \label{GPVoltAmpA}
\end{equation}
Expressing this in terms of the linear response coefficients $\chi_{VV}$
and $\chi_{QQ}$, we obtain an expression
which is completely analogous to Eq.~(\ref{GPDefn})
for the power gain for a position detector:
\begin{equation}
     G_P = \frac{ |\lambda_V|^2}
        {4 \textrm{Im } \chi_{QQ} \cdot \textrm{Im } \chi_{VV}}
     \label{GPVoltAmp}
\end{equation}
Finally, we may again define the effective temperature $T_{\rm
eff}[\omega]$ of the amplifier via Eq.~(\ref{OscTeff}), and define
a quantum-limited voltage amplifier as one which satisfies the
ideal noise condition of Eq.~(\ref{CorrQLCondition}).  For such an
amplifier, the power gain will again be determined by the
effective temperature via Eq.~(\ref{GPTemp}).

Turning to the noise, we can again calculate the total symmetrized
noise at the output port of the amplifier following the same
argument used to the get the output noise of the position detector
(cf.~Eq.~(\ref{SItot})). As we did in the classical approach, we
will again assume that the input impedance of the amplifier is
much larger that source impedance: $\Zin \gg \Zs$; we will
test this assumption for consistency at the end of the
calculation. Focusing only on the amplifier contribution to this
noise (as opposed to the intrinsic noise of the input signal), and
referring this noise back to the amplifier input, we find that the symmetrized
quantum noise spectral density describing the added noise of the amplifier,
$\bS_{VV,{\rm add}}[\omega]$, satisfies the same equation we found for a classical
voltage amplifier, Eq.~(\ref{eq:SVadd}), with each classical spectral density $\clS[\omega]$
being replaced by the corresponding symmetrized quantum spectral density $\bS[\omega]$
as per Eqs.~(\ref{eq:SIsub}) - (\ref{eq:SVIsub}).

It follows that the amplifier noise temperature will again be
given by Eq.~(\ref{eq:TNeqn}), and that the optimal noise
temperature (after optimizing over the source impedance) will be
given by Eq.~(\ref{eq:TNopt}).
 Whereas classically nothing more could be said, quantum mechanically, we now get a further bound from the quantum
 noise constraint of Eq. (\ref{NoiseConstraint}) and the requirement of a large power gain.  The latter requirement
 tells us that the voltage gain $\lambda_V[\omega]$ and the cross-correlator $\bS_{V_{out}Q}$ must be in phase
 (cf.~Sec.~(\ref{subsubsec:NoiseCrossCorr}) and Appendix \ref{app:ImPartsGone}).  This in turn means that
  $\bS_{\VN \IN }$ must be purely imaginary.  In this case, the quantum noise constraint may be
  re-written as:
\begin{eqnarray}
    \bS_{\VN \VN}[\omega] \bS_{\IN \IN}[\omega] - \left[
        \textrm{Im } \bS_{\VN \IN} \right]^2 \geq
        \left( \frac{\hbar \omega}{2} \right)^2
        \label{eq:VAmpNoiseConstraint}
\end{eqnarray}
Using these results in Eq.~(\ref{eq:TNopt}), we find the ultimate
quantum limit on the noise temperature:\footnote{
     Note that our definition of the noise temperature
     $\TN$ conforms with that of \textcite{Devoret00} and most
     electrical engineering texts,
     but is slightly different
     than that of \textcite{Caves82}.  Caves assumes the source is
     initially at zero temperature (i.e. $T_0 = 0$), and consequently
     uses the full quantum expression for its equilibrium noise.
     In contrast, we have assumed that $\kb T_0 \gg \hbar \omega$.
     The different definition of the noise temperature used by Caves leads to the result
     $\kb T_N \geq \hbar \omega / (\ln 3)$ as opposed to our Eq.~(\ref{eq:TNQL}).  
     We stress that the difference between these results
     has nothing to do with physics, but only with how one defines the
     noise temperature.
}
\begin{eqnarray}
    \kb \TN[\omega] \geq \frac{\hbar \omega}{2}
\label{eq:TNQL}
\end{eqnarray}
Similar to the case of the position detector, reaching the quantum
limit here is {\it not} simply a matter of tuning the coupling
(i.e.~tuning the source impedance $\Zs$ to match the noise
impedance, cf.~Eq.~(\ref{ZOpt1}) - (\ref{ZOpt2})); one also needs
to have an amplifier with ``ideal" quantum noise, that is an amplifier 
satisfying Eq.~(\ref{CorrQLCondition}). 

Finally, we need to test our initial assumption that $|\Zs| \ll
|\Zin|$,  taking $|\Zs|$ to be equal to its optimal value $\ZN$.
Using the proportionality condition of Eq.~(\ref{PropCond}) and
the fact that we are in the large power gain limit ($G_P[\omega]
\gg 1$), we find:
\begin{equation}
     \left|
         \frac{\ZN[\omega]}{ \textrm{Re } \Zin [\omega] }
     \right |  =
     \left | \frac{  \alpha }{\textrm{Im } \alpha} \right |
     \frac{\hbar \omega}{4 \kb T_{\rm eff}} =
     \frac{1}{2 \sqrt{G_P[\omega]} } \ll 1
     \label{ZinConstraint}
\end{equation}
It follows that $|\ZN| \ll |\Zin|$ in the large power gain,
large effective temperature regime of interest, thus justifying
the form of Eq.~(\ref{eq:SVadd}). Eq.~(\ref{ZinConstraint}) is
analogous to the case of the displacement detector, where we found
that reaching the quantum limit on resonance required the
detector-dependent damping to be much weaker than the intrinsic
damping of the oscillator (cf.~Eq.~(\ref{OptA2})).

Thus, similar to the situation of the displacement detector, the
linear response approach allows us both to derive rigorously the
quantum limit on the noise temperature $\TN$ of an amplifier, and
to state conditions that must be met to reach this limit.  To
reach the quantum-limited value of $\TN$ with a large power gain,
one needs {\it both} a tuned source impedance $\Zs$, {\it and} an
amplifier which possesses ideal noise properties
(cf.~Eq.~(\ref{CorrQLCondition}) and Eq.~(\ref{PropCond})).

\subsubsection{Role of noise cross-correlations}

Before leaving the topic of a linear voltage amplifier, we pause
to note the role of cross-correlations in current and voltage
noise in reaching the quantum limit.  First, note from
Eq.~(\ref{ZOpt2}) that in both the classical and quantum
treatments, the noise impedance $\ZN$ of the amplifier will have a
reactive part (i.e.~$\textrm{Im } \ZN \neq 0$) if there are
out-of-phase correlations between the amplifier's current and
voltage noises (i.e.~if $\textrm{Im } S_{VI} \neq 0$). Thus, if
such correlations exist, it will not be possible to minimize the
noise temperature (and hence, reach the quantum limit), if one
uses a purely real source impedance $\Zs$.

More significantly, note that the final classical expression for
the noise temperature $\TN$ explicitly involves the real part of
the $S_{VI}$ correlator (cf.~Eq.~(\ref{eq:TNopt})). In contrast,
we have shown that in the quantum case, $\textrm{Re } \bS_{VI}$
{\it must} be zero if one wishes to reach the quantum limit while
having a large power gain (cf.~Appendix \ref{app:ImPartsGone}); as
such, this quantity does not appear in the final expression for
the minimal $\TN$.  It also follows that {\it to reach the quantum
limit while having a large power gain, an amplifier cannot have
significant in-phase correlations between its current and voltage
noise}.

This last statement can be given a heuristic explanation.  If
there are out-of-phase correlations between current and voltage
noise, we can easily make use of these by appropriately choosing
our source impedance. However, if there are in-phase correlations
between current and voltage noise, we cannot use these simply by
tuning the source impedance.  We {\it could} however have used
them by implementing feedback in our amplifier. The fact that we
have not done this means that these correlations represent a kind
of missing information; as a result, we must necessarily miss the
quantum limit. In Sec.~\ref{subsec:MinAmp}, we explicitly give an
example of a voltage amplifier which misses the quantum limit due
to the presence of in-phase current and voltage fluctuations; we
show how this amplifier can be made to reach the quantum limit by
adding feedback in Appendix \ref{subsec:MirrorsFeedback}.

\subsection{Near quantum-limited mesoscopic detectors}

Having discussed the origin and precise definition of the quantum limit 
on the added noise of a linear, phase-preserving amplifier, we now provide
a brief review of work examining whether particular detectors are able (in principle) 
to achieve this ideal limit.  We will focus on the ``op-amp" mode of operation discussed in Sec.~\ref{subsec:TwoKindsAmps},
where the detector is only weakly coupled to the system producing the signal to be amplified.
As we have repeatedly stressed, reaching the quantum limit in this case requires the detector to have
``quantum ideal noise", as defined by Eq.~(\ref{CorrQLCondition}).  
Heuristically, this corresponds to the general
requirement of no wasted information:  there should be no other quantity besides
the detector output that could be monitored to provide information on the input signal
\cite{Clerk03}.  We have already given one simple but relevant example of a detector which reaches the amplifier quantum limit:  the parametric cavity detector, discussed extensively in Sec.~\ref{subsec:CavityDetector}.  Here, we turn to other more complex detectors.

\subsubsection{dc SQUID amplifiers}

The dc SQUID (superconducting quantum interference device) is a detector based on a superconducting ring having two Josephson junctions.  It can in principle be used as a 
near quantum-limited voltage amplifier or flux-to-voltage amplifier.  Theoretically, this was investigated using a quantum Langevin approach \cite{Koch80, Danilov83}, as well as more rigorously by using perturbative techniques \cite{Averin00b} and mappings to quantum impurity problems \cite{Clerk06}.  Experiments on SQUIDS have also confirmed its potential for near quantum-limited operation.  \textcite{Kycia01}  were able to achieve a noise temperature $\TN$ approximately 1.9 times the quantum limited value at an operating frequency of $\omega  = 2\pi \times 519$ MHz.  Working at lower frequencies appropriate to gravitational wave detection applications, \textcite{Vinante01} were able to achieve a $\TN$ approximately 200 times the quantum limited value at a frequency  
$\omega  = 2\pi \times 1.6$ kHz; 
more recently, the same group achieved a $\TN$ approximately 10 times the
quantum limit at a frequency $\omega  = 2\pi \times 1.6$ kHz \cite{Falferi08}.
In practice, it can be difficult to achieve the theoretically-predicted quantum-limited performance due to spurious heating caused by the dissipation in the shunt resistances used in the SQUID.
This effect can be significantly ameliorated, however, by adding cooling fins to the shunts \cite{Wellstood94}.

\subsubsection{Quantum point contact detectors}

A quantum point contact (QPC) is a narrow conducting channel formed in a two-dimensional gas.  The current through the constriction is very sensitive to nearby charges, and thus the QPC acts as a charge-to-current amplifier.  It has been shown theoretically that the QPC can achieve the amplifier quantum limit, both in the regime where transport is due to tunneling \cite{Gurvitz97}, as well as in regimes where the transmission is not small \cite{Levinson97,Aleiner97,Korotkov01c,Pilgram02, Clerk03}.  Experimentally, QPCs are in widespread use as detectors of quantum dot qubits.  The back-action dephasing of QPC detectors was studied in \cite{Buks98, Sprinzak00}; a good agreement was found with the theoretical prediction, confirming that the QPC has quantum-limited back-action noise.      

\subsubsection{Single-electron transistors and resonant-level detectors}

A metallic single-electron transistor (SET) consists of a small metallic island attached via tunnel junctions to larger source and drain electrodes.  Because of Coulomb blockade effects, the conductance of a SET is very sensitive to nearby charges, and hence it acts as a sensitive charge-to-current amplifier.
Considerable work has investigated whether metallic SETs can approach the quantum limit in various different operating regimes.  Theoretically, the performance of a normal-metal SET in the sequential tunneling regime was studied by 
\textcite{Shnirman98,Makhlin00,Devoret00,Aassime01,Johansson02,Johansson03}.  
In this regime, where transport is via a sequence of energy-conserving tunnel events, one is far from optimizing the quantum noise constraint of Eq.~(\ref{CorrQLCondition}), and hence one cannot reach the quantum limit \cite{Shnirman98,Korotkov01b}.  If one instead chooses to work with a normal-metal SET in the  cotunneling regime (a higher-order tunneling process involving a virtual transition), then one can indeed approach the quantum limit \cite{Averin00, MaasenvdBrink02}.  However, by virtue of being a higher-order process, the related currents and gain factors are small, impinging on the practical utility of this regime of operation.  It is worth noting that while most theory on SETs assume a dc voltage bias, to enhance bandwidth, experiments are usually conducted using the rf-SET configuration \cite{Schoelkopf98}, where the SET changes the damping of a resonant LC circuit.  \textcite{Korotkov99b} have shown that this mode of operation for a sequential tunneling SET increases the measurement imprecision noise by approximately a factor of $2$.  The measurement properties of a normal-metal, sequential-tunneling rf-SET (including back-action) were studied experimentally in \textcite{Turek05}. 

Measurement using superconducting SET's has also been studied.  \textcite{Clerk02} have shown that so-called incoherent Cooper-pair tunneling processes in a superconducting SET can have a noise temperature which is approximately a factor of two larger than the quantum limited value.  The measurement properties of a superconducting SET biased at a point of incoherent Cooper-pair tunneling have been probed recently in experiment \cite{Naik06, Rimberg04}.

The quantum measurement properties of phase-coherent, non-interacting resonant level detectors have also been studied theoretically \cite{Averin00b, Mozyrsky04, Gavish06, Clerk04a}.  These systems are similar to metallic SET, except that the central island only has a single level (as opposed to a continuous density of states), and Coulomb-blockade effects are typically neglected.  These detectors can reach the quantum limit in the regime where the voltage and temperature are smaller than the intrinsic energy broadening of the level due to tunneling.  They can also reach the quantum limit in a large-voltage regime that is analogous to the cotunneling regime in a metallic SET \cite{Averin00b, Clerk04a}.  The influence of dephasing processes on such a detector was studied in \textcite{Clerk04a}.

\subsection{Back-action evasion and noise-free amplification}
\label{subsec:BackactionEvasion}
Having discussed in detail quantum limits on phase-preserving
linear amplifiers (i.e.~amplifiers which measure both quadratures
of a signal equally well), we now return to the situation
discussed at the very start of Sec.~\ref{subsec:AmpPrelims}:
imagine we wish only to amplify a {\it single} quadrature of some
time-dependent signal.  For this case, there need not be any added
noise from the measurement.  Unlike the case of amplifying both
quadratures, Liouville's theorem does not require the existence of
any additional degrees of freedom when amplifying a single
quadrature:  phase space volume can be conserved during
amplification simply by contracting the unmeasured quadrature
(cf.~Eq.~\ref{eq:ClassDPA}).  As no extra degrees of freedom are
needed, there need not be any extra noise associated with the
amplification process.

Alternatively, single-quadrature detection can take a form similar
to a QND measurement, where the back-action does not affect the
dynamics of the quantity being measured
\cite{Thorne78,Caves80b,Caves82, Braginsky92, Braginsky80, BockoRMP96}. For concreteness,
consider a high-Q harmonic oscillator with position $x(t)$ and
resonant frequency $\Omega$.  Its motion may be written in terms
of quadrature operators defined as in 
Eq.~(\ref{eq:twoquadraturesdefined}):
\begin{eqnarray}
      \hat{x}(t) & = & \hat{X}_\delta (t) \cos \left(\Omega t   + \delta \right) +
        \hat{Y}_\delta (t) \sin \left( \Omega t   + \delta \right)
        \nonumber \\
        \label{eq:Quadratures}
\end{eqnarray}
Here, $\hat{x}(t)$ is the Heisenberg-picture position operator of the oscillator.  The quadrature
operators can be written in terms of the (Schr\"odinger-picture) oscillator creation and destruction operators as:
\begin{subequations}
\begin{eqnarray}
    \hat{X}_\delta(t) & = & \xrms \left(
         \hat{c}  e^{i \left( \omegam t + \delta \right)} 
         	+  \hat{c}^\dag e^{-i \left( \omegam t + \delta \right)}
    \right) \\
    \hat{Y}_\delta(t) & = & -i \xrms \left(
         \hat{c}  e^{i \left( \omegam t + \delta \right)} -  
         \hat{c}^\dag  e^{-i \left( \omegam t + \delta \right)}
    \right)
\end{eqnarray}
\label{eqs:ExplicitQuadratures}
\end{subequations}
As previously discussed, 
the two quadrature amplitude operators 
$\hat{X}_\delta$ and $\hat{Y}_\delta$ are canonically
conjugate (cf.~Eq.(\ref{eq:QuadsConjugate})).  
Making a measurement of one quadrature amplitude, say
$\hat{X}_\delta$, will thus invariably lead to back-action disturbance of the 
other, conjugate quadrature $\hat{Y}_\delta$.
However, due to the dynamics of a harmonic oscillator, this
disturbance will not affect the measured quadrature at later
times.  One can already see this from the classical equations of
motion.  Suppose our oscillator is driven by a time-dependent
force $F(t)$ which only has appreciable bandwidth near $\Omega$.
We may write this as:
\begin{eqnarray}
    F(t) = F_X(t) \cos(\Omega t + \delta) + F_Y(t) \sin(\Omega t + \delta)
    \label{eq:FQuads}
\end{eqnarray}
where $F_X(t), F_Y(t)$ are slowly varying compared to $\Omega$.
Using the fact that the oscillator has a high-quality factor $Q =
\Omega / \gamma$, one can easily find the equations of motion:
\begin{subequations}
\begin{eqnarray}
    \frac{d}{dt} X_\delta(t) =
        - \frac{\gamma}{2} X_\delta(t) - \frac{F_Y(t)}{2 m \Omega}, \\
    \frac{d}{dt} Y_\delta(t) =
        - \frac{\gamma}{2} Y_\delta(t) + \frac{F_X(t)}{2 m \Omega}.
\end{eqnarray}
    \label{eq:QuadEqns}
\end{subequations}
Thus, as long as $F_Y(t)$ and $F_X(t)$ are uncorrelated and sufficiently slow, 
the dynamics of the two quadratures are completely
independent; in particular, if $Y_\delta$ is subject to a narrow-bandwidth, noisy
force, it is of no consequence to the evolution of $X_\delta$.
An ideal measurement of $X_\delta$ will result in a back-action force having the 
form in Eq.~(\ref{eq:FQuads}) with $F_Y(t) = 0$, implying that 
$X_\delta(t)$ will be completely unaffected by the measurement.

Not surprisingly, if one can measure and amplify $X_{\delta}$ without any back-action, there need not be any added noise due to the amplification.  In such a setup, the only added noise is the measurement-imprecision noise associated with intrinsic fluctuations of the amplifier output.  These may be reduced (in principle) to an arbitrarily small value by simply increasing the amplifier gain (e.g.~by increasing the detector-system coupling):  in an ideal setup, there is no back-action penalty on the measured quadrature associated with this increase.  

The above conclusion can lead to what seems like a contradiction.  Imagine we 
use a back-action evading amplifier to make a ``perfect" measurement of $X_{\delta}$ (i.e. negligible added noise).  We would then have no uncertainty as to the value of this quadrature.  Consequently, we would expect the quantum state of our oscillator to be a squeezed state, where the uncertainty in $X_{\delta}$ is much smaller than $\xrms$.  However, if there is no back-action acting on $X_{\delta}$, how is the amplifier able to reduce its uncertainty?  This seeming paradox can be fully resolved by considering the conditional aspects of an ideal single quadrature measurement, where one considers the state of the oscillator given a particular measurement history \cite{Ruskov05, Clerk08}.  

It is worth stressing that the possibility of amplifying a single quadrature without back-action (and hence, without added noise) relies crucially on our oscillator resembling
a perfect harmonic oscillator: the oscillator $Q$ must be large, and non-linearities (which
could couple the two quadratures) must be small.  In addition, the envelope of the non-vanishing back-action force $F_X(t)$ must have a narrow bandwidth.  One should further note that a very high precision measurement of $X_\delta$ will produce a very large backaction force $F_X$.  If the system is not nearly perfectly harmonic, then
the large amplitude imparted to the conjugate quadrature
$Y_\delta$ will inevitably leak back into $X_\delta$.

Amplifiers or detectors which treat the two signal quadratures
differently are known in the quantum optics literature as `phase
sensitive'; we prefer the designation `phase-non-preserving' since
they do not preserve the phase of the original signal. Such
amplifiers invariably rely on some internal clock (i.e.~an
oscillator with a well defined phase) which breaks
time-translation invariance and picks out the phase of the
quadrature that will be amplified (i.e.~the choice of $\delta$
used to define the two quadratures in
Eq.~(\ref{eq:Quadratures})); we will see this explicitly in what
follows.  This leads to an important caveat:  even in a situation
where the interesting information is in a single signal
quadrature, to benefit from using a phase non-preserving
amplifier, we must know in advance the precise phase of this
quadrature.  If we do not know this phase, we will have either to
revert to a phase-preserving amplification scheme (and thus be
susceptible to added noise) or we would have to develop a
sophisticated and high speed quantum feedback scheme to
dynamically adapt the measurement to the correct quadrature in
real time \cite{MabuchiAdaptivePhase2002}.   In what follows, we will make the above ideas concrete by considering a
few examples of quantum, phase non-preserving
amplifiers.\footnote{%
One could in principle generalize the linear response approach of Sec.~\ref{sec:GenLinResponseTheory} to deal with phase non-preserving detectors.  However, as such detectors are not time-translational invariant, such a description becomes rather cumbersome and is not particularly helpful.  We prefer instead to present concrete examples.}

\subsubsection{Degenerate parametric amplifier}
\label{subsubsec:DPA}


Perhaps the simplest example of a phase non-preserving amplifier
is the degenerate parametric amplifier; the classical version of
this system was described at the start of Sec.~\ref{subsec:AmpPrelims} (cf.\  Eq.~\ref{eq:ClassDPA}).    
The setup is similar to the non-degenerate parametric amplifier discussed in Sec.~\ref{subsec:NDParamps}, 
except that the idler mode is
eliminated, and the non-linearity converts a single pump photon
into two signal photons at frequency $\omega_{\rm S} = \omega_{\rm
P}/2$.  As we now show, the resulting dynamics causes one signal quadrature to be amplified, while the other
is attenuated, in such a way that it is not necessary to add extra noise to preserve the canonical commutation relations. 

The system Hamiltonian is
\begin{eqnarray}
\hat{H}_{\rm sys} &=& \hbar\left(\omegap \ap^\dagger \ap + \omegas \as^\dagger\as \right)\nonumber\\
&+& i\hbar \eta \left(\as^\dagger\as^\dagger\ap -
\as\as\ap^\dagger\right).
\end{eqnarray}
Treating the pump classically as before, the analog of
Eq.~(\ref{eq:nondegintrep}) is 
\be
\hat{V}_{\rm sys} = i\hbar{\lambda \over 2} \left(\as^\dagger\as^\dagger -
\as\as \right) \, ,
\label{eq:degintrep}
\ee
where $\lambda/2=\eta \psi_P$, 
and the analog of Eq.~(\ref{eq:nondegEOMs}) is:
\be
{\dot{\hat{a}}}_{\rm S} = -\frac{\ks}{2}\,\as +\lambda
\as^\dagger -\sqrt
{\ks}\, {\hat b}_{\rm S,in} 
\label{eq:degEOMs}
\ee

The dimensionless quadrature operators corresponding to the signal mode are:
\begin{eqnarray}
{\hat x}_{\rm S} = \frac{1}{\sqrt{2}}\left(\as^\dagger +
\as\right), \quad
{\hat y}_{\rm S} =  \frac{i}{\sqrt{2}}\left(\as^\dagger -
\as\right)
\end{eqnarray}
which obey $[{\hat x}_{\rm S},{\hat y}_{\rm S}]=i$.
We can define quadrature operators $\hat{X}_{S,{\rm in/out}},\hat{Y}_{S,{\rm in/out} }$ 
corresponding to the input and output fields
in a completely analogous manner. 

The steady state solution of Eq.~(\ref{eq:degEOMs}) for the output
fields becomes
\begin{eqnarray}
{\hat X}_{\rm S,out} = \sqrt{G}{\hat X}_{\rm S,in}, \quad
{\hat Y}_{\rm S,out} = \frac{1}{\sqrt{G}}{\hat Y}_{\rm S,in},
\end{eqnarray}
where the number gain $G$ is given by
\be
G = \left[\frac{\lambda + \ks/2}{\lambda-\ks/2}\right]^2.
\ee
We thus see clearly that the amplifier treats the two quadratures differently.  One quadrature is amplified, the other attenuated,
with the result that the commutation relation can be preserved without the necessity of extra degrees of freedom and added noise.
Note  that the large-amplitude pump mode has played the role of a clock in the degenerate paramp:  it is the phase of the 
pump which picks out which quadrature of the signal will be amplified.

Before ending our discussion here, it is important to stress that while the degenerate parametric amplifier is phase-sensitive and has no added noise, it is {\it not} an example of back-action evasion (see \textcite{Caves80b}, footnote on p.~342).  This amplifier is operated in the scattering mode of amplifier operation, a mode where (as discussed extensively in Sec.~\ref{subsec:TwoKindsAmps}) back-action is not at all relevant.  Recall that in this mode of operation, the amplifier input is perfectly impedance matched to the signal source, and the input signal is simply the amplitude of an incident wave on the amplifier input.  This mode of operation necessarily requires a strong coupling between the input signal and the amplifier input (i.e.~$\langle \hb_{\rm S, in}\rangle$).  If one instead tried to weakly couple the degenerate parametric amplifier to a signal source, and operate it in the ``op-amp" mode of operation (cf.~Sec.~\ref{subsec:TwoKindsAmps}), one finds that there is indeed a back-action disturbance of the measured quadrature.  We have yet another example which demonstrates that one must be very careful to distinguish the ``op-amp" and ``scattering" mode of amplifier operation.

\subsubsection{Double-sideband cavity detector}
\label{subsubsec:DoubleSideband}

We now turn to a simple but experimentally-relevant detector  
that is truly back-action evading.  We will take as our input signal the position $\hat{x}$ 
of a mechanical oscillator.  The amplifier setup we consider is almost identical to the cavity position detector discussed in Sec.~\ref{subsec:measurementofoscillatorposition}:  we again have a single-sided resonant cavity whose frequency depends linearly on the oscillator's position, with the Hamiltonian being given by Eq.~(\ref{CavityHam1})
(with $\hat{z} = \hat{x} / \xrms$).
We showed in Sec.~\ref{subsec:measurementofoscillatorposition} and Appendix \ref{subapp:CavityPositionDetector} that by driving the cavity on resonance, it could be used to make a quantum limited position measurement:  one can operate it as a phase-preserving amplifier of the mechanical's oscillators position, and achieve the minimum possible amount of added noise.  To use the same system to make a back-action free measurement of one oscillator quadrature only, one simply uses a different 
cavity drive.  Instead of driving at the cavity resonance frequency $\omegar$, one drives at the two sidebands associated with the mechanical motion (i.e. at frequencies $\omegar \pm \omegam$, where $\Omega$ is as always the frequency of the mechanical resonator).  As we will see, such a drive results in an effective interaction which only couples the cavity to one quadrature of the oscillator's motion.  This setup was first proposed as a means of back-action evasion in \textcite{Braginsky80}; further discussion can be found in \textcite{Caves80b, Braginsky92}, as well as in \textcite{Clerk08}, which gives a fully quantum treatment and considers conditional aspects of the measurement.  In what follows, we sketch the operation of this system following \textcite{Clerk08}; details are provided in Appendix \ref{subapp:SingleQuadrature}. 

We will start by requiring that our system be in the  
``good-cavity" limit, where $\omegar \gg \Omega \gg \kappa$ ($\kappa$ is the damping of the cavity mode); we will
also require the mechanical oscillator to have a high Q-factor.  In this regime, the two sidebands associated with the mechanical motion at $\omegar \pm \Omega$ are well-separated from the main cavity resonance at $\omegar$.  Making a single-quadrature measurement requires that one drives the cavity equally at the two sideband frequencies.  The amplitude of the driving field $\bar{b}_{\rm in}$ entering the cavity will be chosen to have the form:
\begin{eqnarray}
    \bar{b}_{\rm in}(t) & = &
        -\frac{i \sqrt{\dot{N}} }{4}\left( e^{i \delta} e^{-i (\omegar-\omegam) t} - e^{-i \delta} e^{-i (\omegar + \omegam) t} \right) 
        \nonumber \\
        & = &
         \frac{\sqrt{\dot{N} }}{2} \sin(\Omega t + \delta) e^{-i \omegar t}.
         \label{eq:DSBDrive}
\end{eqnarray}
Here, $\dot{N}$ is the photon number flux associated with the cavity drive (see Appendix \ref{app:drivencavity} for more details on how to properly include a drive using input-output theory).
Such a drive could be produced by taking a signal at the cavity resonance frequency, and amplitude modulating it at the mechanical frequency. 

To understand the effect of this drive, note that it sends the cavity both photons with frequency $(\omegar - \Omega)$ and photons with frequency $(\omegar +
\Omega)$.  The first kind of drive photon can be
converted to a cavity photon if a quanta is {\it absorbed from}
the mechanical oscillator; the second kind of drive photon can be
converted to a cavity photon if a quanta is {\it emitted to} the
mechanical oscillator.  The result is that we can create a cavity
photon by either absorbing or emitting a mechanical oscillator
quanta.  Keeping track that there is a well-defined relative phase
of $e^{i 2 \delta}$ between the two kinds of drive photons, we would expect
the double-sideband drive to yield an effective
cavity-oscillator interaction of the form:
\begin{subequations}
\begin{eqnarray}
    V_{\rm eff} & \propto &
    	 \sqrt{\dot{N} } \left[ \hat{a}^{\dag} \left( e^{i \delta} \hat{c}  +
        e^{-i \delta} \hat{c}^\dag \right)
         + \textrm{h.c. }   \right]
    \label{eq:DSBVeff} \\
	& \propto &
	\sqrt{\dot{N} } (\hat{a} + \hat{a}^{\dag}) \hat{X}_\delta
	\label{eq:DSBVeff2}
\end{eqnarray}
\end{subequations}
This is exactly what is found in a full calculation (see Appendix \ref{subapp:SingleQuadrature}).  Note that
we have written the interaction in an interaction picture in which the fast oscillations of the cavity and oscillator operators have been removed. In the second line, we have made use of Eqs.~(\ref{eqs:ExplicitQuadratures}) to show that the effective interaction only involves the $\hat{X}_\delta$ oscillator quadrature.

We thus see from Eq.~(\ref{eq:DSBVeff2}) that the cavity is only coupled to the oscillator $X_{\delta}$ quadrature. As shown rigorously in Appendix \ref{subapp:CavityPositionDetector}, 
the result is that the system only measures and amplifies this quadrature:  the light leaving the cavity has a signature of $\hat{X}_{\delta}$, but not of $\hat{Y}_{\delta}$.
Further, Eq.~(\ref{eq:DSBVeff2}) implies that the cavity operator 
  $\sqrt{\dot{N}} \left( \hat{a} +
\hat{a}^\dag \right)$ will act as a noisy force on the $Y_{\delta}$ quadrature.  While this will cause a back-action heating of $Y_{\delta}$, it will not affect the measured quadrature $X_\delta$.  We thus have a true back-action evading amplifier:  the cavity output light lets one measure $X_{\delta}$ free from any back-action effect.  Note that 
in deriving Eq.~(\ref{eq:DSBVeff}), we have used the fact that the cavity operators have fluctuations in a narrow bandwidth $\sim \kappa \ll \omegam$: the back-action force noise is slow compared to the oscillator frequency.  If this were not the case, we could still have a back-action heating of the measured $X_{\delta}$ quadrature.  Such effects, arising from a non-zero ratio $\kappa / \omegam$, are treated in \textcite{Clerk08}.

Finally, as there is no back-action on the measured $X_{\delta}$ quadrature,
the only added noise of the amplification scheme is measurement-imprecision noise (e.g. shot noise in the light leaving the cavity).  This added noise can be made arbitrarily small by increasing the gain of the detector by, for example, increasing the strength of the cavity drive $\dot{N}$.  In a real system where $\kappa / \omega_M$ is non-zero, the finite-bandwidth of the cavity number fluctuations leads to a small back-action on the $X_{\delta}$.  As a result, one cannot make the added noise arbitrarily small, as too large a cavity drive will heat the measured quadrature.  Nonetheless, for a sufficiently small ratio $\kappa / \omega_M$, one can still beat the standard quantum limit on the added noise \cite{Clerk08}.

\subsubsection{Stroboscopic measurements}

With sufficiently high bandwidth it should be able to do
stroboscopic measurements in sync with the oscillator motion which
could allow one to go below the standard quantum limit in one of
the quadratures of motion \cite{Caves80b,Braginsky92}.  To
understand this idea, imagine an extreme form of phase sensitive
detection in which a Heisenberg microscope makes a strong
high-resolution measurement which projects the oscillator onto a
state of well defined position $X_0$ at time $t=0$:
\be
\Psi_{X_0}(t) = \sum_{n=0}^\infty a_n e^{-i(n+1/2)\Omega t}
\left|n\right\rangle,
\label{eq:stroboscopic1}
\ee
where the coefficients obey $a_n = \langle n|X_0\rangle$.  Because
the position is well-defined the momentum is extremely uncertain.
(Equivalently the momentum kick delivered by the backaction of
the microscope makes the oscillator momentum uncertain.)  Thus the
wave packet quickly spreads out and the position uncertainty
becomes large.  However because of the special feature that the
harmonic oscillator levels are evenly spaced, we can see from
Eq.~(\ref{eq:stroboscopic1}) that the wave packet reassembles
itself precisely once each period of oscillation because $e^{i
n\Omega t}=1$ for every integer $n$.  (At half periods, the packet
reassembles at position $-X_0$.)  Hence stroboscopic measurements
made once (or twice) per period will be backaction evading and
can go below the standard quantum limit. The only limitations will
be the finite anharmonicity and damping of the oscillator.  Note that the possibility of using mesoscopic electron detectors to perform stroboscopic measurements has recently received attention \cite{Ruskov05, Jordan05}.

\section{Bosonic Scattering Description of a Two-Port Amplifier}
\label{sec:ScatteringAmp}

In this section, we return again to the topic of Sec.~\ref{subsec:VoltageAmp},
quantum limits on a quantum voltage amplifier.  We now discuss the physics in terms of the bosonic voltage amplifier
first introduced in Sec.~\ref{subsec:TwoKindsAmps}.  Recall that in that subsection, we demonstrated that the standard Haus-Caves derivation of the quantum limit
was not directly relevant to the usual weak-coupling ``op-amp" mode of amplifier
operation, a mode where the input signal is not simply the amplitude of a wave
incident on the amplifier.  In this section, we will expand upon that discussion, giving
an explicit discussion of the differences between the op-amp description of an amplifier presented in Sec.~\ref{subsec:PositionDetector}, and the scattering description often used in the quantum optics literature \cite{Courty99, Grassia98}.  We will see that what one means by ``back-action" and ``added noise" are not the same in the two descriptions!  Further, even though an amplifier may reach the quantum limit when used in the scattering mode (i.e.~its added noise is as small as allowed by commutation relations), it can nonetheless fail to achieve the quantum limit when used in the op-amp mode.  Finally, the discussion here will also allow us to highlight important aspects of the quantum limit not easily discussed
in the more general context of Sec.~\ref{sec:GenLinResponseTheory}.
\subsection{Scattering versus op-amp representations}
\label{subsec:ScatteringRep}

In the bosonic scattering approach, a generic linear amplifier is
modeled as a set of coupled bosonic modes.  To make matters
concrete, we will consider the specific case of a voltage
amplifier with distinct input and output ports, where each port is
a semi-infinite transmission line (see Fig.~\ref{fig:TwoPortAmp}).
We start by recalling that a quantum transmission line can be described as a set of non-interacting bosonic modes (see Appendix \ref{app:QuantumResistor} for a quick review).
Denoting the input transmission line with an $a$ and the output transmission line with a $b$, the current and voltage operators in these lines may be written:
\begin{subequations}
\begin{eqnarray}
    \hV _q(t) = \int_0^{\infty} \frac{d \omega}{2 \pi}
    \left( \hV_q[\omega]e^{-i \omega t} + h.c. \right)
        \label{eq:VOpdefn}\\
    \hI _q(t) = \sigma_q \int_0^{\infty} \frac{d \omega}{2 \pi}
    \left( \hI_q[\omega] e^{-i \omega t} + h.c. \right)
        \label{eq:IOpdefn}
\end{eqnarray}
\end{subequations}
with
\begin{subequations}
\begin{eqnarray}
    \hV_{q}[\omega] & = &
        \sqrt{\frac{\hbar\omega}{2}Z_{q}}
        \left(\hq_{\rm in}[\omega] + \hq_{\rm out}[\omega] \right)\
    \label{eq:Vdefn}\\
    \hI_{q}[\omega] & = & \sqrt{\frac{\hbar\omega}{2Z_{q}}}
    \left(\hq_{\rm in}[\omega] - \hq_{\rm out}[\omega]\right)
    \label{eq:Idefn}
\end{eqnarray}
\label{eqs:BosonicIV}
\end{subequations}
Here, $q$ can be equal to $a$ or $b$, and we have $\sigma_a=1,
\sigma_b=-1$. The operators $\hain[\omega],\haout[\omega]$
are 
bosonic annihilation operators; $\hain[\omega]$
describes an incoming wave in the input transmission line
(i.e.~incident on the amplifier) having frequency $\omega$, while
$\haout[\omega]$ describes an outgoing wave with frequency
$\omega$. The operators $\hbin[\omega]$ and $\hbout[\omega]$
describe analogous waves in the output transmission line.
We can think of $\hV_{a}$ as the input voltage
to our amplifier, and $\hV_{b}$ as the output voltage. Similarly,
$\hI_{a}$ is the current drawn by the amplifier at the input, and
$\hI_{b}$ the current drawn at the output of the amplifier.
Finally, $Z_a$ ($Z_b$) is the characteristic impedance of the input (output)
transmission line.  

As we have seen, amplification invariably requires additional degrees of freedom.
Thus, to amplify a signal at a particular frequency $\omega$, there will be $2 N$ bosonic modes
involved, where the integer $N$ is necessarily larger than $2$.  
Four of
these modes are simply the frequency-$\omega$ modes in the input
and output lines
(i.e.~$\hain[\omega]$,$\haout[\omega]$,$\hbin[\omega]$ and
$\hbout[\omega]$).  The remaining $2(N-2)$ modes describe
auxiliary degrees of freedom involved in the amplification
process; these additional modes could correspond to frequencies
different from the signal frequency $\omega$.  The auxiliary modes
can also be divided into incoming and outgoing modes.  It is thus
convenient to represent them as additional transmission lines
attached to the amplifier; these additional lines could be
semi-infinite, or could be terminated by active elements.

\subsubsection{Scattering representation}

In general, our generic two-port bosonic amplifier will be
described by a $N\times N$ scattering matrix which determines the
relation between the outgoing mode operators and incoming mode
operators.  The form of this matrix is constrained by the
requirement that the output modes obey the usual canonical
bosonic commutation relations. It is convenient to express the scattering
matrix in a form which \emph{only} involves the input and output
lines explicitly:
\begin{eqnarray}
    \left(\begin{array}{c}
    \haout[\omega]\\
    \hbout[\omega]\end{array}\right) & = & \left(\begin{array}{cc}
    s_{11}[\omega] & s_{12}[\omega]\\
    s_{21}[\omega] & s_{22}[\omega]
    \end{array}\right)
    \left(\begin{array}{c}
    \hain[\omega]\\
    \hbin[\omega]\end{array}\right)
    \nonumber \\
    &&
        +\left(\begin{array}{c}
    \hFF_{a}[\omega]\\
    \hFF_{b}[\omega]\end{array}\right)
    \label{eq:sdefn}
\end{eqnarray}
Here $\hFF_{a}[\omega]$ and $\hFF_{b}[\omega]$ are each an
unspecified linear combination of the incoming auxiliary modes introduced above.
They thus describe noise in the outgoing modes of the
input and output transmission lines which arises from the
auxiliary modes involved in the amplification process. Note the
similarity between Eq.~(\ref{eq:sdefn}) and Eq.~(\ref{eq:fullb})
for the simple one-port bosonic amplifier considered in
Sec.~\ref{subsec:CavesArg}.

In the quantum optics literature, one typically views
Eq.~(\ref{eq:sdefn}) as the defining equation of the amplifier; we
will call this the scattering representation of our amplifier.  The representation is best
suited to the scattering mode of amplifier operation described in Sec.~\ref{subsec:TwoKindsAmps}.  In this mode of operation, the experimentalist ensures
that $\langle
\ha_{in}[\omega] \rangle$ is precisely equal to the signal to be amplified, \emph{irrespective} of what is coming out of the amplifier.
Similarly, the output signal from the
amplifier is the amplitude of the outgoing wave in the output
line, $\langle \hbout[\omega] \rangle$.
If we focus on $\hbout$, we have precisely the same situation as described in Sec.~\ref{CavesQL}, where we presented the Haus-Caves derivation of the quantum limit (cf. Eq.~(\ref{eq:fullb})).  It thus follows that in the scattering mode
of operation, the matrix element $s_{21}[\omega]$ represents the gain of our amplifier at
frequency $\omega$, $|s_{21}[\omega]|^2$ the
corresponding ``photon number gain", and
$\hFF_{b}$ the added noise operator of the amplifier.
The operator $\hFF_{a}$ represents the back-action noise in the scattering mode of
operation; this back-action has no effect on the added noise of the amplifier in the scattering mode.

Similar to Sec.~\ref{subsec:CavesArg}, one can now apply the
standard argument of \textcite{Haus62} and \textcite{Caves82} to
our amplifier. This argument tells us that since the ``out''
operators must have the same commutation relations as the ``in''
operators, the added noise $\hFF_{b}$ cannot be arbitrarily small
in the large gain limit (i.e. $|s_{21}| \gg 1$).  Note that this
version of the quantum limit on the added noise has nothing to do with back-action.
As already discussed, this is perfectly appropriate for the scattering mode of operation, as in this mode, the experimentalist ensures that the signal going into the amplifier is completely independent of whatever is coming out of the amplifier.  This mode of operation could be realized in time-dependent experiments, where a pulse is launched at the amplifier.  This mode is {\it not} realized in most weak-coupling amplification experiments, where the signal to be amplified is not identical to an incident wave
amplitude.

\subsubsection{Op-amp representation}

In the usual op-amp amplifier mode of operation
(described extensively in Sec.~\ref{sec:GenLinResponseTheory}),
the input and output signals are not
simply incoming/outgoing wave amplitudes; thus, the scattering representation
is not an optimal description of our amplifier.  The system we are describing here
is a voltage amplifier:  thus, in the op-amp mode, the experimentalist would ensure that
the voltage at the end of the input line ($ \hV_a $) is equal to the signal to be amplified, and would read out the voltage at the end of the output transmission line ($\hV_b$) as the output of the amplifier.  From Eq.~(\ref{eq:VOpdefn}), we see that this implies that the amplitude of the wave going into the amplifier, $\ain$, will depend on the amplitude of the wave exiting the amplifier, $\aout$.

Thus, if we want to use our amplifier as a voltage amplifier, we
would like to find a description which is more
tailored to our needs than the scattering representation of
Eq.~(\ref{eq:sdefn}).  This can be found by simply re-expressing
the scattering matrix relation of Eq. (\ref{eq:sdefn}) in terms of
voltages and currents. The result will be what we term the ``op
amp" representation of our amplifier, a representation which is standard
in the discussion of classical amplifiers (see, e.g., \textcite{Boylestad06}).  
In this representation, one
views $\hV_a$ and $\hI_b$ as inputs to the amplifier: $\hV_a$ is
set by whatever we connect to the amplifier input, while $\hI_b$
is set by whatever we connect to the amplifier output.  In
contrast, the outputs of our amplifier are the voltage in the
output line, $\hV_b$, and the current drawn by the amplifier at
the input, $\hI_a$.  Note that this interpretation of voltages and
currents is identical to how we viewed the voltage amplifier in
the linear-response/quantum noise treatment of
Sec.~\ref{subsec:VoltageAmp}.

Using Eqs.~(\ref{eq:VOpdefn}) and (\ref{eq:IOpdefn}), and
suppressing frequency labels for clarity, Eq.~(\ref{eq:sdefn}) may
be written explicitly in terms of the voltages and current in the
input ($\hV_a, \hI_a$) and output ($\hV_b, \hI_b$) transmission lines:
\begin{eqnarray}
    \left(\begin{array}{c}
        \hV_{b} \\
        \hI_{a}
    \end{array}\right)
        & = &
    \left(\begin{array}{cc}
        \lambda_{V} & -\Zout\\   
        \frac{1}{\Zin} & \lambda_{I}'
    \end{array}\right)
    \left(\begin{array}{c}
        \hV_{a}\\
        \hI_{b}
    \end{array}\right)  +
    \left(\begin{array}{c}
        \lambda_{V}\cdot \htV \\
        \htI
    \end{array}\right)
    \label{eq:opamprep}
\end{eqnarray}
The coefficients in the above matrix are completely determined by the 
scattering matrix of Eq.~(\ref{eq:sdefn}) (see Eqs.~(\ref{eqs:OpAmpDefs}) below); 
moreover, they are familiar from the
discussion of a voltage amplifier in Sec.~\ref{subsec:VoltageAmp}.
$\lambda_V[\omega]$ is the voltage gain of the amplifier,
$\lambda_{I}'[\omega]$ is the reverse current gain of the
amplifier, $\Zout$ is the output impedance, and $\Zin$ is the
input impedance.  The last term on the RHS of
Eq.~(\ref{eq:opamprep}) describes the two familiar kinds of
amplifier noise.  $\htV$ is the usual voltage noise of the
amplifier (referred back to the amplifier input), while $\htI$ is
the usual current noise of the amplifier.  Recall that in this
standard description of a voltage amplifier
(cf.~Sec.~\ref{subsec:VoltageAmp}), $\tilde{I}$ represents the
back-action of the amplifier: the system producing the input
signal responds to these current fluctuations, resulting in an
additional fluctuation in the input signal going into the
amplifier. Similarly, $\lambda_{V}\cdot\tilde{V}$ represents the
intrinsic output noise of the amplifier: this contribution to the
total output noise does not depend on properties of the input
signal.  Note that we are using a sign convention where a positive
$\langle \hI_a \rangle $ indicates a current flowing {\it into}
the amplifier at its input, while a positive $\langle \hI_b
\rangle$ indicates a current flowing {\it out of} the amplifier at
its output.  Also note that the operators $\hV_a$ and $\hI_b$ on
the RHS of Eq.~(\ref{eq:opamprep}) will have noise; this noise is
entirely due to the systems attached to the input and output of
the amplifier, and as such, should not be included in what we call
the added noise of the amplifier.

Additional important properties of our amplifier follow
immediately from quantities in the op-amp representation.  As
discussed in Sec.~\ref{subsec:PositionDetector}, the most
important measure of gain in our amplifier is the dimensionless
power gain. This is the ratio between power dissipated at the
output to that dissipated at the input, taking the output current
$I_{B}$ to be $V_{B}/ \Zout$:
\begin{eqnarray}
    G_{P} & \equiv &
        \frac{(\lambda_{V})^{2}}{4}\frac{ \Zin }{ \Zout }\cdot
            \left(1+\frac{\lambda_{V}\lambda'_{I}}{2} \frac{ \Zin }{\Zout }\right)^{-1}
    \label{eq:BosonicPowerGain}
\end{eqnarray}

Another important quantity is the loaded input impedance:  what is
the input impedance of the amplifier in the presence of a load
attached to the output?  In the presence of reverse current gain
$\lambda'_I \neq 0$, the input impedance will depend on the output
load.  Taking the load impedance to be $Z_{{\rm load}}$, some simple
algebra yields:
\begin{eqnarray}
\frac{1}{Z_{{\rm in,loaded}}} =
    \frac{1}{\Zin} + \frac{\lambda'_I \lambda_V}{Z_{{\rm load}} + \Zout}
    \label{eq:LoadedZin}
\end{eqnarray}
It is of course undesirable to have an input impedance which
depends on the load.  Thus, we see yet again that it is
undesirable to have appreciable reverse gain in our amplifier
(cf.~Sec.~\ref{subsubsec:GenericLinear}).

\subsubsection{Converting between representations}

Some straightforward algebra now lets us express the op-amp
parameters appearing in Eq.~(\ref{eq:opamprep}) in terms of the
scattering matrix appearing in Eq.~(\ref{eq:sdefn}):
\begin{subequations}
\label{eq:OpAmpParams}
\begin{eqnarray}
    \lambda_{V} & = &
        2\sqrt{\frac{Z_{b}}{Z_{a}}}\frac{s_{21}}{D}
        \label{eq:Gaindefn}\\
    \lambda_{I}' & = &
        2\sqrt{\frac{Z_{b}}{Z_{a}}}\frac{s_{12}}{D}
        \label{eq:RevGaindefn}\\
    \Zout & = &
         Z_{b}\frac{(1+s_{11})(1+s_{22})-s_{12}s_{21}}{D}
         \label{eq:Zoutdefn}\\
    \frac{1}{ \Zin } & = &
        \frac{1}{Z_{a}}\frac{(1-s_{11})(1-s_{22})-s_{12}s_{21}}{D}
        \label{eq:Zindefn}
\end{eqnarray}
\label{eqs:OpAmpDefs}
\end{subequations}
where all quantities are evaluated at the same frequency $\omega$,
and $D$ is defined as:
\begin{eqnarray}
    D & = & (1+s_{11})(1-s_{22})+s_{12}s_{21}\label{eq:Ddefn}
\end{eqnarray}
Further, the voltage and current noises in the op-amp
representation are simple linear combinations of the noises
$\hFF_a$ and $\hFF_b$ appearing in the scattering representation:
\begin{eqnarray}
    \left(\begin{array}{c}
        \htV\\
        Z_{a}\cdot \htI
    \end{array}\right)
    & = &
        \sqrt{2\hbar\omega Z_{a}}
            \left(\begin{array}{cc}
                -\frac{1}{2} & \frac{1+s_{11}}{2s_{21}}\\
                \frac{s_{22}-1}{D} & -\frac{s_{12}}{D}
            \end{array}\right)
        \left(\begin{array}{c}
            \hFF_{a}\\
            \hFF_{b}
        \end{array}\right)
        \nonumber \\
    \label{eq:OpAmpNoises}
\end{eqnarray}
Again, all quantities above are evaluated at frequency $\omega$.

Eq.~(\ref{eq:OpAmpNoises}) immediately leads to an important
conclusion and caveat:  {\it what one calls the ``back-action" and
``added noise" in the scattering representation (i.e. $\FF_{a}$
and $\FF_{b}$ ) are not the same as the ``back-action" and ``added
noise" defined in the usual op-amp representation}. For example,
the op-amp back-action $\htI$ does not in general coincide with
the $\hFF_a$, the back-action in the scattering picture.
If we are indeed interested in using our amplifier as a voltage
amplifier, we are interested in the total added noise of our
amplifier {\it as defined in the op-amp representation}. As we saw
in Sec.~\ref{subsec:VoltageAmp} (cf.~Eq.~(\ref{eq:ClassVNoise})),
this quantity involves both the noises $\htI$ and $\htV$.  We thus see explicitly
something already discussed in Sec.~\ref{subsec:TwoKindsAmps}:
it is very dangerous to make conclusions
about how an amplifier behaves in the op-amp mode of operation based on its properties in the scattering mode of operation.  As we will see, even though an amplifier is ``ideal"
in the scattering mode (i.e.~$\FF_{a}$ as small as possible), it can nonetheless fail to reach the quantum limit in the op-amp mode of operation.


In what follows, we will calculate the op-amp noises $\htV$ and
$\htI$
 in a minimal bosonic voltage amplifier,
and show explicitly how this description is connected to the more
general linear-response treatment of Sec.~\ref{subsec:VoltageAmp}.
However, before proceeding, it is worth noting that Eqs.
(\ref{eq:Gaindefn})-(\ref{eq:Zindefn}) are themselves completely
consistent with linear-response theory.  Using linear-response,
one would calculate the op-amp parameters $\lambda_V, \lambda'_I,
\Zin$ and $\Zout$ using Kubo formulas
(cf.~Eqs.~(\ref{eq:VAmpZin}), (\ref{eq:VAmpZout}) and the
discussion following Eq.~(\ref{eq:VoltAmpHint})).  These in turn
would  involve correlation functions of $\hI_a$ and $\hV_b$
evaluated at zero coupling to the amplifier input and output. Zero
coupling means that there is no input voltage to the amplifier
(i.e.~a short circuit at the amplifier input, $\hV_a=0$) and there
is nothing at the amplifier output drawing current (i.e.~an open
circuit at the amplifier output, $\hI_b=0$).
Eq.~(\ref{eq:opamprep}) tells us that in this case, $\hV_b$ and
$\hI_a$ reduce to (respectively) the noise operators $\lambda_V
\htV$ and $\htI$.  Using the fact that the commutators of $\hFF_a$
and $\hFF_b$ are completely determined by the scattering matrix
(cf.~Eq.~(\ref{eq:sdefn})), we verify explicitly in Appendix
\ref{app:BosonicKubo} that the Kubo formulas yield the same
results for the op-amp gains and impedances as Eqs.
(\ref{eq:Gaindefn})-(\ref{eq:Zindefn}) above.

\subsection{Minimal two-port scattering amplifier}
\label{subsec:MinAmp}


\subsubsection{Scattering versus op-amp quantum limit}
In this subsection we demonstrate that an
amplifier which is ``ideal" and minimally complex when used in the scattering
operation mode fails, when used as a voltage op-amp,
to have a quantum limited noise temperature. The system
we look at is very similar to the amplifier considered by
\textcite{Grassia98}, though our conclusions are somewhat
different than those found there.

In the scattering representation, one might guess that an
``ideal'' amplifier would be one where there are no reflections of
signals at the input and output, and no way for incident signals
at the output port to reach the input.  In this case,
Eq.~(\ref{eq:sdefn}) takes the form:
\begin{eqnarray}
    \left(\begin{array}{c}
    \haout \\
    \hbout \end{array}\right) & = & \left(\begin{array}{cc}
    0 & 0\\
    \sqrt{G} & 0\end{array}\right)\left(\begin{array}{c}
    \hain \\
    \hbin \end{array}\right)+\left(\begin{array}{c}
    \hFF_{a}\\
    \hFF_{b}\end{array}\right)
    \label{eq:sdefnsimp}
\end{eqnarray}
where we have defined $\sqrt{G} \equiv s_{21}$.
All quantities above should be evaluated at the same frequency
$\omega$; for clarity, we will omit writing the explicit $\omega$
dependence of quantities throughout this section.

Turning to the op-amp representation, the above equation implies
that our amplifier has no reverse gain, and that the input and
output impedances are simply given by the impedances of the input
and output transmission lines. From Eqs.~(\ref{eq:OpAmpParams}),
we have:
\begin{subequations}
\begin{eqnarray}
    \lambda_{V} & = & 2\sqrt{\frac{Z_{b}}{Z_{a}} G}
        \label{eq:Gainsimp}\\
    \lambda_{I}' & = & 0\\
    \Zout & = & Z_{b}\\
    \frac{1}{ \Zin} & = & \frac{1}{Z_{a}}
        \label{eq:Zinsimp}
\end{eqnarray}
\end{subequations}
We immediately see that our amplifier looks less ideal as an op-amp.  The input
and output impedances are the same as those of the input and output transmission line.
However, for an ideal op-amp, we would have liked $\Zin \ra \infty$ and $\Zout \ra 0$.

Also of interest are the expressions for the
amplifier noises in the op-amp representation:
\begin{eqnarray}
    \left(\begin{array}{c}
        \htV\\
        Z_{a}\cdot \htI
    \end{array}\right)
         & = &
    -\sqrt{2\hbar\omega Z_{a}}
    \left(\begin{array}{cc}
        \frac{1}{2} & - \frac{1}{2 \sqrt{G}}\\
        1 & 0\end{array}\right)
    \left(\begin{array}{c}
        \hFF_{a}\\
        \hFF_{b}\end{array}
    \right)
    \nonumber \\
    \label{eq:OpAmpNoisesSimp} 
\end{eqnarray}
As $s_{12}=0$, the back-action noise is the same in both the
op-amp and scattering representations: it is determined
completely by the noise operator $\hFF_{a}$. However, the
voltage noise (i.e. the intrinsic output noise) involves {\it
both} $\hFF_{a}$ and $\hFF_{b}$. We thus have the unavoidable
consequence that there will be correlations in $\htI$ and
$\htV$.  Note that from basic linear response theory, we know
that there must be some correlations between $\htI$ and $\htV$
if there is to be gain (i.e. $\lambda_V$ is given by a Kubo
formula involving these operators, cf.~Eq.~(\ref{eq:gain})).

To make further progress, we note again that commutators of the
noise operators $\hFF_a$ and $\hFF_b$ are completely determined by
Eq.~(\ref{eq:sdefnsimp}) and the requirement that the output
operators obey canonical commutation relations. We thus have:
\begin{subequations}
 \begin{eqnarray}
\left[\hFF_{a},\hFF_{a}^{\dagger}\right] & = & 1\\
\left[\hFF_{b},\hFF_{b}^{\dagger}\right] & = & 1-\left|G \right|\\
\left[\hFF_{a},\hFF_{b}\right] = \left[\hFF_{a},\hFF_{b}^{\dagger}\right] & = & 0
\end{eqnarray}
\end{subequations}

\begin{figure}[t]
\begin{center}
\includegraphics[width=3.45in]{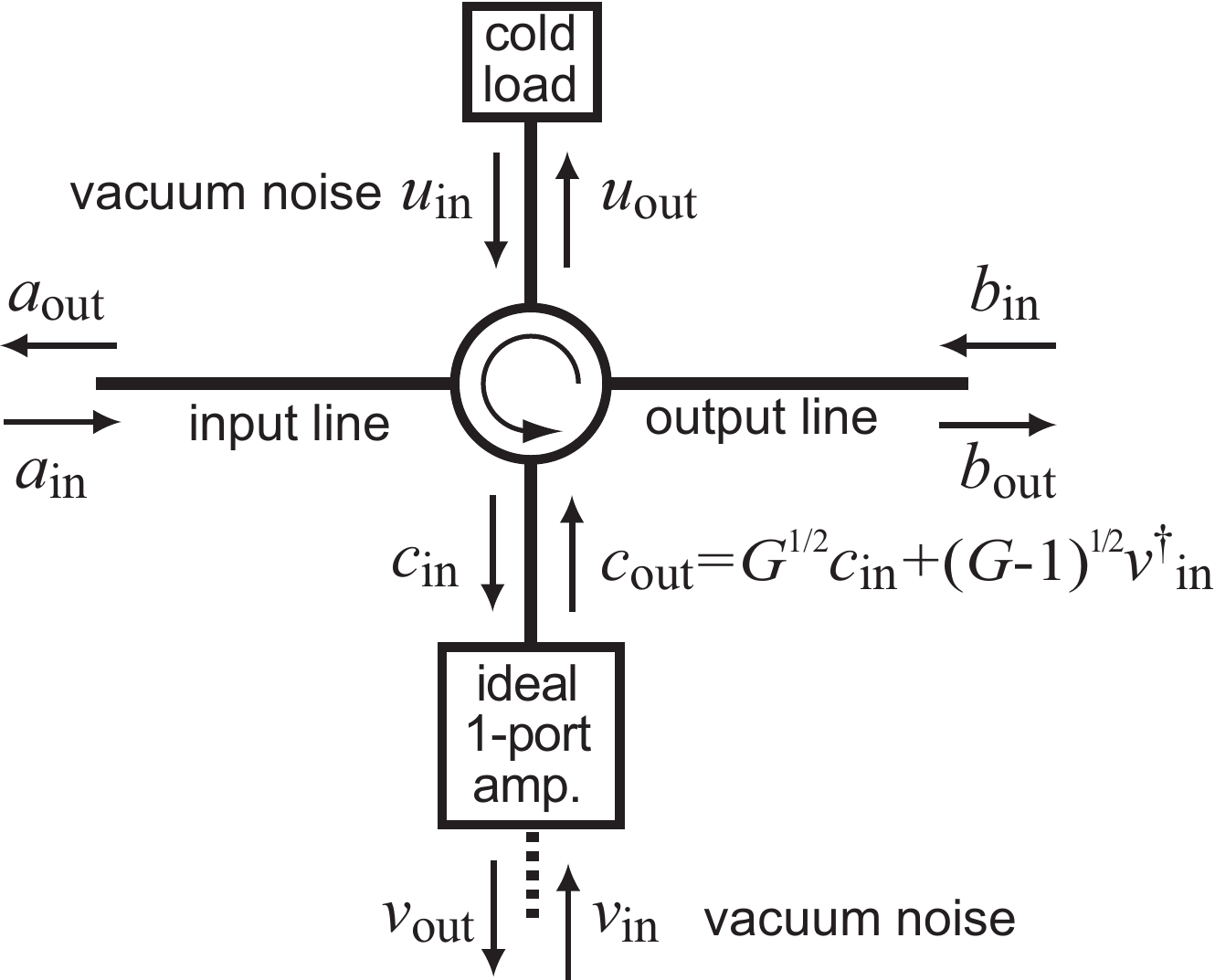}
\caption{Schematic of a ``minimal" two-port amplifier which reaches the
quantum limit in the scattering mode of operation, but
misses the quantum limit when used as a weakly-coupled op-amp.
See text for further description}
\label{fig:IdealTwoPortAmp}
\end{center}
\end{figure}

We will be interested in the limit of a large power gain, which
requires $|G| \gg 1$. A minimal solution to the above
equations would be to have the noise operators determined by two
independent (i.e.~mutually commuting) auxiliary input mode
operators $u_{in}$ and $v_{in}^{\dagger}$:\begin{eqnarray}
\hFF_{a} & = & \huin^{\pd}
    \label{eq:umode}\\
\hFF_{b} & = & \sqrt{|G|-1}\hvin^{\dagger}
    \label{eq:vmode}
\end{eqnarray}
Further, to minimize the noise of the amplifier, we take the
operating state of the amplifier to be the vacuum for both
these modes. With these choices, our amplifier is in exactly
the minimal form described by \textcite{Grassia98}: an input
and output line coupled to a negative resistance box and an
auxiliary ``cold load'' via a four-port circulator (see
Fig.~\ref{fig:IdealTwoPortAmp}).  The negative resistance box
is nothing but the single-mode bosonic amplifier discussed in
Sec.~\ref{subsec:CavesArg}; an explicit realization of this
element would be the parametric amplifier discussed in 
Sec.~\ref{subsec:NDParamps}.
The ``cold load" is a semi-infinite
transmission line which models dissipation due to a resistor at
zero-temperature (i.e.~its noise is vacuum noise, cf.~Appendix
\ref{app:QuantumResistor}).

Note that within the scattering picture, one would conclude that
our amplifier is ideal:  in the large gain limit, the noise added
by the amplifier to $\hbout$ corresponds to a single quantum at
the input:
\begin{eqnarray}
    \frac{
        \left \langle
            \left\{ \hFF^{\pd}_b, \hFF^{\dagger}_b \right\}
        \right \rangle}
        {|G|}
    =
        \frac{ |G|-1 }{|G|}
            \left \langle
                \left\{ \hvin^{\dag}, \hvin^{\pd} \right\}
            \right \rangle
    \ra 1
    \label{eq:ScattAN}
\end{eqnarray}
This however is {\it not} the quantity which interests us:  as we
want to use this system as a voltage op-amp, we would like to
know if the noise temperature {\it defined in the op-amp picture}
is as small as possible.  We are also usually interested in the
case of a signal which is weakly coupled to our amplifier; here,
weak-coupling means that the input impedance of the amplifier is
much larger than the impedance of the signal source
(i.e. $\Zin \gg \Zs$).  In this limit, the amplifier only slightly increases
the total damping of the signal source.

To address whether we can reach the op-amp quantum limit in the
weak-coupling regime, we can make use of the results of the
general theory presented in Sec.~\ref{subsec:VoltageAmp}.  In
particular, we need to check whether the quantum noise constraint
of Eq.~(\ref{eq:VAmpNoiseConstraint}) is satisfied, as this is a
prerequisite for reaching the (weak-coupling) quantum limit.
Thus, we need to calculate the {\it symmetrized} spectral
densities of the current and voltage noises, and their
cross-correlation. It is easy to confirm from the definitions of
Eq. (\ref{eq:VOpdefn}) and (\ref{eq:IOpdefn}) that these
quantities take the form:
\begin{subequations}
\begin{eqnarray}
    \bS_{VV}[\omega] & = &
    \frac{
        \left\langle \left\{ \htV[\omega],\htV^{\dagger}(\omega')\right\}
        \right\rangle
    }{4 \pi \delta(\omega - \omega')} \\
    \bS_{II}[\omega] & = & \frac{
        \left\langle \left\{ \htI[\omega],
        \htI^{\dagger}(\omega') \right\} \right\rangle
    }{4 \pi \delta(\omega-\omega')} \\
    \bS_{VI}[\omega]  & = & \frac{
        \left\langle \left\{ \htV[\omega],
        \htI^{\dagger}(\omega')\right\} \right\rangle
    }{4 \pi \delta(\omega - \omega')}
\end{eqnarray}
\end{subequations}
The expectation values here are over the operating state of the
amplifier; we have chosen this state to be the vacuum for the
auxiliary mode operators $\huin$ and $\hvin$ to minimize the
noise.

Taking $|s_{21}| \gg 1$, and using Eqs.~(\ref{eq:umode}) and
(\ref{eq:vmode}), we have
\begin{subequations}
\label{eq:MinNoises}
\begin{eqnarray}
    \bS_{VV}[\omega] & = &
        \frac{\hbar\omega Z_{a}}{4}\left(\sigma_{uu}+\sigma_{vv}\right)
        = \frac{\hbar\omega Z_{a}}{2}\\
    \bS_{II}[\omega] & = &
        \frac{\hbar\omega}{Z_{a}}\sigma_{uu}
        = \frac{\hbar\omega}{Z_{a}}\\
    \bS_{VI}[\omega] & = &
        \frac{\hbar\omega}{2} \sigma_{uu} = \frac{\hbar\omega}{2}
    \label{eq:SIV}
\end{eqnarray}
\end{subequations}
where we have defined:
\begin{eqnarray}
    \sigma_{ab} & \equiv & \left\langle \ha \hb^{\dagger}+\hb^{\dagger} \ha \right\rangle
\end{eqnarray}
and have used the fact that there cannot be any correlations
between the operators $u$ and $v$ in the vacuum state (i.e.~
$\langle \hu \hv^{\dagger}\rangle=0$).

It follows immediately from the above equations that our minimal
amplifier does not optimize the quantum noise constraint of
Eq.~(\ref{eq:VAmpNoiseConstraint}):
\begin{eqnarray}
    \bS_{VV}[\omega] \bS_{II}[\omega] - \left[
        \textrm{Im } \bS_{VI} \right]^2 = 2 \times
        \left( \frac{\hbar \omega}{2} \right)^2.
        \label{eq:MissedQL}
\end{eqnarray}
The noise product $\bS_{VV} \bS_{II}$ is
precisely twice the quantum-limited value.  As a result, the
general theory of Sec.~\ref{subsec:VoltageAmp} tells us {\it if
one couples an input signal weakly to this amplifier (i.e.~$\Zs
\ll \Zin$), it is impossible to reach the quantum limit on the
added noise}.   Thus, while our amplifier
is ideal in the scattering mode of operation (cf.~Eq.~(\ref{eq:ScattAN})),
it fails to reach the quantum
limit when used in the weak-coupling, op-amp mode of operation.
Our amplifier's failure to have ``ideal" quantum
noise also means that if we tried to use it to do QND qubit
detection, the resulting back-action dephasing would be twice as
large as the minimum required by quantum mechanics
(cf.~Sec.~\ref{subsec:QLimitProof}).

One might object to the above conclusions based on the
classical expression for the minimal noise temperature,
Eq.~(\ref{eq:TNopt}).  Unlike the quantum noise constraint of
Eq.~(\ref{eq:VAmpNoiseConstraint}), this equation also involves
the real part of $\bS_{VI}$, and is optimized by our ``minimal"
amplifier.  However, this does not mean that one can achieve a
noise temperature of $\hbar \omega / 2$ at weak coupling! Recall
from Sec.~\ref{subsec:VoltageAmp} that in the usual process of
optimizing the noise temperature, one starts by assuming the weak
coupling condition that the source impedance $\Zs$ is much
smaller than the amplifier input impedance $\Zin$. One then
finds that to minimize the noise temperature, $|\Zs|$ should be
tuned to match the noise impedance of the amplifier $\ZN \equiv
\sqrt{\bS_{VV}/\bS_{II}}$. However, in our minimal bosonic
amplifier, it follows from Eqs.~(\ref{eq:MinNoises}) that
$\ZN=\Zin / \sqrt{2}\sim \Zin$: the noise impedance is on the
order of the input impedance. Thus, it is {\it impossible} to
match the source impedance to the noise impedance while at the
same time satisfying the weak coupling condition $\Zs \ll
\Zin$. 

Despite its failings, our amplifier can indeed yield a quantum-limited
noise temperature in the op-amp mode of operation {\it if}
we no longer insist on a weak coupling to the input signal. To see this
explicitly, imagine we connected our amplifier to a signal source
with source impedance $Z_{s}.$ The total output noise of the
amplifier, referred back to the signal source, will now have the
form:
\begin{eqnarray}
    \htV_{\rm{tot}} & = &
    -\left(
        \frac{Z_{s}Z_{a}}{Z_{s}+Z_{a}}
    \right) \htI+
    \htV
    \label{eq:vtot}
\end{eqnarray}
Note that this classical-looking equation can be rigorously
justified within the full quantum theory if one starts with a full
description of the amplifier and the signal source (e.g.~a
parallel LC oscillator attached in parallel to the amplifier
input).  Plugging in the expressions for $\htI$ and $\htV$, we
find:
\begin{eqnarray}
    \htV_{\rm{tot}} & = &
        \sqrt{\frac{\hbar\omega}{2}}
        \Bigg[
            \left(\frac{\Zs Z_{a}}{ \Zs+ Z_{a}}\right)
                \left( \frac{2}{ \sqrt{ Z_{a} } } \huin \right)
        -
        \nonumber \\
        && \sqrt{Z_{a}}( \huin -
        \hvin^{\dagger})
        \Bigg] \nonumber \\
    & = &
        \sqrt{\frac{\hbar\omega Z_{a}}{2}}
        \left[\left(
            \frac{\Zs -Z_{a}}{ \Zs +Z_{a}}
            \right)
            \huin  - \hvin^{\dagger}
            \right]
\end{eqnarray}
Thus, if one tunes $\Zs$ to $Z_{a}=\Zin$, the mode $\huin$
does not contribute to the total added noise, and one reaches the
quantum limit. Physically speaking, by matching the signal source
to the input line, the ``back-action'' noise described by $\hFF_a
= \huin$ does not feed back into the input of the amplifier.
Note that achieving this matching explicitly requires one to be
far from weak coupling! Having $Z_{s}=Z_{a}$ means that when we
attach the amplifier to the signal source, we will dramatically
increase the damping of the signal source.

\subsubsection{Why is the op-amp quantum limit not achieved?}

Returning to the more interesting case of a weak amplifier-signal
coupling, one might still be puzzled as to why our seemingly ideal
amplifier misses the quantum limit. While the mathematics behind
Eq.~(\ref{eq:MissedQL}) is fairly transparent, it is also possible
to understand this result heuristically. To that end, note again
that the amplifier noise cross-correlation $\bS_{IV}$ does not
vanish in the large-gain limit (cf.~Eq. (\ref{eq:SIV})).
Correlations between the two amplifier noises represent a kind of
information, as by making use of them, we can improve the
performance of the amplifier.  It is easy to take advantage of
out-of-phase correlations between $\tI$ and $\tV$ (i.e.
$\textrm{Im }\bS_{VI}$) by simply tuning the phase of the source
impedance (cf.~Eq.~(\ref{eq:TNeqn})).  However, one cannot take
advantage of in-phase noise correlations (i.e.~$\textrm{Re
}\bS_{VI}$) as easily.  To take advantage of the information here,
one needs to modify the amplifier itself.  By feeding back some of
the output voltage to the input, one could effectively cancel out
some of the back-action current noise $\tI$ and thus reduce the
overall magnitude of $\bS_{II}$.  Hence, the unused information in
the cross-correlator $\textrm{Re }\bS_{VI}$ represents a kind of
wasted information:  had we made use of these correlations via a
feedback loop, we could have reduced the noise temperature and
increased the information provided by our amplifier.  The presence
of a non-zero $\textrm{Re }\bS_{VI}$ thus corresponds to wasted
information, implying that we cannot reach the quantum limit.
Recall that within the linear-response approach, we were able to
prove rigorously that a large-gain amplifier with ideal quantum
noise {\it must} have $\textrm{Re }\bS_{VI}=0$ (cf.~the discussion
following Eq.~(\ref{eq:GPTeff})); thus, a non-vanishing
$\textrm{Re } \bS_{VI}$ rigorously implies that one cannot be at
the quantum limit.  In Appendix \ref{subsec:MirrorsFeedback}, 
we give an explicit demonstration of how feedback may be used 
to utilize these cross-correlations to reach the quantum limit.

Finally, yet another way of seeing that our amplifier does not
reach the quantum limit (in the weak coupling regime) is to
realize that this system {\it does not have a well defined
effective temperature}. Recall from Sec.\ref{subsec:VoltageAmp}
that a system with ``ideal'' quantum noise (i.e. one that
satisfies Eq. (\ref{eq:VAmpNoiseConstraint}) as an equality)
necessarily has the {\it same} effective temperature at its input
and output ports (cf.~Eq.~(\ref{eq:OneTEff})).  Here, that implies
the requirement:
\begin{eqnarray}
    \frac{|\lambda_{V}|^{2}\cdot \bS_{VV}}{ \Zout }
        & = &
            \Zin \bS_{II} \equiv 2 \kb T_{\rm eff}
\end{eqnarray}
In contrast, our minimal bosonic amplifier has very different
input and output effective temperatures:
\begin{eqnarray}
    2 k_{B}T_{\rm eff,in} & = &
        \Zin \bS_{II}=\frac{\hbar\omega}{2}\\
    2 k_{B}T_{\rm eff,out} & = &
        \frac{|\lambda_{V}|^{2} \cdot \bS_{VV}}{ \Zout }
            =2 |G| \hbar\omega
\end{eqnarray}
This huge difference in effective temperatures means that it is
impossible for the system to possess ``ideal'' quantum noise, and
thus it cannot reach the weak-coupling quantum limit.

While it implies that one is not at the quantum limit, the
fact that $T_{\rm{eff,in}} \ll T_{\rm{eff,out}}$ can nonetheless be viewed as
an asset.  From a practical point of view, a large
$T_{\rm{eff,in}}$ can be dangerous.
Even though the direct effect of the large $T_{\rm{eff,in}}$ is offset by an appropriately weak coupling to the amplifier (see Eq.~(\ref{eq:QLheating}) and following discussion), this large $T_{\rm{eff,in}}$ can also heat up other degrees of freedom if they couple strongly to the back-action noise of the amplifier.  This can in turn lead to unwanted heating of the input system.  As $T_{\rm{eff,in}}$ is usually constant over a broad range of frequencies, this unwanted heating effect can be quite bad.  In the minimal amplifier discussed here, this problem is circumvented by having a small $T_{\rm{eff,in}}$.  The only price that is payed is that the added noise will be $\sqrt{2}$ the quantum limit value.
We discuss this issue further in Sec.~\ref{sec:QLinpractice}.

\section{Reaching the Quantum Limit in Practice}
\label{sec:QLinpractice}

\subsection{Importance of QND measurements}

The fact that QND measurements are repeatable is of fundamental
practical importance in overcoming detector inefficiencies
\cite{Gambetta07}. A prototypical example is the electron shelving
technique \cite{Dehmelt86,Blatt86} used to measure trapped ions. A
related technique is used in present implementations of ion-trap
based quantum computation. Here the (extremely long-lived)
hyperfine state of an ion is read out via state-dependent optical
fluorescence. With properly chosen circular polarization of the
exciting laser, only one hyperfine state fluoresces and the
transition is cycling; that is, after fluorescence the ion almost
always returns to the same state it was in prior to absorbing the
exciting photon. Hence the measurement is QND. Typical
experimental parameters \cite{Wineland1998} allow the cycling
transition to produce $N \sim 10^6$ fluorescence photons. Given
the photomultiplier quantum efficiency and typically small solid
angle coverage, only a very small number ${\bar n}_d$ will be
detected on average. The probability of getting zero detections
(ignoring dark counts for simplicity) and hence misidentifying the
hyperfine state is $P(0) = e^{-{\bar n}_d}$. Even for a very poor
overall detection efficiency of only $10^{-5}$, we still have
${\bar n}_d=10$ and nearly perfect fidelity $F = 1 - P(0)\approx
0.999955$. It is important to note that the total time available
for measurement is not limited by the phase coherence time ($T_2$)
of the qubit or by the measurement-induced dephasing
\cite{Korotkov01a,Makhlin01,Schuster05,Gambetta2006}, but rather
only by the rate at which the qubit makes real transitions between
measurement ($\hat\sigma_z$) eigenstates. In a perfect QND
measurement there is no measurement-induced state mixing
\cite{Makhlin01} and the relaxation rate $1/T_1$ is unaffected by
the measurement process.
%

\subsection{Power matching versus noise matching}
\label{subsec:PowerMatching}

In Sec.~\ref{sec:QLAmplifiers}, we saw that an important part of reaching
the quantum limit on the added noise of an amplifier (when used in the op-amp mode of operation) is to optimize the coupling strength to the amplifier.  For a position detector, this condition corresponds to tuning the strength of the back-action damping $\gamma$ to be much smaller than the intrinsic oscillator damping (cf. Eq.~(\ref{OptA2})).  For a voltage amplifier, this condition corresponds to tuning the impedance of the signal source to be equal to the noise impedance (cf. Eq.~(\ref{ZOpt1})), an impedance which is much smaller than the amplifier's input impedance (cf. Eq.~(\ref{ZinConstraint})).

In this subsection, we make the simple point that optimizing the coupling (i.e.~source impedance) to reach the quantum limit {\it is not} the same as what one would do to optimize the power gain.  To understand this, we need to introduce another measure of power gain commonly used in the engineering community, the available power gain $G_{P,{\rm avail}}$.    For simplicity, we will discuss this quantity in the context of a linear voltage amplifier, using the notations of Sec.~\ref{subsec:VoltageAmp}; it can be analogously defined for the position detector of Sec.\ref{subsec:PositionDetector}.  $G_{P,{\rm avail}}$ tells us how much power we are providing to an optimally matched output load relative to the maximum power we {\it could} in principle have extracted from the source.  This is in marked contrast to the power gain $G_P$, which was calculated using the {\it actual} power drawn at the amplifier input.

For the available power gain, we first consider $P_{\rm in, avail}$.  This is the maximum possible power that could be delivered to the input of the amplifier, assuming we optimized {\it both} the value of the input impedance $Z_{\rm in}$ and the load impedance $Z_{\rm load}$ while keeping $Z_{\rm s}$ fixed.  For simplicity, we will take all impedances to be real in our discussion.
In general, the power drawn at the input of the amplifier is given by 
$	P_{\rm in} = v_{\rm in}^2 \Zin /  \left( Z_{s} + \Zin \right)^2$.
Maximizing this over $\Zin$, we obtain the available input power $P_{\rm in, avail}$:
\begin{eqnarray}
	P_{\rm in, avail} =
		\frac{ v_{\rm in}^2 }{4 Z_{\rm s} }
\end{eqnarray}
The maximum occurs for $\Zin = Z_{\rm s}$.

The output power supplied to the load 
$P_{\rm out} = v_{\rm load}^2 / Z_{\rm load}$
is calculated as before,
keeping $\Zin$ and $Z_{\rm s}$ distinct.  One has:
\begin{eqnarray}
	P_{\rm out} & = &
		\frac{v_{\rm out}^2}{ Z_{\rm load}}
		\left( \frac{Z_{\rm load} }{\Zout + Z_{\rm load} } \right)^2 \nonumber \\
	& = &
		\frac{ \lambda^2 v_{\rm in}^2 }{ \Zout }
		\left( \frac{ \Zin }{\Zin + \Zs} \right)^2
		\frac{ Z_{\rm load} / \Zout }{ \left(1 + Z_{\rm load} / \Zout \right)^2}
\end{eqnarray}
As a function of $Z_{\rm load}$, $P_{\rm out}$ is maximized when $Z_{\rm load} = \Zout$:
\begin{eqnarray}
	P_{\rm out, max} & = &
		\frac{ \lambda^2 v_{\rm in}^2 }{4 \Zout }
		\left( \frac{ \Zin }{\Zin + \Zs} \right)^2
\end{eqnarray}
The available power gain is now defined as:
\begin{eqnarray}
	G_{P, {\rm avail}}  \equiv 
		\frac{ P_{\rm out, max} }{P_{\rm in, avail} } 
		& = &
		\frac{ \lambda^2 Z_{\rm s} }{\Zout }
		\left( \frac{ \Zin }{\Zin + \Zs} \right)^2 \nonumber \\
		& = &
		G_P \frac{ 4 Z_{\rm s} / \Zin }
		{\left( 1 + \Zs / \Zin  \right)^2}
\end{eqnarray}
We see that $G_{P, {\rm avail}}$ is strictly less than or equal to the power gain $G_P$; equality is only achieved when $\Zs = \Zin$ (i.e.~when the source impedance is ``power matched" to the input of the amplifier).  The general situation where
$G_{P, {\rm avail}} < G_P$ indicates that we are not drawing as much power from the source as we could, and hence the actual power supplied to the load is not as large as it could be.

Consider now a situation where we have achieved the quantum limit on the added noise.
This necessarily means that we have ``noise matched", i.e. taken $Z_{\rm s}$ to
be equal to the noise impedance $\ZN$.  The available power gain in this case is:
\begin{eqnarray}
	G_{P, {\rm avail}} \simeq
		\lambda^2 \frac{ Z_{N} }{\Zout}
	\simeq 2 \sqrt{G_P} \ll G_P
\end{eqnarray}
We have used Eq.~(\ref{ZinConstraint}), which tells us that the noise impedance is smaller than the input impedance by a large factor $1/(2 \sqrt{G_P})$.  Thus, as reaching the quantum limit requires the use of a source impedance much smaller than $\Zin$, it results in a dramatic drop in the available power gain compared to the case where we ``power match" (i.e. take $Z_{\rm s} = \Zin$).  In practice, one must decide whether it is more important to minimize the added noise, or maximize the power provided at the output of the amplifier:  one cannot do both at the same time.

\section{Conclusions}

In this review, we have given an introduction to quantum limits for position detection and amplification, limits 
which are tied to fundamental constraints on quantum noise correlators. We wish 
to end by briefly emphasizing notable current developments and pointing out future perspectives in the field.

As we have repeatedly emphasized, much of our discussion has been
directly relevant to the measurement of mechanical nanoresonators, a topic of considerable
recent attention.  These nanoresonators are typically studied by coupling them either to electrical (often superconducting) circuits or to optical cavities.  A key goal is to achieve quantum-limited continuous position detection (cf.~Sec.~\ref{subsec:PositionDetector}); current experiments are coming tantalyzingly close to this limit (cf.~Table \ref{table:QLExperiments}). 
Although being able to follow the nanoresonator's motion with a precision set by the quantum limit is in principle independent from being at low temperatures, it becomes interesting only when the systems are near their ground state; one could then, e.g., monitor the oscillator's
zero-point fluctuations (cf.~Sec.~\ref{subsec:measurementofoscillatorposition}).   Given the comparatively small values of mechanical frequencies (mostly less than GHz), this calls for the application of non-equilibrium cooling techniques which exploit back-action to reduce the effective temperature of the mechanical device, a technique that has been demonstrated recently both in superconducting circuits and optomechanical setups (see \textcite{2009_05_MarquardtGirvin_OptomechanicsReview} for a review).

The ability to do quantum limited position detection will in turn open up many new interesting avenues of research.  Among the most significant, it will allow the possibility of doing quantum feedback control \cite{Wiseman93, Wiseman94}, where one uses the continuously-obtained measurement output to tailor the state of the mechanical resonator.  The relevant theoretical framework is that of quantum conditional evolution and quantum trajectories (see e.g.~\onlinecite{Brun02,JacobsSteck06}), where one tracks the state of a measured quantum system in a particular run of the experiment.  The application of these ideas has only recently been explored in condensed matter contexts \cite{Korotkov99, Korotkov01b, Goan01, Goan01b,Diosi08,Wiseman06,Wiseman08}.  Fully understanding the potential of these techniques, as well as differences that occur in condensed matter versus atomic physics contexts, remains an active area of research.  Other important directions in nanomechanics include the possibility of detecting quantum jumps in the state of a mechanical resonator via QND measurement of its energy \cite{Santamore04, Santamore04b,
Harris08, Jayich08}, as well as the possibility of making backaction evading measurements (cf.~Sec.~\ref{subsec:BackactionEvasion}).  Back-action evasion using a microwave cavity detector coupled to a nanomechanical resonator was recently reported \cite{Schwab09}.  

Another area distinct from nanomechanics where rapid progress is being made is the readout of solid state qubits using microwave signals sent through cavities whose transmission properties are controlled by the qubit. At the moment, one is close to achieving good fidelity single-shot QND readout, which is a prerequisite for a large number of applications in quantum information processing. The gradually growing information about the qubit state is extracted from the measured noisy microwave signal trace, leading to a corresponding collapse of the qubit state. This process can also be described by conditional quantum evolution and quantum trajectories. Ê

A promising method for superconducting qubit readout currently employed is a so-called ``latching measurement'', where the hysteretic behaviour of a strongly driven anharmonic system (e.g. a Josephson junction) is exploited to toggle between two states depending on the qubit state Ê\cite{Siddiqi04,Lupascu06}. Although this is then no longer a linear measurement scheme and therefore distinct from what we have discussed in this review, it can be turned into a linear amplifier for a sufficiently weak input signal.  An interesting and important open question is whether such a setup can reach the quantum limit on linear amplification.

Both qubit detection and mechanical measurements in electrical circuits would benefit from quantum-limited on-chip amplifiers. Such amplifiers are now being developed using the tools of circuit quantum electrodynamics, employing Josephson junctions or SQUIDs coupled to microwave transmission line cavities \cite{ 2008_Devoret_JosephsonRingModulator, Lehnert08}.  Such an amplifier has already been used
to perform continuous position detection with a measurement imprecision below the SQL level \cite{Lehnert09}.

\section*{Acknowledgements} 

This work was supported in part by NSERC,
the Canadian Institute for Advanced Research (CIFAR), NSA under
ARO contract number W911NF-05-1-0365, the NSF under
grants DMR-0653377  and DMR-0603369, the David and Lucile
Packard Foundation, the W.M. Keck Foundation and the Alfred P. Sloan Foundation.
 We
acknowledge the support and hospitality of the Aspen Center for
Physics where part of this work was carried out.
F.M. acknowledges support via DIP, GIF and 
through SFB 631, SFB/TR 12, the Nanosystems Initiative Munich,
and the Emmy-Noether program.
S. M. G. acknowledges the support of the Centre for Advanced Study
at the Norwegian Academy of Science and Letters where part of
this work was carried out.  


\appendix

\section{Basics of Classical and Quantum Noise}
\label{sec:ReviewClassicalNoise}

\subsection{Classical noise correlators}
\label{subsec:classicalnoisecorrelators}

Consider a classical random voltage signal $V(t)$.  The signal is
characterized by zero mean $\langle V(t)\rangle=0$, and
autocorrelation function
\be
G_{VV}(t-t') = \langle V(t) V(t') \rangle
\label{eq:voltageautocorrelator}
\ee
whose sign and magnitude tells us whether the voltage fluctuations
at time $t$ and time $t^\prime$ are correlated, anti-correlated or
statistically independent. We assume that the noise process is
{\em stationary} (i.e., the statistical properties are time
translation invariant) so that $G_{VV}$ depends only on the time
difference. If $V(t)$ is Gaussian distributed, then the mean and
autocorrelation completely specify the statistical properties and
the probability distribution. We will assume here that the noise
is due to the sum of a very large number of fluctuating charges so
that by the central limit theorem, it is Gaussian distributed.  We
also assume that $G_{VV}$ decays (sufficiently rapidly) to zero on
some characteristic correlation time scale $\tau_{\rm c}$ which is
finite.

The spectral density of the noise as measured by a spectrum
analyzer is a measure of the intensity of the signal at different
frequencies. In order to understand the spectral density of a
random signal, it is useful to define its `windowed' Fourier transform as
follows:
\be
V_T[\omega] = \frac{1}{\sqrt{T}}\int_{-T/2}^{+T/2}dt\, e^{i\omega
t}V(t),
\label{eq:FT}
\ee
where $T$ is the sampling time.  In the limit $T\gg\tau_{\rm c}$
the integral is a sum of a large number $N\approx
\frac{T}{\tau_{\rm c}}$ of random uncorrelated terms. We can think
of the value of the integral as the end point of a random walk in
the complex plane which starts at the origin. Because the distance
traveled will scale with $\sqrt{T}$, our choice of normalization
makes the statistical properties of $V[\omega]$ independent of the
sampling time $T$ (for sufficiently large $T$). Notice that
$V_T[\omega]$ has the peculiar units of volts$\sqrt{\rm secs}$ which
is usually denoted volts/$\sqrt{\rm Hz}$.

The spectral density (or `power spectrum') of the noise is defined
to be the ensemble averaged quantity
\be
\clS_{VV}[\omega] \equiv 
	\lim_{T \ra \infty} \langle |V_T[\omega]|^2\rangle
	=  \lim_{T \ra \infty} \langle
V_T[\omega]V_T[-\omega]\rangle
\label{eq:spectraldensity}
\ee
The second equality follows from the fact that $V(t)$ is real
valued. The Wiener-Khinchin theorem (derived in Appendix
\ref{app:WienerKhinchin}) tells us that the spectral density is
equal to the Fourier transform of the autocorrelation function
\be
\clS_{VV}[\omega]=\int_{-\infty}^{+\infty} dt\, e^{i\omega t}
G_{VV}(t).
\label{eq:WienerKhinchin}
\ee
The inverse transform relates the autocorrelation function to the
power spectrum
\be
G_{VV}(t)=\int_{-\infty}^{+\infty} \frac{d\omega}{2\pi}\,
e^{-i\omega t}\clS_{VV}[\omega].
\label{eq:inversetransform}
\ee

We thus see that a short auto-correlation time implies a spectral
density which is non-zero over a wide range of frequencies.  In
the limit of `white noise'
\be
G_{VV}(t)=\sigma^2 \delta(t)
\ee
the spectrum is flat (independent of frequency)
\be
\clS_{VV}[\omega]=\sigma^2
\ee
In the opposite limit of a long autocorrelation time, the signal
is changing slowly so it can only be made up out of a narrow range
of frequencies (not necessarily centered on zero).

Because $V(t)$ is a real-valued classical variable, it naturally
follows that $G_{VV}(t)$ is always real.  Since $V(t)$ is not a
quantum operator, it commutes with its value at other times and
thus, $\langle V(t)V(t')\rangle =\langle V(t')V(t)\rangle$. From
this it follows that $G_{VV}(t)$ is always symmetric in time and
the power spectrum is always symmetric in frequency
\be
\clS_{VV}[\omega]=\clS_{VV}[-\omega].
\ee

As a prototypical example of these ideas, let us consider a simple
harmonic oscillator of mass $M$ and frequency $\Omega$. The
oscillator is maintained in equilibrium with a large heat bath at
temperature $T$ via some infinitesimal coupling which we will
ignore in considering the dynamics.  The solution of Hamilton's
equations of motion are
\begin{eqnarray}
x(t) &=& x(0) \cos(\Omega t) + p(0)\frac{1}{M\Omega}\sin(\Omega
t)\nonumber\\
p(t) &=&p(0) \cos(\Omega t) - x(0){M\Omega}\sin(\Omega t),
\end{eqnarray}
where $x(0)$ and $p(0)$ are the (random) values of the position
and momentum at time $t=0$. It follows that the position
autocorrelation function is
\begin{eqnarray}
G_{xx}(t) &=& \langle x(t) x(0)\rangle \\
&=& \langle x(0)x(0)\rangle\cos(\Omega t)  + \langle p(0)
x(0)\rangle\frac{1}{M\Omega}\sin(\Omega t).\nonumber
\end{eqnarray}
Classically in equilibrium there are no correlations between
position and momentum.  Hence the second term vanishes.  Using the
equipartition theorem $\frac{1}{2}M\Omega^2\langle x^2\rangle =
\frac{1}{2}k_{\rm B}T$, we arrive at
\be
G_{xx}(t) = \frac{\kbt}{M\Omega^2}\cos(\Omega t)
\ee
which leads to the spectral density
\be
\clS_{xx}[\omega] = \pi \frac{\kbt}{M\Omega^2}\left[ \delta(\omega
- \Omega) + \delta(\omega + \Omega)\right]
\label{eq:Sxxclassical}
\ee
which is indeed symmetric in frequency.

\subsection{The Wiener-Khinchin Theorem}
\label{app:WienerKhinchin}

 From the definition of the spectral density in
Eqs.(\ref{eq:FT}-\ref{eq:spectraldensity}) we have
\begin{eqnarray}
\clS_{VV}[\omega]  =
     \frac{1}{T}\int_{0}^{T}dt\int_{0}^{T}dt'\,e^{i\omega(t-t')}\langle
     V(t)V(t')\rangle \nonumber\\
              =
     \frac{1}{T}\int_{0}^{T}dt
     \int_{-2B\left( t\right)}^{+2B\left( t\right) }d\tau \,e^{i\omega
\tau} \left\langle
         V(t+\tau /2)V(t-\tau /2)\right\rangle \nonumber \\
\end{eqnarray}
where
\begin{eqnarray*}
B\left( t\right) &=&t~\mathrm{if~}t<T/2 \\
&=&T-t~\mathrm{if~}t>T/2.
\end{eqnarray*}

If $T$ greatly exceeds the noise autocorrelation time
$\tau_{\rm c}$ then it is a good approximation to extend the
bound $B(t)$ in the second integral to infinity, since the
dominant contribution is from small $\tau$. Using time
translation invariance gives
\begin{eqnarray}
\clS_{VV}[\omega]&=&\frac{1}{T}\int_{0}^{T}dt\int_{-\infty}^{+\infty}d\tau
\,e^{i\omega\tau} \left\langle V(\tau)V(0)\right\rangle\nonumber\\
&=&\int_{-\infty}^{+\infty}d\tau \,e^{i\omega\tau} \left\langle
V(\tau)V(0)\right\rangle.
\end{eqnarray}
This proves the Wiener-Khinchin theorem stated in
Eq.~(\ref{eq:WienerKhinchin}).

A useful application of these ideas is the following.  Suppose
that we have a noisy signal $V(t)=\bar V + \eta(t)$ which we
begin monitoring at time $t=0$.  The integrated signal up to
time $t$ is given by \be I(T) = \int_0^T dt\, V(t) \ee and has mean \be
\langle I(T) \rangle = \bar V T. \ee Provided that the
integration time greatly exceeds the autocorrelation time of
the noise, $I(T)$ is a sum of a large number of uncorrelated
random variables. The central limit theorem tells us in this
case that $I(t)$ is gaussian distributed even if the signal
itself is not.  Hence the probability distribution for $I$ is
fully specified by its mean and its variance \be \langle
(\Delta I)^2\rangle = \int_0^T dt dt'\, \langle
\eta(t)\eta(t')\rangle. \ee From the definition of spectral
density above we have the simple result that the variance of
the integrated signal grows linearly in time with
proportionality constant given by the noise spectral density at
zero frequency
\be
\langle (\Delta I)^2\rangle = \clS_{VV}[0]\, T.
\ee

As a simple application, consider the photon shot noise of a
coherent laser beam.  The total number of photons detected in
time $T$ is \be N(T) = \int_0^T dt\, \dot N(t). \ee The
photo-detection signal $\dot N(t)$ is \emph{not} gaussian, but
rather is a \emph{point process}, that is, a sequence of delta
functions with random Poisson distributed arrival times and
mean photon arrival rate $\Ndotbar$.  Nevertheless at long
times the mean number of detected photons \be \langle
N(T)\rangle = \Ndotbar T \ee will be large and the photon
number distribution will be gaussian with variance \be \langle
(\Delta N)^2\rangle = \SNN\,T. \ee Since we know that for a
Poisson process the variance is equal to the mean \be \langle
(\Delta N)^2\rangle = \langle N(T)\rangle, \ee it follows that
the shot noise power spectral density is \be \SNN(0) =
\Ndotbar. \ee Since the noise is white this result happens to
be valid at all frequencies, but the noise is gaussian
distributed only at low frequencies.


\subsection{Square law detectors and classical spectrum analyzers}
\label{subsec:classicalspectrumanalyzers}

Now that we understand the basics of classical noise, we can
consider how one experimentally measures a
classical noise spectral density.  With modern high speed digital sampling
techniques it is perfectly feasible to directly measure the random
noise signal as a function of time and then directly compute the autocorrelation
function in Eq.~(\ref{eq:voltageautocorrelator}).  This is typically done by first
performing an analog-to-digital conversion of the noise signal, and then numerically 
computing the autocorrelation function.  One can then use Eq.~(\ref{eq:WienerKhinchin}) to calculate the noise spectral density via a numerical Fourier transform.  Note that while Eq.~(\ref{eq:WienerKhinchin}) seems to require an ensemble average, in practice this is not explicitly done.  Instead, one uses a sufficiently long averaging time $T$ (i.e.~much longer than the correlation time of the noise) such that a single time-average is equivalent to an ensemble average.  This approach of measuring a noise spectral density directly from its autocorrelation function is most appropriate for signals at RF frequencies well below 1 MHz.  

For microwave signals with frequencies well above 1 GHz, a very different approach is usually taken.  Here, the standard route to obtain a noise spectral density involves first shifting the signal to a lower intermediate frequency via a technique known as heterodyning (we discuss this more in Sec.~\ref{subsubsec:HeterodyneAnalyzer}).  This intermediate-frequency signal is then sent to a filter which selects a narrow frequency range of interest, the so-called `resolution bandwidth'.  Finally, this filtered signal is sent to a square-law detector (e.g. a diode), and the resulting output is averaged over a certain time-interval (the inverse of the so-called `video bandwidth').  It is this final output which is then taken to be a measure of the noise spectral density.

It helps to put the above into equations.  Ignoring for simplicity the initial heterodyning step,  let  
\be
V_{\rm f}[\omega] = f[\omega] V[\omega]
\ee
be the voltage at the output of the filter and
 the input of the square law detector.  Here, $f[\omega]$ is the (amplitude)
 transmission coefficient of the filter and $V[\omega]$ is the Fourier transform of
 the noisy signal we are measuring.
From Eq.~(\ref{eq:inversetransform}) it follows that the output of
the square law detector is proportional to
\be
\langle I \rangle =
\int_{-\infty}^{+\infty}\frac{d\omega}{2\pi}|f[\omega]|^2
\clS_{VV}[\omega].
\ee
Approximating the narrow band filter centered on frequency $\pm
\omega_0$ as\footnote{A linear passive filter performs a
convolution $V_{\rm out}(t) = \int_{-\infty}^{+\infty} dt'\,
F(t-t') V_{\rm in}(t')$ where $F$ is a real-valued (and causal)
function.  Hence it follows that $f[\omega]$, which is the Fourier
transform of $F$, obeys $f[-\omega]=f^*[\omega]$ and hence
$|f[\omega]|^2$ is symmetric in frequency.}
\be
|f[\omega]|^2 = \delta(\omega-\omega_0) + \delta(\omega+\omega_0)
\label{eq:narrowfilter}
\ee
we obtain
\be
\langle I \rangle = \clS_{VV}(-\omega_0) + \clS_{VV}(\omega_0)
\ee
showing as expected that the classical square law detector
measures the symmetrized noise power.

We thus have two very different basic approaches for the measurement of classical noise spectral densities:  for low RF frequencies, one can directly measure the noise autocorrelation, whereas for high microwave frequencies, one uses a filter and a square law detector.  For noise signals in intermediate frequency ranges, a combination of different methods is generally used.  The whole story becomes even more complicated, as at very high frequencies (e.g.~in the far infrared), devices such as the so-called `Fourier Transform spectrometer' are in fact based on a direct measurement of the equivalent of an auto-correlation function of the signal. In the infrared, visible and ultraviolet, noise spectrometers use gratings followed by a slit acting as a filter.

\section{Quantum Spectrum Analyzers: Further Details}
\label{app:QSAFurtherDetails}

\subsection{Two-level system as a spectrum analyzer}
\label{subsec:TLSanalyzer}

In this sub-appendix, we derive the Golden Rule transition rates Eqs.~(\ref{eqs:FGRRates}) describing a quantum two-level system
coupled to a noise source (cf.~Sec.~\ref{subsec:QuantumSpectrumAnalyzers}).  
Our derivation is somewhat unusual, in that the role of the continuum as a noise
source is emphasized from the outset.
We start by treating the noise $F(t)$ in Eq.~(\ref{eq:TLSCoupling}) as being a classically noisy variable.
We assume that the coupling $A$ is under our control and can be
made small enough that the noise can be treated in lowest order
perturbation theory. We take the state of the two-level system to
be
\begin{equation}
|\psi (t)\rangle =\left(
\begin{array}{c}
\alpha _{g}(t) \\
\alpha _{e}(t)%
\end{array}%
\right) .
\end{equation}%
In the interaction representation, first-order time-dependent
perturbation theory gives
\begin{equation}
|\psi _{\mathrm{I}}(t)\rangle =|\psi (0)\rangle -\frac{i}{\hbar }%
\int_{0}^{t}d\tau \,\,\hat{V}(\tau )|\psi (0)\rangle .
\label{eq:first-order-evolution}
\end{equation}%
If we initially prepare the two-level system in its ground state,
the amplitude to find it in its excited state at time $t$ is from
Eq.~(\ref{eq:first-order-evolution})
\begin{eqnarray}
\alpha _{e} &=&-\frac{iA}{\hbar }\int_{0}^{t}d\tau \,\,\langle e |
\hat{\sigma}_{x}(\tau)|g \rangle F(\tau ), \nonumber\\
&=&-\frac{iA}{\hbar }\int_{0}^{t}d\tau \,\,e^{i\omega _{01}\tau
}F(\tau ).
\label{eq:AmplitudeRandomWalk}
\end{eqnarray}
Since the integrand in Eq.~(\ref{eq:AmplitudeRandomWalk}) is
random, $\alpha_e$ is a sum of a large number of random terms;
i.e.~its value is the endpoint of a random walk in the complex
plane (as discussed above in defining the spectral density of
classical noise). As a result, for times exceeding the
autocorrelation time $\tau_{\rm c}$ of the noise, the integral
will not grow linearly with time but rather only as the square
root of time, as expected for a random walk.  We can now compute
the probability
\begin{equation}
p_{e}(t)\equiv |\alpha _{e}|^{2}=\frac{A^{2}}{\hbar ^{2}}\int_{0}^{t}%
\int_{0}^{t}d\tau _{1}d\tau _{2}\,e^{-i\omega _{01}\left( \tau
_{1}-\tau _{2}\right) }F(\tau _{1})F(\tau _{2})
\end{equation}
which we expect to grow quadratically for short times $t<
\tau_{\rm c}$, but linearly for long times $t> \tau_{\rm c}$.
Ensemble averaging the probability over the random noise yields
\begin{equation}
\bar{p}_{e}(t)=\frac{A^{2}}{\hbar
^{2}}\int_{0}^{t}\int_{0}^{t}d\tau _{1}d\tau _{2}\,e^{-i\omega
_{01}\left( \tau _{1}-\tau _{2}\right) }\left\langle F(\tau
_{1})F(\tau _{2})\right\rangle
\label{pexciteoft}
\end{equation}
  Introducing the noise spectral
density
\begin{equation}
S_{FF}(\omega )=\int_{-\infty }^{+\infty }d\tau \,e^{i\omega \tau
}\langle F(\tau )F(0)\rangle ,
\label{eq:noise-spectral-density}
\end{equation}
and utilizing the Fourier transform defined in Eq.~(\ref{eq:FT})
and the Wiener-Khinchin theorem from Appendix
\ref{app:WienerKhinchin}, we find that the probability to be in
the excited state indeed increases \emph{linearly} with time at
long times,\footnote{Note that for very long times, where there is
a significant depletion of the probability of being in the initial
state, first-order perturbation theory becomes invalid. However,
for sufficiently small $A$, there is a wide range of times
$\tau_{\rm c}\ll t\ll 1/\Gamma$ for which Eq.~\ref{lineart} is
valid. Eqs.~(\ref{gamup}) and (\ref{gamdown}) then yield well-defined
rates which can be used in a master equation to describe the full
dynamics including long times.}
\begin{equation}
\bar{p}_{e}(t)=t\frac{A^{2}}{\hbar ^{2}}S_{FF}(-\omega _{01})
\label{lineart}
\end{equation}
The time derivative of the probability gives the transition rate
from ground to excited states
\begin{equation}
\Gamma _{\uparrow }=\frac{A^{2}}{\hbar ^{2}}S_{FF}(-\omega _{01})
\end{equation}%
Note that we are taking in this last expression the spectral
density on the negative frequency side. If $F$ were a strictly
classical noise source, $\langle F(\tau )F(0)\rangle $ would be
real, and $S_{FF}(-\omega _{01})=S_{FF}(+\omega _{01})$. However,
because as we discuss below $F$ is actually an operator acting on
the environmental degrees of freedom, $\left[ {\hat F}(\tau
),{\hat F}(0)\right] \neq 0
$ and $%
S_{FF}(-\omega _{01})\neq S_{FF}(+\omega _{01})$.

Another possible experiment is to prepare the two-level system in
its excited state and look at the rate of decay into the ground
state. The algebra is identical to that above except that the sign
of the frequency is reversed:
\begin{equation}
\Gamma _{\downarrow }=\frac{A^{2}}{\hbar ^{2}}S_{FF}(+\omega
_{01}).
\end{equation}
We now see that our two-level system does indeed act as a quantum
spectrum analyzer for the noise. Operationally, we prepare the
system either in its ground state or in its excited state, weakly
couple it to the noise source, and after an appropriate interval
of time (satisfying the above inequalities) simply measure whether
the system is now in its excited state or ground state. Repeating
this protocol over and over again, we can find the probability of
making a transition, and thereby infer the rate and hence the
noise spectral density at positive and negative frequencies. 
Naively one imagines that a spectrometers measures the noise
spectrum by extracting a small amount of the signal energy from
the noise source and analyzes it.  This is {\em not} the case
however.  There must be energy flowing in both directions if the
noise is to be fully characterized.

We now rigorously treat the quantity ${\hat F}(\tau )$ as a
quantum Heisenberg operator which acts in the Hilbert space of the
noise source. The previous derivation is unchanged (the ordering
of
  ${\hat F}(\tau_1){\hat F}(\tau_2)$
  having been chosen correctly in anticipation of the quantum treatment), and
Eqs.~(\ref{gamup},\ref{gamdown}) are still valid provided that we
interpret the angular brackets in
Eq.~(\ref{pexciteoft},\ref{eq:noise-spectral-density}) as
representing a quantum expectation value (evaluated in the absence
of the coupling to the spectrometer):
\begin{equation}
S_{FF}(\omega )=\int_{-\infty }^{+\infty }d\tau \,e^{i\omega \tau
}\sum_{\alpha ,\gamma }\rho _{\alpha \alpha }\,\langle \alpha
|{\hat F}(\tau )|\gamma \rangle \langle \gamma |{\hat F}(0)|\alpha
\rangle.
\end{equation}%
Here, we have assumed a stationary situation, where  the density
matrix $\rho$ of the noise source is diagonal in the energy
eigenbasis (in the absence of the coupling to the spectrometer).
However, we do not necessarily assume that it is given by the
equilibrium expression. This yields the standard quantum
mechanical expression for the spectral density:
\begin{eqnarray}
S_{FF}(\omega ) &=&\int_{-\infty }^{+\infty }d\tau \,e^{i\omega
\tau }\sum_{\alpha ,\gamma }\rho _{\alpha \alpha
}\,e^{\frac{i}{\hbar }(\epsilon _{\alpha }-\epsilon _{\gamma })
\tau}\,|\langle \alpha |{\hat F}|\gamma \rangle |
^{2}\nonumber \\
&=&2\pi \hbar \sum_{\alpha ,\gamma }\rho _{\alpha \alpha
}\,|\langle \alpha |{\hat F}|\gamma \rangle |^{2}\delta (\epsilon
_{\gamma }-\epsilon _{\alpha }-\hbar \omega ).
\end{eqnarray}%
Substituting this expression into
Eqs.~(\ref{gamup},\ref{gamdown}), we derive the familiar Fermi
Golden Rule expressions for the two transition rates.

In standard courses, one is not normally taught that the
transition rate of a discrete state into a continuum as described
by Fermi's Golden Rule can (and indeed should!) be viewed as
resulting from the continuum acting as a quantum noise source
which causes the amplitudes of the different components of the
wave function to undergo random walks. The derivation presented
here hopefully provides a motivation for this interpretation.  In
particular, thinking of the perturbation (i.e.~the coupling to the
continuum) as quantum noise with a small but finite
autocorrelation time (inversely related to the bandwidth of the
continuum) neatly explains why the transition probability
increases quadratically for very short times, but linearly for
very long times. 

It it is important to keep in mind that our expressions for the
transition rates are only valid if the autocorrelation time of our
noise is much shorter that the typical time we are interested in;
this typical time is simply the inverse of the transition rate.
The requirement of a short autocorrelation time in turn implies
that our noise source must have a large bandwidth (i.e.~there must
be large number of available photon frequencies in the vacuum) and
must not be coupled too strongly to our system. This is true
despite the fact that our final expressions for the transition
rates only depend on the spectral density at the transition
frequency (a consequence of energy conservation).

One standard model for the continuum is an infinite collection of
harmonic oscillators. The electromagnetic continuum in the
hydrogen atom case mentioned above is a prototypical example. The
vacuum electric field noise coupling to the hydrogen atom has an
extremely short autocorrelation time because the range of mode
frequencies $\omega_\alpha$ (over which the dipole matrix element
coupling the atom to the mode electric field $\vec E_\alpha$ is
significant) is extremely large, ranging from many times smaller
than the transition frequency to many times larger. Thus, the
autocorrelation time of the vacuum electric field noise is
considerably less than $10^{- 15}$s, whereas the decay time of the
hydrogen 2p state is about $10^{- 9}$s. Hence the inequalities
needed for the validity of our expressions are very easily
satisfied.

\subsection{Harmonic oscillator as a spectrum analyzer}
\label{sec:SHOQNOISE}

We now provide more details on the system described in Sec.~\ref{subsec:QuantumSpectrumAnalyzers}, where a harmonic
oscillator acts as a spectrometer of quantum noise. We start with the coupling
Hamiltonian givein in Eq.~(\ref{eq:SHOCoupling}).
In analogy to the TLS spectrometer, noise in $\hat{F}$ at the oscillator
frequency $\Omega$ can cause transitions between its eigenstates.
We assume both that $A$ is small, and that our noise source
has a short autocorrelation time, so  we may again use
perturbation theory to derive rates for these transitions.  There
is a rate for increasing the number of quanta in the oscillator by
one, taking a state $|n\rangle$ to $|n+1\rangle$:
\begin{equation}
     \Gamma_{n \ra n+1} = \frac{A^2}{\hbar^2}
         \left[ (n+1) \xrms^2 \right]
         S_{FF}[-\Omega]
         \equiv (n+1) \Gamma_{\ua}
\end{equation}
As expected, this rate involves the noise at $-\Omega$, as energy
is being {\it absorbed from} the noise source.  
Similarly, there is a rate for decreasing the number of quanta in the oscillator by one:
\begin{equation}
     \Gamma_{n \ra n-1} = \frac{A^2}{\hbar^2}
         \left( n \xrms^2 \right)
         S_{FF}[\Omega]
         \equiv n \Gamma_{\da}
\end{equation}
This rate involves the noise at $+\Omega$, as energy is being {\it
emitted to} the noise source.

Given these transition rates, we may immediately write a simple master
equation for the probability
$p_n(t)$ that there are $n$ quanta in the oscillator:
\begin{eqnarray}
    \frac{d}{dt} p_n  &= &        \left[ n \Gamma_{\ua} p_{n-1} +
(n+1) \Gamma_{\da} p_{n+1}
         \right]\nonumber\\
         &-& \left[ n \Gamma_{\da} + (n+1) \Gamma_{\ua} \right] p_n
         \label{OscMaster}
\end{eqnarray}
The first two terms describe transitions into the state $| n
\rangle$ from the states $|n +1\rangle$ and $|n -1 \rangle$, and
hence increase $p_n$.  In contrast, the last two terms describe
transitions out of the state $| n \rangle$ to the states $|n
+1\rangle$ and $|n -1 \rangle$, and hence decrease $p_n$.
The stationary state of the oscillator is given by solving 
Eq.~(\ref{OscMaster}) for $\frac{d}{dt} p_n = 0$, yielding:
\begin{equation}
     p_n = e^{-n \hbar \Omega / (\kb T_{\rm eff})}
         \left(1 - e^{-\hbar \Omega / (\kb T_{\rm eff})} \right)
         \label{OscTherm}
\end{equation}
where the effective temperature $T_{\rm eff}[\Omega]$ is defined in 
Eq.~(\ref{OscTeff}).
Eq.~(\ref{OscTherm}) describes a thermal equilibrium distribution
of the oscillator, with an effective oscillator temperature
$T_{\rm eff}[\Omega]$ determined by the quantum noise spectrum of
$\hat{F}$. This is the same effective temperature that emerged in
our discussion of the TLS spectrum analyzer.  As we have seen,
if the noise source is in thermal equilibrium at a temperature $T_{\rm eq}$, then $T_{\rm eff}[\Omega] = T_{\rm eq}$.
In the more general case where the noise source is not
in thermal equilibrium,  $T_{\rm eff}$ only serves to characterize the
asymmetry of the quantum noise, and will vary with frequency
\footnote{Note that the effective temperature can become negative
if the noise source prefers emitting energy versus absorbing it;
in the present case, that would lead to an instability.}.

We can learn more about the quantum noise spectrum of $\hat{F}$ by
also looking at the dynamics of the oscillator.
In particular, as the average energy
$\langle E \rangle$ of the oscillator is just given by
$     \langle E(t) \rangle = \sum_{n=0}^{\infty} \hbar \Omega \left(
         n + \frac{1}{2} \right) p_n(t)$,
we can use the master equation Eq.~(\ref{OscMaster}) to derive an
equation for its time dependence.  One thus finds Eq.~(\ref{OscEnergy}).
By demanding $d\langle E \rangle/dt=0$ in this equation, we find that the combination of damping and heating effects causes
the energy to reach a steady state mean value of
$\langle E \rangle= P/\gamma$.
This implies that the finite ground state
  energy $\langle E \rangle = \hbar\Omega/2$ of the
  oscillator is determined via the balance between the
  `heating' by the
  zero-point fluctuations of the environment (described by the
  symmetrized correlator at $T=0$) and the dissipation. It is possible
  to take an alternative but equally correct viewpoint, where only the deviation
  $\langle\delta E\rangle=\langle E \rangle-\hbar \Omega/2$ from the
  ground state energy is considered. Its evolution equation
\begin{equation}
\frac{d}{dt} \langle
  \delta E\rangle = \langle \delta E \rangle
  (\Gamma_{\ua}-\Gamma_{\da}) + \Gamma_{\ua} \hbar \Omega
\label{alternativeOscEnergy}
\end{equation}
only contains a decay term at $T=0$, leading to $\langle \delta E
 \rangle \rightarrow 0$.

\subsection{Practical quantum spectrum analyzers}
\label{subsec:practicalquantumspectrometers}

As we have seen, a `quantum spectrum analyzer' can in principle be
constructed from a two level system (or a harmonic oscillator) in
which we can separately measure the up and down transition rates
between states differing by some precise energy $\hbar\omega>0$
given by the frequency of interest.  The down transition rate
tells us the noise spectral density at frequency $+\omega$ and the
up transition rate tells us the noise spectral density at
$-\omega$.  While we have already discussed experimental implementation of these ideas
using two-level systems and oscillators, similar schemes have been implemented in other systems.
A number of recent experiments have made use of 
superconductor-insulator-superconductor junctions \cite{Deblock03, Pierre06, Onac06b} to measure quantum noise, as the current-voltage characteristics of such junctions are very sensitive to the absorption or emission of energy (so-called photon-assisted transport processes).  It has also been suggested that tunneling of flux in a SQUID can be used to measure quantum noise \cite{Amin08}.

In this subsection, we discuss additional methods for the detection of 
quantum noise.  Recall from Sec.~\ref{subsec:classicalspectrumanalyzers}  that one of the most basic classical noise spectrum analyzers consists of a 
linear narrow band filter and a square law detector such as a
diode.  In what follows, we will consider a  simplified
quantum treatment of such a device where we do not explicitly model a diode,
but instead focus on the energy of the filter circuit.
We then turn to various noise detection schemes making use of a photomultiplier.  We will show that depending on the detection scheme used, one can measure either the symmetrized quantum noise spectral density $\bS[\omega]$, or the non-symmetrized spectral density $S[\omega]$.

\subsubsection{Filter plus diode}

Using the simple treatment we gave of a harmonic oscillator as a quantum
spectrum analyzer in Sec.~\ref{sec:SHOQNOISE}, one can attempt to provide
a quantum treatment of the classical `filter plus diode' spectrum analyzer discussed in Sec.~\ref{subsec:classicalspectrumanalyzers}.  This approach is due to \textcite{Lesovik97} and \textcite{Gavish00}.  The analysis starts by modeling the spectrum analyzer's resonant filter circuit as a harmonic oscillator of frequency $\Omega$ weakly coupled to some equilibrium dissipative bath.  The oscillator thus has an intrinsic damping rate $\gamma_0 \ll \Omega$, and is initially at a finite temperature $T_{\rm eq}$.  One then drives this damped oscillator (i.e.~the filter circuit) with the noisy quantum force $\hF(t)$ whose spectrum at frequency $\Omega$ is to be measured.  

In the classical `filter plus diode' spectrum analyzer, the output of the filter circuit was sent to a square law detector, whose time-averaged output was then taken as the measured spectral density.  To simplify the analysis, we can instead consider how the noise changes the average energy of the resonant filter circuit, taking this quantity as a proxy for the output of the diode.  Sure enough, if we subject the filter circuit to purely classical noise, it would cause the average energy of the circuit $\langle E \rangle$ to increase an amount directly proportional to the classical spectrum 
$\clS_{FF}[\Omega]$.  We now consider 
$\langle E \rangle$ in the case of a quantum noise source, and ask how it relates to the 
quantum noise spectral density $S_{FF}[\Omega]$.

The quantum case is straightforward to analyze using the approach of Sec.~\ref{sec:SHOQNOISE}.  Unlike the classical case, the noise will both lead to additional
fluctuations of the filter circuit {\it and} increase its damping rate by an amount $\gamma$ (c.f.~Eq.~(\ref{OscGamma})). 
To make things quantitative, we let $n_{\rm eq}$ denote the average number of quanta in the filter circuit prior to coupling to $\hF(t)$, i.e.
\begin{eqnarray}
	n_{\rm eq} = \frac{1}{\exp \left( \frac{ \hbar \Omega}{k_B T_{\rm eq}} \right) -1},
\end{eqnarray}
and let $n_{\rm eff}$ represent the Bose-Einstein factor associated with the effective temperature $T_{\rm eff}[\Omega]$ of the noise source $\hF(t)$,
\begin{eqnarray}
	n_{\rm eff} = \frac{1}{\exp \left( \frac{ \hbar \Omega}{k_B T_{\rm eff}[\Omega] } \right) -1}.
\end{eqnarray}
One then finds \cite{Lesovik97, Gavish00}:
\begin{eqnarray}
	\Delta \langle E \rangle & = &
		\hbar \Omega \cdot \frac{ \gamma}{\gamma_0 + \gamma} 
		(n_{\rm eff} - n_{\rm eq}) 
		\label{eq:Deltanosc1}
\end{eqnarray}
This equation has an extremely simple interpretation:  the first term results from the expected heating effect of the noise, while the second term results from the noise source having increased the circuit's damping by an amount $\gamma$.  Re-expressing this result in terms of the symmetric and anti-symmetric in frequency parts of the quantum noise spectral density $S_{FF}[\Omega]$, we have:
\begin{eqnarray}
	\Delta \langle E \rangle & = &
		\frac{ 
				\bS_{FF}(\Omega) - 
				\left(n_{\rm eq} + \frac{1}{2} \right) \left(
			 	S_{FF}[\Omega] - S_{FF}[-\Omega] \right)    
			}
			{	2 m \left( \gamma_0 + \gamma \right) } \nonumber 
				\\	\label{eq:Deltanosc2}
\end{eqnarray}
We see that $\Delta \langle E \rangle$ is in general {\it not} simply proportional to the symmetrized noise $\bS_{FF}[\Omega]$.  Thus, the `filter plus diode' spectrum analyzer does not simply measure the symmetrized quantum noise spectral density.  We stress that there is nothing particularly quantum about this result.  The extra term on the RHS of  Eq.~(\ref{eq:Deltanosc2}) simply reflects the fact that coupling the noise source to the filter circuit could change the damping of this circuit; this could easily happen in a completely classical setting.  As long as this additional damping effect is minimal, the second term in Eq.~(\ref{eq:Deltanosc2}) will be minimal, and our spectrum analyzer will (to a good approximation) measure the symmetrized noise.  Quantitatively, this requires:
\begin{eqnarray}
	n_{\rm eff} \gg n_{\rm eq}.
\end{eqnarray}
We now see where quantum mechanics enters:  if the noise to be measured is close to being zero point noise (i.e. $n_{\rm eff} \ra 0$), the above condition can never be satisfied, and thus it is {\it impossible} to ignore the damping effect of the noise source on the filter circuit.  In the zero point limit, this damping effect (i.e.~second term in Eq.~(\ref{eq:Deltanosc2}))  will always be greater than or equal to the expected heating effect of the noise (i.e.~first term in Eq.~(\ref{eq:Deltanosc2})).

\subsubsection{Filter plus photomultiplier}

We now turn to quantum spectrum analyzers involving a square law detector
we can accurately model-- a photomultiplier.  As a first example of 
such a system, consider
a photomultiplier with a narrow band
filter placed in front of it.  The mean photocurrent is then given
by
\be
\langle I \rangle = \int_{-\infty}^{+\infty}d\omega\,
|f[\omega]|^2 r[\omega] \SVV[\omega],
\label{eq:photocurrent}
\ee
where $f$ is the filter (amplitude) transmission function defined
previously and $r[\omega]$ is the response of the photodetector at
frequency $\omega$, and $\SVV$ represents the electric field
spectral density incident upon the photodetector. Naively one
thinks of the photomultiplier as a square law detector with the
square of the electric field representing the optical power.
 However, according to the Glauber theory of (ideal) photo-detection
\cite{Glauber2006,Walls94,Gardiner00}, photocurrent is produced
if, and only if, a photon is absorbed by the detector, liberating
the initial photo-electron.  Glauber describes this in terms of
normal ordering of the photon operators in the electric field
autocorrelation function.  In our language of noise power at
positive and negative frequencies, this requirement becomes simply
that $r[\omega]$ vanishes for $\omega>0$.  Approximating the
narrow band filter centered on frequency $\pm \omega_0$ as in
Eq.~(\ref{eq:narrowfilter}), we obtain
\be
\langle I \rangle = r[-\omega_0] \SVV[-\omega_0]
\ee
which shows that this particular realization of a quantum
spectrometer only measures electric field spectral density at
negative frequencies since the photomultiplier never emits energy
into the noise source. Also one does not see in the output any
`vacuum noise' and so the output (ideally) vanishes as it should
at zero temperature.  Of course real photomultipliers suffer from
imperfect quantum efficiencies and have non-zero dark current.  Note
that we have assumed here that there are no additional fluctuations associated with the filter circuit.  Our result thus coincides with what we found in the previous subsection 
for the `filter plus diode' spectrum analyzer (c.f.~Eq.~(\ref{eq:Deltanosc2}), in the limit where the filter circuit is initially at zero temperature (i.e.~$n_{\rm eq} = 0$).


\subsubsection{Double sideband heterodyne power spectrum}
\label{subsubsec:HeterodyneAnalyzer}

At RF and microwave frequencies, practical spectrometers often
contain heterodyne stages which mix the initial frequency down to
a lower frequency $\omega_{\rm IF}$ (possibly in the classical regime).  Consider
a system with a mixer and local oscillator at frequency
$\omega_{\rm LO}$ that mixes both the upper sideband input at
$\omega_{\rm u}= \omega_{\rm LO}+\omega_{\rm IF}$ and the lower
sideband input at $\omega_{\rm l}= \omega_{\rm LO}-\omega_{\rm
IF}$ down to frequency $\omega_{\rm IF}$.  This can be achieved by
having a Hamiltonian with a 3-wave mixing term which (in the
rotating wave approximation) is given by
\be
V=\lambda[
	\aIF\al\aLO^\dagger + \aIF^\dagger\al^\dagger\aLO] +
\lambda[\aIF^\dagger\au\aLO^\dagger + \aIF\au^\dagger\aLO]
\ee
The interpretation of this term is that of a Raman process. Notice
that there are two energy conserving processes that can create an
IF photon which could then activate the photodetector.  First, one
can absorb an LO photon and \emph{emit} two photons, one at the IF
and one at the lower sideband.  The second possibility is to
\emph{absorb} an upper sideband photon and create IF and LO
photons.  Thus we expect from this that the power in the IF
channel detected by a photomultiplier would be proportional to the
noise power in the following way
\be
I\propto S[+\omega_{\rm l}] + S[-\omega_{\rm u}]
\label{eq:dblsidenband}
\ee
since creation of an IF photon involves the signal source either
absorbing a lower sideband photon from the mixer or the signal
source emitting an upper sideband photon into the mixer. In the
limit of small IF frequency this expression would reduce to the
symmetrized noise power
\be
I\propto S[+\omegaLO]+S[-\omegaLO] = 2 \bS[ \omegaLO ]
\ee
which is the same as for a `classical' spectrum analyzer with a
square law detector (c.f.~Appendix~\ref{subsec:classicalspectrumanalyzers}).  
For equilibrium noise
spectral density from a resistance $R_0$ derived in Appendix
\ref{app:QuantumResistor} we would then have
\be \SVV[\omega]+\SVV[-\omega] = 2\r0\hbar|\omega|
[2\nb(\hbar|\omega|)  +1] ,
\label{eq:SVVresistor}
\ee
Assuming our spectrum analyzer has high input impedance so that it
does not load the noise source, this voltage spectrum will
determine the output signal of the analyzer.  This symmetrized
quantity does \emph{not} vanish at zero temperature and the output
contains the vacuum noise from the input.  This vacuum noise has
been seen in experiment.  \cite{Schoelkopf97}

\section{Modes, Transmission Lines and Classical Input/Output Theory}
\label{app:classicalinputoutput}

In this appendix we introduce a number of important classical
concepts about electromagnetic signals which are essential to
understand before moving on to the study of their quantum
analogs.  A signal at carrier frequency $\omega$ can be
described in terms of its amplitude and phase or equivalently
in terms of its two quadrature amplitudes
\be
s(t) = X \cos(\omega t) + Y \sin(\omega t).
\label{eq:signalquads}
\ee
We will see in the following that the physical oscillations of
this signal in a transmission line  are precisely the
sinusoidal oscillations of a simple harmonic oscillator.
Comparison of Eq.~(\ref{eq:signalquads}) with
$x(t) = x_0 \cos \omega t + (p_0/M \omega) \sin \omega t$
shows that we can identify the
quadrature amplitude $X$ with the coordinate of this oscillator
and thus the quadrature amplitude $Y$ is proportional to the
momentum conjugate to $X$. Quantum mechanically, $X$ and $Y$
become operators $\hat X$ and $\hat Y$ which do not commute.
Thus their quantum fluctuations obey the Heisenberg
uncertainty relation.

Ordinarily (e.g., in the absence of squeezing), the phase
choice defining the two quadratures is arbitrary and so their
vacuum (i.e. zero-point) fluctuations are equal
\be
X_{\rm ZPF} = Y_{\rm ZPF}.
\ee
Thus the canonical commutation relation becomes
\be
[\hat X,\hat Y] = i X_{\rm ZPF}^2.
\ee
We will see that the fact that $X$ and $Y$ are canonically
conjugate has profound implications both classically and
quantum mechanically. In particular, the action of any circuit
element (beam splitter, attenuator, amplifier, etc.) must
preserve the Poisson bracket (or in the quantum case, the
commutator) between the signal quadratures.  This places strong
constraints on the properties of these circuit elements and in
particular, forces every amplifier to add noise to the signal.

\subsection{Transmission lines and classical input-output theory}
\label{subsec:ClassicalTransmissionLines}

\begin{figure}[ht]
\begin{center}
\includegraphics[width=3.45in]{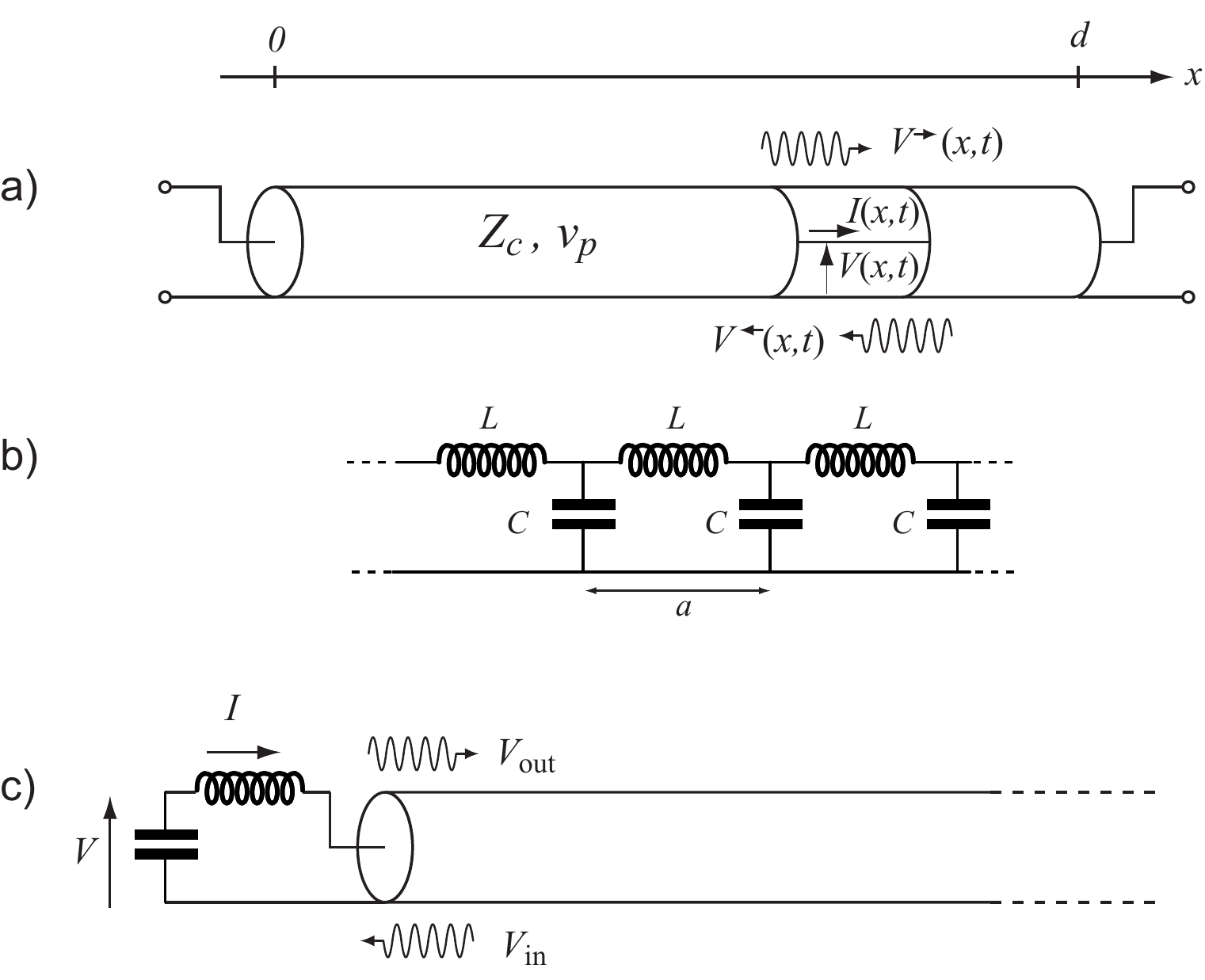}
\caption{
a) Coaxial transmission line, indicating voltages and currents as defined in the main text.
b) Lumped element representation of a transmission line with capacitance per unit length $c=C/a$ and
inductance per unit length $\ell = L/a$. c) Discrete LC resonator terminating a transmission line.}
\label{fig:transmissionline}
\end{center}
\end{figure}

We begin by considering a coaxial transmission line modeled as
a perfectly conducting wire with inductance per unit length of
$\ell$ and capacitance to ground per unit length $c$ as shown
in Fig.~\ref{fig:transmissionline}. If the voltage at
position $x$ at time $t$ is $V(x,t)$, then the charge density
is $q(x,t) = cV(x,t)$. By charge conservation the current $I$
and the charge density are related by the continuity equation
\be
\partial_t q + \partial_x I = 0.
\label{eq:continuityeq}
\ee
The constitutive relation (essentially Newton's law) gives the
acceleration of the charges
\be \ell\partial_t I = -\partial_x V.
\label{eq:constitutiveeq}
\ee
We can decouple Eqs.~(\ref{eq:continuityeq}) and
(\ref{eq:constitutiveeq}) by introducing left and right
propagating modes
\begin{eqnarray}
V(x,t) &=& [V^\rightarrow + V^\leftarrow]\label{eq:Vsum}\\
I(x,t) &=& \frac{1}{\Zc}[V^\rightarrow - V^\leftarrow]\label{eq:Idiff}
\end{eqnarray}
where $\Zc\equiv\sqrt{\ell/c}$ is called the characteristic
impedance of the line.  In terms of the left and right
propagating modes, Eqs.~(\ref{eq:continuityeq}) and
\ref{eq:constitutiveeq} become
\begin{eqnarray}
\vp \partial_x V^\rightarrow + \partial_t V^\rightarrow &=& 0 \label{eq:Vright}\\
\vp \partial_x V^\leftarrow - \partial_t V^\leftarrow &=& 0 \label{eq:left}
\end{eqnarray}
where $\vp\equiv 1/\sqrt{\ell c}$ is the wave phase velocity.
These equations have solutions which propagate by uniform
translation without changing shape since the line is
dispersionless
\begin{eqnarray}
V^\rightarrow(x,t) &=& \Vout(t-\frac{x}{\vp})\label{eq:Voutdef}\\
V^\leftarrow(x,t) &=& \Vin(t+\frac{x}{\vp})\label{eq:Vindef},
\end{eqnarray}
where $\Vin$ and $\Vout$ are \emph{arbitrary} functions of
their arguments.  For an infinite transmission line, $\Vout$
and $\Vin$ are completely independent.  However for the case of
a semi-infinite line terminated at $x=0$ (say) by some system
$S$, these two solutions are not independent, but rather
related by the boundary condition imposed by the system.  We
have
\begin{eqnarray}
V(x=0,t) &=& [\Vout(t) + \Vin(t)] \label{eq:Vx0} \\
I(x=0,t) &=& \frac{1}{\Zc}[\Vout(t) - \Vin(t)],\label{eq:Ix0}
\end{eqnarray}
from which we may derive
\be
\Vout(t) = \Vin(t) + \Zc I(x=0,t).
\label{eq:inputoutputIV}
\ee

If the system under study is just an open circuit so that
$I(x=0,t)=0$, then $\Vout=\Vin$, meaning that the outgoing wave
is simply the result of the incoming wave reflecting from the
open circuit termination.  In general however, there is an
additional outgoing wave radiated by the current $I$ that is
injected by the system dynamics into the line.  In the absence
of an incoming wave we have \be V(x=0,t) = \Zc I(x=0,t), \ee
indicating that the transmission line acts as a simple resistor
which, instead of dissipating energy by Joule heating, carries
the energy away from the system as propagating waves.  The fact
that the line can dissipate energy despite containing only
purely reactive elements is a consequence of its infinite
extent.  One must be careful with the order of limits, taking
the length to infinity \emph{before} allowing time to go to
infinity.  In this way the outgoing waves never reach the far
end of the transmission line and reflect back.  Since this is a
conservative Hamiltonian system, we will be able to quantize
these waves and make a quantum theory of resistors
\cite{caldeiraleggett} in Appendix \ref{app:QuantumResistor}.
The net power flow carried to the right by the line is \be P =
\frac{1}{\Zc}[\Vout^2(t) - \Vin^2(t)]. \label{eq:Poynting} \ee

The fact that the transmission line presents a dissipative
impedance to the system means that it causes damping of the
system.  It also however opens up the possibility of
controlling the system via the input field which partially
determines the voltage driving the system.  From this point of
view it is convenient to eliminate the output field by writing
the voltage as \be V(x=0,t) = 2\Vin(t) + \Zc I(x=0,t).
\label{eq:Vis2Vin} \ee As we will discuss in more detail below,
the first term drives the system and the second damps it.  From
Eq.~(\ref{eq:inputoutputIV}) we see that measurement of the
outgoing field can be used to determine the current $I(x=0,t)$
injected by the system into the line and hence to infer the
system dynamics that results from the input drive field.

As a simple example, consider the system consisting of an LC
resonator shown in Fig.~(\ref{fig:transmissionline} c). This
can be viewed as a simple harmonic oscillator whose coordinate
$Q$ is the charge on the capacitor plate (on the side connected
to $L_0$). The current $I(x=0,t)=\dot Q$ plays the role of the
velocity of the oscillator.  The equation of motion for the
oscillator is readily obtained from
\be Q = C_0[-V(x=0^+,t)-L_0 \dot I(x=0^+,t)].
\ee Using Eq.~(\ref{eq:Vis2Vin}) we obtain a harmonic
oscillator damped by the transmission line and driven by the
incoming waves
\be
\ddot Q = -\Omega_0^2 Q - \gamma \dot Q
-\frac{2}{L_0} \Vin(t),
\label{eq:SHOLCeqnmotion}
\ee
where the
resonant frequency is $\Omega_0^2\equiv 1/\sqrt{L_0C_0}$. Note
that the term $\Zc I(x=0,t)$ in Eq.~(\ref{eq:Vis2Vin}) results
in the linear viscous damping rate $\gamma \equiv \Zc/L_0$.

If we solve the equation of motion of the oscillator, we can
predict the outgoing field.  In the  present instance of a
simple oscillator we have a particular example of the general
case where the system responds linearly to the input field.  We
can characterize any such system by a complex, frequency
dependent impedance $Z[\omega]$ defined by \be Z[\omega] = -
\frac{V(x=0,\omega)}{I(x=0,\omega)}. \label{eq:systemimpedance}
\ee Note the peculiar minus sign which results from our
definition of positive current flowing to the right (out of the
system and into the transmission line). Using
Eqs.~(\ref{eq:Vx0}, \ref{eq:Ix0}) and
Eq.~(\ref{eq:systemimpedance}) we have
\be
\Vout[\omega] =
r[\omega] \Vin[\omega],
\ee
where the reflection coefficient
$r$ is determined by the impedance mismatch between the system
and the line and is given by the well known result \be
r[\omega] = \frac{Z[\omega] - \Zc}{Z[\omega] + \Zc}.
\label{eq:reflimpedmismatch} \ee

If the system is constructed from purely reactive (i.e.
lossless) components, then $Z[\omega]$ is purely imaginary and
the reflection coefficient obeys $|r|=1$ which is consistent
with Eq.~(\ref{eq:Poynting}) and the energy conservation
requirement of no net power flow into the lossless system. For
example, for the series $LC$ oscillator we have been
considering, we have \be Z[\omega] = \frac{1}{j\omega C_0} +
j\omega L_0, \label{eq:ZomegaLC} \ee where, to make contact
with the usual electrical engineering sign conventions, we have
used $j=-i$.  If the damping $\gamma$ of the oscillator induced
by coupling it to the transmission line is small, the quality factor 
of the resonance will be high and we need only consider frequencies near the
resonance frequency $\Omega_0\equiv 1/\sqrt{L_0C_0}$ where the
impedance has a zero.  In this case we may approximate \be
Z[\omega] \approx \frac{2}{jC_0\Omega_0^2}[\Omega_0-\omega]
= 2j L_0 (\omega-\Omega_0) \label{eq:ZomegaLCapprox} \ee which yields for the reflection
coefficient \be r[\omega] = \frac{\omega-\Omega_0 +
j\gamma/2}{\omega-\Omega_0 - j\gamma/2} \ee showing that indeed
$|r|=1$ and that the phase of the reflected signal winds by
$2\pi$ upon passing through the resonance.  \footnote{For the
case of resonant \emph{transmission} through a symmetric
cavity, the phase shift only winds by $\pi$.}

Turning to the more general case where the system also contains lossy elements,
one finds that $Z[\omega]$ is no longer purely imaginary, but has a real part satisfying $\textrm{Re }Z[\omega] > 0$.  This in turn implies via Eq.~(\ref{eq:reflimpedmismatch}) 
that $|r|<1$.  In the special case of impedance matching $Z[\omega]=\Zc$,  all the incident power is dissipated in the system and none is reflected.
  The other two limits of interest are open circuit termination with $Z=\infty$
  for which $r=+1$ and short circuit termination $Z=0$ for which $r=-1$.

Finally, if the system also contains an active device which has energy being pumped into it from a separate external source, it may under the right conditions be described by an effective {\it negative} resistance $\mathrm{Re}\,Z[\omega]<0$ over a certain frequency range.  Eq.~(\ref{eq:reflimpedmismatch}) then gives $|r| \geq 1$, implying $|\Vout| > |\Vin|$.  Our system will thus act like the one-port amplifier discussed in Sec.~\ref{subsec:TwoKindsAmps}:  it amplifies signals incident upon it.  We will discuss this idea of negative resistance further in Sec.~\ref{subsec:NegativeResistance}; a physical realization is provided by the two-port reflection parametric amplifier discussed in Appendix \ref{subsec:NDParamps}.

\subsection{Lagrangian, Hamiltonian, and wave modes for a transmission line}

 Prior to moving on to the case of quantum noise it is useful to
 review the classical statistical mechanics of transmission lines.
 To do this we need to write down the Lagrangian and then determine the
 canonical momenta and the Hamiltonian.  Very conveniently, the
system is simply a large collection of harmonic oscillators
(the normal modes) and hence can be readily quantized.  This
representation of a physical resistor is essentially the one
used by Caldeira and Leggett \cite{caldeiraleggett} in their
seminal studies of the effects of dissipation on tunneling. The
only difference between this model and the vacuum fluctuations
in free space is that the relativistic bosons travel in one
dimension and do not carry a polarization label. This changes
the density of states as a function of frequency, but has no
other essential effect.

It is convenient to define a flux variable
\cite{DevoretLesHouches} \be \varphi(x,t) \equiv
\int_{-\infty}^td\tau\, V(x,\tau), \ee where
$V(x,t)=\partial_t\varphi(x,t)$ is the local voltage on the
transmission line at position $x$ and time $t$.  Each segment
of the line of length $dx$ has inductance $\ell\,dx$ and the
voltage drop along it is $-dx\,
\partial_x\partial_t\varphi(x,t)$.  The flux through this
inductance is thus $-dx\,\partial_x \varphi(x,t)$ and the local
value of the current is given by the constitutive equation \be
I(x,t)=-\frac{1}{\ell}\,\partial_x\varphi(x,t).
\label{eq:current} \ee

The Lagrangian for the system is 
\be L_g \equiv \int_0^\infty
dx\, {\cal L}(x,t)= \int_0^\infty
dx\, \left( \frac{c}{2}(\partial_t\varphi)^2 -
\frac{1}{2\ell}\,(\partial_x\varphi)^2 \right), \ee The Euler-Lagrange
equation for this Lagrangian is simply the wave equation
\be
\vp^2\partial_x^2 \varphi - \partial_t^2\varphi = 0.
\label{waveeqnEulerLagrange}
\ee
The
momentum conjugate to $\varphi(x)$ is simply the charge density
\be
q(x,t) \equiv \frac{\delta {\cal L}}{\delta \partial_t \varphi}
= c\partial_t\varphi = cV(x,t)
\label{eq:qconjtophi}
\ee
and so the Hamiltonian is given by \be H = \int dx\,
\left\{\frac{1}{2c} q^2 + \frac{1}{2\ell} (\partial_x\varphi)^2
\right\}. \ee

We know from our previous results that the charge density
consists of left and right moving solutions of arbitrary fixed
shape.  For example we might have for the right moving case
\be
q(t-x/\vp) = \alpha_k \cos[k(x-\vp t)] + \beta_k \sin[k(x-\vp
t)].
\ee
A confusing point is that since $q$ is real valued, we see that
it necessarily contains both $e^{ikx}$ and $e^{-ikx}$ terms
even if it is only right moving.  Note however that for $k>0$
and a right mover, the $e^{ikx}$ is associated with the
positive frequency term $e^{-i\omega_k t}$ while the $e^{-ikx}$
term is associated with the negative frequency term
$e^{+i\omega_k t}$ where $\omega_k \equiv \vp |k|$.  For left
movers the opposite holds. We can appreciate this better if we
define
\be
A_k \equiv \frac{1}{\sqrt{L}} \int dx\, e^{-ikx}
\left\{ \frac{1}{\sqrt{2c}} q(x,t) - i
\sqrt{\frac{k^2}{2\ell}}\varphi(x,t) \right\}
\label{eq:modeamplitudes}
\ee
where for
simplicity we have taken the fields to obey periodic boundary
conditions on a length $L$. Thus we have (in a form which
anticipates the full quantum theory)
\be
H = \frac{1}{2}\sum_k\left(A_k^* A_k + A_k A_k^*\right).
\label{eq:anticipatesquantumH}
\ee
The classical equation of motion (\ref{waveeqnEulerLagrange})
yields the simple result
\be
\partial_t A_k = -i\omega_k A_k.
\label{eq:anticipatesquantumEOM}
\ee Thus
\begin{eqnarray}
&&q(x,t)\nonumber\\
&=& \sqrt{\frac{c}{2L}}\sum_k e^{ikx} \left[ A_k(0)e^{-i\omega_k t} + A_{-k}^*(0) e^{+i\omega_k t}\right]\\
&=& \sqrt{\frac{c}{2L}}\sum_k  \left[ A_k(0)e^{+i(kx-\omega_k t)} + A_{k}^*(0) e^{-i(kx-\omega_k t)}\right].\nonumber\\
\end{eqnarray}
We see that for $k>0$ ($k<0$) the wave is right (left) moving,
and that for right movers the $e^{ikx}$ term is associated with
positive frequency and the $e^{-ikx}$ term is associated with
negative frequency.  We will return to this in the quantum case
where positive (negative) frequency will refer to the
destruction (creation) of a photon.  Note that the right and
left moving voltages are given by
\begin{widetext}
\begin{eqnarray}
V^\rightarrow &=& \sqrt{\frac{1}{2Lc}}\sum_{k>0}
\left[ A_k(0)e^{+i(kx-\omega_k t)} + A_{k}^*(0) e^{-i(kx-\omega_k t)}\right] \label{eq:VrightB38}\\
V^\leftarrow &=& \sqrt{\frac{1}{2Lc}}\sum_{k<0}
\left[ A_k(0)e^{+i(kx-\omega_k t)} + A_{k}^*(0) e^{-i(kx-\omega_k t)}\right]\label{eq:VleftB39}
\end{eqnarray}
The voltage spectral density for the right moving waves is thus
\begin{eqnarray}
S_{VV}^\rightarrow[\omega] &=& \frac{2\pi}{2Lc} \sum_{k>0} \left\{ \langle  A_kA_{k}^*\rangle \delta(\omega - \omega_k)+ \langle  A_{k}^*A_k\rangle \delta(\omega + \omega_k) \right\}
\label{eq:voltagespecdensright}
\end{eqnarray}
The left moving spectral density has the same expression but
$k<0$.

Using Eq.~(\ref{eq:Poynting}), the above results lead to a net
power flow (averaged over one cycle) within a frequency band
defined by a pass filter $G[\omega]$ of
\be
P = P^\rightarrow - P^\leftarrow = \frac{\vp}{2L}\sum_k
\mathrm{sgn}(k)  \left[ G[\omega_k]\langle  A_k A_k^*\rangle +
G[-\omega_k]\langle A_k^* A_k \rangle \right].
\label{eq:powerrightminusleft}
\ee
\end{widetext}

\subsection {Classical statistical mechanics of a transmission line}
\label{subsec:TLClassicalStatMech}

Now that we have the Hamiltonian, we can consider the classical
statistical mechanics of a transmission line in thermal
equilibrium at temperature $T$.  Since each mode $k$ is a
simple harmonic oscillator we have from
Eq.~(\ref{eq:anticipatesquantumH}) and the equipartition
theorem
\be
\langle A_k^* A_k\rangle = \kb T.
\label{eq:equipart}
\ee
Using this, we see from Eq.~(\ref{eq:voltagespecdensright}) 
that the right moving voltage signal has a simple white
noise power spectrum.
 Using Eq.~(\ref{eq:powerrightminusleft}) we have for the
right moving power in a bandwidth $B$ (in Hz rather than
radians/sec) the very simple result
\begin{eqnarray}
P^\rightarrow &=& \frac{\vp}{2L}\sum_{k>0}
 \left\langle G[\omega_k]A_k^* A_k + G[-\omega_k] A_k A_k^*\right\rangle\nonumber\\
&=& \frac{\kb T}{2} \int_{-\infty}^{+\infty}\frac{d\omega}{2\pi}\, G[\omega] \nonumber\\
&=& \kb T B.
\label{eq:powerperunitbandwidth}
\end{eqnarray}
where we have used the fact mentioned in connection with
Eq.~(\ref{eq:narrowfilter}) and the discussion of square law
detectors that all passive filter functions are symmetric in
frequency.

One of the basic laws of statistical mechanics is Kirchhoff's
law stating that the ability of a hot object to emit radiation
is proportional to its ability to absorb.  This follows from
very general thermodynamic arguments concerning the thermal
equilibrium of an object with its radiation environment and it
means that the best possible emitter is the black body.  In
electrical circuits this principle is simply a form of the
fluctuation dissipation theorem which states that the
electrical thermal noise produced by a circuit element is
proportional to the dissipation it introduces into the circuit.
Consider the example of a terminating resistor at the end of a
transmission line.  If the resistance $R$ is matched to the
characteristic impedance $\Zc$ of a transmission line, the
terminating resistor acts as a black body because it absorbs
100\% of the power incident upon it.  If the resistor is held
at temperature $T$ it will bring the transmission line modes
into equilibrium at the same temperature (at least for the case
where the transmission line has finite length).  The rate at
which the equilibrium is established will depend on the
impedance mismatch between the resistor and the line, but the
final temperature will not.

\begin{figure}[ht]
\begin{center}
\includegraphics[width=3.45in] {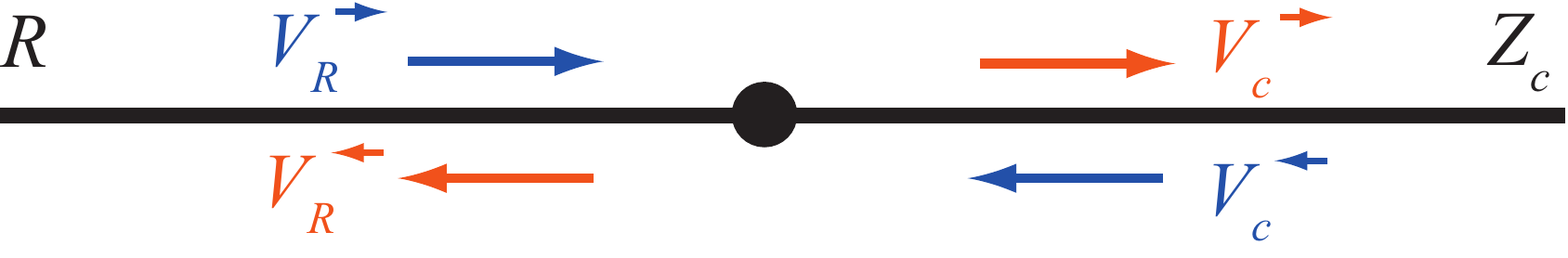}
\caption{(Color online) Semi-infinite transmission line of impedance $\Zc$ terminated by a resistor $R$ which is
represented as a second semi-infinite
transmission line.}
\label{fig:resistor+transmissionline}
\end{center}
\end{figure}

\begin{figure}[ht]
\begin{center}
\includegraphics[width=3.45in] {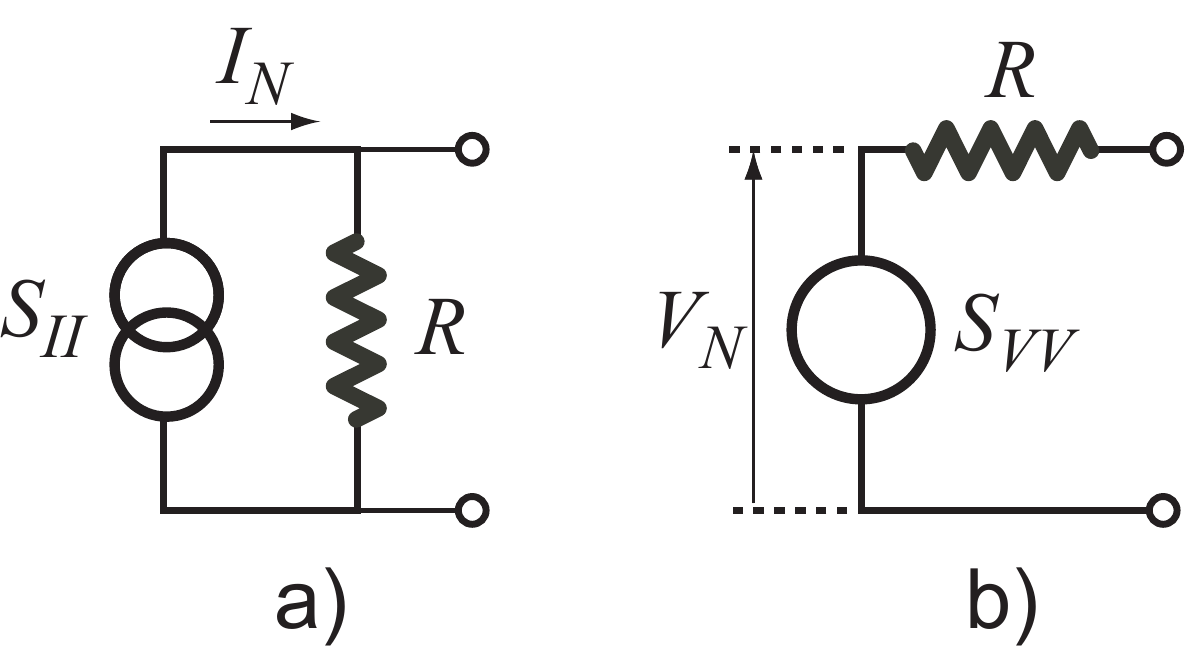}
\caption{Equivalent circuits for noisy resistors.}
\label{fig:resistornoises}
\end{center}
\end{figure}

A good way to understand the fluctuation-dissipation theorem is
to represent the resistor $R$ which is terminating the $\Zc$
line in terms of a second semi-infinite transmission line of
impedance $R$ as shown in
Fig.~(\ref{fig:resistor+transmissionline}). First consider the
case when the $R$ line is not yet connected to the $\Zc$ line.
Then according to Eq.~(\ref{eq:reflimpedmismatch}), the open
termination at the end of the $\Zc$ line has reflectivity
$|r|^2=1$ so that it does not dissipate any energy.
Additionally of course, this termination does not transmit any
signals from the $R$ line into the $\Zc$.  However when the two
lines are connected the reflectivity becomes less than unity
meaning that incoming signals on the $\Zc$ line see a source of
dissipation $R$ which partially absorbs them. The absorbed
signals are not turned into heat as in a true resistor but are
partially transmitted into the $R$ line which is entirely
equivalent.  Having opened up this port for energy to escape
from the $\Zc$ system, we have also allowed noise energy
(thermal or quantum) from the $R$ line to be transmitted into
the $\Zc$ line. This is completely equivalent to the effective
circuit shown in Fig.~(\ref{fig:resistornoises} a) in which a
real resistor has in parallel a random current generator
representing thermal noise fluctuations of the electrons in the
resistor.    This is the essence of the fluctuation dissipation
theorem.

In order to make a quantitative analysis in terms of the power
flowing in the two lines, voltage is not the best variable to
use since we are dealing with more than one value of line
impedance.  Rather we define incoming and outgoing fields via
\begin{eqnarray}
\Ain &=& \frac{1}{\sqrt{\Zc}}\Vc^\leftarrow\label{eq:Ain}\\
\Aout&=& \frac{1}{\sqrt{\Zc}}\Vc^\rightarrow\label{eq:Aout}\\
\Bin &=& \frac{1}{\sqrt{R}}\VR^\rightarrow\label{eq:Bin}\\
\Bout &=& \frac{1}{\sqrt{R}}\VR^\leftarrow\label{eq:Bout}
\end{eqnarray}
Normalizing by the square root of the impedance allows us to
write the power flowing to the right in each line in the simple
form
\begin{eqnarray}
P_{\rm c}&=&(\Aout)^2 - (\Ain)^2\\
P_R &=& (\Bin)^2 - (\Bout)^2
\end{eqnarray}
The out fields are related to the in fields by the $\smat$ matrix
\be
\left(
\begin{array}{cc}
\Aout\\
\Bout
\end{array}
\right) = \smat \left(
\begin{array}{cc}
\Ain\\
\Bin
\end{array}
\right)
\ee

Requiring continuity of the voltage and current at the
interface between the two transmission lines, we can solve for
the scattering matrix $\smat$:
\be
\smat = \left(
\begin{array}{cc}
+r & t\\
t&-r
\end{array}
\label{eq:2x2Smatrix}
\right)
\ee
where
\begin{eqnarray}
r &=&\frac{R-\Zc}{R+Zc}\\
t &=&\frac{2\sqrt{R\Zc}}{R+Zc}.
\end{eqnarray}
Note that $|r|^2 + |t|^2=1$ as required by energy conservation
and that $\smat$ is unitary with det~$(\smat)=-1$.  By
moving the point at which the phase of the $\Bin$ and $\Bout$
fields are determined one-quarter wavelength to the left, we
can put $\smat$ into different standard form
\be
{\smat}' = \left(
\begin{array}{cc}
+r & it\\
it&+r
\end{array}
\right)
\label{eq:2x2Sprime}
\ee
which has det~$(\smat')=+1$.

As mentioned above, the energy absorbed from the $\Zc$ line by
the resistor $R$ is not turned into heat as in a true resistor
but is is simply transmitted into the $R$ line, which is
entirely equivalent.
Kirchhoff's law is now easy to understand.  The energy absorbed
{\it from} the $\Zc$ line by $R$, and the energy transmitted {\it into} it by
thermal fluctuations in the $R$ line are both proportional to
the absorption coefficient
\be
A = 1- |r|^2=|t|^2=\frac{4R\Zc}{(R+\Zc)^2}.
\ee

The requirement that the transmission line $\Zc$ come to
equilibrium with the resistor allows us to readily compute the
spectral density of current fluctuations of the random current
source shown in Fig.~(\ref{fig:resistornoises} a).
%
%
The power dissipated in $\Zc$ by the current source attached to
$R$ is
\begin{eqnarray}
P &=& \int_{-\infty}^{+\infty} \frac{d\omega}{2\pi}\,S_{II}[\omega] \frac{R^2\Zc}{(R+\Zc)^2}\nonumber\\
\end{eqnarray}
For the special case $R=\Zc$ we can equate this to the right
moving power $P^\rightarrow$ in
Eq.~(\ref{eq:powerperunitbandwidth}) because left moving waves
in the $\Zc$ line are not reflected and hence cannot contribute
to the right moving power.  Requiring $P=P^\rightarrow$ yields
the classical Nyquist result for the current noise of a
resistor
\be
\clS_{II}[\omega] = \frac{2}{R}\kb T
\label{eq:nyquist1}
\ee
or in the electrical engineering convention
\be
\clS_{II}[\omega]+\clS_{II}[-\omega] = \frac{4}{R}\kb T.
\ee

We can derive the equivalent expression for the voltage noise
of a resistor (see Fig.~\ref{fig:resistornoises} b) by
considering the voltage noise at the open termination of a
semi-infinite transmission line with $\Zc=R$.  For an open
termination $V^\rightarrow = V^\leftarrow$ so that the voltage
at the end is given by
\be
V=2V^\leftarrow = 2V^\rightarrow
\label{eq:B59}
\ee
and thus using Eqs.~(\ref{eq:voltagespecdensright}) and
(\ref{eq:equipart}) we find
\be
\clS_{VV} = 4\clS_{VV}^\rightarrow = 2R\kb T
\label{eq:Johnson1}
\ee
which is equivalent to Eq.~(\ref{eq:nyquist1}).

\subsection{Amplification with a transmission line and a negative resistance}
\label{subsec:NegativeResistance}

We close our discussion of transmission lines by further expanding upon the idea mentioned at the end of App.~\ref{subsec:ClassicalTransmissionLines} 
that one can view a one-port amplifier as a transmission line terminated
by an effective negative resistance.  The discussion here will be very general:  we will explore what can be learned about amplification by simply extending the results we have obtained on transmission lines to the case of an effective negative resistance.  Our general discussion will not address the important issues of {\it how} one achieves an effective negative resistance over some appreciable frequency range:  for such questions, one must focus on a specific physical realization, such as the parametric amplifier discussed in Sec.~\ref{subsec:NDParamps}.

We start by noting that for the case $-\Zc<R<0$ the power gain $G$ is given by
\be
G=|r|^2 >1,
\ee
and the ${\smat}'$ matrix introduced in Eq.~(\ref{eq:2x2Sprime}) becomes
\be
{\smat}' = -\left(
\begin{array}{cc}
\sqrt{G} & \pm \sqrt{G-1}\\
\pm \sqrt{G-1}&\sqrt{G}
\end{array}
\right)
\label{eq:2x2Smatrixamplifier}
\ee
where the sign choice depends on the branch cut chosen in the
analytic continuation of the off-diagonal elements. This
transformation is clearly no longer unitary (because there is
no energy conservation since we are ignoring the work done by
the amplifier power supply). Note however that we still have
det~$({\smat'})=+1$. It turns out that this naive analytic
continuation of the results from positive to negative
resistance is not strictly correct.  As we will show in the
following, we must be more careful than we have been so far in
order to insure that the transformation from the in fields
to the out fields must be canonical.

In order to understand the canonical nature of the
transformation between input and output modes, it is necessary
to delve more deeply into the fact that the two quadrature
amplitudes of a mode are canonically conjugate.  Following the
complex amplitudes defined in
Eqs.~(\ref{eq:Ain}-\ref{eq:Bout}), let us define a vector of
real-valued quadrature amplitudes for the incoming and outgoing
fields
\be
{\vec q}^{\, \rm in} = \left(
\begin{array}{c}
           \XAin\\ \XBin\\ \YBin\\ \YAin
\end{array}
\right), \,\, {\vec q}^{\, \rm out} = \left(
\begin{array}{c}
           \XAout\\ \XBout\\ \YBout\\ \YAout
\end{array}.
\label{eq:quadampvectorq}
\right)
\ee
The Poisson brackets amongst the different quadrature
amplitudes is given by
\be
\{q_i^{\rm in},q_j^{\rm in}\} \propto J_{ij},
\ee
or equivalently the quantum commutators are
\be
[q_i^{\rm in},q_j^{\rm in}]=i X_{\rm ZPF}^2 J_{ij},
\ee
where
\be
J \equiv \left(
\begin{array}{cccc}
0&0&0&+1\\
0&0&+1&0\\
0&-1&0&0\\
-1&0&0&0
\end{array}
\right).
\ee
In order for the transformation to be canonical, the same
Poisson bracket or commutator relations must hold for the
outgoing field amplitudes
\be
[q_i^{\rm out},q_j^{\rm out}]=i X_{\rm ZPF}^2 J_{ij}.
\label{eq:commutatoroutfields}
\ee
  In the case of a non-linear device
these relations would apply to the small fluctuations in the
input and output fields around the steady state solution.
Assuming a linear device (or linearization around the steady
state solution) we can define a $4\times 4$ real-valued 
scattering matrix $\ts$
in analogy to the $2\times 2$ complex-valued scattering matrix $\smat$
in Eq.~(\ref{eq:2x2Smatrix}) which relates the output fields to
the input fields
\be
q_i^{\rm out} = {\ts}_{ij}q_j^{\rm in}.
\label{eq:4x4Smatrix}
\ee
Eq.~(\ref{eq:commutatoroutfields}) puts a powerful constraint
on on the $\ts$ matrix, namely that it must be symplectic.
That is, $\ts$ and its transpose must obey
\be
\ts J \ts^{\rm T} = J.
\label{eq:symplectic}
\ee
From this it follows that
\be
\det {\ts} = \pm 1.
\ee
This in turn immediately implies Liouville's theorem that
Hamiltonian evolution preserves phase space volume (since $\det
{\ts}$ is the Jacobian of the transformation which
propagates the amplitudes forward in time).

Let us further assume that the device is phase preserving, that
is that the gain or attenuation is the same for both
quadratures.  One form for the $\ts$ matrix consistent with
all of the above requirements is
\be
{\ts} = \left(
\begin{array}{cccc}
+\cos\theta&\sin\theta&0&0\\
\sin\theta&-\cos\theta&0&0\\
0&0&-\cos\theta&\sin\theta\\
0&0&\sin\theta&+\cos\theta\\
\end{array}
\right).
\ee
This simply corresponds to a beam splitter and is the
equivalent of Eq.~(\ref{eq:2x2Smatrix}) with $r=\cos\theta$. As
mentioned in connection with Eq.~(\ref{eq:2x2Smatrix}), the
precise form of the scattering matrix depends on the choice of
planes at which the phases of the various input and output
waves are measured.

Another allowed form of the scattering matrix is:
\be
{\ts}' = -\left(
\begin{array}{cccc}
+\cosh\theta&+\sinh\theta&0&0\\
+\sinh\theta&+\cosh\theta&0&0\\
0&0&+\cosh\theta&-\sinh\theta\\
0&0&-\sinh\theta&+\cosh\theta\\
\end{array}
\right).
\ee
If one takes $\cosh\theta=\sqrt{G}$, this scattering matrix
is essentially the canonically correct formulation of the negative-resistance
scattering matrix we tried to write in Eq.~(\ref{eq:2x2Smatrixamplifier}).
Note that the off-diagonal terms have changed sign for the $Y$ quadrature relative to the naive expression in Eq.~(\ref{eq:2x2Smatrixamplifier}) (corresponding
to the other possible analytic continuation choice). This is
necessary to satisfy the symplecticity condition and hence make
the transformation canonical.  The scattering matrix $\ts'$ can describe amplification.
Unlike the beam splitter scattering matrix $\ts$ above, $\ts'$ is
not unitary (even though $\det {\ts}' = 1$).  Unitarity would correspond to power conservation.  Here, power is not conserved, as we are not explicitly tracking the power source supplying our active system.

The form of the negative-resistance amplifier scattering matrix $\ts'$ confirms
many of the general statements we made about phase-preserving amplification in 
Sec.~\ref{subsec:CavesArg}.  First, note that the requirement of finite gain $G>1$ and phase preservation
makes all the diagonal elements of $\ts'$ (i.e.~$\cosh\theta$ ) equal.  We see that
to amplify the $A$ mode, it is impossible to avoid
coupling to the $B$ mode (via the $\sinh\theta$ term) because
of the requirement of symplecticity.  We thus see that it is
impossible classically or quantum mechanically to build a
linear phase-preserving amplifier whose only effect is to
amplify the desired signal.  The presence of the $\sinh\theta$
term above means that the output signal is always contaminated
by amplified noise from at least one other degree of freedom
(in this case the $B$ mode). If the thermal or quantum noise in
$A$ and $B$ are equal in magnitude (and uncorrelated), then in
the limit of large gain where $\cosh\theta\approx\sinh\theta$,
the output noise (referred to the input) will be doubled.  This
is true for both classical thermal noise and quantum vacuum
noise. 

The negative resistance model of an amplifier here gives us another way
to think about the noise added by an amplifier:  crudely speaking, we can view it as being
directly analogous to the fluctuation-dissipation theorem
simply continued to the case of negative dissipation.  Just as
dissipation can occur only when we open up a new channel and
thus we bring in new fluctuations, so amplification can occur
only when there is coupling to an additional channel.  Without
this it is impossible to satisfy the requirement that the
amplifier perform a canonical transformation.

\section{Quantum Modes and Noise of a Transmission Line}
\label{app:QuantumResistor}


\subsection{Quantization of a transmission line}

Recall from Eq.~(\ref{eq:qconjtophi}) and the discussion in
Appendix \ref{app:classicalinputoutput} that the momentum
conjugate to the transmission line flux variable $\varphi(x,t)$
is the local charge density $q(x,t)$.  Hence in order to
quantize the transmission line modes we simply promote these
two physical quantities to quantum operators obeying the
commutation relation
\be
[\hat q(x),\hat\varphi(x')]=-i\hbar\delta(x-x')
\ee
from which it follows that the mode amplitudes defined in
Eq.~(\ref{eq:modeamplitudes}) become quantum operators obeying
\be
[\hat A_{k'},\hat A_k^\dagger] = \hbar\omega_k \delta_{kk'}
\ee
and we may identify the usual raising and lowering operators by
\be
\hat A_k =  \sqrt{\hbar\omega_k}\,\hat b_k
\ee
where $\hat b_k$ destroys a photon in mode $k$.  The quantum
form of the Hamiltonian in Eq.~(\ref{eq:anticipatesquantumH})
is thus
\be
H = \sum_k \hbar\omega_k \left[\hat b_k^\dagger \hat{b}_k +
\frac{1}{2}\right].
\ee
For the quantum case the thermal equilibrium expression
 then becomes
\be
\langle \hat A_k^\dagger \hat A_k\rangle = \hbar\omega_k
\nb(\hbar\omega_k),
\ee
which reduces to Eq.~(\ref{eq:equipart}) in the classical limit
$\hbar\omega_k \ll \kb T$.

We have seen previously in Eqs.~(\ref{eq:Vsum}) that the
voltage fluctuations on a transmission line can be resolved
into right and left moving waves which are functions of a
combined space-time argument
\be
V(x,t) = V^\rightarrow(t-\frac{x}{\vp}) +
V^\leftarrow(t+\frac{x}{\vp}).
\ee
Thus in an infinite transmission line, specifying
$V^\rightarrow$ everywhere in space at $t=0$ determines its
value for all times.  Conversely specifying $V^\rightarrow$ at
$x=0$ for all times fully specifies the field at all spatial
points.  In preparation for our study of the quantum version of
input-output theory in Appendix \ref{app:drivencavity}, it is
convenient to extend
Eqs.~(\ref{eq:VrightB38}-\ref{eq:VleftB39}) to the quantum case ($x=0$):
\begin{eqnarray}
	\hV^{\ra}(t) & = &
		 \sqrt{ \frac{1}{2 L c} } \sum_{k > 0} \sqrt{\hbar \omega_k}
			\left[ \hb^{\pd}_k e^{-i \omega_k t} + h.c. \right] 
			\nonumber \\
		& = &
		\int_0^{\infty} \frac{d \omega}{2 \pi} 
			\sqrt{
				\frac{\hbar \omega \Zc}{2}
			}
			\left[
				\hb^\ra[\omega] e^{-i \omega t} + h.c.
			\right]
			\label{eq:QuantVRight}
\end{eqnarray}
In the second line, we have defined:
\begin{eqnarray}
	\hb^{\ra}[\omega] & \equiv &
		2 \pi \sqrt{ \frac{v_p}{L} } \sum_{k>0} \hb_k  \delta(\omega - \omega_k)
		\label{eq:QuantBRight}
\end{eqnarray}
In a similar fashion, we have:
\begin{eqnarray}
	\hV^{\leftarrow}(t) 		& = &
		\int_0^{\infty} \frac{d \omega}{2 \pi} 
			\sqrt{
				\frac{\hbar \omega \Zc}{2}
			}
			\left[
				\hb^{\leftarrow}[\omega] e^{-i \omega t} + h.c.
			\right] 
						\label{eq:QuantVLeft} \\
	\hb^{\leftarrow}[\omega] & \equiv &
		2 \pi \sqrt{ \frac{v_p}{L} } \sum_{k<0}  \hb_k  \delta(\omega - \omega_k)
					\label{eq:QuantBLeft}
\end{eqnarray}
One can easily verify that among the $\hb^\ra[\omega], \hb^\leftarrow[\omega]$ operators and their conjugates, the only non-zero commutators are given by:
\be
	\left[ \hb^\ra[\omega] , \left(\hb^{\ra} [\omega'] \right)^{\dag} \right] = 
	\left[ \hb^\la[\omega] , \left( \hb^{\la}[\omega'] \right)^{\dag} \right] = 
	2 \pi \delta(\omega - \omega')
\ee
We have taken the continuum limit $L \ra \infty$ here, allowing us to change sums on $k$ to integrals.  We have thus obtained the description of a quantum transmission line in terms of left and right-moving frequency resolved modes, as used in our discussion
of amplifiers in Sec.~\ref{sec:ScatteringAmp} (see Eqs.~\ref{eqs:BosonicIV}).  Note that if the right-moving modes are further taken to be in thermal equilibrium, one finds (again, in the continuum limit):
\begin{subequations}
\begin{eqnarray}
	\left \langle \left( \hbr[\omega] \right)^\dag \hbr[\omega'] \right \rangle &=& 
		2 \pi \delta(\omega - \omega') 
			\nb(\hbar \omega) \\
	\left \langle \hbr[\omega] \left( \hbr[\omega'] \right)^\dag \right \rangle &=& 
		2 \pi \delta(\omega - \omega') 
			\left[1+\nb(\hbar \omega)\right]	 \nonumber \\		  .
\end{eqnarray}
\end{subequations}
	
We are typically interested in a relatively narrow band of frequencies centered on some characteristic drive or resonance frequency $\Omega_0$.  In this case, it is useful to work in the time-domain, in a frame rotating at $\Omega_0$.  
Fourier transforming
\footnote{As in the main text, we use in this appendix 
a convention which differs from the one
commonly used in quantum optics: $\ha[\omega] =
\int_{-\infty}^{+\infty}dt\,e^{+i\omega t}\ha(t)$ and
$\ha^\dagger[\omega]=[\ha[-\omega]]^\dagger=\int_{-\infty}
^{+\infty}dt\,e^{+i\omega t}\ha^\dagger(t)$.}
Eqs.~(\ref{eq:QuantBRight}) and (\ref{eq:QuantBLeft}), one finds:
\begin{subequations}
\begin{eqnarray}
	\hbr(t) & = & 
		\sqrt{\frac{\vp}{L}} \sum_{k>0} e^{-i(\omega_k
		-\Omega_0)t}\hat{b}_k(0),
	\label{eq:hbr}	\\
	\hb^{\la}(t) & = & \sqrt{\frac{\vp}{L}} \sum_{k<0} e^{-i(\omega_k
		-\Omega_0)t}\hat{b}_k(0).
	\label{eq:hbl}	
\end{eqnarray}
	\label{eqs:hbtime}
\end{subequations}
These represent temporal right and left moving modes.  
Note that the normalization factor in
Eqs.~(\ref{eqs:hbtime}) has been chosen so that the right moving
photon flux at $x=0$ and time $t$ 
is given by
\be
\langle\dot N\rangle= 
\langle \hbrd(t) \hbr(t) \rangle
\ee

In the same rotating frame, and within the approximation that all relevant frequencies are near $\Omega_0$, 
Eq.~(\ref{eq:QuantVRight}) becomes simply:
\begin{eqnarray}
\hV^\ra(t) &\approx& 
 	\sqrt{\frac{\hbar\Omega_0\Zc}{2}}
\left[ \hbr(t) + \hbrd(t)\right]
\label{eq:VrightII}
\end{eqnarray}
We have already seen that using classical statistical mechanics,
the voltage noise in equilibrium is white.  The
corresponding analysis of the temporal modes using
Eqs.~(\ref{eqs:hbtime}) shows that the quantum commutator obeys
\begin{eqnarray}
[\hbr(t),\hbrd(t')] &=& \delta(t-t').
\label{eq:noisecommu}
\end{eqnarray}
In deriving this result, we have converted summations over mode index to  
integrals over frequency.  Further, because (for finite time resolution at least) the integral is dominated by frequencies near $+\Omega_0$ we can, within the Markov (Wigner Weisskopf) approximation, extend the lower limit of frequency integration to minus infinity and thus arrive at a delta function in time.
If we further take the right moving modes to be in thermal equilibrium, then we 
may similarly approximate:
\begin{subequations}
\begin{eqnarray}
\langle \hbrd(t')\hbr(t)\rangle &=& \nb(\hbar\Omega_0)\delta(t-t')\label{eq:C10}\\
\langle \hbr(t)\hbrd(t')\rangle &=& \left[1+\nb(\hbar\Omega_0)\right]\delta(t-t').\label{eq:C11}
\end{eqnarray}
\end{subequations}
Equations (\ref{eq:VrightII}) to (\ref{eq:C11}) indicate that  $\hV^\ra(t)$  can
be treated as the quantum operator equivalent of white noise; a similar line of reasoning applies
\emph{mutatis mutandis} to the left moving modes.  We stress that these results rely crucially on our assumption that we are dealing with a relatively narrow band of frequencies in the vicinity of $\Omega_0$; the resulting approximations we have made are known as the Markov approximation.
As one can already see from the form of 
Eqs.~(\ref{eq:QuantVRight},\ref{eq:QuantVLeft}), and as will be discussed further, the actual spectral density of vacuum noise on
a transmission line is not white, but is linear in frequency.  The
approximation made in Eq.~(\ref{eq:noisecommu}) treats it as a
constant within the narrow band of frequencies of interest.  If
the range of frequencies of importance is large then the Markov
approximation is not applicable.

\subsection{Modes and the windowed Fourier transform}
\label{subapp:WindowedFT}

While delta function correlations can make the quantum
noise relatively easy to deal with in both the time and
frequency domain, it is sometimes the case that it is easier to
deal with a `smoothed' noise variable.  The introduction of an
ultraviolet cutoff regulates the mathematical singularities in
the noise operators evaluated at equal times and is physically
sensible because every real measurement apparatus has finite
time resolution.  A second motivation is that real spectrum
analyzers output a time varying signal which represents the
noise power in a certain frequency interval (the `resolution
bandwidth') averaged over a certain time interval (the inverse
`video bandwidth').  The mathematical tool of choice for
dealing with such situations in which time and frequency both
appear is the `windowed Fourier transform'.  The windowed transform
uses a kernel which is centered on some frequency window and
some time interval.  By summation over all frequency and time
windows it is possible to invert the transformation.  The
reader is directed to \cite{Mallat98} for the mathematical
details.



For our present purposes where we are interested in just a
single narrow frequency range centered on $\Omega_0$, a
convenient windowed transform kernel for smoothing the quantum
noise is simply a box of width $\Delta t$ representing the
finite integration time of our detector. In the frame rotating
at $\Omega_0$ we can define
\be
\hBr_j = \frac{1}{\sqrt{\Delta
t}}\int_{t_j}^{t_{j+1}}d\tau\,\hbr(\tau)
\label{eq:C13}
\ee
where $t_j=j(\Delta t)$ denotes the time of arrival of the
$j$th temporal mode at the point $x=0$.  Recall that $\hbr$ has
a photon flux normalization and so $\hBr_j$ is dimensionless.
From Eq.~(\ref{eq:noisecommu}) we see that these smoothed
operators obey the usual bosonic commutation relations
\be
[\hBr_j,\hBrd_k]=\delta_{jk}.
\ee

\begin{figure}[t]
\begin{center}
\includegraphics[width= 1.00 \columnwidth]{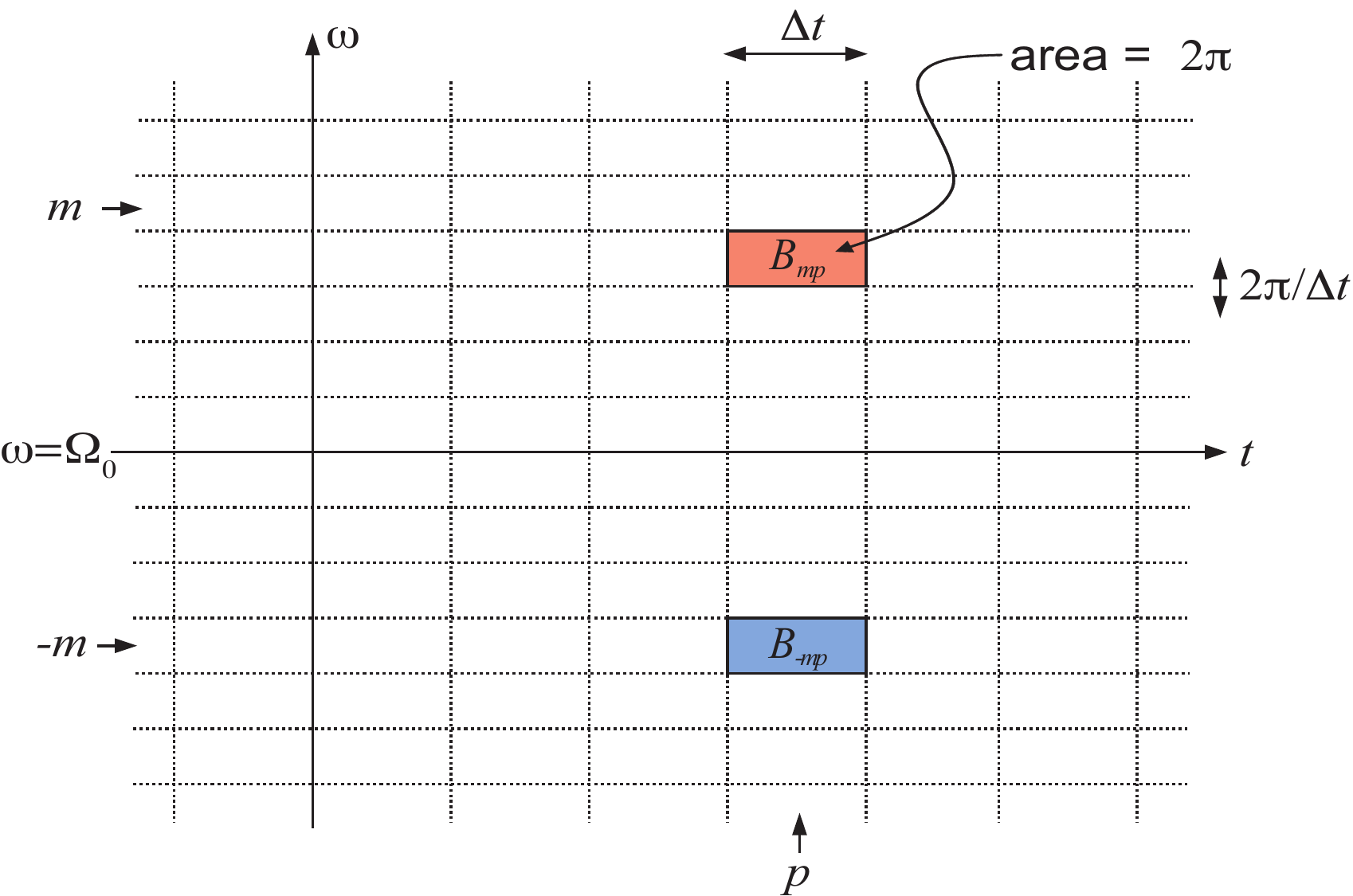}
\caption{(Color online) Schematic figure indicating how the various
modes defined by the windowed Fourier transform tile the time-frequency plane.
Each individual cell corresponds to a different mode, and has an area $2 \pi$.  
\label{fig:MichelsMusicalScoreTilingFigure}}
\end{center}
\end{figure}

The state $B_j^\dagger|0\rangle$ has a single photon occupying
basis mode $j$, which is centered in frequency space at
$\Omega_0$ and in time space on the interval $j\Delta
t<t<(j+1)\Delta t$ (i.e.~this temporal mode passes the
point $x=0$ during the $j$th time interval.)  This basis mode
is much like a note in a musical score:  it has a certain
specified pitch and occurs at a specified time for a specified
duration. Just as we can play notes of different frequencies
simultaneously, we can define other temporal modes on the same
time interval and they will be mutually orthogonal provided the
angular frequency spacing is a multiple of $2\pi/\Delta t$.
The result is a set of modes $B_{m,p}$ labeled by both a frequency index 
$m$ and a time index $p$.  $p$ labels the time interval as before, while $m$ labels 
the angular frequency:
\begin{eqnarray}
	\omega_{m} = \Omega_0 + m \frac{2 \pi}{\Delta t}
\end{eqnarray}
The result is, as illustrated in
Fig.~(\ref{fig:MichelsMusicalScoreTilingFigure}), a
complete lattice of possible modes tiling the frequency-time
phase space, each occupying area $2\pi$ corresponding to the
time-frequency uncertainty principle.

We can form other modes of arbitrary shapes centered on
frequency $\Omega_0$ by means of linear superposition of our
basis modes (as long as they are smooth on the time scale
$\Delta t$).  Let us define
\be
\Psi=\sum_j \psi_j \hBr_j.
\ee
This is also a canonical bosonic mode operator obeying
\be
[\Psi,\Psi^\dagger]=1
\ee
provided that the coefficients obey the normalization condition
\be
\sum_j |\psi_j|^2 = 1.
\ee
We might for example want to describe a mode which is centered
at a slightly higher frequency $\Omega_0+\delta\Omega$ (obeying
$(\delta\Omega)(\Delta t)<<1$) and spread out over a large time
interval $T$ centered at time $T_0$.  This could be given for
example by
\be
\psi_j = {\mathcal N} e^{-\frac{(j \Delta t-T_0)^2}{4T^2}}e^{-i(\delta\Omega) (j \Delta t)}
\ee
where ${\mathcal N}$ is the appropriate normalization constant.

The state having $n$ photons in the mode is simply
\be
\frac{1}{\sqrt{n!}} \left(\Psi^\dagger\right)^n|0\rangle.
\ee

The concept of `wave function of the photon' is fraught with dangers. 
In the very special case where we restrict attention solely to the subspace of single photon Fock states, 
we can usefully think of the amplitudes $\{\psi_j\}$ as the `wave function of the photon' \cite{CTPhotonsAtoms}
since it tells us about the spatial mode which is excited.  In the general case however it is essential to keep in mind that the transmission line is a collection of coupled LC oscillators with an infinite number of degrees of freedom.  Let us
simplify the argument by considering a single LC oscillator.
We can perfectly well write a wave function for the system as a
function of the coordinate (say the charge $q$ on the capacitor).  The ground state wave function $\chi_0(q)$ is a gaussian function of the coordinate.  The one photon
state created by $\Psi^\dagger$ has a wave function $\chi_1(q)\sim q\chi_0(q)$ proportional to the
coordinate times the same gaussian.  In the general case $\chi$ is a \emph{wave functional} of the charge distribution $q(x)$ over the entire transmission line.


Using Eq.~(\ref{eq:C10}) we have
\be
\langle \hBrd_j\hBr_k\rangle = \nb(\hbar\Omega_0)\delta_{jk}
\label{eq:WindowedModeOccupancy}
\ee
independent of our choice of the coarse-graining  time window
$\Delta t$.  This result allows us to give meaning to the
phrase one often hears bandied about in descriptions of
amplifiers that `the noise temperature corresponds to a mode
occupancy of $X$ photons'. This simply means that the photon flux
\emph{per unit bandwidth} is $X$.  Equivalently the flux in
bandwidth $B$ is
\be
\overline{{\dot N}} = \frac{X}{\Delta t} (B\Delta t)=XB.
\label{eq:PhotonFluxXB}
\ee
The interpretation of this is that $X$ photons in a temporal
mode of duration $\Delta t$ pass the origin in time $\Delta t$.
Each mode has bandwidth $\sim\frac{1}{\Delta t}$ and so there
are $B \Delta t$ independent temporal modes in
bandwidth $B$ all occupying the same time interval $\Delta t$.
The longer is $\Delta t$ the longer it takes a given mode to
pass the origin, but the more such modes fit into the frequency
window.

As an illustration of these ideas, consider the following
elementary question:  What is the mode occupancy of a laser
beam of power $P$ and hence photon flux $\overline{{\dot
N}}=\frac{P}{\hbar\Omega_0}$?   We cannot answer this without
knowing the coherence time or equivalently the bandwidth.  The
output of a good laser is like that of a radio frequency
oscillator--it has essentially no amplitude fluctuations.  The
frequency is nominally set by the physical properties of the
oscillator, but there is nothing to pin the phase which
consequently undergoes slow diffusion due to unavoidable noise
perturbations.  This leads to a finite phase coherence time
$\tau$ and corresponding frequency spread $1/\tau$ of the laser
spectrum.  (A laser beam differs from a thermal source that has
been filtered to have the same spectrum in that it has smaller
amplitude fluctuations.)  Thus we expect that the mode
occupancy is $X=\overline{\dot N}\tau$.  A convenient
approximate description in terms of temporal modes is to take
the window interval to be $\Delta t=\tau$.  Within the $j$th
interval we take the phase to be a (random) constant
$\varphi_j$ so that  (up to an unimportant normalization
constant) we have the coherent state
\be
\prod_j e^{\sqrt{X}e^{i\varphi_j}\hBrd_j}|0\rangle
\ee
which obeys
\be
\langle \hBr_k\rangle = \sqrt{X}e^{i\varphi_k}
\ee
and
\be
\langle \hBrd_k \hBr_k\rangle = X.
\ee



%


\subsection{Quantum noise from a resistor}
\label{subapp:ResistorNoise}

Let us consider the quantum equivalent to
Eq.~(\ref{eq:Johnson1}), $\clS_{VV} = 2R\kb T$,  for the case of a semi-infinite
transmission line with open termination, representing a
resistor. \null From Eq.~(\ref{eq:current}) we see that the
proper boundary condition for the $\varphi$ field is
$\partial_x\varphi(0,t) =
\partial_x\varphi(L,t)=0$. (We have temporarily made the transmission line have
a large but finite length $L$.) The normal mode expansion that
satisfies these boundary conditions is \be \varphi(x,t) =
\sqrt{\frac{2}{L}}\sum_{n=1}^\infty \varphi_n(t) \cos(k_nx),
\ee where $\varphi_n$ is the normal coordinate and $k_n \equiv
\frac{\pi n}{L}$. Substitution of this form into the Lagrangian
and carrying out the spatial integration yields a set of
independent harmonic oscillators representing the normal modes.
\be L_g = \sum_{n=1}^\infty \left( \frac{c}{2} \dot\varphi_n^2 -
\frac{1}{2\ell} k_n^2 \varphi_n^2 \right). \ee From this we can find
the momentum operator $\hat p_n$ canonically conjugate to the
coordinate operator $\hat\varphi_n$ and quantize the system to
obtain an expression for the operator representing the voltage
at the end of the transmission line in terms of the mode
creation and destruction operators \be \hat{V} =
\sum_{n=1}^\infty \sqrt{\frac{\hbar\Omega_n}{Lc}}
i(\hat{b}^\dagger_n - \hat{b}_n). \ee The spectral density of
voltage fluctuations is then found to be
\begin{eqnarray}
\SVV[\omega] = \frac{2\pi}{L} \sum_{n=1}^\infty
\frac{\hbar\Omega_n}{c} \big\{ \nb
(\hbar\Omega_n)\delta(\omega+\Omega_n)\nonumber\\
  +
[\nb(\hbar\Omega_n)+1] \delta(\omega-\Omega_n) \big\},
\end{eqnarray}
  where $\nb(\hbar\omega)$ is the Bose occupancy factor
for a photon with energy $\hbar\omega$. Taking the limit
$L\rightarrow\infty$ and converting the summation to an
integral yields
\be
\SVV(\omega) = 2\Zc\hbar|\omega| \big\{ \nb(\hbar|\omega|)
\Theta(-\omega) + [\nb(\hbar|\omega|)+1] \Theta(\omega) \big\},
\label{eq:transmissionlinenoise}
\ee
where $\Theta$ is the step function. We see immediately that at
zero temperature there is no noise at negative frequencies
because energy can not be extracted from zero-point motion.
However there remains noise at positive frequencies indicating
that the vacuum is capable of absorbing energy from another quantum system.
The voltage spectral density at both zero and non-zero
temperature is plotted in Fig.~(\ref{fig:resistornoiseplot}).


Eq.~(\ref{eq:transmissionlinenoise}) for this `two-sided'
spectral density of a resistor can be rewritten in a more
compact form
\be \SVV[\omega] = \frac{2\Zc\hbar\omega}
{1-e^{-{\hbar \omega / \kbt}} } ,
\ee
which reduces to the more familiar expressions in various
limits. For example, in the classical limit $k_{\rm B}T \gg
\hbar\omega$ the spectral density is equal to the Johnson noise
result\footnote{Note again that in the engineering convention
this would be $\SVV[\omega] = 4\Zc k_{\rm B}T$.}
\be
\SVV[\omega] = 2\Zc k_{\rm B}T,
\ee
 in agreement with Eq.~(\ref{eq:Johnson1}).
In the quantum limit it reduces to
\be \SVV[\omega] = 2\Zc\hbar\omega \Theta(\omega).
\ee Again, the step function tells us that the resistor can
only absorb energy, not emit it, at zero temperature.

If we use the engineering convention and add the noise at
positive and negative frequencies we obtain \be \SVV[\omega] +
\SVV[-\omega] = 2\Zc\hbar\omega \coth\frac{\hbar\omega}{2k_{\rm
B}T} \label{ResistorDensitySymm} \ee
  for the symmetric part of the
noise, which appears in the quantum fluctuation-dissipation
theorem (cf.~Eq.~(\ref{eq:FDFinal})). The antisymmetric part
of the noise is simply \be \SVV[\omega] - \SVV[-\omega] =
2\Zc\hbar\omega, \label{ResistorDensityAsymm} \ee yielding
\be
\frac{\SVV[\omega] - \SVV[-\omega]}{\SVV[\omega] +
\SVV[-\omega]} = \tanh\frac{\hbar\omega}{2k_{\rm B}T}.
\ee

This quantum treatment can also be applied to any arbitrary
dissipative network
\cite{DevoretLesHouches,BurkhardANDDivencenzo}. If we have a
more complex circuit containing capacitors and inductors, then
in all of the above expressions, $\Zc$ should be replaced by
${\rm Re}\, Z[\omega]$ where $Z[\omega]$ is the complex
impedance presented by the circuit.

In the above we have explicitly quantized the standing wave
modes of a finite length transmission line.  We could instead
have used the running waves of an infinite line and recognized
that, as the in classical treatment in Eq.~(\ref{eq:B59}), the
left and right movers are not independent.  The open boundary
condition at the termination requires $V^\leftarrow =
V^\rightarrow$ and hence $b^\rightarrow=b^\leftarrow$.  We then
obtain
\be
S_{VV}[\omega] = 4S_{VV}^\rightarrow[\omega]
\ee
and from the quantum analog of
Eq.~(\ref{eq:voltagespecdensright}) we have
\begin{eqnarray}
S_{VV}[\omega] &=&
\frac{4\hbar|\omega|}{2c\vp}\left\{\Theta(\omega)(\nb+1)+\Theta(-\omega)\nb\right\}\nonumber\\
&=&2Z_c \hbar|\omega|\left\{\Theta(\omega)(\nb+1)+\Theta(-\omega)\nb\right\} \nonumber \\
\end{eqnarray}
in agreement with Eq.~(\ref{eq:transmissionlinenoise}).

\section{Back Action and Input-Output Theory for Driven Damped Cavities}
\label{app:drivencavity}

A high $Q$ cavity whose resonance frequency can be parametrically
controlled by an external source can act as a very simple quantum
amplifier, encoding information about the external source in the
phase and amplitude of the output of the driven cavity.  For
example, in an optical cavity, one of the mirrors could be
moveable and the external source could be a force acting on that
mirror. This defines the very active field of optomechanics, which also deals
with microwave cavities coupled to nanomechanical systems and other related setups \cite{1985_11_Meystre_RadiationPressureDrivenInterferometers,%
2005_02_MarquardtHarrisGirvin_Cavity,2004_12_HoehbergerKarrai_CoolingMicroleverNature,%
Arcizet2006,Gigan2006,%
2006_11_Kippenberg_RadPressureCooling,2006_05_Harris_MicrocantileverMirror,%
Harris08,2007_Wineland_RFcircuitCooling,%
2007_01_Marquardt_CantileverCooling,2007_02_WilsonRae_Cooling,KonradCooling2008}.
In the case of a microwave cavity containing a qubit, the
state-dependent polarizability of the qubit acts as a source
which shifts the frequency of the cavity
\cite{Blais04,Wallraff04,Schuster05}.

The dephasing of a qubit in a microwave cavity and the
fluctuations in the radiation pressure in an optical cavity
both depend on the quantum noise in the number of photons
inside the cavity.  We here use a simple equation of motion
method to exactly solve for this quantum noise in the
perturbative limit where the dynamics of the qubit or mirror
degree of freedom has only a weak back action effect on the
cavity.

In the following, we first give a basic discussion of
the cavity field noise spectrum, deferring
the detailed microscopic derivation to subsequent subsections.
We then provide a review of the input-output theory
for driven cavities, and employ this theory to analyze the important example of a
dispersive position measurement, where we demonstrate how the standard quantum limit can
be reached. Finally, we analyze an example where a modified
dispersive scheme is used to detect only one quadrature of a harmonic oscillator's motion,
such that this quadrature does not feel any back-action.

\subsection{Photon shot noise inside a cavity and back action}

Consider a degree of freedom $\hat{z}$ coupled parametrically
with strength $A$ to the cavity oscillator
\be
{\hat H}_{\rm int} = \hbar\omegar(1 + A \hat z)\, [{\hat
a}^\dagger \hat{a}-\langle{\hat a}^\dagger \hat{a}\rangle]
\label{eq:paramcoupling}
\ee
where following Eq.~(\ref{CavityHam1}), we have taken $A$ to be
dimensionless, and use $\hat z$ to denote the dimensionless 
system variable that we wish to probe.  For example, 
$\hat{z}$ could represent the dimensionless position
of a mechanical oscillator
\be
\hat z \equiv \frac{\hat x}{x_{\rm ZPF}}.
\ee
 We have subtracted the $\langle\hat a^\dagger \hat
a\rangle$ term so that the mean force on the degree of freedom
is zero. To obtain the full Hamiltonian, we would have to add
the cavity damping and driving terms, as well as the
Hamiltonian governing the intrinsic dynamics of the system
$\hz$. From Eq.~(\ref{eq:1sidedcavityforcenoise}) we know
that the back action noise force acting on $\hz$ is
proportional to the quantum fluctuations in the number of
photons ${\hat n}={\hat a}^{\dagger} {\hat a}$ in the cavity,
\be
S_{nn}(t) = \langle \hat{a}^\dagger(t) \hat{a}(t) \hat{a}^
\dagger(0) \hat{a}(0)\rangle - \langle \hat{a}^\dagger(t)
\hat{a}(t)\rangle^2.
\label{eq:Snoise}
\ee
For the case of continuous wave driving at frequency $\omegal=\omegar+\Delta$
detuned by $\Delta$ from the resonance, the cavity is in a
coherent state $|\psi\rangle$ obeying
\be
\hat{a}(t) = e^{-i\omegal t} [\bar a + \hat{d}(t)]
\label{eq:coherent}
\ee
where the first term is the `classical part' of the mode amplitude
$\psi(t) = \bar a e^{-i\omegal t}$ determined by the strength of
the drive field, the damping of the cavity and the detuning
$\Delta$, and $d$ is the quantum part. By definition,
\be
\hat{a}|\psi\rangle=\psi|\psi\rangle
\ee
so the coherent state is annihilated by ${\hat d}$:
\be
\hat{d}|\psi\rangle=0.
\label{eq:dvacuum}
\ee
That is, in terms of the operator $\hat{d}$, the coherent state
looks like the undriven quantum ground state .  The
displacement transformation in Eq.~(\ref{eq:coherent}) is
canonical since
\be
[\hat{a},\hat{a}^\dagger]=1\, \Rightarrow \,
[\hat{d},\hat{d}^\dagger] =1.
\ee

Substituting the displacement transformation into
Eq.~(\ref{eq:Snoise}) and using Eq.~(\ref{eq:dvacuum}) yields
\be
 S_{nn}(t) = \bar n \langle
\hat{d}(t)\hat{d}^\dagger(0)\rangle,
\label{eq:Snnt}
\ee
where $\bar n=|\bar a|^2$ is the mean cavity photon number.
   If we set the cavity energy damping
rate to be $\kappa$, such that the amplitude damping rate is $\kappa/2$,
then the undriven state obeys
\be
\langle \hat{d}(t) \hat{d}^\dagger(0) \rangle = e^{+i\Delta
t}e^{-\frac{\kappa}{2}|t|}.
\ee
This expression will be justified formally in the subsequent subsection, 
after introducing input-output theory.
We thus arrive at the very simple result
\be
S_{nn}(t) = \bar n e^{i\Delta t-\frac{\kappa}{2}|t|}.
\label{eq:autocorrelation}
\ee

\begin{figure}[t]
\begin{center}
\includegraphics[width=\columnwidth]{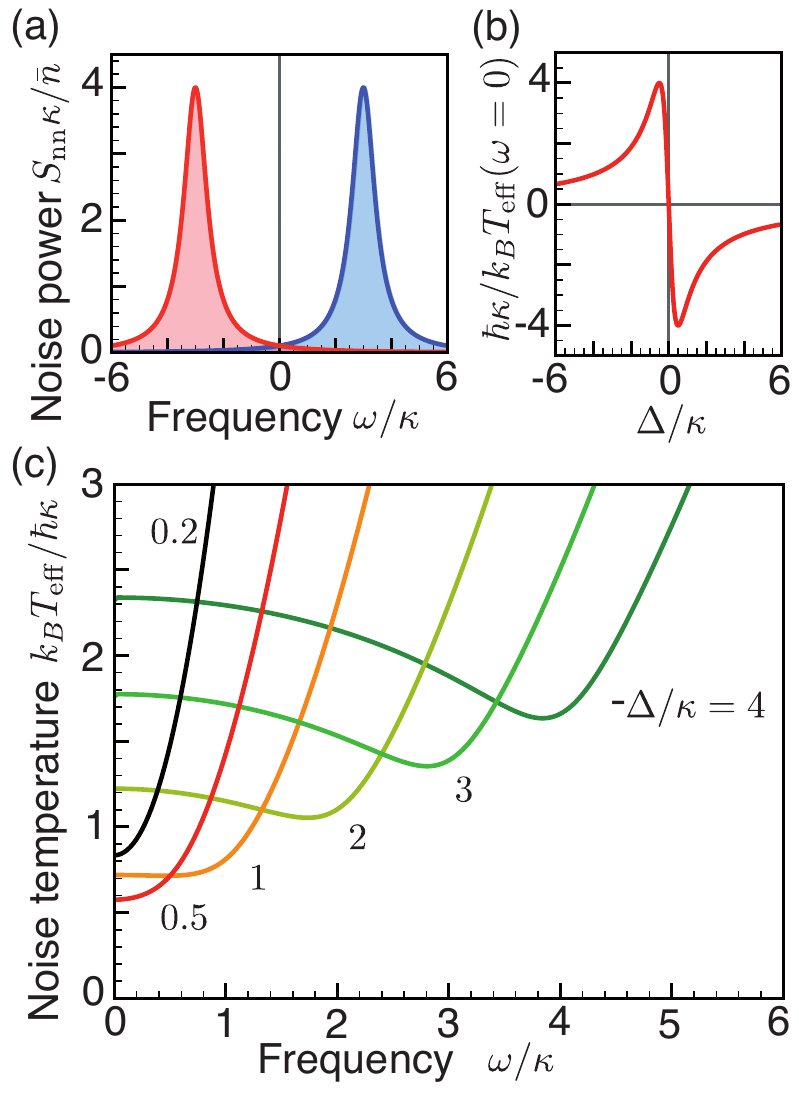}
\caption{ (Color online) (a) Noise spectrum of the photon number in a
driven cavity as a function of frequency when the cavity drive
frequency is detuned from the cavity resonance by $\Delta=+3\kappa$
(left peak) and $\Delta=-3\kappa$ (right peak). (b)
Effective temperature $T_{\rm eff}$ of the low frequency noise, $\omega\rightarrow 0$,
as a function of the detuning $\Delta$ of the drive from the cavity resonance.
(c) Frequency-dependence of the effective noise temperature, for different
values of the detuning.
\label{fig:CavityTeff}}
\end{center}
\end{figure}

The power spectrum of the noise is, via the Wiener-Khinchin
theorem (Appendix \ref{app:WienerKhinchin}), simply the Fourier
transform of the autocorrelation function given in
Eq.~(\ref{eq:autocorrelation})
\be
S_{nn}[\omega]=\int_{-\infty}^{+\infty} dt\, e^{i \omega t}
S_{nn}(t)=\bar n \frac{\kappa}{(\omega+\Delta)^2 +
(\kappa/2)^2}.
\label{eq:Snnomega}
\ee
As can be seen in Fig.~\ref{fig:CavityTeff}a, for positive
detuning $\Delta=\omega_L-\omegar>0$, i.e. for a drive that is
blue-detuned with respect to the cavity, the noise peaks at
{\em negative} $\omega$. This means that the noise tends to
pump energy into the degree of freedom $\hat z$ 
(i.e.~it contributes negative damping). For negative detuning the noise
peaks at positive $\omega$ corresponding to the cavity
absorbing energy from $\hat z$. Basically, the interaction with
$\hat z$ (three wave mixing) tries to Raman scatter the drive
photons into the high density of states at the cavity
frequency. If this is uphill in energy, then $\hat z$ is
cooled.

As discussed  in Sec.~\ref{sec:SHOQNOISE} (c.f.~Eq.~(\ref{OscTeff})), 
at each frequency $\omega$, we can use
detailed balance to assign the noise an effective temperature $T_{\rm eff}[\omega]$:
\begin{eqnarray}
      \frac{S_{nn}[\omega]}{S_{nn}[-\omega]}&=&e^{\hbar \omega / k_B T_{\rm eff}[\omega]} \Leftrightarrow \nonumber \\
      \kb T_{\rm eff}[\omega] &\equiv&
      \frac{\hbar \omega}
      {\log \left[\frac{S_{nn}[\omega]}{S_{nn}[-\omega]} \right]}
      \label{eq:CavityTeff}
\end{eqnarray}
or equivalently
\be
\frac{S_{nn}[\omega]-S_{nn}[-\omega]}
{S_{nn}[\omega]+S_{nn}[-\omega]}=\tanh(\beta\hbar\omega/2).
\ee
If ${\hat z}$ is the coordinate of a harmonic oscillator of frequency $\omega$ (or some
non-conserved observable of a qubit with level splitting $\omega$), then that system will acquire
a temperature $T_{\rm eff}[\omega]$ in the absence of coupling to any other environment. In particular,  if the characteristic oscillation frequency of the system ${\hat z}$ is much
smaller than $\kappa$, then we have the simple result
\begin{eqnarray}
\frac{1}{k_{\rm B}T_{\rm eff}}&=&\lim_{\omega\rightarrow 0^+}
\frac{2}{\hbar\omega}\frac{S_{nn}[\omega]-S_{nn}[-\omega]}
{S_{nn}[\omega]+S_{nn}[-\omega]}\nonumber \\ &=& 2\frac{d\ln
S_{nn}[\omega]}{d\hbar\omega}\nonumber\\ &=&
\frac{1}{\hbar}\frac{-4\Delta}{\Delta^2 + (\kappa/2)^2}.
\end{eqnarray}
As can be seen in Fig.~\ref{fig:CavityTeff}, the asymmetry in
the noise changes sign with detuning, which causes the effective
temperature to change sign.

First we discuss the case of a positive $T_{\rm eff}$, where this mechanism can be used
to laser cool  an oscillating mechanical cantilever, provided $T_{\rm eff}$ is lower than the
intrinsic equilibrium  temperature of the cantilever.
\cite{2004_12_HoehbergerKarrai_CoolingMicroleverNature,Arcizet2006,%
Gigan2006,2006_11_Kippenberg_RadPressureCooling,2006_05_Harris_MicrocantileverMirror,%
Harris08,2007_Wineland_RFcircuitCooling,2007_01_Marquardt_CantileverCooling, 2007_02_WilsonRae_Cooling}.
A simple classical argument helps us understand this cooling
effect.   Suppose that the moveable mirror is at the right hand
end of a cavity being driven below the resonance frequency.  If
the mirror moves to the right, the resonance frequency will
fall and the number of photons in the cavity will rise.  There
will be a time delay however to fill the cavity and so the
extra radiation pressure will not be fully effective in doing
work on the mirror. During the return part of the oscillation
as the mirror moves back to the left, the time delay in
emptying the cavity will cause the mirror to have to do extra
work against the radiation pressure. At the end of the cycle it
ends up having done net positive work on the light field and
hence is cooled.  The effect can therefore be understood as
being due to the introduction of some extra optomechanical
damping.

The signs reverse (and $T_{\rm eff}$ becomes negative) if the
cavity is driven above resonance, and consequently the
cantilever motion is heated up. In the absence of intrinsic
mechanical losses, negative values of the effective temperature
indicate a dynamical instability of the cantilever (or
population inversion in the case of a qubit), where the
amplitude of motion grows until it is finally stabilized by
nonlinear effects. This can be interpreted as negative damping
introduced by the optomechanical coupling and can be used to
create parametric amplification of mechanical forces acting on
the oscillator.

Finally, we mention that cooling towards the quantum ground
state of a mechanical oscillator (where phonon numbers become
much less than one), is only possible
\cite{2007_01_Marquardt_CantileverCooling,2007_02_WilsonRae_Cooling}
in the ``far-detuned regime'', where $-\Delta=\omega \gg
\kappa$ (in contrast to the $\omega \ll \kappa$ regime
discussed above).

\subsection{Input-output theory for a driven cavity}

\label{subsec:inputoutput}

The results from the previous section can be more formally and
rigorously derived in a full quantum theory of a cavity driven
by an external coherent source.  The theory relating the drive,
the cavity and the outgoing waves radiated by the cavity is
known as input-output theory and the classical description was
presented in Appendix \ref{app:classicalinputoutput}.  The present
quantum discussion closely follows standard references on the
subject \cite{Walls94,YurkeInputOutput,Yurke84}. The crucial
feature that distinguishes such an approach from many other
treatments of quantum-dissipative systems is the goal of
keeping the bath modes instead of tracing them out. This is
obviously necessary for the situations we have in mind, where
the output field emanating from the cavity contains the
information acquired during a measurement of the system coupled
to the cavity.  As we learned from the classical treatment, we
can eliminate the outgoing waves in favor of a damping term for
the system.  However we can recover the solution for the
outgoing modes completely from the solution of the equation of
motion of the damped system being driven by the incoming waves.

In order to drive the cavity we must partially open one of its
ports which exposes the cavity both to the external drive and
to the vacuum noise outside which permits energy in the cavity
to leak out into the surrounding bath.  We will formally
separate the degrees of freedom into internal cavity modes and
external bath modes. Strictly speaking, once the port is open,
these modes are not distinct and we only have `the modes of the
universe'
\cite{ScullyOpenCavityLaser,ScullyModesofUniverse1,ScullyModesofUniverse2}.
However for high $Q$ cavities, the distinction is well-defined
and we can model the decay of the cavity in terms of a
spontaneous emission process in which an internal boson is
destroyed and an external bath boson is created.  We assume a
single-sided cavity. For a high $Q$ cavity, this
physics is accurately captured in the following Hamiltonian
\be
\hat{H} = \hat{H}_{\rm sys} + \hat{H}_{\rm bath} + \hat{H}_{\rm
int}.
\label{eq:cavitypluscantileverH}
\ee
    The bath Hamiltonian is
\be
\hat{H}_{\rm bath} = \sum_q\hbar\omega_q \hat{b}^\dagger_q
\hat{b}_q
\ee
where $q$ labels the quantum numbers of the independent
harmonic oscillator bath modes obeying
\be
[\hat{b}_q,\hat{b}^\dagger_{q'}]=\delta_{q,q'}.
\ee
Note that since the bath terminates at the system, there is no
translational invariance, the normal modes are standing not
running waves, and the quantum numbers $q$ are not necessarily
wave vectors.

The coupling Hamiltonian is (within the rotating wave
approximation)
\be
\hat{H}_{\rm int} = -i\hbar\sum_q\left[f_q \hat{a}^\dagger
\hat{b}_q - f^*_q \hat{b}^\dagger_q \hat{a}\right].
\label{eq:couplingHam}
\ee
   For the moment we will leave the system (cavity) Hamiltonian to
be completely general, specifying only that it consists of a
single degree of freedom (i.e.\   we concentrate on only a
single resonance of the cavity with frequency $\omegar$)
obeying the usual bosonic commutation relation
\be
[\hat{a},\hat{a}^\dagger] = 1.
\ee
(N.B. this does not imply that it is a harmonic oscillator.  We
will consider both linear and non-linear cavities.) Note that the
most general linear coupling to the bath modes would include terms
of the form $\hat{b}^\dagger_q \hat{a}^\dagger$ and $\hat{b}_q a$
but these are neglected within the rotating wave approximation
because in the interaction representation they oscillate at high
frequencies and have little effect on the dynamics.

The Heisenberg equation of motion (EOM) for the bath variables is
\be
\dot{\hat{b}}_q=\frac{i}{\hbar}[\hat{H},\hat{b}_q]=-i\omegaq
\hat{b}_q + f_q^*\hat{a}
\label{eq:bathEOM}
\ee
We see that this is simply the EOM of a harmonic oscillator driven by
a forcing term due to the motion of the cavity degree of freedom.
    Since this is a linear system, the
EOM can be solved exactly.  Let $t_0<t$ be a time in the distant
past before any wave packet launched at the cavity has reached it.
The solution of Eq.~(\ref{eq:bathEOM}) is
\be
\hat{b}_q(t)=e^{-i\omegaq(t-t_0)}\hat{b}_q(t_0) + \int_{t_0}^t
d \tau\, e^{-i\omegaq(t-\tau)} f^*_q\, \hat{a}(\tau).
\label{eq:bathEOMsolution}
\ee
The first term is simply the free evolution of the bath while the
second represents the waves radiated by the cavity into the bath.

The EOM for the cavity mode is
\be
\dot{\hat{a}} = \frac{i}{\hbar}[\hat{H}_{\rm sys},\hat{a}] -\sum_q
f_q \hat{b}_q.
\label{eq:cavityEOM}
\ee
Substituting Eq.~(\ref{eq:bathEOMsolution}) into the last term
above yields
\begin{eqnarray}
\sum_qf_q\hat{b}_q = \sum_q f_q
e^{-i\omegaq(t-t_0)}\hat{b}_q(t_0)\nonumber\\
+ \sum_q|f_q|^2 \int_{t_0}^t d\tau\, e^{-i(\omegaq-\omegar)(t-\tau)}
[e^{+i\omegar(\tau-t)}\hat{a}(\tau)],\,\,\,\,
\end{eqnarray}
where the last term in square brackets is a slowly varying
function of $\tau$. To simplify our result, we note that if the
cavity system were a simple harmonic oscillator of frequency
$\omegar$ then the decay rate from the $n=1$ single photon
excited state to the $n=0$ ground state would be given by the
following Fermi Golden Rule expression
\be
\kappa(\omegar)
=2\pi\sum_q|f_q|^2\delta(\omegar-\omegaq).
\ee
\null   From this it follows that
\be
\int_{-\infty}^{+\infty}\frac{d\nu}{2\pi} \kappa(\omegar+ \nu)
e^{-i\nu(t-\tau)} =\sum_q|f_q|^2
e^{-i(\omegaq-\omegar)(t-\tau)}.
\label{eq:markov0707}
\ee
We now make the Markov approximation which assumes that
$\kappa(\nu)=\kappa$ is a constant over the range of
frequencies relevant to the cavity so that
Eq.~(\ref{eq:markov0707}) may be represented as
\be
\sum_q|f_q|^2 e^{-i(\omegaq-\omegar)(t-\tau)} = \kappa
\delta(t-\tau).
\ee
Using
\be
\int_{-\infty}^{x_0}dx\,\delta(x-x_0) = \frac{1}{2}
\ee
we obtain for the cavity EOM
\be
\dot{\hat{a}} = \frac{i}{\hbar}[\hat{H}_{\rm sys},\hat{a}] -
\frac{\kappa}{2} \hat{a} -\sum_q f_q
e^{-i\omegaq(t-t_0)}\hat{b}_q(t_0) .
\label{eq:cavityEOM2}
\ee
The second term came from the part of the bath motion representing
the wave radiated by the cavity and, within the Markov
approximation, has become a simple linear damping term for the
cavity mode. Note the important factor of 2.  The amplitude decays
at half the rate of the intensity (the energy decay rate
$\kappa$).

Within the spirit of the Markov approximation it is further
convenient to treat $f\equiv\sqrt{|f_q|^2}$ as a constant and
define the density of states (also taken to be a constant) by
\be
\rho = \sum_q \delta(\omegar - \omegaq)
\ee
so that the Golden Rule rate becomes
\be
\kappa = 2\pi f^2\rho.
\ee
   We can now define the so-called `input
mode'
\be
\hbin(t) \equiv \frac{1}{\sqrt{2\pi\rho}}\sum_q
e^{-i\omegaq(t-t_0)}\hat{b}_q(t_0) \, .
\label{eq:inputmode}
\ee
For the case of a transmission line treated in Appendix \ref{app:QuantumResistor}, this
coincides with the field ${\hat b}^{\rightarrow}$ moving towards the cavity [see Eq.~(\ref{eq:hbr})]. 
We finally have for the cavity EOM
\be
\dot{\hat{a}} = \frac{i}{\hbar}[\hat{H}_{\rm sys},\hat{a}] -
\frac{\kappa}{2} \hat{a} -\sqrt{\kappa}\, \hbin(t).
\label{eq:cavityEOM3}
\ee
Note that when a wave packet is launched from the bath towards
the cavity, causality prevents it from knowing about the
cavity's presence until it reaches the cavity.  Hence the input
mode evolves freely as if the cavity were not present until the
time of the collision at which point it begins to drive the
cavity.  Since $\hbin$(t) evolves under the free bath
Hamiltonian and acts as the driving term in the cavity EOM, we
interpret it physically as the input mode.
Eq.~(\ref{eq:cavityEOM3}) is the quantum analog of the
classical equation (\ref{eq:SHOLCeqnmotion}), for our previous example of an LC-oscillator driven
by a transmission line.  The latter would
also have been first order in time if as in
Eq.~(\ref{eq:anticipatesquantumEOM}) we had worked with the
complex amplitude $A$ instead of the coordinate $Q$.

Eq.~(\ref{eq:inputmode}) for the input mode contains a time
label just as in the interaction representation.  However it is
best interpreted as simply labeling the particular linear
combination of the bath modes which is coupled to the system at
time $t$.  Some authors even like to think of the bath modes as
non-propagating while the cavity flies along the bath (taken to be 1D) 
at a velocity $v$.  The system then only interacts briefly
with the local mode positioned at $x=vt$ before moving on and interacting with the
next local bath mode. 
We will elaborate on this view further at the end of this subsection.

The expression for the power $P_{\rm in}$ (energy per time)
impinging on the cavity depends on the normalization chosen in
our definition of $\hbin$. It can be obtained, for example, by
imagining the bath modes ${\hat b}_q$ to live on a
one-dimensional waveguide with propagation velocity $v$ and
length $L$ (using periodic boundary conditions). In that case
we have to sum over all photons to get the average power
flowing through a cross-section of the waveguide, $P_{\rm in} =
\sum_q {\hbar \omega_q} (v_p/L) \left\langle {\hat b}_q^{\dagger}
{\hat b}_q \right\rangle$. Inserting the definition for
$\hbin$, Eq.~(\ref{eq:inputmode}), the expression for the input
power carried by a monochromatic beam at frequency $\omega$ is
\be
P_{\rm in}(t) = \hbar \omega \left\langle \hbin^{\dagger}(t)
\hbin(t) \right\rangle
\label{eq:inputpower}
\ee
Note that this has the correct dimensions due to our choice of
normalization for $\hbin$ (with dimensions $\sqrt{\omega}$). In
the general case, an integration over frequencies is needed (as
will be discussed further below). An analogous formula holds
for the power radiated by the cavity, to be discussed now.


The output mode $\hbout(t)$ is radiated into the bath and evolves freely
after the system interacts with $\hbin(t)$.   If the cavity did not respond at all,
then the output mode would simply be the input mode reflected
off the cavity mirror.  If the mirror is partially transparent
then the output mode will also contain waves radiated by the
cavity (which is itself being driven by the input mode
partially transmitted into the cavity through the mirror) and
hence contains information about the internal dynamics of the
cavity. To analyze this output field, let $t_1>t$ be a time in
the distant future after the input field has interacted with
the cavity. Then we can write an alternative solution to
Eq.~(\ref{eq:bathEOM}) in terms of the final rather than the
initial condition of the bath
\be
\hat{b}_q(t) = e^{-i\omega_q(t-t_1)}\hat{b}_q(t_1) -
\int_t^{t_1}d\tau\, e^{-i\omegaq(t-\tau)}f_q^*\hat{a}(\tau).
\ee
Note the important minus sign in the second term associated
with the fact that the time $t$ is now the lower limit of
integration rather than the upper as it was in
Eq.~(\ref{eq:bathEOMsolution}).

Defining
\be
\hbout(t) \equiv \frac{1}{\sqrt{2\pi\rho}}\sum_q
e^{-i\omegaq(t-t_1)}\hat{b}_q(t_1),
\label{eq:boutfuture}
\ee
we see that this is simply the free evolution of the bath modes
from the distant future (after they have interacted with the cavity)
back to the present, indicating that it is indeed appropriate to
interpret this as the outgoing field.  Proceeding as before we
obtain
\be
\dot{\hat{a}} = \frac{i}{\hbar}[\hat{H}_{\rm sys},\hat{a}] +
\frac{\kappa}{2} \hat{a} -\sqrt{\kappa}\, \hbout(t).
\label{eq:cavityEOM4}
\ee
Subtracting  Eq.~(\ref{eq:cavityEOM4}) from
Eq.~(\ref{eq:cavityEOM3}) yields
\be
\hbout(t) = \hbin(t) + \sqrt{\kappa}\, \hat{a}(t)
\label{eq:boutput}
\ee
which is consistent with our interpretation of the outgoing field
as the reflected incoming field plus the field radiated by the
cavity out through the partially reflecting mirror.

The above results are valid for any general cavity Hamiltonian.
The general procedure is to solve Eq.~(\ref{eq:cavityEOM3}) for
$\hat a(t)$ for a given input field, and then solve
Eq.~(\ref{eq:boutput}) to obtain the output field. For the case
of an empty cavity we can make further progress because the
cavity mode is a harmonic oscillator
\be
\hat{H}_{\rm sys} = \hbar \omegar \hat{a}^\dagger \hat{a}.
\ee
In this simple case, the cavity EOM becomes
\be
\dot{\hat{a}} = -i\omegar \hat{a} - \frac{\kappa}{2} \hat{a}
-\sqrt{\kappa}\, \hbin(t).
\label{eq:cavityEOM5}
\ee
Eq.~(\ref{eq:cavityEOM5}) can be solved by Fourier
transformation, yielding
\begin{eqnarray}
{\hat{a}}[\omega] &=& -\frac{\sqrt{\kappa}}{i(\omegar-\omega)
+ \kappa/2}\, {{\hat{b}}}_{\rm in }[\omega] \\
&=& -\sqrt{\kappa} \chiR[\omega-\omegar] \hbin[\omega]
\label{eq:asolution}
\end{eqnarray}
and
\be
{{\hat{b}}}_{\rm out}[\omega]=\frac{\omega-\omegar
-i\kappa/2}{\omega - \omegar +i\kappa/2}\, {{\hat{b}}}_{\rm in
}[\omega]
\label{eq:outputsolution}
\ee
which is the result for the reflection coefficient quoted in
Eq.~(\ref{eq:onesidedcavityreflection}). For brevity, here
and in the following, we will sometimes use the susceptibility
of the cavity, defined as
\be
	\chiR[\omega-\omegar] \equiv \frac{1}{-i(\omega-\omegar) +
	{\kappa / 2}}
	\label{eq:CavitySusceptibility}
\ee
For the case of steady  driving on resonance where
$\omega=\omegar$, the above equations yield
\be
{{\hat{b}}}_{\rm
out}[\omega]=\frac{\sqrt{\kappa}}{2}{\hat{a}}[\omega].
\ee
In steady state, the incoming power equals the outgoing power,
and both are related to the photon number inside the
single-sided cavity by
\be
P=\hbar \omega \left\langle \hbout^\dagger(t) \hbout(t)
\right\rangle = \hbar \omega \frac{\kappa}{4} \left\langle
{\hat a}^\dagger(t) {\hat a}(t) \right\rangle
\label{eq:D44}
\ee
Note that this does not coincide with the naive expectation,
which would be $P=\hbar \omega \kappa \left\langle {\hat
a}^{\dagger} {\hat a}\right\rangle$. The reason for this
discrepancy is the
 the interference between the part of the incoming
wave which is promptly reflected from the cavity and the field
radiated by the cavity. The naive expression becomes correct
after the drive has been switched off (where ignoring the
effect of the incoming vacuum noise, we would have $\hbout =
\sqrt{\kappa} {\hat a}$). We note in passing that for a driven
two-sided cavity with coupling constants $\kappa_L$ and
$\kappa_R$ (where $\kappa=\kappa_L+\kappa_R$), the incoming
power sent into the left port is related to the photon number
by
\be
P=\hbar \omega {\kappa^2 / ( 4 \kappa_L ) } \left\langle {\hat
a}^{\dagger} {\hat a}\right\rangle.
\ee
Here for $\kappa_L=\kappa_R$ the interference effect completely
eliminates the reflected beam and we have in contrast to
Eq.~(\ref{eq:D44})
\be
P=\hbar\omega\frac{\kappa}{2}\left\langle {\hat a}^{\dagger}
{\hat a}\right\rangle.
\ee

Eq.~(\ref{eq:cavityEOM5}) can also be solved in the time domain to
obtain
\begin{eqnarray}
\hat{a}(t) &=& e^{-(i\omegar+\kappa/2)(t-t_0)}\hat{a}(t_0)\nonumber\\
&-&\sqrt{\kappa}\int_{t_0}^td\tau\,
e^{-(i\omegar+\kappa/2)(t-\tau)}\hbin(\tau).
\label{eq:cavityEOMsolution1}
\end{eqnarray}

If we take the input field to be a coherent drive at frequency
$\omegal = \omegar+\Delta$ so that its amplitude has a
classical and a quantum part
\be
{\hat b}_{\rm in}(t) = e^{-i\omegal t}[\barbin + \hat\xi(t)]
\label{eq:barbinplusnoise}
\ee
and if we take the limit $t_0\rightarrow\infty$ so that the
initial transient in the cavity amplitude has damped out, then the
solution of Eq.~(\ref{eq:cavityEOMsolution1}) has the form
postulated in Eq.~(\ref{eq:coherent})
with
\be
\bar a = -\frac{\sqrt{\kappa}}{-i\Delta+\kappa/2}\barbin
\ee
and (in the frame rotating at the drive frequency)
\be
{\hat d}(t)= -\sqrt{\kappa} \int_{-\infty}^td\tau\,
e^{+(i\Delta-\kappa/2)(t-\tau)}\hat\xi(\tau).
\label{eq:hatdoft}
\ee

Even in the absence of any classical drive, the input field
delivers vacuum fluctuation noise to the cavity.  Notice that from
Eqs.~(\ref{eq:inputmode}, \ref{eq:barbinplusnoise})
\begin{eqnarray}
[\hbin(t),\hbin^\dagger(t')] &=&
[\hat\xi(t),\hat\xi^\dagger(t')]\nonumber\\
&=&\frac{1}{2\pi \rho} \sum_q 
	e^{-i (\omegaq -\omegal) (t-t')}\nonumber\\
&=& \delta(t-t'),
\label{eq:vacuumnoise1}
\end{eqnarray}
which is similar to Eq.~(\ref{eq:noisecommu}) for a quantum transmission line. 
This is the
operator equivalent of white noise. Using
Eq.~(\ref{eq:cavityEOMsolution1}) in the limit $t_0 \rightarrow
-\infty$ in Eqs.~(\ref{eq:coherent},\ref{eq:hatdoft}) yields
\begin{eqnarray}
[\hat{a}(t),\hat{a}^\dagger(t)] &=&
[\hat{d}(t),\hat{d}^\dagger(t)]\nonumber\\
&=& \kappa \int_{-\infty}^t d\tau \int_{-\infty}^t d\tau'\,
e^{-(-i\Delta+\kappa/2)(t-\tau)}\nonumber\\
&&e^{-(+i\Delta+\kappa/2)(t-\tau')}\delta(\tau-\tau')\nonumber\\
&=& 1
\end{eqnarray}
as is required for the cavity bosonic quantum degree of freedom.
We can interpret this as saying that the cavity zero-point
fluctuations arise from the vacuum noise that enters through the
open port.  We also now have a simple physical interpretation of
the quantum noise in the number of photons in the driven cavity in
Eqs.~(\ref{eq:Snoise},\ref{eq:Snnt},\ref{eq:Snnomega}). It is due
to the vacuum noise which enters the cavity through the same ports
that bring in the classical drive.  The interference between the
vacuum noise and the classical drive leads to the photon number
fluctuations in the cavity.

In thermal equilibrium, ${\hat \xi}$ also contains thermal radiation.  If the bath is
being probed only over a narrow range of frequencies centered on
$\omegar$ (which we have assumed in making the Markov
approximation) then we have to a good
approximation (consistent with the above commutation relation)
\begin{eqnarray}
\langle\hat\xi^\dagger(t)\hat\xi(t')\rangle&=&N\delta(t-t')\label{eq:xinoise1}\\
\langle\hat\xi(t)\hat\xi^\dagger(t')\rangle&=&(N+1)\delta(t-t')\label{eq:xinoise2}
\end{eqnarray}
where $N=\nb(\hbar\omegar)$ is the thermal equilibrium
occupation number of the mode at the frequency of interest. We
can gain a better understanding of Eq.~(\ref{eq:xinoise1}) by
Fourier transforming it to obtain the spectral density
\be
S[\omega] = \int_{-\infty}^{+\infty}
dt\,\langle\hat\xi^\dagger(t)\hat\xi(t')\rangle e^{i \omega (t-t')} = N.
\ee
As mentioned previously, this dimensionless quantity is the
spectral density that would be measured by a photomultiplier:
it represents the number
of thermal photons passing a given point per unit time per unit
bandwidth. Equivalently the thermally radiated power in a
narrow bandwidth $B$ is
\be
P=\hbar\omega N B.
\ee
One often hears the confusing statement that the noise added by
an amplifier is a certain number $N$ of photons ($N=20$, say
for a good cryogenic HEMT amplifier operating at 5 GHz).  This
means that the excess output noise (referred back to the input by dividing by the power gain) 
produces a flux of N photons per second in a 1 Hz bandwidth, or $10^6 N$ photons per second in 1
MHz of bandwidth (see also Eq.~(\ref{eq:PhotonFluxXB})).



We can gain further insight into input-output theory by using
the following picture.  The operator $\hbin(t)$ represents the
classical drive plus vacuum fluctuations which are just about
to arrive at the cavity.  We will be able to show that the
output field is simply the input field a short while later
after it has interacted with the cavity. Let us consider the
time evolution over a short time period $\Delta t$ which is
very long compared to the inverse bandwidth of the vacuum noise
(i.e., the frequency scale beyond which the vacuum noise cannot
be treated as constant due to some property of the environment)
but very short compared to the cavity system's slow dynamics.
In this circumstance it is useful to introduce the quantum
Wiener increment related to Eq.~(\ref{eq:C13})
\be
\dW \equiv \int_t^{t+\Delta t}d\tau\, \hat\xi(\tau)
\ee
which obeys
\be
[\dW,\dWd]=\Delta t.
\ee

In the interaction picture (in a displaced frame in which the
classical drive has been removed) the Hamiltonian term that
couples the cavity to the quantum noise of the environment is
from Eq.~(\ref{eq:couplingHam})
\be
\hat V=-i\hbar\sqrt{\kappa} (\hat a^\dagger\hat \xi - \hat
a\hat \xi^\dagger).
\ee
Thus the time evolution operator (in the interaction picture)
on the $j$th short time interval $[t_j,t_j+\Delta t]$ is
\be
\hat U_j=e^{\sqrt{\kappa}(\hat a\,\dWd - \hat a^\dagger\,\dW)}
\ee
Using this we can readily evolve the incoming temporal mode
forward in time by a small step $\Delta t$
\be
\dW' = \hat U^\dagger \dW\hat U \approx \dW+\sqrt{\kappa}\Delta
t\, \ha.
\label{eq:outfieldHeis}
\ee

Recall that in input-output theory we formally defined the
outgoing field as the bath field far in the future propagated
back (using the free field time evolution) to the present,
which yielded
\be
\hbout = \hbin + \sqrt{\kappa} \ha.
\ee
Eq.~(\ref{eq:outfieldHeis}) is completely equivalent to this.
Thus we confirm our understanding that the incoming field is
the bath temporal mode just before it interacts with the cavity
and the outgoing field is the bath temporal mode just after it
interacts with the cavity.

This leads to the following picture which is especially useful
in the quantum trajectory approach to conditional quantum
evolution of a system subject to weak continuous measurement
\cite{GardinerParkinsZoller92,Walls94}.
On top of the classical drive ${\bar b}_{\rm in}(t)$, the bath
supplies to the system a continuous stream of ``fresh''
harmonic oscillators, each in their ground state (if $T=0$).
Each oscillator with its quantum fluctuation $\dW$ interacts
briefly for a period $\Delta t$ with the system and then is
disconnected to propagate freely thereafter, never interacting
with the system again.  Within this picture it is useful to
think of the oscillators arrayed in an infinite stationary line
and the cavity flying over them at speed $\vp$ and touching
each one for a time $\Delta t$.

\subsection{Quantum limited position measurement using a cavity detector}
\label{subapp:CavityPositionDetector}
We will now apply the input-output formalism introduced in the
previous section to the important example of a dispersive
position measurement, which employs a cavity whose resonance
frequency shifts in response to the motion of a harmonic
oscillator.  This physical system was considered heuristically
in Sec.~\ref{subsec:measurementofoscillatorposition}.  Here
we will present a rigorous derivation using the (linearized)
equations of motion for the coupled cavity and oscillator
system.

Let the dimensionless position operator
\be
\hat z=\frac{1}{x_{\rm ZPF}}\hx =[\hat c^\dagger + \hat c]
\label{eq:xdefn}
\ee
be the coordinate of a harmonic oscillator whose energy is
\be
H_{\rm M} = \hbar\omega_{\rm M} {\hat c}^\dagger {\hat c}
\ee
and whose position uncertainty in the quantum ground state is
$x_{\rm ZPF}=\sqrt{\langle 0|\hat x^2|0\rangle}$.

This Hamiltonian could be realized for example by mounting one of the
cavity mirrors on a flexible cantilever (see the discussion above).

When the mirror moves, the cavity resonance frequency shifts,
\be
\tomegar = \omegar [1+A\hz(t)]
\ee
where for a cavity of length $L$, $A=-x_{\rm ZPF}/L$.

 Assuming that the mirror moves slowly enough for the
cavity to adiabatically follow its motion (i.e.
$\omegam\ll\kappa$), the outgoing light field suffers a phase
shift which follows the changes in the mirror position.  This
phase shift can be detected in the appropriate homodyne set up
as discussed in Sec.~\ref{subsec:CavityDetector}, and from
this phase shift we can determine the position of the
mechanical oscillator. In addition to the actual zero-point
fluctuations of the oscillator, our measurement will suffer
from shot noise in the homodyne signal and from additional
uncertainty due to the back action noise of the measurement
acting on the oscillator.  All of these effects will appear
naturally in the derivation below.

 We begin by considering the optical cavity equation of motion
 based on Eq.~(\ref{eq:cavityEOM3}) and the optomechanical
 coupling Hamiltonian in Eq.~(\ref{eq:paramcoupling}).  These
 yield
\be
\dot\ha = -i \omegar(1 +A\hz)\ha -\frac{\kappa}{2}\ha
-\sqrt{\kappa}\hbin.
\ee
Let the cavity be driven by a laser at a frequency  $\omegal
=\omegar+\Delta$ detuned from the cavity by $\Delta$. Moving to
a frame rotating at $\omegal$ we have
\be
\dot\ha=+i(\Delta-A \omegar \hz)\ha -
\frac{\kappa}{2}\ha-\sqrt{\kappa}\hbin.
\ee
and we can write the incoming field as a constant plus white
noise vacuum fluctuations (again, in the rotating frame)
\be
\hbin= 
	{\bar b}_{\rm in} +\hat\xi 
\ee
and similarly for the cavity field following
Eq.~(\ref{eq:coherent})
\be
\ha = 
	\bar a + \hd. 
\label{eq:newcoherent} 
\ee
Substituting these expressions into the equation of motion, we
find that the constant classical fields obey
\be
\bar a = -\frac{\sqrt{\kappa}}{\kappa/2 - i\Delta}{\bar b}_{\rm
in}
\ee
and the new quantum equation of motion is, after neglecting a small term
${\hat d}{\hat z}$: 
\be
\dot\hd = +i\Delta\hd -iA\omegar\bar a \hz -\frac{\kappa}{2}\hd
-\sqrt{\kappa}\hxi.
\ee
The quantum limit for position measurement will be reached only
at zero detuning, so we specialize to the case $\Delta=0$.  We
also choose the incoming field amplitude and phase to obey
\be
\barbin =-i\sqrt{\Ndotbar},
\ee
so that
\be
\bar a = +2i\sqrt{\frac{\Ndotbar}{\kappa}},
\ee
 where $\Ndotbar$ is the incoming photon number flux.
The quantum equation of motion for the cavity then becomes
\be
\dot\hd = + g \hz -\frac{\kappa}{2}\hd -\sqrt{\kappa}\hxi,
\label{eq:cavquantflucEOM}
\ee
where the opto-mechanical coupling constant is proportional to
the laser drive amplitude
\be
g\equiv
2A\omegar\sqrt{\frac{\Ndotbar}{\kappa}}=A\omegar\sqrt{\bar n}.
\ee
and
\be
\bar n=|\bar a|^2=4\frac{\Ndotbar}{\kappa}
\ee
 is the mean cavity
photon number. Eq.~(\ref{eq:cavquantflucEOM}) is easily solved
by Fourier transformation
\be
\hd[\omega] = \frac{1}{[\kappa/2 -i\omega]}\left\{
g\hz[\omega]-\sqrt{\kappa}\hxi[\omega] \right\}.
\ee

Let us assume that we are in the limit of low mechanical
frequency relative to the cavity damping, $\omegam \ll \kappa$,
so that the cavity state adiabatically follows the motion of
the mechanical oscillator. Then we obtain to a good
approximation
\begin{eqnarray}
\hd[\omega] &=& \frac{2}{\kappa}\left\{
g\hz[\omega]-\sqrt{\kappa}\hxi[\omega] \right\}
\label{eq:ddot}\\
\hddag[\omega] &=& \frac{2}{\kappa}\left\{
g\hz[\omega]-\sqrt{\kappa}\hxidag[\omega] \right\}
\label{eq:ddagdot}
\end{eqnarray}

The mechanical oscillator equation of motion which is identical
in form to that of the optical cavity
\be
\partial_t \hc = -[\frac{\gamma_0}{2} + i\omegam]\hc -
\sqrt{\gamma_0}{\hat \eta}(t) + \frac{i}{\hbar}[\hat H_{\rm
int},\hc(t)],
\label{eq:OscillatorEOM}
\ee
where $\hat H_{\rm int}$ is the Hamiltonian in
Eq.~(\ref{eq:paramcoupling}) and ${\hat \eta}$ is the
mechanical vacuum noise from the (zero temperature) bath which
is causing the mechanical damping at rate $\gamma_0$. Using
Eq.~(\ref{eq:newcoherent}) 
and expanding to first order in small
fluctuations yields the equation of motion linearized about the
steady state solution
\be
\partial_t \hc = -[\frac{\gamma_0}{2} + i\omegam]\hc -
\sqrt{\gamma_0}{\hat \eta}(t)
+2\frac{g}{\sqrt\kappa}[\hat\xi(t)-\hat\xi^\dagger(t)].
\label{eq:partialtofc}
\ee

It is useful to consider an equivalent formulation in which we
expand the Hamiltonian in Eq.~(\ref{eq:paramcoupling}) to
second order in the quantum fluctuations about the classical
solution
\be
{\hat H}_{\rm int} \approx \hbar\omegar \hddag\hd + \hx\hF ,
\label{eq:paramcouplinglinearized}
\ee
where the force (including the coupling $A$)  is (up to a sign) 
\be
\hF = -i \frac{\hbar g}{\xrms}[\hd -\hddag].
\ee

Note that the radiation pressure fluctuations (photon shot
noise) inside the cavity provide a forcing term.  The state of
the field inside the cavity in general depends on the past
history of the cantilever position.  However for this special
case of driving the cavity on resonance, the dependence of the
cavity field on the cantilever history is such that the latter
drops out of the radiation pressure.  To see this explicitly,
consider the equation of motion for the force obtained from
Eq.~(\ref{eq:cavquantflucEOM})
\be
\dot\hF = -\frac{\kappa}{2} \hF +i\frac{\hbar
g}{\xrms}\sqrt{\kappa}[\hxi-\hdxi].
\label{eq:forcedot}
\ee
Within our linearization approximation, the position of the
mechanical oscillator has no effect on the radiation pressure
(photon number in the cavity), but of course it does affect the
\emph{phase} of the cavity field (and hence the outgoing field) which
is what we measure in the homodyne detection.

\begin{figure}[t]
\begin{center}
\includegraphics[width=3.45in]{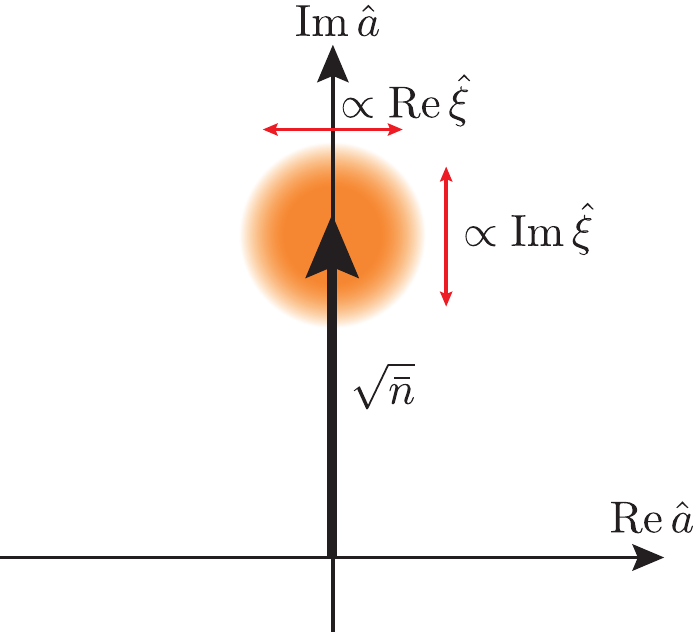}
\caption{(Color online) Phasor diagram for the cavity amplitude showing that (for our choice of parameters) the
imaginary quadrature of the vacuum noise ${\hat \xi}$ interferes with the classical drive to produce
photon number fluctuations while the real quadrature produces phase fluctuations which lead to measurement imprecision. 
The quantum fluctuations are illustrated in the usual fashion, depicting the Gaussian Wigner density of the coherent state in terms of color intensity.}
\label{fig:cavityphasordiagram}
\end{center}
\end{figure}

Thus for this special case $\hat z$ does not appear on the RHS
of either Eq.~(\ref{eq:forcedot}) or
Eq.~(\ref{eq:partialtofc}), which means that there is no
optical renormalization of the cantilever frequency (`optical
spring') or optical damping of the cantilever. The lack of
back-action damping in turn implies that the effective
temperature $T_{{\rm eff}}$ of the cavity detector is infinite
(cf.~Eq.~(\ref{OscTeff})).  For this special case of zero
detuning the back action force noise is controlled by a single
quadrature of the incoming vacuum noise (which interferes with
the classical drive to produce photon number fluctuations).
This is illustrated in the cavity amplitude phasor diagram of
Fig.~(\ref{fig:cavityphasordiagram}).  We see that the vacuum
noise quadrature ${\hat \xi}+{\hat \xi}^{\dagger}$ conjugate to $\hF$ controls the phase noise
which determines the measurement imprecision (shot noise in the
homodyne signal).  This will be discussed further below.

The solution for the cantilever position can again be obtained
by Fourier transformation.  For frequencies small on the scale
of $\kappa$ the solution of Eq.~(\ref{eq:forcedot}) is
\be
\hF[\omega] = \frac{2i\hbar
g}{\xrms\sqrt{\kappa}}\left\{\hxi[\omega]-\hdxi[\omega]\right\}
\ee
and hence the back action force noise spectral density is at
low frequencies
\be
S_{FF}[\omega] =  \frac{4\hbar^2 g^2}{\xrms^2\kappa}
\label{eq:backactionforceinagreement}
\ee
in agreement with Eq.~(\ref{eq:1sidedcavityforcenoise}).

Introducing a quantity proportional to the cantilever
(mechanical) susceptibility (within the rotating wave
approximation we are using)
\be
	\chiM[\omega-\omegam] \equiv \frac{1}{-i(\omega-\omegam) +
	\frac{\gamma_0}{2}},
	\label{eq:MechanicalSusceptibility}
\ee
we find from Eq.~(\ref{eq:partialtofc})
\be
\hz[\omega]=\hz_0[\omega] -\frac{i}{\hbar}\xrms\left\{
\chiM[\omega-\omegam]-\chiM[\omega+\omegam]
\right\}\hF[\omega],
\label{eq:xsolution}
\ee
where the equilibrium fluctuations in position are given by
\be
\hz_0[\omega]\equiv -\sqrt{\gamma_0}\left\{
\chiM[\omega-\omegam] {{\hat\eta}}[\omega]
+\chiM[\omega+\omegam] {{\hat\eta}}^\dagger[\omega]\right\}.
\label{eq:x0solution}
\ee

We can now obtain the power spectrum $S_{zz}$ describing the
total position fluctuations of the cantilever driven by the
mechanical vacuum noise plus the radiation pressure shot noise.
From Eqs.~(\ref{eq:xsolution}, \ref{eq:x0solution}) we find
\begin{eqnarray}
\frac{S_{xx}[\omega]}{\xrms^2} &=& S_{zz}[\omega] \nonumber  \\
&=& \gamma_0 |\chi[\omega-\omegam]|^2
 \\
&&
+
\frac{\xrms^2}{\hbar^2}\left|\chiM[\omega-\omegam] - \chiM[\omega+\omegam]\right|^2
S_{FF}. 
\nonumber
\end{eqnarray}
Note that (assuming high mechanical $Q$, i.e. $\gamma_0\ll
\omegam$) the equilibrium part has support only at positive
frequencies while the back action induced position noise is
symmetric in frequency reflecting the effective infinite
temperature of the back action noise. Symmetrizing this result
with respect to frequency (and using $\gamma_0\ll \omegam$) we
have
\be
{\bar S}_{xx}[\omega] \approx \Sxxsym^0[\omega]\left(1 +
\frac{\Sxxsym^0[\omegam]}{\hbar^2}{\bar S}_{FF}\right),
%
%
\ee
where $\Sxxsym^0[\omega]$ is the symmetrized spectral density
for position fluctuations in the ground state given by
Eq.~(\ref{eq:zeroTsymmspecden}).

Now that we have obtained the effect of the back action noise
on the position fluctuations, we must turn our attention to the
imprecision of the measurement due to shot noise in the output.
The appropriate homodyne quadrature variable to monitor to be
sensitive to the output phase shift caused by position
fluctuations is
\be
\hI = \hbout + \hdbout,
\ee
which, using the input-output results above, can be written
\be
\hI = -(\hxi+\hdxi)+\lambda \hx.
\label{eq:outputsignalI}
\ee
We see that the cavity homodyne detector system acts as a
position transducer with gain
\be
\lambda = \frac{4g}{\xrms\sqrt{\kappa}}.
\label{eq:lambdagaincoeff}
\ee
The first term in Eq.~(\ref{eq:outputsignalI}) represents the
vacuum noise that mixes with the homodyne local oscillator to
produce the shot noise in the output.  The resulting
measurement imprecision (symmetrized) spectral density referred
back to the position of the oscillator is
\be
{\bar S}^{\rm I}_{xx} = \frac{1}{\lambda^2}.
\ee
Comparing this to Eq.~(\ref{eq:backactionforceinagreement}) we
see that we reach the quantum limit relating the imprecision
noise to the back action noise
\be
{\bar S}^{\rm I}_{xx}{\bar S}_{FF}=\frac{\hbar^2}{4}
\ee
in agreement with Eq.~(\ref{eq:SFFSIxxproduct}).

 Notice also from 
Eq.~(\ref{eq:outputsignalI}) that the quadrature of the vacuum
noise which leads to the measurement imprecision is conjugate
to the one which produces the back action force noise as
illustrated previously in Fig.~(\ref{fig:cavityphasordiagram}).
Recall that the two
quadratures of motion of a harmonic oscillator in its ground
state have no classical (i.e., symmetrized) correlation. Hence
 the symmetrized cross correlator
\be
{\bar S}_{IF}[\omega]=0
\ee
vanishes. Because there is no correlation between the output
imprecision noise and the forces controlling the position
fluctuations, the total output noise referred back to the
position of the oscillator is simply
\begin{eqnarray}
\bSxtot[\omega] &=& {\bar S}_{xx}[\omega] + {\bar S}^{\rm I}_{xx}\\
&=&
\Sxxsym^0[\omega]\left(1 +
\frac{\Sxxsym^0[\omegam]}{\hbar^2}{\bar S}_{FF}\right)+\frac{\hbar^2}{4{\bar S}_{FF}}.\nonumber
\label{eq:totaloutnoiserefback}
\end{eqnarray}
This expression again clearly illustrates the competition
between the back action noise proportional to the drive laser
intensity and the measurement imprecision noise which is
inversely proportional. We again emphasize that all of the
above relations are particular to the case of zero detuning of
the cavity drive field from the cavity.

The total output noise at some particular frequency will be a
minimum at some optimal drive intensity.  The precise optimal
value depends on the frequency chosen.  Typically this is taken
to be the mechanical resonance frequency where we find that the
optimal coupling leads to an optimal back action noise
\be
{\bar S}_{FF, \rm
opt}=\frac{\hbar^2}{2\Sxxsym^0[\omegam]}=\frac{\hbar^2\gammam}{4\xrms^2}.
\ee
This makes sense because the higher the damping the less
susceptible the oscillator is to back action forces.  At this
optimal coupling the total output noise spectral density at
frequency $\omegam$ referred to the position  is simply twice
the vacuum value
\be
\bSxtot[\omegam]=2\Sxxsym^0[\omegam],
\ee
in agreement with Eq.~(\ref{eq:twicevacnoise}).  Evaluation of
Eq.~(\ref{eq:totaloutnoiserefback}) at the optimal coupling
yields the graph shown in Fig.~(\ref{fig:zeroTSHOlineshape}).
The background noise floor is due to the frequency independent
imprecision noise with value $\frac{1}{2}\Sxxsym^0[\omegam]$.
The peak value at $\omega=\omegam$ rises a factor of three
above this background.

We derived the gain $\lambda$ in Eq.~(\ref{eq:lambdagaincoeff})
by direct solution of the equations of motion.  With the
results we have derived above, it is straightforward to show
that the Kubo formula in Eq.~(\ref{eq:gain}) yields equivalent
results.  We have already seen that the classical (i.e.
symmetrized) correlations between the output signal $\hI$ and
the force $\hF$ which couples to the position vanishes. However
the Kubo formula evaluates the \emph{quantum} (i.e.
antisymmetric) correlations for the uncoupled system ($A=g=0$).
Hence we have
\begin{widetext}
\be
\chi_{IF}(t) = -\frac{i}{\hbar}\theta(t)\left\langle
\left[-\big(\hxi(t-\delta t) +\hdxi(t-\delta
t)\big),\frac{2i\hbar g}{\xrms
\sqrt{\kappa}}\big(\hxi(0)-\hdxi(0)\big) \right]
\right\rangle_{0},
\ee
\end{widetext}
where $\delta t$ is a small (positive) time representing the
delay between the time when the vacuum noise impinges on the
cavity and when the resulting outgoing wave reaches the
homodyne detector.  (More precisely it also compensates for
certain small retardation effects neglected in the limit
$\omega\ll\kappa$ used in several places in the above
derivations.)  Using the fact that the commutator between the
two quadratures of the vacuum noise is a delta function,
Fourier transformation of the above yields (in the limit
$\omega\,\delta t\ll 1$ the desired result
\be
\chi_{IF}[\omega]=\lambda.
\ee
Similarly we readily find that the small retardation causes the
reverse gain to vanish.  Hence all our results are consistent
with the requirements needed to reach the standard quantum
limit.

Thus with this study of the specific case of an oscillator
parametrically coupled to a cavity, we have reproduced all of
the key results in Sec.~\ref{subsec:PositionDetector} derived
from completely general considerations of linear response
theory.


\subsection{Back-action free single-quadrature detection}
\label{subapp:SingleQuadrature}

We now provide details on the cavity single-quadrature detection
scheme discussed in Sec.~\ref{subsubsec:DoubleSideband}.  We again
consider a high-$Q$ cavity whose resonance frequency is modulated by
a high-$Q$ mechanical oscillator with co-ordinate $\hat{x}$
(cf.~Eqs.~(\ref{eq:paramcoupling}) and (\ref{eq:xdefn})).  To use
this system for amplification of a single quadrature, we will
consider the typical case of a fast cavity ($\omegar \gg \Omega$), and
take the ``good cavity" limit, where $\Omega \gg \kappa$.  As explained in
the main text, the crucial ingredient for single-quadrature detection is to
take an amplitude-modulated cavity drive 
described by the classical input field $\bar{b}_{\rm in}$ given in
Eq.~(\ref{eq:DSBDrive}).
As before (cf.~Eq.~(\ref{eq:coherent})), we may write the cavity
annihilation operator $\hat{a}$ as the sum of a classical piece
$\bar{a}(t)$ and a quantum piece $\hat{d}$; only $\hat{d}$ is influenced by the
mechanical oscillator. $\bar{a}(t)$ is easily found from the
classical (noise-free) equations of motion for the isolated
cavity; making use of the conditions $\omegar \gg \Omega \gg
\kappa$, we have
\begin{eqnarray}
    \bar{a}(t) & \simeq &
        \frac{\sqrt{ \dot{N} \kappa}}{2 \Omega}  \cos \left( \Omega t
            + \delta \right) e^{-i \omegar t}
\end{eqnarray}


To proceed with our analysis, we work in an interaction picture
with respect to the uncoupled cavity and oscillator Hamiltonians.
Making standard rotating-wave approximations, the Hamiltonian in
the interaction picture takes the simple form corresponding to Eq.~(\ref{eq:DSBVeff2}):
\begin{eqnarray}
    H_{\rm int} & = &
        \hbar \tilde{A} 
            \left(\hat{d} + \hat{d}^\dag \right)
        		\left(
            		e^{i \delta }\hat{c}  + e^{-i \delta }  \hat{c}^\dag
            	\right)
            \nonumber \\
     & = &
        \hbar \tilde{A} \left( \hat{d} + \hat{d}^{\dag} \right)
        \frac{ \hat{X}_\delta}{ \xrms},
\end{eqnarray}
where 
\begin{eqnarray}
	\tilde{A} & = & A \cdot \omegar  
		\frac{ \sqrt{\dot{N} \kappa} }{ 4 \Omega},
\end{eqnarray}   
and in the second line, we have made use of the definition of the quadrature operators 
$\hat{X}_{\delta}, \hat{Y}_\delta$  given in Eqs.~(\ref{eqs:ExplicitQuadratures}).  The form of $H_{\rm int}$ was discussed heuristically in the main text in terms 
of Raman processes where photons are removed from the classical drive $\bar{b}_{\rm in}$ and either up or down converted to the cavity frequency via absorption or emission of a mechanical phonon.  Alternatively, we can think of the drive yielding a time-dependent cavity-oscillator coupling which ``follows" the $X_\delta$ quadrature.  
Note that we made crucial of use of the good cavity limit ($\kappa \ll \Omega$)
to drop terms in ${\hat H}_{int}$ which oscillate at frequencies $\pm 2 \Omega$.
These terms represent Raman sidebands which are away from the cavity resonance
by a distance $\pm 2\Omega$.  In the good cavity limit, the
density of photon states is negligible so far off resonance and
these processes are suppressed.

Similar to Eqs.~(\ref{eq:cavityEOM5}) and
(\ref{eq:OscillatorEOM}), the Heisenberg equations of motion (in
the rotating frame) follow directly from $H_{\rm int}$ and the
dissipative terms in the total Hamiltonian:
\begin{subequations}
\begin{eqnarray}
    \partial_t \hat{d} & = &
        - \frac{\kappa}{2} \hat{d} - \sqrt{\kappa} \hat{\xi}(t) e^{i \omegar t}
        - i \tilde{A} \left( e^{i \delta} \hc + e^{-i \delta} \hc^{\dagger} \right)\label{eq:onequadriving} \\
    \partial_t \hat{c} & = &
        - \frac{\gamma_0}{2} \hat{c} - \sqrt{\gamma_0} \hat{\eta}(t) e^{i \Omega t}
        - i e^{-i \delta} \left[
          \tilde{A} \left( \hd + \hd^{\dagger} \right) - f(t)\right]
        \nonumber \\
\end{eqnarray}
\end{subequations}
As before, $\hat{\xi}(t)$ represents the unavoidable noise in
the cavity drive, and $\hat{\eta}(t)$, $\gamma_0$ are the noisy
force and damping resulting from an equilibrium bath coupled to
the mechanical oscillator.  Note from
Eq.~(\ref{eq:onequadriving}) that as anticipated, the cavity is
only driven by one quadrature of the oscillator's motion. We
have also included a driving force $F(t)$ on the mechanical
oscillator which has some narrow bandwidth centered on the
oscillator frequency; this force is parameterized as:
\begin{eqnarray}
    F(t) = \frac{2 \hbar}{\xrms} \textrm{Re} \left[ 
    	f(t) e^{-i \Omega t} e^{- i \delta}
	\right]
\end{eqnarray}
where $f(t)$ is a complex function which is slowly varying on the
scale of an oscillator period.


The equations of motion are easily solved upon Fourier
transformation, resulting in:
\begin{subequations}
\begin{eqnarray}
     \hat{X}_{\delta }[\omega] & = &
            - \xrms \cdot \chi_M[\omega] 
             \Bigg[
                i \left( f^*[-\omega] - f[\omega] \right)
            \label{eq:xslowanswer} \\
        && + \sqrt{\gamma_0}
                 \left(  e^{i \delta} \hat{\eta}(\omega+\Omega)
             +  e^{- i \delta} \hat{\eta}^{\dag}(\omega-\Omega) \right)
            \Bigg]
            \nonumber \\
    \hat{Y}_{ \delta }[\omega] & = &
        i  \xrms \cdot \chi_M[\omega]  \Bigg[
            (-i) \left( f[\omega] + f^*[-\omega] \right)
            \label{eq:pslow}     \\
        &&
            +  \sqrt{\gamma_0}  \left(
             e^{i \delta} \hat{\eta}(\omega+\Omega)
        - e^{-i \delta} \hat{\eta}^{\dag}(\omega-\Omega) \right)
            \nonumber \\
        &&
            -2 i \tilde{A} \chiR[\omega] \sqrt{\kappa}
              \left(
             \hat{\xi}(\omega+\omegar)
            +  \hat{\xi}^{\dagger}(\omega-\omegar)
        \right)
           \Bigg]      \nonumber
\end{eqnarray}
\end{subequations}
where the cavity and mechanical susceptibilities $\chiR, \chiM$ are defined in 
Eqs.~(\ref{eq:CavitySusceptibility}) and (\ref{eq:MechanicalSusceptibility}).


As anticipated, the detected quadrature $\hat{X}_\delta$ is {\it
completely} unaffected by the measurement:
Eq.~(\ref{eq:xslowanswer}) is identical to what we would have if
there were no coupling between the oscillator and the cavity. In
contrast, the conjugate quadrature $\hat{Y}_{\delta}$ experiences
an extra stochastic force due to the cavity: this is the
measurement back-action.

Turning now to the output field from the cavity
$\hat{b}_{\textrm{out}}$, we use the input-output relation
Eq.~(\ref{eq:boutput}) to find in the {\it lab} (i.e.~non-rotating) frame:
\begin{eqnarray}
    \hbout[\omega] & = &
        \bar{b}_{\rm out}[\omega]+
        \left[
            \frac{ -i (\omega-\omegar) - \kappa/2}
            { -i (\omega-\omegar) + \kappa/2} \right] \hat{\xi}[\omega]
    \nonumber \\
    &&
        - i 
        \frac{\tilde{A} \sqrt{\kappa} }{\xrms}
         \chiR[\omega-\omegar] \cdot
        \hat{X}_{\delta}(\omega-\omegar)
                    \nonumber \\ \label{eq:DSBbout}
\end{eqnarray}
The first term on the RHS simply represents the output field from
the cavity in the absence of the mechanical oscillator and any
fluctuations.  It will yield sharp peaks at the two sidebands
associated with the drive, $\omega = \omegar \pm \Omega$.
The second term on the RHS of Eq.~(\ref{eq:DSBbout}) represents
the reflected noise of the incident cavity drive.  This noise will
play the role of the ``intrinsic ouput noise" of this amplifier.

Finally, the last term on the RHS of Eq.~(\ref{eq:DSBbout}) is the
amplified signal: it is simply the amplified quadrature
$\hat{X}_\delta$ of the oscillator.  This term will result in a
peak in the output spectrum at the resonance frequency of the
cavity, $\omegar$.  As there is no back-action on the measured
$\hat{X}_\delta$ quadrature, the added noise can be made
arbitrarily small by simply increasing the drive strength $\dot{N}$ (and
hence $\tilde{A}$).



\section{Information Theory and Measurement Rate}
\label{app:measurementrateinformationtheory}

Suppose that we are measuring the state of a qubit via the phase
shift $\pm\theta_0$ from a one-sided cavity.  Let $I(t)$ be the
homodyne signal integrated up to time $t$ as in
Sec.~\ref{subsec:CavityDetector}.  We would
like to understand the relationship between the signal-to-noise
ratio defined in Eq.~(\ref{eq:cavitySNR}), and the rate at which
information about the state of the qubit is being gained. The
probability distribution for $I$ conditioned on the state of the
qubit $\sigma=\pm 1$ is
\be
p(I|\sigma) = \frac{1}{\sqrt{2\pi \Sthetatheta t}} \exp\left[
\frac{- (I-\sigma\theta_0 t)^2}{2\Sthetatheta t}   \right].
\label{eq:conditionalprob}
\ee
Based on knowledge of this conditional distribution, we now
present two distinct but equivalent approaches to giving an
information theoretic basis for the definition of the measurement
rate.

\subsection{Method I}
\label{subsec:methodI}

Suppose we start with an initial qubit density matrix
\be
\rho_0 = \left(
\begin{array}{cc}
\frac{1}{2}&0\\
0&\frac{1}{2}
\end{array}
\right).
\ee
After measuring for a time $t$, the new density matrix conditioned
on the results of the measurement is
\be
\rho_1 = \left(
\begin{array}{cc}
p_+&0\\
0&p_-
\end{array}
\right)
\ee
where it will be convenient to parameterize the two probabilities by
the polarization $m\equiv {\rm Tr} (\sigma_z \rho_1)$ by
\be
\p_\pm = \frac{1\pm m}{2}.
\ee

The information gained by the measurement is the entropy
loss\footnote{It is important to note that we use throughout here
the physicist's entropy with the natural logarithm rather than the
log base 2 which gives the information in units of bits.} of the
qubit
\be
{\cal I} = {\rm Tr}(\rho_1\ln\rho_1 - \rho_0\ln\rho_0).
\ee
We are interested in the initial rate of gain of information at
short times $\theta_0^2 t \ll \Sthetatheta$ where $m$ will be
small.  In this limit we have
\be
{\cal I}\approx \frac{m^2}{2}.
\ee
We must now calculate $m$ conditioned on the measurement result
$I$
\be
m_I \equiv \sum_\sigma \sigma p(\sigma|I).
\ee

From Bayes theorem we can express this in terms of $p(I|\sigma)$,
which is the quantity we know,
\be
p(\sigma|I)=\frac{p(I|\sigma)p(\sigma)}{\sum_{\sigma'}p(I|\sigma')p(\sigma')}.
\ee
Using Eq.~(\ref{eq:conditionalprob}) the polarization is easily
evaluated
\be
m_I = \tanh\left( \frac{I\theta_0}{\Sthetatheta} \right).
\ee
The information gain is thus
\be
{\cal I}_I = \frac{1}{2}\tanh^2\left(
\frac{I\theta_0}{\Sthetatheta}
\right)\approx\frac{I^2}{2}\left(\frac{\theta_0}{\Sthetatheta}\right)^2
\ee
where the second equality is only valid for small $|m|$.  Ensemble
averaging this over all possible measurement results yields the
mean information gain at short times
\be
{\cal I} \approx \frac{1}{2}\frac{\theta_0^2}{\Sthetatheta} t
\label{eq:infogain}
\ee
 which justifies the definition of
the measurement rate given in
Eq.~(\ref{eq:cavitymeasurementrate}).

\subsection{Method II}
\label{subsec:methodII}

An alternative information theoretic derivation is to consider the
qubit plus measurement device to be a signaling channel. The two
possible inputs to the channel are the two states of the qubit.
The output of the channel is the result of the measurement of $I$.
By toggling the qubit state back and forth, one can send
information through the signal channel to another party.  The
channel is noisy because even for a fixed state of the qubit, the
measured values of the signal $I$ have intrinsic fluctuations.
Shannon's noisy channel coding theorem \cite{CoverandThomas} tells
us the maximum rate at which information can be reliably sent down
the channel by toggling the state of the qubit and making
measurements of $I$.  It is natural to take this rate as defining
the measurement rate for our detector.

The reliable information gain by the receiver on a noisy channel
is a quantity known as the `mutual information' of the
communication channel \cite{Clerk03, CoverandThomas}
\be
R = -\int_{-\infty}^{+\infty}dI \left\{ p(I)\ln p(I)-\sum_\sigma
p(\sigma) \left[p(I|\sigma)\ln p(I|\sigma)\right] \right\}
\ee
The first term is the Shannon entropy in the signal $I$ when we do
not know the input signal (the value of the qubit).  The second
term represents the entropy given that we do know the value of the
qubit (averaged over the two possible input values).  Thus the
first term is signal plus noise, the second is just the noise.
Subtracting the two gives the net information gain. Expanding this
expression for short times yields
\begin{eqnarray}
R &=& \frac{1}{8}\frac{(\langle I(t)\rangle_+ - \langle
I(t)\rangle_-)^2}{\Sthetatheta t}\nonumber\\
&=& \frac{\theta_0^2}{2\Sthetatheta}t\nonumber\\
&=& \Gamma_{\rm meas} t
\end{eqnarray}
exactly the same result as Eq.~(\ref{eq:infogain}).  (Here
$\langle I(t)\rangle_\sigma$ is the mean value of $I$ given that
the qubit is in state $\sigma$.)

\section{Number Phase Uncertainty}
\label{app:numberphase}
In this appendix, we briefly review the number-phase uncertainty relation, and from it we derive the relationship between the spectral densities describing the photon number fluctuations and the phase fluctuations. Consider a coherent state labeled by its classical amplitude
$\alpha$
\be
|\alpha\rangle =
\exp\left\{-\frac{|\alpha|^2}{2}\right\}\exp\{\alpha \hat
a^\dagger\}|0\rangle.
\ee
This is an eigenstate of the destruction operator
\be
\hat a|\alpha\rangle = \alpha|\alpha\rangle.
\ee
It is convenient to make the unitary displacement transformation
which maps the coherent state onto a new vacuum state and the
destruction operator onto
\be
\hat a=\alpha + \hat d
\ee
where $d$ annihilates the new vacuum.  Then we have
\be
\bar N=\langle \hat N \rangle = \langle 0|(\alpha^* +\hat
d^\dagger)(\alpha + \hat d)|0\rangle = |\alpha|^2,
\ee
and
\be
(\Delta N)^2 = \langle (\hat N-\bar N)^2 \rangle =
|\alpha|^2\langle 0|\hat d \hat d^\dagger|0\rangle = \bar N.
\label{eq:Nuncertainty}
\ee

Now define the two quadrature amplitudes
\begin{eqnarray}
\hat X&=&\frac{1}{\sqrt{2}}(\hat a+\hat a^\dagger)\\
\hat Y&=&\frac{i}{\sqrt{2}}(\hat a^\dagger -\hat a).
\end{eqnarray}
Each of these amplitudes can be measured in a homodyne
experiment.
For convenience, let
us take $\alpha$ to be real and positive. Then
\be
\langle \hat X \rangle = \sqrt{2}\alpha
\ee
and
\be
\langle \hat Y\rangle = 0.
\ee
If the phase of this wave undergoes a small modulation due for
example to weak parametric coupling to a qubit then one can
estimate the phase by
\be
\langle \theta \rangle = \frac{\langle \hat Y\rangle}{\langle \hat
X\rangle}.
\ee
This result is of course only valid for small angles, $\theta \ll 1$.
For $\bar{N} \gg 1$, the uncertainty will be
\be
(\Delta\theta)^2=\frac{\langle \hat Y^2 \rangle}{(\langle \hat
X\rangle)^2}=\frac{\frac{1}{2}\langle 0|\hat d \hat
d^\dagger|0\rangle}{2\bar N}=\frac{1}{4\bar N}.
\ee
Thus using Eq.~(\ref{eq:Nuncertainty}) we arrive at the
fundamental quantum uncertainty relation
\be
\Delta\theta\Delta N = \frac{1}{2}.
		\label{eq:AppNumberPhase}
\ee

Using the input-output theory described in
Appendix~\ref{app:drivencavity} we can restate the results
above in terms of noise spectral densities.  Let the amplitude
of the field coming in to the homodyne detector be
\be
\hbin = {\bar b}_{\rm in} + {\hat \xi}(t)
\ee
where ${\hat \xi}(t)$ is the vacuum noise obeying
\be
[{\hat \xi}(t),{\hat \xi}^\dagger(t')]=\delta(t-t').
\ee
We are using a flux normalization for the field operators so
\be
\Ndotbar = \langle \hbin^\dagger\hbin\rangle = |{\bar b}_{\rm in}|^2
\ee
and
\bew
\langle \dot N(t) \dot N(0)\rangle -\Ndotbar^2= \langle 0| ({\bar
b}_{\rm in}^* + {\hat \xi}^\dagger(t))({\bar b}_{\rm in} + {\hat \xi}(t))({\bar
b}_{\rm in}^* + {\hat \xi}^\dagger(0))({\bar b}_{\rm in} + {\hat \xi}(0))
|0\rangle - \left| {\bar b}_{\rm in} \right|^4=\Ndotbar \delta(t).
\eew
From this it follows that the shot noise spectral density is
\be
\SNN=\Ndotbar.
\ee

Similarly the phase can be estimated from the quadrature operator
\be
\hat\theta=\frac{i(\hbin^\dagger-\hbin)}{\langle \hbin^\dagger +
\hbin\rangle} = \langle\hat\theta\rangle +i\frac{({\hat \xi}^\dagger -
{\hat \xi})}{2{\bar b}_{\rm in}}
\ee
which has noise correlator
\be
\langle\delta{\hat \theta}(t)\delta{\hat \theta}(0)\rangle=\frac{1}{4\Ndotbar}\delta(t)
\ee
corresponding to the phase imprecision spectral density
\be
\Sthetatheta =\frac{1}{4\Ndotbar}.
\ee
We thus arrive at the fundamental quantum limit relation
\be
\sqrt{\Sthetatheta \SNN} = \frac{1}{2}.
	\label{eq:AppNumberPhaseNoise}
\ee

\section{Using feedback to reach the quantum limit}
\label{subsec:MirrorsFeedback}

In Sec.~(\ref{subsec:MinAmp}), we demonstrated that any two port
amplifier whose scattering matrix has $s_{11}=s_{22} = s_{12}=0$
will fail to reach quantum limit when used as a weakly coupled op-amp;
at best, it will miss optimizing the quantum noise constraint of
Eq.~(\ref{eq:VAmpNoiseConstraint}) by a factor of two.  Reaching
the quantum limit thus requires at least one of $s_{11},s_{22}$
and $s_{12}$ to be non-zero.  In this subsection, we demonstrate
how this may be done.  We show that by introducing a form of
negative feedback to the ``minimal" amplifier of the previous
subsection, one can take advantage of noise correlations to 
reduce the back-action current noise $S_{II}$ by a factor of two.  As a result,
one is able to reach the weak-coupling (i.e.~op-amp) quantum limit.
Note that quantum amplifiers with feedback are also treated in
\textcite{Grassia98, Courty99}.

On a heuristic level, we can understand the need for either
reflections or reverse gain to reach the quantum limit. A problem
with the ``minimal" amplifier of the last subsection was that its
input impedance was too low in comparison to its noise impedance
$\ZN \sim Z_a$.  From general expression for the input impedance,
Eq.~(\ref{eq:Zindefn}), we see that having non-zero reverse gain
(i.e.~$s_{12} \neq 0$) and/or non-zero reflections (i.e.~$s_{11}
\neq 0$ and/or $s_{22} \neq 0$) could lead to $\Zin \gg Z_{a}$.
This is exactly what occurs when feedback is used to reach the
quantum limit.  Keep in mind that having non-vanishing reverse
gain is dangerous: as we discussed earlier, an appreciable
non-zero $\lambda'_I$ can lead to the highly undesirable
consequence that the amplifier's input impedance depends on the
impedance of the load connected to its output
(cf.~Eq.~(\ref{eq:LoadedZin})).


\subsection{Feedback using mirrors}

\begin{figure}[t]
\begin{center}
\includegraphics[width=3.45in]{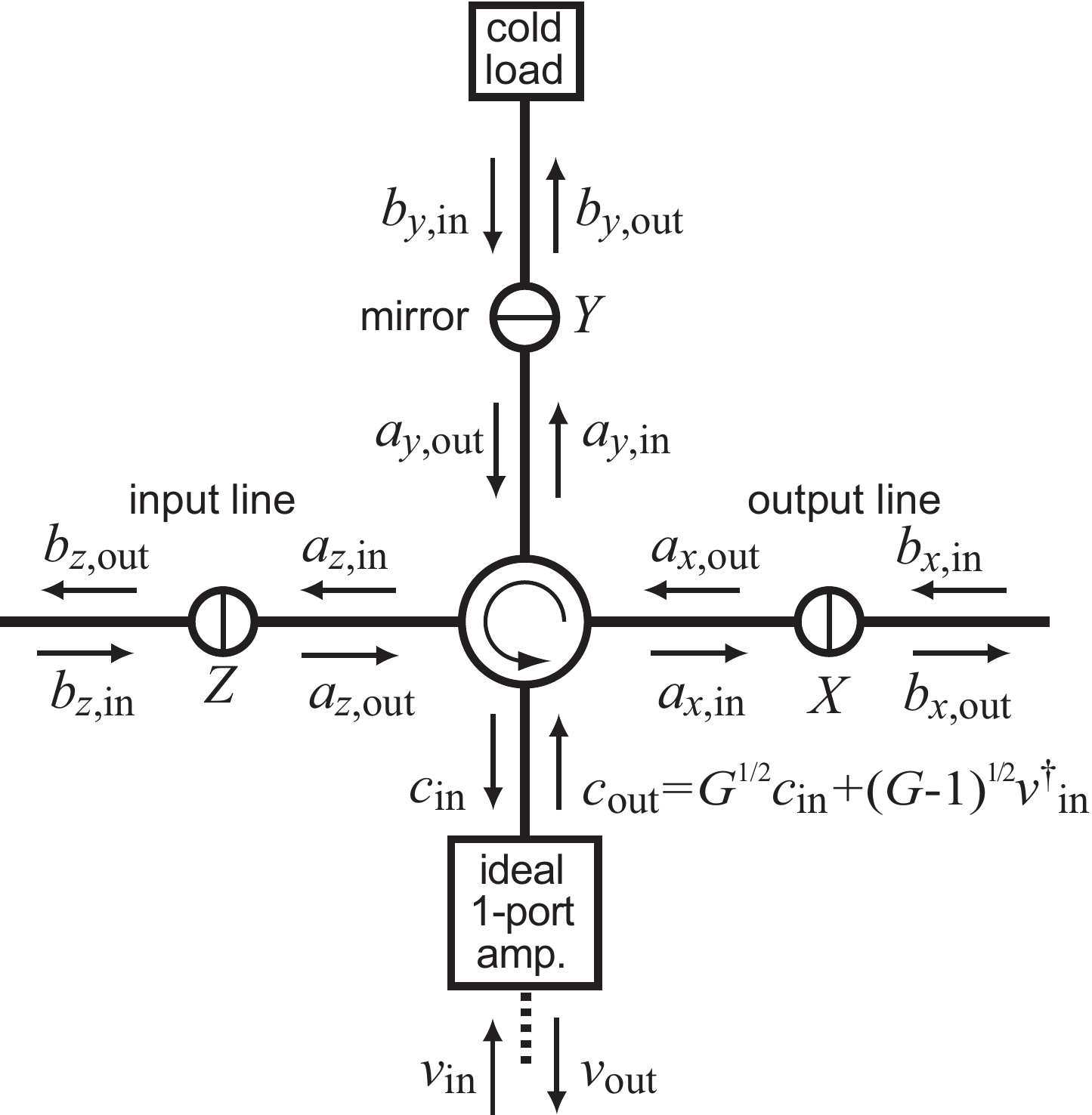}
\caption{Schematic of a modified minimal two-port amplifier, where partially
reflecting mirrors have been inserted in the input and output
transmission lines, as well as in the line leading to the cold load.  By tuning
the reflection coefficient of the mirror in the cold load arm (mirror $Y$), we
can induce negative feedback which takes advantage of correlations between
current and voltage noise.  This then allows this system to reach the quantum limit
as a weakly coupled voltage op amp.
See text for further description}
\label{fig:MirrorsAmp}
\end{center}
\end{figure}

To introduce reverse gain and reflections into the ``minimal"
two-port bosonic amplifier of the previous subsection, we will
insert mirrors in three of the four arms leading from the
circulator: the arm going to the input line, the arm going to the
output line, and the arm going to the auxiliary {}``cold load''
(Fig.~\ref{fig:MirrorsAmp}).
Equivalently, one could imagine that each of these lines is not
perfectly impedance matched to the circulator. Each mirror will be
described by a $2\times2$ unitary scattering
matrix:\begin{eqnarray} \left(\begin{array}{c}
    \ha_{j,{\rm out}}\\
    \hb_{j,{\rm out}}\end{array}\right) & = & U_{j}\cdot\left(\begin{array}{c}
    \hb_{j,{\rm in}}\\
    \ha_{j,{\rm in}}\end{array}\right)\\
    U_{j} & = & \left(\begin{array}{cc}
    \cos\theta_{j} & -\sin\theta_{j}\\
    \sin\theta_{j} & \cos\theta_{j}\end{array}\right)
    \label{eq:MirrorsU}
\end{eqnarray}
Here, the index $j$ can take on three values: $j=z$ for the mirror
in the input line, $j=y$ for the mirror in the arm going to the
cold load, and $j=x$ for the mirror in the output line. The mode
$a_{j}$ describes the {}``internal'' mode which exists between the
mirror and circulator, while the mode $b_{j}$ describes the
{}``external'' mode on the other side of the mirror. We have taken
the $U_{j}$ to be real for convenience. Note that $\theta_{j}=0$
corresponds to the case of no mirror (i.e. perfect transmission).

It is now a straightforward though tedious exercise to construct
the scattering matrix for the entire system. From this, one can
identify the reduced scattering matrix $s$ appearing in
Eq.~(\ref{eq:sdefn}), as well as the noise operators $\FF_{j}$.  These
may then in turn be used to obtain the op-amp description of the
amplifier, as well as the commutators of the added noise
operators.  These latter commutators determine the usual noise
spectral densities of the amplifier.  Details and intermediate
steps of these calculations may be found in Appendix
\ref{app:MirrorsDetail}.

As usual, to see if our amplifier can reach the quantum limit
when used as a (weakly-coupled) op-amp , we need to see if it
optimizes the quantum noise constraint of
Eq.~(\ref{eq:VAmpNoiseConstraint}). We consider the optimal
situation where both the auxiliary modes of the amplifier
($\huin$ and $\hvin^{\dagger}$) are in the vacuum state. The
surprising upshot of our analysis (see Appendix
\ref{app:MirrorsDetail}) is the following: {\it if we include a
small amount of reflection in the cold load line with the
correct phase, then we can reach the quantum limit,
irrespective of the mirrors in the input and output lines}. In
particular, if $\sin\theta_{y}=-1/\sqrt{G}$, our amplifier
optimizes the quantum noise constraint of
Eq.~(\ref{eq:VAmpNoiseConstraint}) in the large gain
(i.e.~large $G$) limit, independently of the values of
$\theta_x$ and $\theta_y$. Note that tuning $\theta_y$ to reach
the quantum limit does not have a catastrophic impact on other
features of our amplifier.  One can verify that this tuning only
causes the voltage gain $\lambda_V$ and power gain $G_P$ to
decrease by a factor of two compared to their $\theta_y=0$
values (cf.~Eqs.~(\ref{eq:Gainmirror}) and
(\ref{eq:Gpowermirror})). This choice for $\theta_y$ also leads
to $\Zin \gg Z_a \sim \ZN$ (cf.~(\ref{eq:ZinMirror})), in
keeping with our general expectations.

Physically, what does this precise tuning of $\theta_y$ correspond
to? A strong hint is given by the behaviour of the amplifier's
cross-correlation noise $\bS_{VI}[\omega]$
(cf.~Eq.~(\ref{eq:SIVmirrors})).  In general, we find that
$\bS_{VI}[\omega]$ is real and non-zero.  However, the tuning {\it
$\sin \theta_y = -1/\sqrt{G}$ is exactly what is needed to have
$\bS_{VI}$ vanish}.  Also note from Eq.~(\ref{eq:SImirrors}) that
this special tuning of $\theta_y$ decreases the back-action
current noise precisely by a factor of two compared to its value
at $\theta_y=0$.  A clear physical explanation now emerges.  Our
original, reflection-free amplifier had correlations between its
back-action current noise and output voltage noise
(cf.~Eq.~(\ref{eq:SIV})).  By introducing negative feedback of the
output voltage to the input current (i.e.~via a mirror in the
cold-load arm), we are able to use these correlations to decrease
the overall magnitude of the current noise (i.e.~the voltage
fluctuations $\tV$ partially cancel the original current
fluctuations $\tI$). For an optimal feedback (i.e. optimal choice
of $\theta_y$), the current noise is reduced by a half, and the
new current noise is not correlated with the output voltage noise.
Note that this is indeed negative (as opposed to positive)
feedback-- it results in a reduction of both the gain and the
power gain. To make this explicit, in the next section we will map
the amplifier described here onto a standard op-amp with negative
voltage feedback.

\subsection{Explicit examples}

To obtain a more complete insight, it is useful to go back and
consider what the reduced scattering matrix of our system looks
like when $\theta_y$ has been tuned to reach the quantum limit.
From Eq.~(\ref{eq:smirrors}), it is easy to see that at the
quantum limit, the matrix $s$ satisfies:
\begin{subequations}
\begin{eqnarray}
    s_{11} & = & -s_{22} \\
    s_{12} & = & \frac{1}{G} s_{21}
\end{eqnarray}
\end{subequations}
The second equation also carries over to the op-amp picture; at
the quantum limit, one has:
\begin{eqnarray}
    \lambda'_I = \frac{1}{G} \lambda_V
\end{eqnarray}

One particularly simple limit is the case where there are no
mirrors in the input and output line ($\theta_{x}=\theta_{z}=0$),
only a mirror in the cold-load arm.  When this mirror is tuned to
reach the quantum limit (i.e. $\sin\theta_{y}=-1/\sqrt{G}$), the
scattering matrix takes the simple form:
\begin{eqnarray}
s & = & \left(\begin{array}{cc}
0 & 1/\sqrt{G}\\
\sqrt{G} & 0\end{array}\right)\end{eqnarray} In this case, the principal
effect of the weak mirror in the cold-load line is to introduce a
small amount of reverse gain.  The amount of this reverse gain is
exactly what is needed to have the input impedance diverge
(cf.~Eq.~(\ref{eq:Zindefn})). It is also what is needed to achieve
an optimal, noise-canceling feedback in the amplifier. To see
this last point explicitly, we can re-write the amplifier's
back-action current noise ($\tI$) in terms of its original noises
$\tI_0$ and $\tV_0$ (i.e. what the noise operators would have been
in the absence of the mirror).  Taking the relevant limit of small
reflection (i.e. $\tilde{r} \equiv \sin \theta_y$ goes to zero as
$|G| \ra \infty$), we find that the modification of the current
noise operator is given by:
\begin{eqnarray}
    \tI & \simeq &
              \tI_0 +
            \frac{2 \sqrt{G} \tilde{r}} {1 - \sqrt{G} \tilde{r} } \frac{\tV_0}{Z_a}
\end{eqnarray}
As claimed, the presence of a small amount of reflection
$\tilde{r}\equiv \sin \theta_y$ in the cold load arm ``feeds-back"
the original voltage noise of the amplifier $\tV_0$ into the
current. The choice $\tilde{r} = -1/ \sqrt{G} $ corresponds to a negative
feedback, and optimally makes use of the fact that $\tI_0$ and
$\tV_0$ are correlated to reduce the overall fluctuations in
$\tI$.

While it is interesting to note that one can reach the quantum
limit with no reflections in the input and output arms, this case
is not really of practical interest.  The reverse current gain in
this case may be small (i.e. $\lambda'_I \propto 1/ \sqrt{G}$), but it is
not small enough:  one finds that because of the non-zero
$\lambda'_I$, the amplifier's input impedance is strongly reduced
in the presence of a load (cf.~Eq.~(\ref{eq:LoadedZin})).

There is a second simple limit we can consider which is more
practical.  This is the limit where reflections in the input-line
mirror and output-line mirror are both strong. Imagine we take
$\theta_{z}=-\theta_{x}=\pi/2-\delta/G^{1/8}$. If again we set
$\sin\theta_{y}=-1/\sqrt{G}$ to reach the quantum limit, the scattering
matrix now takes the form (neglecting terms which are order
$1/\sqrt{G}$):
\begin{eqnarray}
    s & = &
        \left(\begin{array}{cc}
            +1 & 0\\
            \frac{\delta^2 G^{1/4} }{2} & -1
        \end{array}\right)
        \label{eq:StrongRsmat}
\end{eqnarray}
In this case, we see that at the quantum limit, the reflection
coefficients $s_{11}$ and $s_{22}$ are exactly what is needed to
have the input impedance diverge, while the reverse gain
coefficient $s_{12}$ plays no role. For this case of strong
reflections in input and output arms, the voltage gain is reduced
compared to its zero-reflection value:
\begin{eqnarray}
    \lambda_V \ra
        \sqrt{\frac{Z_b}{Z_a}}  \left( \frac{\delta}{2} \right)^2 G^{1/4}
\end{eqnarray}
The power gain however is independent of $\theta_x,\theta_z$, and
is still given by $G/2$ when $\theta_{y}$ is tuned to be at the
quantum limit.

\subsection{Op-amp with negative voltage feedback}

\begin{figure}[t]
\begin{center}
\includegraphics[width=3.45in]{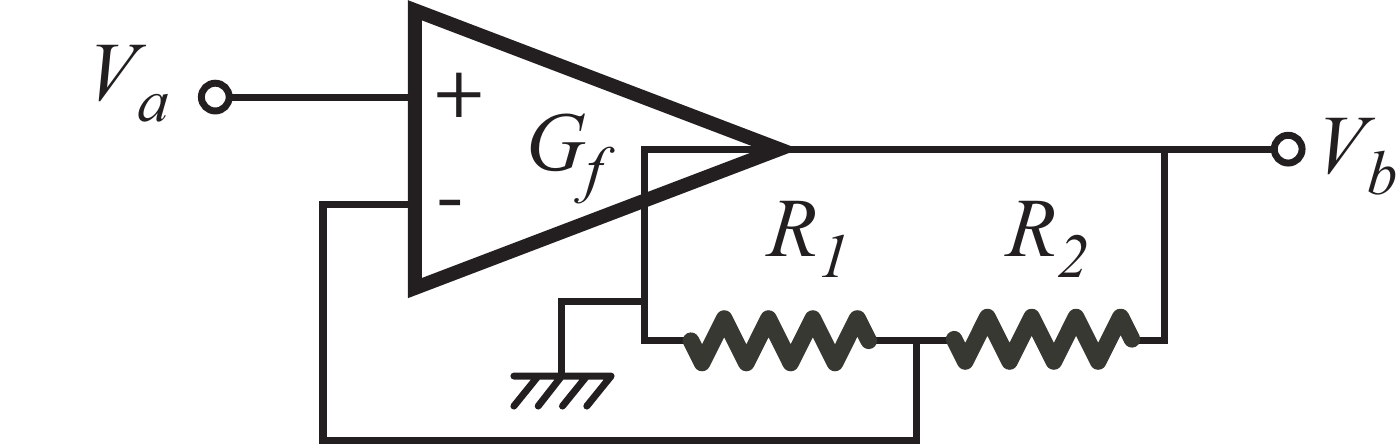}
\caption{Schematic of a voltage op-amp with negative feedback.}
\label{fig:FeedbackOpAmp}
\end{center}
\end{figure}

We now show that a conventional op-amp with feedback can be mapped
onto the amplifier described in the previous subsection.  We will
show that tuning the strength of the feedback in the op-amp
corresponds to tuning the strength of the mirrors, and that an
optimally tuned feedback circuit lets one reach the quantum limit.
This is in complete correspondence to the previous subsection,
where an optimal tuning of the mirrors also lets one reach the
quantum limit.

More precisely, we consider a scattering description of a
non-inverting op-amp amplifier having negative voltage feedback.
The circuit for this system is shown in Fig.~\ref{fig:FeedbackOpAmp}.
A fraction $B$ of the output voltage of the amplifier is fed back to the negative
input terminal of the op-amp. In practice, $B$ is determined by
the two resistors $R_1$ and $R_2$ used to form a voltage divider
at the op-amp output.  The op-amp with zero feedback is described
by the ``ideal" amplifier of Sec.~\ref{subsec:MinAmp}: at zero
feedback, it is described by Eqs.~(\ref{eq:Gainsimp})-
(\ref{eq:Zinsimp}).  For simplicity, we consider the relevant case
where:
\begin{eqnarray}
    Z_b \ll R_1,R_2 \ll Z_a
\end{eqnarray}
In this limit,  $R_1$ and $R_2$ only play a role through the
feedback fraction $B$, which is given by:
\begin{eqnarray}
    B = \frac{R_2}{R_1 + R_2}
\end{eqnarray}

Letting $G_f$ denote the voltage gain at zero feedback ($B=0$), an
analysis of the circuit equations for our op-amp system yields:
\begin{subequations}
\begin{eqnarray}
    \lambda_{V} & = &  \frac{G_{f}}{1+B\cdot G_{f} }\\
    \lambda'_{I} & = & \frac{B}{1+B\cdot G_{f}}\\
    \Zout & = & \frac{Z_{b}}{1+B\cdot G_{f}}\\
    \Zin & = & (1+B\cdot G_{f})Z_{a}\\
    G_{P} & = & \frac{G_{f}^{2} / 2}{B\cdot G_{f}+2 Z_b/Z_a}
\end{eqnarray}
\end{subequations}
Again, $G_{f}$ represents the gain of the amplifier in the absence
of any feedback, $Z_{a}$ is the input impedance at zero feedback,
and $Z_{b}$ is the output impedance at zero feedback.

Transforming this into the scattering picture yields a scattering
matrix $s$ satisfying:
\begin{subequations}
\begin{eqnarray}
    s_{11} = -s_{22} & = & -
        \frac{B G_f \left( Z_a - (2+B G_f) Z_b \right)}
            {B G_f Z_a + (2+ B G_f)^2 Z_b}
        \\
    s_{21} & = & -
        \frac{2 \sqrt{Z_a Z_b} G_f \left(1+ B G_f  \right)}
            {B G_f Z_a + (2+ B G_f)^2 Z_b}
        \\
    s_{12} & = & \frac{B}{G_f} s_{21}
\end{eqnarray}
\end{subequations}
Note the connection between these equations and the necessary form
of a quantum limited s-matrix found in the previous subsection.

Now, given a scattering matrix, one can always find a minimal
representation of the noise operators $\FF_{a}$ and $\FF_{b}$
which have the necessary commutation relations. These are given in
general by:
\begin{eqnarray}
    \hFF_{a} & = &
        \sqrt{1-|s_{11}|^{2}-|s_{12}|^{2}+|l|^{2}}\cdot \hu_{in}
            +l\cdot \hv_{in}^{\dagger}\\
    \hFF_{b} & = & \sqrt{|s_{21}|^{2}+|s_{22}|^{2}-1}\cdot \hv_{in}^{\dagger}\\
    l & = & \frac{s_{11}s_{21}^{*}+s_{12}s_{22}^{*}}{\sqrt{|s_{21}|^{2}+|s_{22}|^{2}-1}}
\end{eqnarray}
Applying this to the s matrix for our op-amp, and then taking the
auxiliary modes $\huin$ and $\hvin^{\dagger}$ to be in the
vacuum state, we can calculate the minimum allowed $\bS_{VV}$ and
$\bS_{II}$ for our non-inverting op-amp amplifier. One can then
calculate the product $\bS_{VV}\bS_{II}$ and compare against the
quantum-limited value ($\bS_{IV}$ is again real). In the case of
zero feedback (i.e. $B=0$), one of course finds that this product
is twice as big as the quantum limited value. However, if one
takes the large $G_f$ limit while keeping \textbf{$B$} non-zero
but finite, one obtains:\begin{eqnarray} \bS_{VV}\bS_{II} &
\rightarrow &
(\hbar\omega)^{2}\left(1-\frac{2B}{G_f}+O\left(\frac{1}{G_f}\right)^{2}\right)\end{eqnarray}
Thus, for a fixed, non-zero feedback ratio $B$, it is possible to
reach the quantum limit. Note that if $B$ does not tend to zero as
$G_f$ tends to infinity, the voltage gain of this amplifier will
be finite. The power gain however will be proportional to $G_f$
and will be large. If one wants a large voltage gain, one could
set $B$ to go to zero with $G_f$ i.e.
$B\propto\frac{1}{\sqrt{G_f}}$. In this case, one will still reach
the quantum limit in the large $G_f$ limit, and the voltage gain
will also be large (i.e. $\propto\sqrt{G_f}$). Note that in all
these limits, the reflection coefficients $s_{11}$ and $s_{22}$
tend to $-1$ and 1 respectively, while the reverse gain tends to
$0$. This is in complete analogy to the amplifier with mirrors
considered in the previous subsection, in the case where we took
the reflections to be strong at the input and at the output
(cf.~Eq.~(\ref{eq:StrongRsmat})). We thus see yet again how the
use of feedback allows the system to reach the quantum limit.


\section{Additional Technical Details}

This appendix provides further details of calculations presented in the main text.

\subsection{Proof of quantum noise constraint}
\label{app:noiseineq}

Note first that we may write the symmetrized $\hI$ and $\hF$ noise
correlators defined in Eqs.~(\ref{SFSymm}) and (\ref{SISymm}) as
sums over transitions between detector energy eigenstates:
\begin{widetext}
\begin{eqnarray}
     \bS_{FF}[\omega] & = & \pi \hbar \sum_{i,f}
         \langle i | \hat{\rho}_0 | i \rangle
         \cdot
         | \langle f | \hat{F} | i \rangle |^2
         \left[
             \delta(E_f-E_i + \hbar\omega) +
             \delta(E_f-E_i - \hbar\omega)
         \right]
          \label{LehmanRep} \\
     \bS_{II}[\omega] & = & \pi \hbar \sum_{i,f}
         \langle i | \hat{\rho}_0 | i \rangle
         \cdot
         | \langle f | \hat{I} | i \rangle |^2
         \left[
             \delta(E_f-E_i + \hbar\omega) +
             \delta(E_f-E_i - \hbar\omega)
         \right]          \label{LehmanRep2}
\end{eqnarray}
\end{widetext}
Here, $\hat{\rho}_0$ is the stationary density matrix describing
the state of the detector, and $| i \rangle$ ($ | f \rangle$) is a
detector energy eigenstate with energy $E_{i}$ ($E_{f}$).
Eq.~(\ref{LehmanRep}) expresses the noise at frequency $\omega$ as
a sum over transitions. Each transition starts with an an initial
detector eigenstate $|i \rangle$, occupied with a probability
$\langle i | \rho_0 | i \rangle$, and ends with a final detector
eigenstate $|f \rangle$, where the energy difference between the
two states is either $+\hbar \omega$ or $-\hbar \omega$ . Further,
each transition is weighted by an appropriate matrix element.

To proceed, we fix the frequency $\omega>0$, and let the index
$\nu$ label each transition $| i \rangle \ra | f \rangle$
contributing to the noise.  More specifically, $\nu$ indexes each
ordered pair of detector energy eigenstates states $\{|i \rangle,
| f \rangle\}$ which satisfy $E_f - E_i \in \pm \hbar [\omega,\omega+d\omega]$ 
and $\langle i | \rho_0 | i \rangle \neq 0$.  
We can now consider the
matrix elements of $\hI$ and $\hF$ which contribute to
$\bS_{II}[\omega]$ and $\bS_{FF}[\omega]$ to be complex vectors
$\vec{v}$ and $\vec{w}$.   Letting $\delta$ be any real number,
let us define:
\begin{eqnarray}
     \left[ \vec{w} \right]_{\nu} & = &
         \langle f(\nu) | \hF | i(\nu) \rangle \\
     \left[ \vec{v} \right]_{\nu} & = &
        \begin{cases}
             e^{-i \delta} \langle f(\nu) | \hI | i(\nu) \rangle
            & \text{if } E_{f(\nu)} - E_{i(\nu)} = +\hbar \omega,  \\
             e^{i \delta} \langle f(\nu) | \hI | i(\nu) \rangle
            & \text{if } E_{f(\nu)} - E_{i(\nu)} = -\hbar \omega.
    \end{cases}
    \nonumber \\
\end{eqnarray}
Introducing an inner product $\langle \cdot , \cdot
\rangle_{\omega}$ via:
\begin{equation}
     \langle \vec{a}, \vec{b} \rangle_\omega =
        \pi  \sum_{\nu} \langle i(\nu) | \hat{\rho}_0 | i(\nu) 
         \rangle \cdot
             \left( a_{\nu} \right)^* b_{\nu},
\end{equation}
we see that the noise correlators $\bS_{II}$ and $\bS_{FF}$ may be
written as:
\begin{eqnarray}
     \bS_{II}[\omega] d \omega & = &
         \langle \vec{v} , \vec{v} \rangle_{\omega}
         \label{SIvec}\\
     \bS_{FF}[\omega] d \omega & = &
         \langle \vec{w} , \vec{w} \rangle_{\omega}
         \label{SFvec}
\end{eqnarray}

We may now employ the Cauchy-Schwartz inequality:
\begin{equation}
    \langle \vec{v} , \vec{v} \rangle_{\omega}
    \langle \vec{w} , \vec{w} \rangle_{\omega}
         \geq
    \left|
        \langle \vec{v} , \vec{w} \rangle_{\omega}
     \right|^2
     \label{Cauchy}
\end{equation}
A straightforward manipulation shows that the real part of
$\langle \vec{v} , \vec{w} \rangle_{\omega}$ is determined by the
symmetrized cross-correlator $\bS_{IF}[\omega]$ defined in
Eq.~(\ref{SIFSymm}):
\begin{equation}
         \Re \langle \vec{v} , \vec{w} \rangle_{\omega} =
     \Re \left[ e^{i \delta} \bS_{IF}[\omega] \right]
        d \omega
         \label{SIFre}
\end{equation}
In contrast, the imaginary part of $\langle \vec{v} , \vec{w}
\rangle_{\omega}$ is independent of $\bS_{IF}$; instead, it is
directly related to the gain $\chiIF$ and reverse gain $\chiFI$
of the detector:
\begin{equation}
    \textrm{Im } \langle \vec{v} , \vec{w} \rangle_{\omega}
        =
        \frac{\hbar}{2} \Re \left[
        e^{i \delta} \left(
                \chiIF[\omega] -
        \left[\chiFI[\omega]\right]^*
        \right) \right] d \omega
    \label{SIFim}
\end{equation}
Substituting Eqs. (\ref{SIFim}) and (\ref{SIFre}) into
Eq.~(\ref{Cauchy}), one immediately finds the quantum noise
constraint given in Eq.~(\ref{DeltaNoiseConstraint}).  As in the main text,
we let $\tchiIF = \chiIF - \chiFI^*$.   Maximizing
the RHS of this inequality with respect to the phase $\delta$, one
finds that the maximum is achieved for $\delta = \delta_0 =
-\arg(\tchiIF) + \dtild_0$ with 
\begin{eqnarray}
    \tan 2 \dtild_0 =
        - \frac{ | \bS_{IF} | \sin 2 \phi }
        { (\hbar/2) |\tchiIF| + |\bS_{IF} | \cos 2 \phi }
        \label{eq:tan2delta}
\end{eqnarray}
where $\phi = \arg ( \bS_{IF} \tchiIF* )$.  At $\delta =
\delta_0$,  Eq.~(\ref{DeltaNoiseConstraint}) becomes the final
noise constraint of Eq.~(\ref{NoiseConstraint}).

The proof given here also allows one to see what must be done in
order to achieve the ``ideal" noise condition of
Eq.~(\ref{CorrQLCondition}): one must achieve equality in the
Cauchy-Schwartz inequality of Eq.~(\ref{Cauchy}). This requires
that the vectors $\vec{v}$ and $\vec{w}$ be proportional to
one another; there must exist a complex factor $\alpha$ (having
dimensions $[I] / [F]$) such that:
\begin{equation}
     \vec{v} = \alpha \cdot \vec{w}
\end{equation}
Equivalently, we have that
\begin{equation}
     \langle f | I | i \rangle =
     \begin{cases}
      e^{i \delta} \alpha \langle f | F | i \rangle
        & \text{if } E_f - E_i = +\hbar \omega \\
      e^{-i \delta} \alpha \langle f | F | i \rangle
        & \text{if }  E_f - E_i = -\hbar \omega.
\end{cases}
     \label{PropCond}
\end{equation}
for {\it each} pair of initial and final states $|i \rangle$, $|f
\rangle$ contributing to $\bS_{FF}[\omega]$ and $\bS_{II}[\omega]$
(cf.~Eq.~(\ref{LehmanRep})).  Note that this {\it not} the same as
requiring Eq.~(\ref{PropCond}) to hold for all possible states
$|i\rangle$ and $| f \rangle$.  
This proportionality condition in turn implies a proportionality between the input
and output (unsymmetrized) quantum noise spectral densities:
\begin{eqnarray}
	S_{II}[\omega] = |\alpha|^2 S_{FF}[\omega]
\end{eqnarray}
It thus also follows that the imaginary parts of the input and output susceptibilities are proportional:
\begin{eqnarray}
	\textrm{Im } \chi_{II}[\omega] = |\alpha|^2 \textrm{Im } \chi_{FF}[\omega],
	\label{eq:ChiProp}
\end{eqnarray}
as well as the symmetrized input and output noise (i.e. Eq.~(\ref{AlphaDefn})).
Finally, one can also use Eq.~(\ref{PropCond}) to relate the {\it unsymmetrized}
$I$-$F$ quantum noise correlator $S_{IF}[\omega]$ to $S_{FF}[\omega]$:
(cf.~Eq.~(\ref{eq:SIFunsymm})):
\begin{eqnarray}
    S_{IF}[\omega] =
    \begin{cases}
            e^{-i \delta} \alpha^* S_{FF}[\omega] & \text{if } \omega > 0 , \\
            e^{i \delta} \alpha^* S_{FF}[\omega] & \text{if } \omega < 0
    \end{cases}
    \label{eq:SIProf}
\end{eqnarray}
Note that $S_{FF}[\omega]$ is necessarily real and positive.

Finally, for a detector with quantum-ideal noise properties, the
magnitude of the constant $\alpha$ can be found from Eq.
(\ref{AlphaDefn}). The phase of $\alpha$ can also be determined
from:
\begin{eqnarray}
    \frac{- \textrm{Im } \alpha}{|\alpha|} & = &
        \frac{ \hbar | \tchiIF | / 2}{\sqrt{\bS_{II} \bS_{FF} }}
        \cos \tilde{\delta}_0
        \label{eq:AlphaPhase}
\end{eqnarray}
For zero frequency or for a large detector effective temperature,
this simplifies to:
\begin{eqnarray}
    \frac{- \textrm{Im } \alpha}{|\alpha|} & = &
        \frac{ \hbar  \tchiIF  / 2}{\sqrt{\bS_{II} \bS_{FF} }}
        \label{eq:AlphaPhase2}
\end{eqnarray}

Note importantly that to have a non-vanishing gain and power gain,
one needs $\textrm{Im } \alpha \neq 0$. This in turn places a very
powerful constraint on a quantum-ideal detectors: {\it all
transitions contributing to the noise must be to final states $| f
\rangle$ which are completely unoccupied}.  To see this, imagine a
transition taking an initial state $| i \rangle = | a \rangle$ to
a final state $| f \rangle = | b \rangle$ makes a contribution to
the noise.  For a quantum-ideal detector, Eq.~(\ref{PropCond})
will be satisfied:
\begin{equation}
     \langle b | \hI | a \rangle = e^{\pm i \delta}\alpha
         \langle b | \hF | a \rangle
         \label{PropAsymm1}
\end{equation}
where the plus sign corresponds to $E_b > E_a$, the minus to $E_a
> E_b$. If now the final state $| b \rangle$ was also occupied
(i.e. $\langle b | \hat{\rho}_0 | b \rangle \neq 0$), then the
reverse transition $| i=b \rangle \ra | f=a \rangle$) would also
contribute to the noise.  The proportionality condition of
Eq.~(\ref{PropCond}) would now require:
\begin{equation}
     \langle a | \hI | b \rangle = e^{\mp i \delta} \alpha
         \langle a | \hF | b \rangle
         \label{PropAsymm2}
\end{equation}
As $\hI$ and $\hF$ are both Hermitian operators, and as $\alpha$
must have an imaginary part in order for there to be gain, we have
a contradiction: Eq.~(\ref{PropAsymm1}) and (\ref{PropAsymm2})
cannot both be true.  It thus follows that the final state of a
transition contributing to the noise {\it must} be unoccupied in
order for Eq.~(\ref{PropCond}) to be satisfied and for the
detector to have ideal noise properties.  Note that this necessary
asymmetry in the occupation of detector energy eigenstates
immediately tells us that {\it a detector or amplifier cannot
reach the quantum limit if it is in equilibrium}.

\subsection{Proof that a noiseless detector does not amplify}
\label{subapp:NoAmplification}
With the above results in hand, we can now prove assertions made in Sec.~\ref{subsubsec:NoiseCrossCorr} that detectors which evade the quantum noise
constraint of Eq.~(\ref{NoiseConstraint}) and simply satisfy
\begin{eqnarray}
	 \bS_{FF} \bS_{II} = |\bS_{IF}|^2
	 \label{eq:ClassicalNoiseConstraint}
\end{eqnarray}
are at best transducers, as their power gain is limited to being at most one.

The first way to make the RHS of  Eq.~(\ref{NoiseConstraint}) vanish is to have $\chiIF = \chiFI^*$.  We have already seen
that whenever this relation holds, the detector power gain cannot be any larger than one (c.f.~Eq.~(\ref{eq:GPReverse})).
Now, imagine that the detector also has a minimal amount of noise, 
i.e.~Eq.~(\ref{eq:ClassicalNoiseConstraint}) also holds.  This latter fact implies
that the proportionality condition of Eq.~(\ref{PropCond}) also must hold.  In this situation, 
the detector {\it must} have a power gain of unity, and is thus a transducer.
There are two possibilities to consider here.  First,  $\bS_{FF}$ and $\bS_{II}$ 
could both be non-zero, but perfectly correlated:  $|\bS_{IF}|^2 = \bS_{FF} \bS_{II}$.  In this case,
the proportionality constant $\alpha$ must be real (c.f.~Eq.~(\ref{eq:AlphaPhase})).  Using this fact along with 
Eqs.~(\ref{eq:SIProf}) and (\ref{eq:LambdaSIF}), one immediately finds that $S_{FF}[\omega]
= S_{FF}[-\omega]$.  This implies the back-action damping $\gamma$ associated with the detector input vanishes (c.f.~Eq.~(\ref{OscGamma})). 
It thus follows immediately from Eq.~(\ref{GPDefn}) and Eq.~(\ref{eq:GPReverse}) that the power gain $G_{P,{\rm rev}}$ (defined in a way that
accounts for the reverse gain) is exactly one.  The detector is thus simply a transducer.  The other possibility here is that $\chiIF = \chiFI^*$ and one or both of $\bS_{II},\bS_{FF}$ are equal zero.  Note that
if the symmetrized noise vanishes, then so must the asymmetric part of the noise.  Thus, it follows that either the
damping induced by the detector input, $\gamma$, or that induced by the output, $\gamma_{\rm out}$ (c.f. Eq.~(\ref{gammaout}))
(or both) must be zero.  Eqs.~(\ref{GPDefn}) and (\ref{eq:GPReverse}) then again yield a power gain $G_{P,{\rm rev}}=1$.
We thus have shown that any detector which has $\chiFI = \chiIF^*$ and satisfies
$	 \bS_{FF} \bS_{II} = |\bS_{IF}|^2$
must necessarily be a transducer, with a power gain precisely equal to one.

A second way to make the RHS of  Eq.~(\ref{NoiseConstraint}) vanish is to have $\bS_{IF} / \tchiIF$ be purely imaginary and larger in magnitude than $\hbar/2$.  Suppose this is the case, and that the detector also satisfies the minimal noise requirement of Eq.~(\ref{eq:ClassicalNoiseConstraint}).  Without loss of generality, we take $\tchiIF$ to be real, implying that $\bS_{IF}$ is purely imaginary.
Eqs.~(\ref{SIFim}) and (\ref{eq:tan2delta}) then imply that the phase factor $e^{i \delta}$ appearing in the proportionality relation of 
Eq.~(\ref{eq:SIProf}) is purely imaginary, while the constant $\alpha$ is purely real.  Using this proportionality relation in Eq.~(\ref{eq:LambdaSIF}) for $\tchiIF$ yields:
\begin{eqnarray}
	 \tchiIF  & = &   \frac{\alpha}{\hbar} \left( S_{FF}[\omega] - S_{FF}[-\omega] \right) 
	 \nonumber \\
	 & = & 
	 		2 \alpha \left[ - \textrm{Im }\chi_{FF}[\omega] \right]
\end{eqnarray}
Using this result and the relation between $\chi_{FF}$ and $\chi_{II}$ in Eq.~(\ref{eq:ChiProp}), we can write the power gain in the absence of reverse gain, $G_P$ (c.f.~Eq.~(\ref{GPDefn})), as
\begin{eqnarray}
 	G_P =1/ \left| 1 - \chiFI / \chiIF \right|^2
\end{eqnarray}
If the reverse gain vanishes (i.e.~$\chiFI=0$), we immediately find that $G_P = 1$:  the detector has a power gain of one, and is thus simply a transducer.  If the reverse gain is non-zero, we must take the expression for $G_P$ above and plug it into Eq.~(\ref{eq:GPReverse}) for the power gain with reverse gain, $G_{P,{\rm rev}}$.  Some algebra again yields that the full power gain is at most unity.  We again have the conclusion that the detector does not amplify.


\subsection{Simplifications for a quantum-limited detector}
\label{app:ImPartsGone}

In this appendix, we derive the additional constraints on the
property of a detector that arise when it satisfies the quantum
noise constraint of Eq.~(\ref{CorrQLCondition}).  We focus on the ideal case
where the reverse gain $\chiFI$ vanishes.
To start, we substitute Eq.~(\ref{eq:SIProf}) into
Eqs.~(\ref{eq:LambdaSIF})-(\ref{eq:SIFBar}); writing $S_{FF}[\omega]$
in terms of the detector effective temperature $T_{\rm eff}$
(cf.~Eq.~(\ref{OscTeff})) yields:
\begin{eqnarray}
     \frac{\hbar \lambda[\omega]}{2} & = &
            -  e^{-i \delta} \hbar
        \left[-\textrm{Im }\chi_{FF}[\omega] \right]
        \\
        &&
            \left[
                     \left( \textrm{Im }\alpha \right) \coth \left( \frac{\hbar \omega}{2 \kb T_{\rm eff}} \right)
             + i  \left( \textrm{Re } \alpha \right)
            \right]
         \nonumber \\
     \bS_{I F}[\omega] & = &
            e^{-i \delta} \hbar
        \left[-\textrm{Im }\chi_{FF}[\omega] \right]
        \\
        &&
            \left[
                     \left( \Re  \alpha \right)
             \coth \left( \frac{\hbar \omega}{2 \kb T_{\rm eff}} \right)
              - i \left( \textrm{Im } \alpha \right)
            \right]
        \nonumber
\end{eqnarray}

To proceed, let us write:
\begin{eqnarray}
    e^{-i \delta} = \frac{\lambda}{|\lambda|} e^{-i \dtild}
\end{eqnarray}
The condition that $|\lambda|$ is real yields the condition:
\begin{eqnarray}
    \tan \dtild = \frac{ \Re \alpha}{ \rm{Im } \alpha}
        \tanh\left( \frac{\hbar \omega}{2 \kb T_{\rm eff} } \right)
\end{eqnarray}

We now consider the relevant limit of a large detector power gain
$G_P$.  $G_P$ is determined by Eq.~(\ref{GPTemp}); the only way
this can become large is if $\kb T_{\rm eff} / (\hbar \omega) \ra
\infty$ while $\textrm{Im } \alpha$ does not tend to zero.  We
will thus take the large $T_{\rm eff}$ limit in the above
equations while keeping both $\alpha$ and the phase of $\lambda$
fixed.  Note that this means the parameter $\dtild$ must evolve;
it tends to zero in the large $T_{\rm eff}$ limit.  In this limit,
we thus find for $\lambda$ and $\bS_{IF}$:
\begin{eqnarray}
     \frac{\hbar \lambda[\omega]}{2} & = &
            -  2 e^{-i \delta} \kb T_{\rm eff} \gamma[\omega]
                     \left( \textrm{Im }\alpha \right)
             \left[ 1
                + O\left[
                \left( \frac{\hbar \omega}{ \kb T_{\rm eff} } \right)^2 \right]
            \right]
        \nonumber \\
        \\
     \bS_{I F}[\omega] & = &
            2 e^{-i \delta} \kb T_{\rm eff} \gamma[\omega]
                     \left( \Re  \alpha \right)
             \left[ 1
                + O\left( \frac{\hbar \omega}{ \kb T_{\rm eff} } \right)
            \right]
            \nonumber \\
\end{eqnarray}

Thus, in the large power-gain limit (i.e.~large $T_{\rm eff}$
limit), the gain $\lambda$ and the noise cross-correlator
$\bS_{IF}$ have the same phase:  $\bS_{IF} / \lambda$ is purely
real.



\subsection{Derivation of non-equilibrium Langevin equation}
\label{app:Langevin}

In this appendix, we prove that an oscillator weakly coupled to an
arbitrary out-of-equilibrium detector is described by the Langevin
equation given in Eq.~(\ref{Langevin}), an equation which
associates an effective temperature and damping kernel to the
detector.  The approach taken here is directly related to the pioneering work
of Schwinger \cite{Schwinger61}.

We start by defining the oscillator matrix Keldysh green function:
\begin{equation}
     \check{G}(t) = \left(
         \begin{array}{cc}
           G^K(t) & G^R(t) \\
           G^A(t) & 0 \\
         \end{array}
     \right)
\end{equation}
where $G^R(t-t') =  - i \theta(t-t') \langle
\left[\hat{x}(t),\hat{x}(t')\right] \rangle$, $G^A(t-t')  =   i
\theta(t'-t) \langle \left[\hat{x}(t),\hat{x}(t')\right] \rangle$,
and    $G^K(t-t') = - i \langle
         \{\hat{x}(t),\hat{x}(t')\} \rangle$.
At zero coupling to the detector ($A=0$), the oscillator is only
coupled to the equilibrium bath, and thus $\check{G}_0$ has the
standard equilibrium form:
\begin{equation}
     \check{G}_0[\omega] = \frac{\hbar}{m}
     \left(
         \begin{array}{cc}
             - 2 \textrm{Im } g_0[\omega]
          \coth \left(\frac{\hbar \omega}{ 2 \kb T_{{\rm bath}}}
\right) & g_0[\omega] \\
           g_0[\omega]^* & 0 \\
         \end{array}
     \right)
\end{equation}
where:
\begin{equation}
     g_0[\omega] = \frac{1}{\omega^2 - \Omega^2 + i \omega
     \gamma_0 / m}
\end{equation}
and where $\gamma_0$ is the intrinsic damping coefficient, and
$T_{{\rm bath}}$ is the bath temperature.

We next treat the effects of the coupling to the detector in
perturbation theory.  Letting $\check{\Sigma}$ denote the
corresponding self-energy, the Dyson equation for $\check{G}$ has
the form:
\begin{equation}
\left[ \check{G}[\omega] \right]^{-1} =
     \left[ \check{G}_0[\omega] \right]^{-1} -
     \left(
         \begin{array}{cc}
           0 & \Sigma^A[\omega] \\
           \Sigma^R[\omega] & \Sigma^K[\omega] \\
         \end{array}
     \right)
     \label{Dyson}
\end{equation}
To lowest order in $A$, $\check{\Sigma}[\omega]$ is given by:
\begin{eqnarray}
     \check{\Sigma}[\omega] & = & A^2 \check{D}[\omega] \\
     & \equiv & \frac{A^2}{\hbar} \int dt \hspace{3pt} e^{i \omega t}
     \\
     && \left(
         \begin{array}{cc}
           0 & i \theta(-t) \langle [ \hF(t), \hF(0) ] \rangle \\
           -i \theta(t) \langle [ \hF(t), \hF(0) ] \rangle
             & -i  \langle \{ \hF(t), \hF(0) \} \rangle \\
         \end{array}
     \right) \nonumber
\end{eqnarray}
Using this lowest-order self energy, Eq.~(\ref{Dyson}) yields:
\begin{eqnarray}
     G^R[\omega] & = &  \frac{\hbar}
         { m(\omega^2 - \Omega^2) - A^2 \textrm{Re } D^R[\omega] + i
\omega (\gamma_0 +
         \gamma[\omega]) } \nonumber \\
         \label{GR} \\
     G^A[\omega]  & = & \left[G^R[\omega] \right]^* \\
     G^K[\omega] & = & -2 i \textrm{Im } G^R[\omega] \times
         \nonumber \\
         && \frac{
         \gamma_0 \coth \left(\frac{\hbar \omega}{ 2 \kb T_{{\rm
bath}}} \right)
         + \gamma[\omega] \coth \left(\frac{\hbar \omega}{ 2 \kb T_
{\rm eff}}
         \right)}
         { \gamma_0 +
         \gamma[\omega]} \nonumber \\
         \label{GK}
\end{eqnarray}
where $\gamma[\omega]$ is given by Eq.~(\ref{OscGamma}), and
$T_{\rm eff}[\omega]$ is defined by Eq.~(\ref{OscTeff}).  The main
effect of the real part of the retarded $\hat{F}$ Green function
$D^R[\omega]$ in Eq.~(\ref{GR}) is to renormalize the oscillator
frequency $\Omega$ and mass $m$; we simply incorporate these
shifts into the definition of $\Omega$ and $m$ in what follows.

If $T_{\rm eff}[\omega]$ is frequency independent, then Eqs.
(\ref{GR}) - (\ref{GK}) for $\check{G}$ corresponds exactly to an
oscillator coupled to two equilibrium baths with damping kernels
$\gamma_0$ and $\gamma[\omega]$.  The correspondence to the
Langevin equation Eq.~(\ref{Langevin}) is then immediate.  In the
more general case where $T_{\rm eff}[\omega]$ has a frequency
dependence, the correlators $G^R[\omega]$ and $G^K[\omega]$ are in
exact correspondence to what is found from the Langevin equation
Eq.~(\ref{Langevin}):  $G^K[\omega]$ corresponds to symmetrized
noise calculated from Eq.~(\ref{Langevin}), while $G^R[\omega]$
corresponds to the response coefficient of the oscillator
calculated from Eq.~(\ref{Langevin}).  This again proves the
validity of using the Langevin equation Eq.~(\ref{Langevin}) to
calculate the oscillator noise in the presence of the detector to
lowest order in $A$.



\subsection{Linear-response formulas for a two-port bosonic amplifier}
\label{app:BosonicKubo}

In this appendix, we use the standard linear-response Kubo
formulas of Sec.~\ref{subsec:VoltageAmp} to derive expressions for
the voltage gain $\lambda_V$, reverse current gain $\lambda'_I$,
input impedance $Z_{in}$ and output impedance $Z_{out}$ of a
two-port bosonic voltage amplifier
(cf.~Sec.~\ref{sec:ScatteringAmp}).  We recover the same
expressions for these quantities obtained in
Sec.~\ref{sec:ScatteringAmp} from the scattering approach.  We
stress throughout this appendix the important role played by the
causal structure of the scattering matrix describing the
amplifier.

In applying the general linear response formulas, we must bear in
mind that these expressions should be applied to the {\it
uncoupled} detector, i.e. nothing attached to the detector input
or output.  In our two-port bosonic voltage amplifier, this means
that we should have a short circuit at the amplifier input
(i.e.~no input voltage, $V_a = 0$), and we should have open circuit at
the output (i.e. $I_b=0$, no load at the output drawing current).
These two conditions define the uncoupled amplifier.  Using the
definitions of the voltage and current operators
(cf.~Eqs.~(\ref{eq:Vdefn}) and (\ref{eq:Idefn})), they take the
form:
\begin{subequations}
\begin{eqnarray}
    \hain[\omega] & = & -\haout[\omega]
        \label{eq:Kuboain1} \\
    \hbin[\omega] & = & \hbout[\omega]
        \label{eq:Kubobin1}
\end{eqnarray}
\end{subequations}
The scattering matrix equation Eq.~(\ref{eq:sdefn}) then allows us
to solve for $\ha_{in}$ and $\ha_{out}$ in terms of the added
noise operators $\hFF_a$ and $\hFF_b$.
\begin{subequations}
\begin{eqnarray}
    \hain[\omega] & = & -\frac{1-s_{22}}{D} \hFF_a[\omega] - \frac{s_{12}}{D} \hFF_b[\omega]
        \label{eq:Kuboain2} \\
    \hbin[\omega] & = & -\frac{s_{21}}{D} \hFF_a[\omega] + \frac{1+s_{11}}{D} \hFF_b[\omega]
        \label{eq:Kubobin2}
\end{eqnarray}
\end{subequations}
where $D$ is given in Eq.~(\ref{eq:Ddefn}), and we have omitted
writing the frequency dependence of the scattering matrix.
Further, as we have already remarked, the commutators of the added
noise operators is completely determined by the scattering matrix
and the constraint that output operators have canonical
commutation relations.  The non-vanishing commutators are thus
given by:
\begin{subequations}
\begin{eqnarray}
    \left[ \hFF^{\pd}_a[\omega], \hFF^{\dagger}_a(\omega') \right] & = &
        2 \pi \delta(\omega - \omega')
        \left(
            1 - |s_{11}|^2 - |s_{12}|^2 \right)
        \nonumber \\
            \\
    \left[ \hFF^{\pd}_b[\omega], \hFF^{\dagger}_b(\omega') \right] & = &
        2 \pi \delta(\omega - \omega')
        \left(
            1 - |s_{21}|^2 - |s_{22}|^2 \right)
        \nonumber \\
            \\
    \left[ \hFF^{\pd}_a[\omega], \hFF^{\dagger}_b(\omega') \right] & = &
        -2 \pi \delta(\omega - \omega')
        \left(
            s_{11} s_{21}^* + s_{12} s_{22}^*  \right)
            \nonumber \\
\end{eqnarray}
\label{eq:KuboFCommutators}
\end{subequations}
The above equations, used in conjunction with Eqs.~(\ref{eq:Vdefn})
and (\ref{eq:Idefn}), provide us with all all the information
needed to calculate commutators between current and voltage
operators.  It is these commutators which enter into the
linear-response Kubo formulas. As we will see, our calculation
will crucially rely on the fact that the scattering description
obeys causality: disturbances at the input of our system must take
some time before they propagate to the output.  Causality
manifests itself in the energy dependence of the scattering
matrix: as a function of energy, it is an analytic function in the
upper half complex plane.

\subsubsection{Input and output impedances}

Eq.~(\ref{eq:VAmpZin}) is the linear response Kubo formula for the
input impedance of a voltage amplifier. Recall that the input
operator $\hQ$ for a voltage amplifier is related to the input
current operator $\hI_a$ via $- d \hQ /dt = \hI_a$
(cf.~Eq.~(\ref{eq:VAmpCoupling})).  The Kubo formula for the input
impedance may thus be re-written in the more familiar form:
\begin{equation}
    Y_{\rm in,Kubo}[\omega] \equiv \frac{i}{\omega}
        \left( - \frac{i}{\hbar}  \int_0^{\infty} dt  \langle [\hI_a(t), \hI_a(0)] \rangle e^{i \omega t} \right)
\end{equation}
where $Y_{\rm in}[\omega] = 1/ \Zin [\omega]$,

Using the defining equation for $\hI_a$ (Eq.~(\ref{eq:IOpdefn}))
and Eq.~(\ref{eq:Kuboain1}) (which describes an uncoupled
amplifier), we obtain:
\begin{eqnarray}
    && Y_{\rm in,Kubo}[\omega]  =       \label{eq:KuboYin1}
    \\
    &&
        \frac{2}{Z_a}
        \int_0^{\infty} dt e^{i \omega t}
        \int_0^{\infty} \frac{d \omega'}{2 \pi} \frac{\omega'}{\omega}
        \Lambda_{aa}(\omega')
        \left(
            e^{-i \omega' t} - e^{i \omega' t}
        \right) \nonumber
\end{eqnarray}
where we have defined the real function $\Lambda_{aa}[\omega]$ for
$\omega>0$ via:
\begin{eqnarray}
    \left[ \ha^{\pd}_{\rm in}[\omega], \ha^{\dag}_{\rm in}(\omega') \right]
    = 2 \pi \delta(\omega - \omega') \Lambda_{aa}[\omega]
    \label{eq:KuboLambdaAA}
\end{eqnarray}

It will be convenient to also define $\Lambda_{aa}[\omega]$ for
$\omega < 0$ via $\Lambda_{aa}[\omega] = \Lambda_{aa}[-\omega]$.
Eq.~(\ref{eq:KuboYin1}) may then be written as:
\begin{eqnarray}
    Y_{\rm in,Kubo}[\omega] & = &
        \frac{2}{Z_a}
        \int_0^{\infty} dt
        \int_{-\infty} ^ {\infty} \frac{d \omega'}{2 \pi} \frac{\omega'}{\omega}
        \Lambda_{aa}(\omega')
            e^{i (\omega - \omega') t}
        \nonumber \\
        & = &
        \frac{\Lambda_{aa}[\omega]}{Z_a} +
        \frac{i}{\pi \omega }
         \mathcal{P} \int_{-\infty} ^ {\infty} d \omega'
        \frac{\omega '\Lambda_{aa}(\omega')/Z_a}{\omega - \omega'}
        \nonumber \\
        \label{eq:KuboYin2}
\end{eqnarray}

Next, by making use of Eq.~(\ref{eq:Kuboain2}) and
Eqs.~(\ref{eq:KuboFCommutators}) for the commutators of the added
noise operators, we can explicitly evaluate the commutator in
Eq.~(\ref{eq:KuboLambdaAA}) to calculate $\Lambda_{aa}[\omega]$.
Comparing the result against the result Eq.~(\ref{eq:Zindefn}) of
the scattering calculation, we find:
\begin{equation}
    \frac{\Lambda_{aa}[\omega]}{Z_a} =
        \textrm{Re } Y_{\rm in,scatt}[\omega]
\end{equation}
where $Y_{\rm in,scatt}[\omega]$ is the input admittance of the
amplifier obtained from the scattering approach.  Returning to
Eq.~(\ref{eq:KuboYin2}), we may now use the fact that
$Y_{in,scatt}[\omega]$ is an analytic function in the upper half
plane to simplify the second term on the RHS, as this term is
simply a Kramers-Kronig integral:
\begin{eqnarray}
     && \frac{1}{\pi \omega }
        \mathcal{P} \int_{-\infty} ^ {\infty} d \omega'
        \frac{\omega '\Lambda_{aa}(\omega')/Z_a}{\omega - \omega'}
        \nonumber \\
     &&=
        \frac{1}{\pi \omega }
        \mathcal{P} \int_{-\infty} ^ {\infty} d \omega'
        \frac{\omega' \textrm{Re } Y_{\rm in,scatt}(\omega') }{\omega - \omega'}
        \nonumber \\
    && =
        \textrm{Im } Y_{\rm in,scatt}[\omega]
\end{eqnarray}
It thus follows from Eq.~(\ref{eq:KuboYin2}) that input impedance
calculated from the Kubo formula is equal to what we found
previously using the scattering approach.

The calculation for the output impedance proceeds in the same
fashion, starting from the Kubo formula given in
Eq.~(\ref{eq:VAmpZout}).  As the steps are completely analogous to
the above calculation, we do not present it here. One again
recovers Eq.~(\ref{eq:Zoutdefn}), as found previously within the
scattering approach.

\subsubsection{Voltage gain and reverse current gain}

Within linear response theory, the voltage gain of the
amplifier ($\lambda_V$) is determined by the commutator between
the ``input operator" $\hQ$ and $\hV_b$
(cf.~Eq.~(\ref{eq:gain}); recall that $\hQ$ is defined by $d
\hQ /dt = - \hI_a$. Similarly, the reverse current gain
($\lambda'_I$) is determined by the commutator between
$\hI_{a}$ and $\hat{\Phi}$, where $\hat{\Phi}$ is defined via
$d \hat{\Phi} /dt = - \hV_b$ (cf.~Eq.~(\ref{ReverseGain})).
Similar to the calculation of the input impedance, to properly
evaluate the Kubo formulas for the gains, we must make use of
the causal structure of the scattering matrix describing our
amplifier.

Using the defining equations of the current and voltage operators
(cf.~Eqs.~(\ref{eq:VOpdefn}) and (\ref{eq:IOpdefn})),  as well as
Eqs.~(\ref{eq:Kuboain1}) and (\ref{eq:Kubobin1}) which describe
the uncoupled amplifier, the Kubo formulas for the voltage gain
and reverse current gain become:
\begin{eqnarray}
    \lambda_{V,{\rm Kubo}} [\omega] & = &
            4 \sqrt{ \frac{Z_b}{Z_a}}
                    \label{eq:KuboVGain1}   \\
            && \times
            \int_0^{\infty} dt \phantom{\cdot}
            e^{i \omega t}
                \textrm{Re } \left[
                    \int_0^{\infty} \frac{d \omega' }{2 \pi} \Lambda_{ba}(\omega') e^{-i \omega' t}
                        \right] \nonumber \\
    \lambda'_{I,{\rm Kubo}}[\omega] & = &
            - 4 \sqrt{ \frac{Z_b}{Z_a}}
                        \label{eq:KuboIGain1} \\
            && \times
            \int_0^{\infty} dt \phantom{\cdot}
            e^{i \omega t}
                \textrm{Re }  \left[
                    \int_0^{\infty} \frac{d \omega' }{2 \pi} \Lambda_{ba}(\omega')
                     e^{i \omega' t}
                        \right] \nonumber
\end{eqnarray}
where we define the complex function $\Lambda_{ba}[\omega]$ for
$\omega>0$ via:
\begin{eqnarray}
    \left[ \hb_{in}^{\pd}[\omega], \ha_{in}^{\dagger} (\omega') \right]
        & \equiv & \left( 2 \pi \delta(\omega - \omega') \right) \Lambda_{ba}[\omega]
\end{eqnarray}
We can explicitly evaluate $\Lambda_{ba}[\omega]$ by using
Eqs.~(\ref{eq:Kuboain2})-(\ref{eq:KuboFCommutators}) to evaluate
the commutator above.  Comparing the result against the scattering
approach expressions for the gain and reverse gain
(cf.~Eqs.~(\ref{eq:Gaindefn}) and (\ref{eq:RevGaindefn})), one
finds:
\begin{eqnarray}
    \Lambda_{ba}[\omega] = \lambda_{V, {\rm scatt} }[\omega] -
        \left[ \lambda'_{I,{\rm scatt} }[\omega] \right]^*
        \label{eq:KuboLambdaBA}
\end{eqnarray}
Note crucially that the two terms above have different analytic
properties:
 the first is analytic in the upper half plane, while the second is analytic
 in the lower half plane.  This follows directly from the fact that the scattering matrix is causal.

At this stage, we can proceed much as we did in the calculation of
the input impedance. Defining $\Lambda_{ba}[\omega]$ for $\omega <
0$ via $\Lambda_{ba}[-\omega] =  \Lambda_{ba}^*[\omega]$, we can
re-write Eqs.~(\ref{eq:KuboVGain1}) and (\ref{eq:KuboIGain1}) in
terms of principle part integrals.
\begin{eqnarray}
    \lambda_{V,{\rm Kubo}}[\omega] & = &
        \frac{Z_b}{Z_a}
            \left(
                \Lambda_{ba}[\omega] +
                    \frac{i}{\pi}
                    \mathcal{P} \int_{-\infty} ^ {\infty} d \omega'
                    \frac{\Lambda_{ba}(\omega')}{\omega - \omega'}
            \right)
        \nonumber \\
    \lambda'_{I,{\rm Kubo} }[\omega] & = &
        -\frac{Z_b}{Z_a}
            \left(
                \Lambda_{ba}[\omega] +
                    \frac{i}{\pi}
                    \mathcal{P} \int_{-\infty} ^ {\infty} d \omega'
                    \frac{\Lambda_{ba}(\omega')}{\omega + \omega'}
            \right)
        \nonumber \\
\end{eqnarray}
Using the analytic properties of the two terms in
Eq.~(\ref{eq:KuboLambdaBA}) for $\Lambda_{ba}[\omega]$, we can
evaluate the principal part integrals above as Kramers-Kronig
relations.  One then finds that the Kubo formula expressions for
the voltage and current gain coincide precisely with those
obtained from the scattering approach.

While the above is completely general, it is useful to go through
a simpler, more specific case where the role of causality is more
transparent.  Imagine that all the energy dependence in the
scattering in our amplifier arises from the fact that there are
small transmission line ``stubs" of length $a$ attached to both
the input and output of the amplifier (these stubs are matched to
the input and output lines).  Because of these stubs, a wavepacket
incident on the amplifier will take a time $\tau = 2 a /v$ to be
either reflected or transmitted, where $v$ is the characteristic
velocity of the transmission line.  This situation is  described
by a scattering matrix which has the form:
\begin{eqnarray}
    s[\omega] =  e^{2 i \omega a / v} \cdot \bar{s}
\end{eqnarray}
where $\bar{s}$ is frequency-independent and real.  To further
simplify things, let us assume that $\bar{s}_{11} = \bar{s}_{22} =
\bar{s}_{12}=0$.  Eqs.~(\ref{eq:KuboLambdaBA}) then simplifies to
\begin{eqnarray}
    \Lambda_{ba}[\omega] =
        & = & s_{21}[\omega] = \bar{s}_{21} e^{i \omega \tau}
\end{eqnarray}
where the propagation time $\tau = 2 a / v$ We then have:
\begin{eqnarray}
    \lambda_V[\omega_0] & = &
            2 \sqrt{ \frac{Z_b}{Z_a}} \bar{s}_{21} \int_0^{\infty} dt \phantom{\cdot} e^{i \omega_0 t}
                  \delta(t - \tau) \\
    \lambda_I[\omega_0] & = &
            -2 \sqrt{ \frac{Z_b}{Z_a}} \bar{s}_{21} \int_0^{\infty} dt \phantom{\cdot} e^{i \omega_0 t}
                 \delta(t + \tau)
\end{eqnarray}
If we now do the time integrals and then take the limit $\tau \ra
0^+$, we recover the results of the scattering approach
(cf.~Eqs.~(\ref{eq:Gaindefn}) and (\ref{eq:RevGaindefn})); in
particular, $\lambda_I = 0$.  Note that if we had set $\tau=0$
from the outset of the calculation, we would have found that both
$\lambda_V$ and $\lambda_I$ are non-zero!


\subsection{Details for the two-port bosonic voltage amplifier with feedback}
\label{app:MirrorsDetail}


In this appendix, we provide more details on the calculations for
the bosonic-amplifier-plus-mirrors system discussed in
Sec.~\ref{subsec:MirrorsFeedback}.  Given that the scattering
matrix for each of the three mirrors is given by
Eq.~(\ref{eq:MirrorsU}), and that we know the reduced scattering
matrix for the mirror-free system (cf.~Eq.~(\ref{eq:sdefnsimp})),
we can find the reduced scattering matrix and noise operators for
the system with mirrors.  One finds that the reduced scattering
matrix $s$ is now given by:
\begin{eqnarray}
    s & = &
        \frac{1}{M} \times \\
        \label{eq:smirrors}
    &&
        \left(\begin{array}{cc}
            \sin\theta_{z}+\sqrt{G} \sin\theta_{x}\sin\theta_{y} &
                - \cos\theta_{x}\cos\theta_{z}\sin\theta_{y}\\
        \sqrt{G} \cos\theta_{x}\cos\theta_{z} &
            \sin\theta_{x}+
            \sqrt{G} \sin\theta_{y}\sin\theta_{z}\end{array}\right)
            \nonumber
\end{eqnarray}
where the denominator $M$ describes multiple reflection
processes:
\begin{eqnarray}
	M & = &
	1+\sqrt{G}\sin\theta_{x}\sin\theta_{z}\sin\theta_{y}
\end{eqnarray}
Further, the noise operators are given by:
\begin{widetext}
\begin{eqnarray}
    &&
    \left(\begin{array}{c}
     \FF_{a}\\
    \FF_{b}\end{array}\right)  =
         \frac{1}{M}
          \left(\begin{array}{cc}
        \cos\theta_{y}\cos\theta_{z} &
        \sqrt{G-1}\cos\theta_{z}\sin\theta_{x}\sin\theta_{y}\\
        -\sqrt{G}\cos\theta_{x}\cos\theta_{y}\sin\theta_{z} & \sqrt{G-1}\cos\theta_{x}
    \end{array}\right)
    \left(\begin{array}{c}
        u_{in}\\
        v_{in}^{\dagger}
    \end{array}\right)
\end{eqnarray}
\end{widetext}

The next step is to convert the above into the op-amp
representation, and find the gains and impedances of the
amplifier, along with the voltage and current noises. The voltage
gain is given by:
\begin{eqnarray}
    \lambda_{V} & = &
         \sqrt{\frac{Z_{B}}{Z_{A}}}
        \frac{2\sqrt{G}}{1-\sqrt{G}\sin\theta_{y}}\cdot
        \frac{1+\sin\theta_{x}}{\cos\theta_{x}}
        \frac{1-\sin\theta_{z}}{\cos\theta_{z}}
        \label{eq:Gainmirror}
\end{eqnarray}
while the reverse gain is related to the voltage gain by the
simple relation:
\begin{eqnarray}
    \lambda'_{I} & = & -\frac{\sin\theta_{y}}{\sqrt{G}}\lambda_V
\end{eqnarray}
The input impedance is determined by the amount of reflection in
the input line and in the line going to the cold load:
\begin{eqnarray}
    \Zin & = &
        Z_{a} \frac{1-\sqrt{G}\sin\theta_{y}}{1+\sqrt{G}\sin\theta_{y}}\cdot
        \frac{1+\sin\theta_{z}}{1-\sin\theta_{z}}
        \label{eq:ZinMirror}
\end{eqnarray}
Similarly, the output impedance only depends on the amount of
reflection in the output line and the in the cold-load line:
\begin{eqnarray}
    \Zout & = &
        Z_{b} \frac{1+ \sqrt{G} \sin\theta_{y}}{1-\sqrt{G} \sin\theta_{y}} \cdot
        \frac{1 + \sin\theta_{x}}{1 - \sin\theta_{x}}
\end{eqnarray}
Note that as $\sin\theta_{y}$ tends to $-1/\sqrt{G}$ , both the input
admittance and output impedance tend to zero.

Given that we now know the op-amp parameters of our amplifier, we
can use Eq.~(\ref{eq:BosonicPowerGain}) to calculate the
amplifier's power gain $G_{P}$. Amazingly, we find that the
power gain is completely independent of the mirrors in the input
and output lines:
\begin{eqnarray}
    G_{P} & = & \frac{G}{1 + G \sin^2 \theta_y}
    \label{eq:Gpowermirror}
\end{eqnarray}
Note that at the special value $\sin \theta_y = -1 \sqrt{G}$ (which allows one to
reach the quantum limit), the power gain is reduced by a factor of two compared to the reflection free case (i.e.~$\theta_y=0$).

Turning to the noise spectral densities, we assume the optimal
situation where both the auxiliary modes $\hu_{in}$ and
$\hv_{in}^{\dagger}$ are in the vacuum state.   We then find that
both $\htI$ and $\htV$ are independent of the amount of reflection
in the output line (e.g. $\theta_{x}$):
\begin{subequations}
\begin{eqnarray}
    \bS_{II} & = &
        \frac{2\hbar\omega}{Z_{a}}
        \left[
        \frac{1 - \sin\theta_{z}}{1 + \sin\theta_{z}} \right]
            \times \nonumber \\
        &&
        \left(
            \frac{G \sin^{2}\theta_{y}+\cos(2\theta_{y})}
            {\left(\sqrt{G} \sin\theta_{y}-1\right)^{2}}
        \right)
            \label{eq:SImirrors} \\
    \bS_{VV} & = &
        \hbar\omega Z_{a}
        \left[
            \frac{1 + \sin\theta_{z}}{1 - \sin\theta_{z}}
        \right]  \times \nonumber \\
        &&
        \left(
            \frac{3+\cos(2\theta_{y})}{4} - \frac{1}{2 G}
        \right)
                \label{eq:SVmirrors} \\
    \bS_{VI} & = &
        \frac{\sqrt{G}(1-1/G) \sin \theta_y + \cos^2\theta_y  }{1 - \sqrt{G} \sin \theta_y}
        \label{eq:SIVmirrors}
\end{eqnarray}
\end{subequations}
As could be expected, introducing reflections in the input line
(i.e.~$\theta_z \neq 0$) has the opposite effect on $\bS_{II}$ versus
$\bS_{VV}$: if one is enhanced, the other is suppressed.

It thus follows that the product of noise spectral densities
appearing in the quantum noise constraint of
Eq.~(\ref{eq:VAmpNoiseConstraint}) is given by (taking the
large-$G$ limit):
\begin{eqnarray}
    \frac{ \bS_{II} \bS_{VV} }{\left( \hbar \omega \right)^2} & = &
        \left( 2 - \sin^2\theta_y \right)
        \cdot\frac{1+G \sin^{2}\theta_{y}}{\left(1- \sqrt{G} \sin
        \theta_{y}\right)^{2}}
        \label{eq:TNmirrors}
\end{eqnarray}
Note that somewhat amazingly, this product (and hence the
amplifier noise temperature) is completely independent of the
mirrors in the input and output arms (i.e. $\theta_z$ and
$\theta_x$). This is a result of both $\bS_{VV}$ and $\bS_{II}$ having no
dependence on the output mirror ($\theta_x$), and their having
opposite dependencies on the input mirror ($\theta_z$).  Also note
that Eq.~(\ref{eq:TNmirrors}) does indeed reduce to the result of
the last subsection: if $\theta_{y}=0$ (i.e. no reflections in the
line going to the cold load), the product $\bS_{II} \bS_{VV}$ is equal to
precisely twice the quantum limit value of $(\hbar\omega)^{2}$.
For $\sin(\theta_y) = -1/\sqrt{G}$, the RHS above reduces to one,
implying that we reach the quantum limit for this tuning of the
mirror in the cold-load arm.


\bibliographystyle{apsrmp}
\bibliography{RMPRefs}
\newpage


\end{document}